**Information Path from Randomness and Uncertainty to Information, Thermodynamics, and Intelligence of Observer**

Vladimir S. Lerner, USA

The introduced path connects uncertainty of random process to certainty of observer information process.
The path integrates multiple acts of impulse observation emerging in interactive process.
Each inter-action is a discrete Yes-No action of impulse ↓↑modeling potential information Bit, while multiple interactions generate a random process of potential multiple bits along the path.
The problem is to reveal the emergent Observer during the impulse observation creating the path.
The random process probabilities emerge from the probability field of the random events starting observation.
Markov chains models the random process of the interacting impulses.
The impulse observations sequentially cut the random process correlations disclosing entropy hidden in correlation, as a measure of process uncertainty, and reveal hidden information Bits. The entropy of the impulses and time interval of observations integrates entropy functional (EF) along the observing process.
The impulse of the interactive No-action cuts the maximum entropy, while its Yes-action transfers a minimum cut to the next impulse, thus creating the maxmin-minimax principle, decreasing the uncertainty along the observing process.
The revealing hidden information integrates information path functional (IPF). The minimax variation principle, applying to the EF-IPF, determines information dynamic equations formally describing the path of interactive observation.
The EF extremals describe reversible symmetrical uncertain dynamics of the path. Between the EF and IPF extremals locate bridges, whose overcoming continues the path with irreversible asymmetrical information dynamics.
Along the path evolves hierarchical multilevel self-organized interactive dynamics, beginning with probabilistic virtual observing process and probabilistic causality.
The process correlations temporary memorize *time intervals* of the impulse-observation trajectory. Random merging of bordered impulses on the trajectory might generate a microprocess of the superposing merging conjugated entropy fractions.
The fractions entangle during the time interval within which the space interval is formed. The impulse interactions curve the time-space impulse geometry whose curvature creates asymmetry of the impulses. Such interaction logically erases the entangled entropy removing the causal probabilistic entropy with symmetrical reversible logic and bringing asymmetrical information logic. That entropy-information gap connects the asymmetrical logic with logical Bits starting formation in the microprocess. Real local gap reveals physical Markov diffusion whose energy memorizes logical Bit or two qubits. The Bits conserve the information logic. Jarzynski's equality confirms thermodynamic conversion of the entropy to the information Bit at probability approaching one.
A flow of moving Bits self–forms a unit of the information macroprocess (UP).The UP size limits the unit's starting maximal and ending minimal Information speeds, attracting a new UP through its free Information.
The anatomy and density of information Bit depend on nature of each Yes-No action and their interactive result.
A minimum of three self-connected Bits assembles the optimal UP-basic triplet. Its free information requests and binds a new UP triplet that joins in a knot memorizing the triplet's Information. During macro-movement, multiple UP triples adjoin the time space hierarchical network (IN) whose free Information's request produces new UP at higher level's knot-node and encodes it in triple code logic. Each UP has a unique position in the IN hierarchy, which defines the exact location of each code's logical structures. The IN node hierarchical level classifies the quality of the assembled Information, while the currently ending IN node integrates the Information enfolding all IN levels. Multiple INs binds their ending triplets, enclosing Observer Information, logic, cognition, and intelligence. The Observer cognition assembles common units through multiple attractions in resonances loops forming IN triplet hierarchical logic. The cognitive logic self-controls the process encoding the intelligence in a double helix coding structure (DSS).
The clock time intervals open access to external energy at each specific level of the IN multiple hierarchy, enabling the memorization and encoding the hierarchy of these Bits. The intelligent observer, self-reflective to the DSS, enables reading and understanding the message meaning. The Interactive Integrated Information Dynamics describe the emerging doublet-triplet logic, the IN formation, integration the logic, memory, cognition, encoding the intelligent observer in triplet code, which further encodes physical information, cognition, intelligence including consciousness, and enables create AI. Each elementary interaction, modeling binary 1-0 value, a Bit, connects phenomenon of interaction with the phenomenon of information, emerging in the impulse' observations.

## INTRODUCTION

Uncertainty about some facts initiates a search, for example on the Web, certain facts-information creating a path from uncertainty to needed information. A potential observer of this information, a participator within the path, sends probing impulses, activating the brain neuron's yes-no impulses. During such observation, many unsubstantial temporal images, which the observer evaluates with low probability, could emerge, until searching high probable image becoming most informative actual image appears for the observer, and temporal observing images, as a virtual for this observer, will disappear. An actual image emerges as information is registered.

*This path becomes information creating information observer with real actions.*

Similar examples are in many scientific researches, searching certain facts-information by multiple experiments-probes, or observing unknown particles, planets in a yet unknown Galaxy, tracking their *probable or real interaction*.

Like an astronomer traces an unobserving planet image measuring its probabilistic trajectory until it become most probable and informative, gets its information fact by copying-registering the most probable image, which needs spending energy, for example for a shot. Or a physicist traces trajectory interactive particles in an Accelerator.

These identify interaction is a primary indicator of a potential probabilistic object during an observation.

*The notions of object, particle, and image replace the probability of event observation. Starting event with its probability originates observation.* Beginning of this process is hidden in *uncertainty* regarding the facts and reality.

What is the scientific way to uncover a path from uncertainty to certainty-as the fact of reality, focusing not on physics of the observing process but on its information-theoretical essence?

Up to now such *information process* has not scientific definition, neither conclusive definition of information.

Diverse forms of interaction describe elementary yes-no ($\downarrow\uparrow$) inter-active actions of the impulse which presents well known Bit-a unit of Information.

The formal recurring *inter-actions*, modeling elementary bits, are *independent of its physical nature* from elementary particles up to the multiple interacting objects. Multiple interactive actions produce a manifold of random elementary 1-0 events-process described by *the probability path of observation.*

The path formalizes the impulse process, observing by objective probabilities along the path.

Even though the discrete interactions are fundamental in Universe, their unified description was unknown and is problematic. Moreover, as the unification reveals information nature of the inter-active phenomena, the observation path creating an information observer is unidentified and unscripted.

It requires integrating the probabilities along the path and measuring the path information by information path integral. The path united information approach shows how:

-information emerges from observing random process probabilities of interacting impulses, bringing a probabilistic causality of the observations;

-removing the random process' uncertainty converts it to equivalent certainty-information;

-assembling and moving information creates observer's information process, with information dynamics, information network, its logic and code;

-raising the observer's structures with information regularities, differentiation, evolution, cognition, and intelligence;

-such path formalizes *Interactive Integrated Information Dynamics* (IIID) which join interactive observation with emergence the certainty, information, information process, and information macrodynamics–information form of irreversible thermodynamics. The IIID theory unifies the observed random path with the IMD multiple information and physical processes dynamics.

Revealing nature of *Information* in various interactive processes, including multiple physical interactions, human observations, human-machine communications, biological, social, economic, among others, and integrated that nature into an information observer, becomes an important scientific task.

Multiple physical interactions start with four fundamental interactions (gravitation, electromagnetic, weak, and strong) toward different chemical, thermodynamic, biological, human observation, up brain neuron inter-active actions, cognitive, intellectual, communication, all forms of life, and various substances of our world.

In such a manifold of interactions, the physical processes interact with energies of different qualities (from light to heat dissipation) where each process with a high quality compensates for the entropy of a lesser quality. Transferring entropy during the interaction unifies the physical processes, human observations and communications, neuronal and cognitive processes, forming superimposing processes [27]. This sequential chain of diverse interactions builds a path which unites the information processes along the path. Any such interaction naturally "observes" the transferring information, becoming an Observer of this information. The multiple inter-actions generalize Yes-No $\downarrow\uparrow$ impulse elementary delta-actions along the path. Each $\downarrow$ action cuts maximal entropy from the observed path, which decreases uncertainty of the path, while each $\uparrow$ action transfers a minimum of the delta-impulse' cutting entropy back to the path.

Such multiple impulse actions decrease the uncertainty of the observing process, imposing maximin-minimax uncertainty principle along the path. The principle formalizes variation principle whose minimax determines Information Path Functional (IPF). The IPF integrates the emerging "observing" information process which builds the optimal minimax structure of Information Observer along the path. The path integrates the superimposing process in the Observer.

*Physical approach* to the observer, developed in Copenhagen interpretation of quantum mechanics [1-5], requires an act of observation, as a physical carrier of the observer knowledge. But this observer' role not describes the formalism of quantum mechanics.

As N.Bohr believed, *probability itself a fundamental nature of reality.*

Wheeler and Feynman developed the time-symmetric direct interaction theory [6] referring to all phenomena captured by classical electrodynamics [7] that "the curves of action and reaction cross".

According to D. Bohm ontological interpretation of quantum physics [8]: physical processes are determined by information, which "is a difference of form that makes a difference of content, i.e., meaning”. Bohm believed that “meaning unfolds into intention, intention into actions”; and “intention generally arises out of a previous perception of meaning or significance of a certain total situation.”.” That observer entails mental processes”.

The quantum approach of J. C. Eccles [9] “is to find a way for the ‘self’ to control its brain.”

J.A. Wheeler introduces physical theory [10-14] of information-theoretic origin of an observer. Wheeler hypothesized that the Bit participates in the origin of all physical processes. Summarizing his physical theory [9-14] of an Observer-Participator, he introduced the doctrine “It from Bit”. In his memoir [13], Wheeler divided his life into three themes or periods which reflect the historical development of modern physics. The first period he called “everything is particles.” In the second period, “everything is fields.” And in the third period, “everything is Information.” These three stages represent an increasing generality of worldview.”

But Wheeler’s theory does not explain *how* the Bit self-creates.

Previously, many physicists [1-14], including Einstein [15], Penrose [16], and others, defined the Observer as having a separate and *physical* origin.

The problem of probability in quantum mechanics, writes Weinberg [17], “is that in quantum mechanics, the way that wave functions change with time is governed by an equation, the Schrödinger equation, *that does not involve probabilities*. It is just as deterministic as Newton’s equations of motion and gravitation... So if we regard the whole process of measurement as being governed by the equations of quantum mechanics, and these equations are perfectly deterministic, how do probabilities get into quantum mechanics?”

D. Tong [18] argues that Quantum Fields are the real building blocks of the universe. The origin of physical particles is the natural probabilities of the vacuum. The comparative review of Wheeler theory and contemporary physics [19] shows that “Everything is From Field”. Due to the quantum origins, “The elementary act of observer-participatorship transcends the category of time (delayed-choice double slit)” [19].

Still problem consists in unification classical and quantum physics.

But since Information originates in quantum processes, its study should focus not on the physics of the observing process’s interacting particles, but on its Information-theoretical essence.

That leads to possibility of such unification using *Information formalism.*

A.N. Kolmogorov [20] established Probability Theory as the foundation of Information Theory and logic. Kolmogorov defined *random* simply as “the absence of periodicity” [21, p. 664].

C.E. Shannon’s *Mathematical Theory of Communication* [22] measures relative entropy, which applies to the random states of an Information process. Kullback-Leibler’s divergence [23], also known as relative entropy, measures the relative Information connections between the states of an observed process. The probabilistic origin of Information is well established [21-23, others], along with its unit, the Bit.

There are many studies of Information mechanisms employing various physical phenomena to account for intelligence. E. T. Jaynes [24] applied Bayesian probabilities to propose a plausible reasoning mechanism [25] whose rules of deductive logic connects maximum Bayes Information (entropy) to human mental activities, as a subjective observer.

The Observer is an *interactant*, present in all phenomena.

Understanding all these starting with definition and role information developments in our Information Age have become a critical task for scientific researchers, and economic institutions.

Wikipedia defines Information though its universal action: "Information is any entity or form that provides the answer to a question of some kind or resolves uncertainty"[38].

The cited references, along with many others, studying information mechanisms in intelligence, explain these through various physical phenomena, whose specifics are still mostly unknown.

Science knows that interactions have built structure of the Universe as its fundamental phenomena.

There have been many studies these interactions specifics; however, no one approach has unified the study of all their common information origins, regularities, and differentiation.

The first approach unifying these studies was published in [27-29], and extended results were published in [30-37]. This unified approach focuses on observations as interactions producing the Observer itself.

The Information Observer emerges by observing a random interactive process.

The essence is the probabilistic tracing of interacting events, which an Information path measures.

During the observation, the uncertainty of the random interactive process is converted into certainty. Thus, certainty is a source of Information. Any single certain interaction is a “Yes-No” action which identifies a Bit, the elementary unit of Information. Multiple observations generate the Bit-moving dynamics, or Informational dynamics.

Bits organize themselves in triplets, which logically self-organize and assemble an Informational network. In the process of assembling the network, the triplets merge and interact with each other. Triplet interactions are memorized and become nodes of the Informational network. Then, the nodes themselves organize logically. A sequence of the logically organized triplet nodes defines a code of the network. This code integrates and carries all prior observations in the emerging Information Observer. The Information Observer emerges from probabilistic observation without any pre-existing physical law.

The well-known Shannon approach defines entropy as probability measures of the uncertainty of the observation. If the entropy of the observation decreases, uncertainty disappears, instead appearing as an equal certainty. Revealing certainty from uncertainty is the scientific path which determines the facts of reality. The physical erasure of entropy by observation creates a certainty. This certainty, or Information, is a physical entity that contains physical energy equivalent to the energy spent to erase the entropy. By this process, a Bit, the elementary unit of Information, is created.

Expressed as a mathematical formalism, the Bit evolves from the abstract probability of the observation as an elementary observer itself. Every step of this approach is substantiated here through a unified formalism of mathematics and logic.

Shannon’s Communications Theory [22] shows the following:

1. Shannon H-entropy measures the set of probabilities of symbols of a message as a source of signals.
2. Maximum H is the most uncertain situation.
3. Minimum H entropy measures the maximal probability.
4. Channel capacity measures its entropy.
5. H is maximized when H is equal to the Channel capacity entropy.
6. Encoding the source message in Bit equalizes the entropy of channel capacity with the entropy of the source (thereby maximizing the equal uncertainty). According to Landauer [26], encoding requires the expenditure of energy quantified by this maximal entropy. (Therefore, the energy of encoding erases the maximal entropy-uncertainty to zero, reaching maximal probability.)
7. “H=0 if and only when we are certain of the outcome does H vanish.”
8. H measures the amount of Information bits encoding the entropy source of message.

These principles of Shannon’s theory of communications agree with the main principle of our approach: Entropy, as measure of uncertainty, erases energy, converting it to the equal Information, measuring certainty. The approach’s main contribution is to extend these principles to any observing random process whose entropy and Information integrate the related path functionals. The integral encoding structures an observer of this information.

Shannon wrote [22]: “A physical system, or a mathematical model of a system which produces a sequence of symbols governed by a set of probabilities, is known as a stochastic process... Conversely, any stochastic process which produces a discrete sequence of symbols chosen from a finite set may be considered a discrete source... Stochastic processes of the type described above are known mathematically as discrete Markoff processes and have been extensively studied in the literature.”

In our formal model of the observation, the impulse observation runs axiomatic probabilities of a random field, linking Kolmogorov 0-1 law and Markov process probabilities [20]. The field connects sets of possible and actual events with their probabilities. The field’s energy covers actual events. This triad specifies the observation.

*The Kolmogorov 0-1 probabilities' act of observation generates the Markov process within the field.*
The Markov process models the arising observer's process, collecting observations which change its measure of probabilities similar to the sequence of *a priori-a posteriori* Bayes probabilities.
These objective probabilities, being an immanent part of the process, virtually observe and measure the Markov correlation connecting states-events, discretely changing the entropy of correlation which generates probabilistic impulses.
Each such impulse virtually cuts the observing entropy-uncertainty hidden in the cutting correlation. The cutting entropy decreases the initial Markov entropy, and increases the entropy of the cutting impulse.
Such multiple interactions minimize the uncertainty of the Markov process and maximize the entropy of each subsequent observing impulse. That runs the minimax principle for each observing impulse along the Markov interactive impulses.
When the observing probability approaches 1, the impulse cutting entropy converts to Information. The merging impulse curves and rotates the interactive yes-no conjugated entropies of the microprocess.
The entropy entanglement starts within the impulse time interval before its space forms, and ends at the beginning of the space during the reversible relative time interval $0.15625\pi$ part of impulse entropy measure $\pi$.
The opposite curvature, enclosing the entropy of the interacting impulses, lowers the potential energy of an external process that converts entropy to a Bit of the interacting process, memorizing the Bit by delivering Landauer's energy [26]. Sequential interactive cuts along the process integrate the cutoff Hidden Information in the Information macroprocess, with a irreversible time course. Each memorized Information binds the reversible microprocess within an impulse with the irreversible Information macroprocess along the multi-dimensional process. The impulse observation consecutively converts entropy to Information in the emerging Information observer, conveying Information causality, certain logic, and complexity. The curving interaction is main information mechanism connecting the emerging information structures of Observer, analogously to gravitation in physics.
Multiple interacting Bits self-organize the Information process, encoding Information causality, logic, and complexity. The trajectory of the observation process carries the wave function both probabilistic and certain, which self-builds the Information macrounits-triplets. Macrounits logically self-organize Information Networks (IN), encoding the units in geometrical structures which enclose the triplet code. Multiple INs bind their ending triplets, enclosing the Observer's Information cognition and intelligence.

The Observer cognition assembles common units through multiple attractions and resonances as it forms the IN triplet hierarchy. The maximal number of accepted triplet levels in multiple INs measures the Observer maximum comparative Information intelligence.
The intelligent Observer recognizes and encodes digital images in message transmission. Being self-reflective, this enables it to understand the meaning of the message. The cognitive logic self-controls the process encoding the intelligence in a double helix coding structure.
Integrating the process entropy in the Entropy Functional and the Bits in the Information Path Integral's measures formalizes the variation problem in the minimax law, determining all regularities of the processes. Solving the problem mathematically describes the micro-macro processes, the IN, and the invariant conditions of the Observer's self-organization and self-replication.
*Observing information becomes equivalent of Observer time.*
These functional regularities create a united Information mechanism whose integral logic self-operates, transforming interacting uncertainties to physical reality-matter. This enables the exchange of human Information and the design of Artificial Intelligences.
Both Information and Information processes emerge as phenomena of natural interactions. Each specific field triad generates an Information process, creating its Observer.
The Information equations described here finalize the main results, validate them numerically, and present Information models of many interactive physical processes.
The approach focuses on formal Information mechanisms in an Observer, without reference to the specific physical processes which originate these mechanisms in the Observations. The Information formalism describes a self -building information machine which creates both Humans and Nature.
*Comments. A bridge connecting of physical results* [6, 7] *and* [39] *with our approach.*
Many years passed since Schrödinger introduced his equation of Quantum Mechanics as a new physical microscopic theory of interacting particles.
However, up to now, the scientific origin of the connection between Quantum Mechanics and classical physics has not been established That must include linking a wave function to a probabilistic field, and connecting Quantum Mechanics to Quantum Information Theory.
Resuts [6,7] have shown that "theory of direct inter-particle interaction, associated with a particle acting upon itself, derives from the motion of a system of charged particles under the influence of electromagnetic forces."
However, the inter-particle interaction in the electrodynamics Maxwell field deals with problem that action and symmetric (adjunct) reaction should merge. Satisfaction of this requirement allows the connection of Maxwell's equations with the equations for atomic particles using the variation principle for total energy as the equivalent of a conservation law for such adjunct interactions. Solution of the obtained equation leads to a *discrete* action crossing reaction.

Study [39] obtains Schrödinger's equation in Quantum Mechanics from Maxwell equations. The equations for energy, momentum, frequency and wavelength of the electromagnetic wave in the atom are derived using the model of atom by analogy with the transmission line. The balance of electromagnetic energy in the atom satisfies the structural constant for the atom $so = 8.277\ 56$.

This constant connects to the physical structure constant $1/h_{\alpha}^{o*} \cong 137.036$ (the updated value) by relation $so = (1/2h_{\alpha}^{o*})^{1/2}$.

Results (Sec.2.2.3) identify a *bridge between minimal uncertainty and a certainty* measured by the entropy invariant $S_{\mp a}^{*} = 2h_{\alpha}^{o}$ which enables creation of an initial Information macrounit—a triplet with probability $p_{\pm a} = \exp(-2h_{\alpha}^{o}) = 0.98555075021 \to 1$ approximating the certainty.

This is the *bridge between micro-and macroprocesses* emerging along the path of observing the impulse interactions from maximal uncertainty to Information certainty [40].

The invariant connects this microprocess, which arises at the merge of interactive action and reaction (Sec. 2.2), with the motion of the interactive adjunct charged particle in a Maxwell field.

This proves the requirement for the *discrete action merging reaction,* which leads to impulse interaction rising the microprocess. Since the merging microprocess emanates for random field, it indicates that equations of the electromagnetic wave in the atom also originate in random field.

Moreover, the Schrödinger equation, describing the microprocess, emerges from the *initial random impulses* of the merging actions and reactions, while both references [6, 7] and [39] have studied the *deterministic* processes.

The invariant constant also binds the emerging micro-macroprocess with Maxwell equations extended to an equation of the interacting atom particles. In addition, the extended model of the atoms, covering three of the four fundamental interactions (electro-magnetic, weak and strong interactions), allows the *Information description*, which confirms "It from bit". The merging impulses 1-0 and 1-0 also explain *creation of qubit* $|0\rangle$ and $|1\rangle$ *in the emerging microprocess* during the entanglement. (Sec.5.4.1.). •

Study [41] has shown "that gravity, just as electromagnetism in Wheeler-Feynman's time symmetric electrodynamics, also be an "adjunct field" instead of an independent entity".

In [40] and Sec.4.4 we calculate a weak information forces' analogy with the gravitational force. •

Reference [42] has revealed that "the entangling space-time works just like a quantum error-correcting code, protecting information in jittery qubits to store it not in individual qubits, but in patterns of entanglement among many", starting with a triple.

Sec. 5.4.1 details the mechanism of *assembling the entangled* qubits triple in an information unit, with emerging a space interval during reversible time interval at the entanglement. •

## REFERENCES TO INTRODUCTION

# I. INFORMATION PATH INITIAL FORMALISM

## 1. OBSERVING PROCESS AND ITS INTEGRAL MEASURE

### 1.1. The Initial Points

1. Multiple interactive processes build the Universe. Reality is built from emerging interactions. Natural interactions unify a sequence of interactive impulses Yes-No or No-Yes actions. Multiple interactions form a random field.
2. Each real (that is, certain) interaction is composed of opposite Yes-No actions of discrete impulse $\downarrow\uparrow$ which model an elementary Bit of information.
3. Observation of a random process uncovers a real Bit and/or multiple Bits.
4. Observation is a series of formal acts emerging from changing the probability of interactive events in random impulses of the random process.
5. Observation defines a random sequence of the interacting impulses (Yes-No or No-Yes actions), which formally models a sequence of Kolmogorov's 0-1 law events. That is, objective probabilities measure the random impulses in a formal random field.
6. An axiomatic Kolmogorov field formally connects the sets of possible events, the sets of actual events, and the probability function. This triad conserves energy of actual (real) events. It models the occurrence of specific events and starts each sequence in a probabilistic observation of multiple interacting events in a random process.
7. The random process models a Markov diffusion process of probabilistic impulses emerging from the random field. The diffusion encloses a set of event-holding correlations.
8. The sequence of the probabilistic 0-1 (No-Yes) impulses, acting on the Markov diffusion process, initiates Bayes probabilities that virtually self-observe the Markov process.
9. These objective probabilities link the Kolmogorov law's axiomatic probabilities with the Bayesian probabilities, bringing discrete Bayes probabilities into the process of the observer. A particular objective Bayes probability observes specific set of events, and its correlation holds the entropy measure.
10. By virtually cutting the Markov correlation, the discrete impulse probabilities observe the entropy-uncertainty hidden in the correlation.
11. Sequentially cutting the entropy of correlation decreases the uncertainty of Markov diffusion, and increases the probabilities of the observing process. Each impulse $\downarrow\uparrow$ action $\downarrow$ cuts the Markov maximal probability, opening a path to certainty, while the following interaction $\uparrow$ cutting the maximal entropy carries an equivalent unit of Information. Such interactive impulses $\downarrow\uparrow$ enclose the certainty of real observation, bringing the energy field to the actual interacting cut, which *converts* the maximal entropy to the Information unit.
12. The information Bit uncovers informative events of the observing process hidden under the cutting correlations.
13. Multiple interacting Bits self-organize the information process, which creates the Information Observer. *Finally, both Information and the Information Process emerge as phenomena of natural interactions, while each random field triad generates an Observer from the process.*

### 1.2. The Initial Formalism and Probabilistic Model

#### *1.2.1. Probability Field, Kolmogorov Law, Interacting Markov process, and Entropy Functional*

Multiple interactive actions are random events-variables $\omega$ in a surrounding random probability field. The probability field defines a mathematical triad [1]: $\Phi = (\Omega, F, P)$, where $\Omega$ is the set of all possible $\omega$, $F$ is Borel's $\sigma$-algebra subsets from sets $\Omega$, and probability $\mathrm{P}$ is a non-negative function of the sets, defined on $F$ at condition $\mathrm{P}(\Omega)=1$. This triad formally connects the sets of possible events, the sets of actual events, and their probability function.

*Example:*

If an experiment consists of one flip of a fair coin, then the result is either heads H or tails T (or none of them), called tail events. Then $\Omega=(H,T)$, and the $\sigma$-algebra $F=2^{\Omega}$ contains $2^2=4$, these tail events with the probability measures $\mathrm{P}(\bullet)=0, \mathrm{P}(\mathrm{H})=0.5, \mathrm{P}(\mathrm{T})=0.5, \mathrm{P}(\mathrm{H},\mathrm{T})=1$. For three-dimensional spaces, from the probability field emerges a potential observer with eight possible probability measures in the form of a random cube.

In the infinite sequence of random variables $\omega$, distributed in the field, a discrete $\mathrm{P}(\bullet)=0$ can happen among $P(\omega)$. •

Let $F$ be the $\sigma$-algebra generated by $\omega$, and $F_o$ be the $\sigma$-algebra generated by the sequence of mutually independent variables $\varpi$ and its function $f(\varpi)$.

Then, the Kolmogorov 0-1 law is fulfilled when "the conditional probability $P_{\bar{\varpi}}[f(\varpi)=0]$ of the relation $f(\varpi)=0$ remains, and the first $n$ variables (of $\varpi$) equals to the absolute probability $P_{\bar{\varpi}}[f(\varpi)=0]=0$ or $P_{\bar{\varpi}}[f(\varpi)=0]=1$ for every $n$." The assumptions of the law are satisfied if random variables $\varpi$ are mutually independent and if the value of function $f(\varpi)$ remains unchanged when only a finite variable changes [1:69].

The probability of the independent events is an equivalent to the probability in the Big Number law [1:69].

Let us also have a Markov diffusion process in the field [1]: $X_t=X(\omega)$ of event $\omega=(x,t)$ including states $x$ and their time moment $t$. The Markov process is $n$-dimensional and all dimensions start in a probability field with different triads associated with local random frequencies of the initial events $\omega$ attribute. Generally $n\rightarrow\infty$.

Trajectories of the Markov diffusion process are defined on $n$-dimensional probability distribution $P_n=P_n[X(\omega)]$ with transitional probabilities $P(s,\tilde{x},t,B)$, $\sigma$ - algebra $F(s,t)$ created by events $\{\tilde{x}(\tau)\in B\}$ at $s\le\tau\le t$, and conditional probability distributions $P_{s,x}$ on $F(s,t)$.

Relation [2]:

$$\tilde{P}_{s,x}(d\omega)=p(\omega)P_{s,x}(d\omega) \tag{1.0}$$

measures transformation of this probability on trajectories of Markov process $(\tilde{x}_t,P_{s,x})$ which holds distributions $\tilde{P}_{s,x}=\tilde{P}_{s,x}(A)$ on extensive $\sigma$-algebra $F(s,\infty)$ with density measure:

$$p(\omega)=\frac{\tilde{P}_{s,x}(d\omega)}{P_{s,x}(d\omega)} \tag{1.1}$$

Applying definition of conditional entropy [3] to mathematical expectation of the logarithmic probability functional density measure (1.1), we introduce *the Entropy Functional (EF) on trajectories* of a Markov diffusion process [4]:

$$S=E\{-\ln[p(\omega)]\}=\int\limits_{\tilde{x}(t)\in B}-\ln[p(\omega)]P_{s,x}(d\omega), \tag{1.2}$$

where $E=E_{x,s,\tilde{x}_t}$ is the conditional mathematical expectation, taken along the trajectories of process $\tilde{x}_t$ at a varied $(\tilde{x},s)$ (by analogy with Kac [5]).

Markov diffusion process describes a drift function $a=a(t,x)$ and a function of diffusion $\sigma=\sigma(t,x)$, which together define additive functional [6-9]:

$$\varphi_s^T = 1/2\int_s^T a(t,\tilde{x})^T (2b(t,\tilde{x}))^{-1} a(t,\tilde{x}_t)dt + \int_s^T \sigma(t,\tilde{x})^{-1} a(t,\tilde{x})d\xi(t), 2b(t,\tilde{x}) = \sigma(t,\tilde{x})\sigma^T(t,\tilde{x}) > 0. \quad (1.3)$$

Functional (1.3) describes the transformation of the Markov processes' random time as it traverses the trajectory of the process.

The additive functional (1.3) measures the probability density (1.1), which also can measure $\varphi_s^t(\omega)$ :

$$p(\omega) = \exp\{-\varphi_s^t(\omega)\}, \text{ or } \varphi_s^t(\omega) = -\ln p(\omega). \quad (1.4)$$

Transitional probabilities for a diffusion process $\varsigma_t$, transforming density (1.4) with aid of additive functional (1.3), satisfy relation (1.0) in form

$$\tilde{P}(s,\varsigma,t,B) = \int_{\tilde{x}(t)\in B} \exp\{-\varphi_s^t(\omega)\}P_{s,x}(d\omega). \quad (1.5)$$

Applying the definition of the Entropy Functional (1.2) to process $\tilde{x}_t$ conditional to process $\varsigma_t$, we get the Entropy Functional measure expressed via the additive functional $\varphi_s^t(\omega)$ on the trajectories of the diffusion processes:

$$S[\tilde{x}_t / \varsigma_t] = E[\varphi_s^t(\omega)]. \quad (1.6)$$

The minimum of this functional, depending on the additive functional, measures the closeness of the above distributions in the form:

$$\min_{\varphi_s^t} S[\tilde{x}_t / \varsigma_t] = S^o \quad (1.7)$$

Let the transformed process be

$$\varsigma_t = \int_s^t \sigma(v,\xi_v)d\xi_v \quad (1.8)$$

having the same diffusion as the initial process $\tilde{x}_t$ but zero drift, then $\varsigma_t$ models standard perturbations -"white noise."

Process $\varsigma_t$ is a transformed version of process $\tilde{x}_t$ whose transition probability satisfies (1.1).

Transformed probability $\tilde{P}_{s,x}$ for this process evaluates the Feller kernel measure [10, 11].

Since transformed process $\varsigma_t$ (1.8) has the same diffusion matrix but zero drift, the right part of additive functional in (1.3) satisfies relation.

$$E[\int_s^T (\sigma(t,\tilde{x})^{-1} a(t,\tilde{x})d\xi(t)] = 0. \quad (1.9)$$

It brings the integral measure of the Entropy Functional expressed via parameters of Ito stochastic equation [34] in the form:

$$\Delta S[\tilde{x}_t]|_s^T = 1/2E_{s,x}\{\int_s^T a(t,\tilde{x}_t)^T (2b(t,\tilde{x}_t))^{-1} a(t,\tilde{x}_t)dt\} = \int_{\tilde{x}(t)\in B} -\ln[p(\omega)]P_{s,x}(d\omega) = -E_{s,x}[\ln p(\omega)]. \quad (1.10)$$

Formulas (1.2-1.5), (1.6), (1.7), (1.8) and (1.10) are in [4, 12] with related citations and references.

The Entropy Functional (EF) in forms (1.6, 1.10) is an *information indicator* of distinction between the probability measures of processes $\tilde{x}_t$ and $\varsigma_t$; it measures the *quantity of information* of process $\tilde{x}_t$ conditional to process $\varsigma_t$.

When the EF in (1.10) equals zero, the observing process's Bayes impulses are indistinct from the noise.

Since the time, following from the zero EF, is also equal to zero, the observation has *not* begun.
It is an indicator of an observation starting.
The right side of (1.10) is the EF equivalent formula, expressing it via probability density $p(\omega)$ of random events $\omega$, integrated with the probability measure $P_{s,x}(d\omega)$ along the process trajectories $\tilde{x}(t) \in B$, which are defined at the set $B$.
*Formula (1.10) directly connects the probabilities, defining the EF and the Markov process function's drift and diffusion, without the need to the implement probabilities measurements for the given process.*
For the processes with the equivalent probability measures, quantity (1.10) is zero, and it is positive for nonequivalent measures of the process. The Variation problem (1.7) for the EF functional was solved in [12].
Mathematical expectation (1.10) on the *process's trajectories,* conditional to the transformed probability measure of Feller kernel $\tilde{P}_{s,x}$, defines an invariant measure at Markov transformations along the trajectories. The invariant measures the Radon-Nikodym probability density measure (1.1) where both $P_{s,x}$ and $\tilde{P}_{s,x}$ are defined. Thus, integral (1.10) is the Markov process's $\tilde{x}_t$ functional entropy measure conditional to the kernel probability measure. The Entropy measure in (1.2) is conditional to any transformed Markov diffusion process, not necessarily satisfying (1.10).
The probability of each process is local for each random ensemble as a part of the $n$-dimensional process ensemble.
The field of events, having probabilities $P_{\bar{\varpi}} =\in 0,1$, may interact with Markov process probability $P_{s,x}(\omega)$ of events $\omega$ satisfying the relation for joint probability of independent events:
$P = P_{\bar{\varpi}} \times P_{s,x}(d\omega), P_{\bar{\varpi}} =\in 0,1$.
From that follows the sequence of probabilities: at $P_{\bar{\varpi}} =0, P = P_{s,x}(d\omega) = 0$ and at $P_{\bar{\varpi}} = 1, P = P_{s,x}(d\omega)$.
Changing probability $P_{s,x}(d\omega) = 0$ to probability $P_{s\to t,x}(d\omega)$ of the interacting events initiates transitional Markov probability $P(s,\tilde{x},t,B)$. Probability $P_{s\to t,x}(d\omega)$, acting via the additive functional in (1.5), switches the transitional probability $P(s,\tilde{x},t,B)$ to the transitional probability $\bar{P}(\bar{s},\bar{\tilde{x}},t,B)$ and density $p(\omega)$. The density, defined via the process additive functional (1.4), links with it drift function in (1.3). Transformation (1.5) determines the transition probabilities with the same diffusion but different drift. Transitional probability $\bar{P}(\bar{s},\bar{\tilde{x}},t,B)$ changes related Markov probability $\bar{P}_{s,x}(\omega_\alpha)$ at $(\bar{s},\bar{\tilde{x}}) = \omega_\alpha$, which, changes its drift functions: $a^\omega = a(\omega) \xrightarrow{\bar{P}_{s,x}(\omega_\alpha)} \bar{a}^\omega = a(\omega_\alpha)$.
That switches the Markov movement from its current drift $a^\omega = a(\omega)$ to Markov movement under another drift $\bar{a}^\omega = a(\omega_\alpha)$. The random sequence 0-1-0-1-0-0-1-…of the probabilistic actions ↓↑ affects the probabilities $\bar{P}_{s,x}(d\omega), \bar{P}_{s,x}(d\omega_\alpha)$, which through Markov density $p(\omega)$ randomly switch the process movement accordingly.
Changing the probabilities implements the act of observation (of the initial point 4.) For example, an infinite sequence of coin-toss tail events hit a table which randomly moves it when each tail interacts with the table.
The law 0-1 interacting probabilities initiate the interactive random impulse movement within the initial (primary) Markov process, which we call a secondary Markov process.

*1.1.2.2. The Bayes' Probability Rules*

For each $i,k$ random event $A_i, B_k$ along the observing events, each conditional a priori probability $P(A_i / B_k)$ follows the conditional a posteriori probability: $P(B_k / A_i) = P(A_i / B_k) / P(A_i)$.

Substituting $P(A_i / B_k) = P(A_i \cup B_k) / P(B_k)$ we get $P(B_k / A_i) = P(A_i \cup B_k) / P(A_i)P(B_k)$,
where average probability of expecting events along the observing events is:

$$P(B_k) = \sum_{i=1}^{n} P(B_k / A_i)P(A_i) . \qquad (1.11)$$

Ratio *a priori* to *a posteriori* probabilities:

$$P(A_i / B_k) / P(B_k / A_i) = P(A_i)$$

determines the current observing probability $P(A_i)$ which may include some observations of previous events $A_{i-1}, A_{i-2},...,$.

*1.2.3. Applying the Bayes Formulas and the Link to the Kolmogorov Law for the Markov Diffusion Process*

Defining *a priori* probability $P^a_{s,x}(d\omega) = P_{s,x}(d\omega)$ and *a posteriori* $P^p_{s,x}(d\omega)$ probabilities by Bayes's rule, we *link* the Kolmogorov 0-1 Law and Bayes probabilities through the Markov diffusion process.

The switch of Markov drifts $a^{\omega} \downarrow \bar{a}^{\omega}$ models No-Yes actions 0-1, while the switch of the drifts $\bar{a}^{\omega} \uparrow a^{\omega}$ models Yes-No interacting actions 1-0 of these impulses.

Both impulses, acting on the Markov process, change its *a priori* probability to *a posteriori* probability.

The impulse $\downarrow\uparrow$ replicates a random bit which the 0-1 probabilities cover, while the switches reveal the interactive link of the Kolmogorov 0-1 Law to the Bayes probabilities *within* the Markov diffusion process.

The Markov process probabilities $P^a_{s,x}(d\omega_\alpha)$ and $P^p_{s,x}(d\omega_\alpha)$ of events $d\omega_\alpha$ correlate under its drift function and diffusion. Increasing $\bar{P}(\bar{s}, \bar{\bar{x}}, t, B)$ raises the probability density $p(d\omega_\alpha) = P^p_{s,x}(d\omega_\alpha) / P^a_{s,x}(d\omega_\alpha)$, which grows the correlations and increases the closeness of these probabilities to density $p(\omega)$, and finally to $p(d\omega_\alpha) \to 1$.

The rising $p(d\omega_\alpha)$ grows a posterior probability, revealing an actual drift $\bar{a}^{\omega} = a(\omega_\alpha)$ moving the process. That increases probability of impulse $\downarrow\uparrow$ up to a certain Bit whose action $\uparrow$ on $\bar{a}^{\omega} = a(\omega_\alpha)$ processes real movement.

Each considered probability is an abstract axiomatic of Kolmogorov's probability.

It predicts the probability measurement on the experiment whose probability distributions, tested by the relative frequencies of occurrences of events, satisfy the condition of symmetry of the equally probable events [13].

In the theory of randomness, the probability of each event is *virtual*, or, at every instance, prescribed to such an imaginary event. Many of its potential probabilities might occur simultaneously. However, an explicit probability describes a physical possibility of some of them.

Multiple random actions describe the probability distributions on the observing sequence of a specific set of events, which formally define the observing triad in the probability field.

Processing interactions $\varpi$ along $X_t$ possess a common time course in the field, which, for each field ensemble with the fractions of random events, is a part of the time in the common time course.

We assume the random field conceals a randomly distributed energy that the events hold.

The probability field of the concealed energy is timeless, reversible, symmetrical, and scalar.

Random interactions, disturbing the field, randomly reveal the energy which the interaction or its measurement acquires from the field for actual events.

## 1.2. The Notions of Observation, Observing Process, Virtual Observer, and Uncertainty

The objective Kolmogorov 0-1 probabilities quantify the idealized (virtual) impulses whose No-Yes actions represent an act of a virtual observation of Markov *a priori* probability $P^a_{s,x}(d\omega) = P_{s,x}(d\omega)$ shifting to Markov *a posteriori* probability $P^p_{s,x}(d\varpi)$ during the impulse finite time interval $\delta\tau$ .

Each observation measures a probability of possible events for a potential observer.

Thus, linking the *a priori* and *a posteriori* Bayes probabilities of the impulses virtually observes the Markov diffusion process on its trajectory as it moves under the drift.

The transitional probabilities on the trajectories of the Markov diffusion process transit multiple virtual observations along a *virtual observing process* $\bar{x}_t$ of interacting impulses. Each such interactive No-Yes actions provide a step-down action (0) and step up action (1) within the impulse No-Yes probabilities.

(The secondary Markov process under the Bayes umpulse observation is a virtual observing process).

The objective probabilities measure the virtual observing probabilities of the potential *(virtual) observer.* (The term virtual applies only to a probable observation by an observer).

Let us substantiate the probabilistic measurement and the entropy measure of the virtual observer.

For each $i$ random events $A_i$ of the observing process and its current *a priori* probability $\bar{P}_{s,x}(A_i)$ includes the previously observed probabilities tracking back to $(n-1)$ transitional Markov probability.

For the Markov diffusion process with the $i$ -events' transitional probability $P_i(s,\tilde{x},t,B)$, a generalized (integral) *a priori* probability, which integrates all accessible *a priori* probabilities has the form

$$\bar{P}_{s,x}(A_i) = \int\limits_{s,t,\tilde{x},B,} \prod_{i=1}^{n-1} P_i(s,\tilde{x},t,B), \qquad (2.1)$$

where $\bar{P}_{s,x}(A_i) = P^a_{s,x}(d\omega) = P_{s,x}(d\omega)$ and generalized distribution (2.1) are defined on $\sigma$ -algebra of events $A_i = A_i(s,\tilde{x},t)$ [9].

Each of the equal probabilities is defined on the same Markov random states-events and is *observing* in the Markov process $\bar{x}_t$ . The observing automatically includes No-Yes actions linking to the emerging Markov observing process.

The ratio of *a posteriori* to *a priori* Bayes probabilities defines *a posterior probability density*

$$\bar{p}(\omega) = \frac{P^p_{s,x}(d\omega)}{P^a_{s,x}(d\omega)} = \frac{P^p_{s,x}(d\omega)}{\bar{P}_{s,x}(A_i)} \ . \qquad (2.2)$$

Each following observation updates *a posteriori* density $\bar{p}(d\omega)$ as well as the previous observations.

Substituting $\tilde{P}_{s,x}(d\omega) = P_{s,x}(d\omega)p(d\omega)$ to $P^p_{s,x}(d\omega) = \bar{p}(d\omega)P_{s,x}(d\omega)$ leads to

$$P^p_{s,x}(d\omega) = \tilde{P}_{s,x}(d\omega)\bar{p}(d\omega) / p(d\omega) \qquad (2.3)$$

at $\bar{p}(d\omega) / p(d\omega) = P^p_{s,x}(d\omega) / \tilde{P}_{s,x}(d\omega)$ . (2.4)

In relations (2.4), all probabilities including $P^p_{s,x}(d\omega)$ are defined on the same Markov process. The observation of the moving observing process concurrently allows updating both *a priori* and *a posteriori* probabilities.

However, the Markov process defines the current probabilities of two time-steps, one with a head up and one with a head down, which in (2.1) limits the number of each current observation $i$ by $i=1$ and $i=2$ , or $i-1, i, i+1$ . That holds two current impulses also in (1.11). Observation of these events within a fixed random interval determines a threshold of starting correlations [18:32]. Thus, notion of observation connects minimum of three current events by a triple correlations. Each previous observation decreases a current *a priori* probability $\bar{P}_{s,x}(A_i)$ in (2.1) at each fixed $i$, since multiplication of the probabilities is less than 1, decreasing the total multiplied probability in (2.1) and $\bar{P}_{s,x}(A_i)$.
Integration of each of these small triple-process increments in (2.1) for each $i$ does not change the decrease.
Such integration at a given $P^{p}_{s,x}(d\omega)$ increases $\bar{p}(d\omega)$, updating it through a prior observation.

The posterior observation also updates *a posteriori* density $\bar{p}(d\omega)$ growing with rising $P^{p}_{s,x}(d\omega)$.

*Example:* Suppose a process stops at each $P=0$ and the process starts at each $P=1$, and let step $i=1$ have

$$\bar{P}_{s,x}(A_{i=1}) = P^{a}_{s,x}(d\omega) = P_{s,x}(d\omega), P_{s,x}(d\omega) \xrightarrow{P=0} \bar{P}_{s,x}(d\omega) = 0.5, p(\omega) = 0.8, \tilde{P}_{s,x}(d\omega) = P^{p}_{s,x}(d\omega) = 0.4.$$

At step $i=2$, it holds $P_{s,x}(d\omega) \xrightarrow{P=1} \bar{P}_{s,x}(d\omega_\alpha) = P^{a}_{s,x}(d\omega_\alpha) = 0.55, \bar{p}(\omega_\alpha) = 0.81, \tilde{\bar{P}}_{s,x}(d\omega_\alpha) = P^{p}_{s,x}(d\omega_\alpha) = 0.4455,$

where $P^{a}_{s,x}(d\omega_\alpha)$ updates $\bar{P}_{s,x}(A_{i+1})$, which includes $\bar{P}_{s,x}(A_{i=1}) = 0.5$.

That changes the density in (2.4). Integral (2.1) from discrete $i=1$ to $i=2$ approximates the sum of multiplications:

$\bar{P}_{s,x}(A_{i+1}) = P^{a}_{s,x}(d\omega) \times \bar{P}_{s,x}(d\omega_\alpha) + P^{a}_{s,x}(d\omega) \times \bar{P}_{s,x}(d\omega_\alpha)$, which brings

$\bar{P}_{s,x}(A_{i+1}) = 0.5 \times 0.55 + 0.5 \times 0.55 = 0.55$, and

$$\bar{p}(\omega_\alpha) = P^{p}_{s,x}(d\omega_\alpha) / \bar{P}_{s,x}(A_{i+1}) = 0.4455 / 0.55 = 0.81. \quad (2.5)$$

Here probability $P^{p}_{s,x}(d\omega_\alpha)$ is updating the probability of Markov drift $a(\omega_\alpha)$ as well as $P^{a}_{s,x}(d\omega_\alpha)$ is doing it. •

Results [14] prove that cutting the impulse correlation increases the observing Markov probability density, which turns to the next impulse cutting off. Each cutoff analytically runs Dirac delta-action, or Kronicker 0-1 action.

Let us define Bayes entropy on the observing process in the form

$$S_B = E_{s,x}[-\ln \bar{p}(\omega)], \quad (2.6)$$

where $E_{s,x}$ is the conditional mathematical expectation moving along this process. Applying to (2.6) formula (2.2) brings

$$S_B = -\int_{\tilde{x} \in B} \ln \bar{p}(\omega) P^{a}_{s,x}(d\omega). \quad (2.7)$$

This formula also concurs with Bayes (conditional) probability in the form (1.11) applied to (2.6).
Entropy $S_B$ measures the uncertainty of the observing process.

For the virtual observer observing this process, $S_B$ measures the integral of the virtual observer's uncertainty.

Maximal uncertainty measures non-correlating *a priori-a posteriori* probabilities when their connection approaches zero. Such a theoretical uncertainty has infinite entropy. Its conditional entropy and time do not exist.

The finite uncertainty measure has non-zero correlating finite *a priori–a posteriori* probabilities with a finite time interval and the subsequent finite conditional entropy.

An example of a finite uncertainty process is "white noise." Using it allows us to measure the uncertainty of the observing process relative to the uncertainty-entropy of white noise (Sec.1.1).

Measuring conditional entropy (1.10) relative to diffusion process $\varsigma_t$ is useful, since average $\varsigma_t$ models zero of the observing process, from which a potential observation begins. When a Yes–No probable impulse affects such a process, the impulse of the observing process starts a potential observer.

## 1.3. CERTAINTY, INFORMATION, CUTTING ENTROPY

Information, as a notion of certainty opposed to uncertainty, originates through a reduction of the observing uncertainty toward maximal posteriori probability 1, which, we assume, evaluates an observing probabilistic fact with certainty.

If an elementary Kronicker impulse increases each observing Bayes a posteriori probability, it concurrently increases the probability of each virtual impulse (up to a real impulse with posteriori probability 1) and decreases the related uncertainty. Each $\downarrow\uparrow$ impulse carries the secondary Markov process and interacts with the primary Markov diffusion.

Each observing action $\downarrow$ cuts entropy of the interacting Markov diffusion process decreasing its uncertainty. The cutting action transfers the cutting entropy, reduced in interaction, to the impulse action $\uparrow$, which increases the entropy of the observing impulse's Bayes probabilities along the secondary Markov process. Multiple cutting actions $\downarrow$ minimize the uncertainty of the primary Markov process, and interacting actions $\uparrow$ maximize entropy in each following impulse of the observing process. The observing impulse accumulates maximum entropy when the Markov process entropy reaches the minimum and maximum of its probability. Thus, each action $\downarrow$ cuts maximal entropy, while the following interacting action $\uparrow$ transfers the reduced maximal entropy to the observing process. An interactive impulse $\downarrow\uparrow$, enclosing the certainty of a real observation, brings energy to actual interacting impulse $\uparrow\downarrow$ cut, which *converts* the its maximal entropy to the Information unit. Measuring the maximal Markov transitional probability indicates the *rapprochement* of the observing impulse to the moment of cutting its maximal uncertainty, thus extracting the certainty of the Observing bit. (When the potential (virtual) observer obtains a Bit through cutting the maximal entropy of the observing Bayes impulse, it automatically measures minimal $S_B$ of the Markov diffusion process, approaching probability 1.)

Finally, erasure of the collected entropy-uncertainty during observation of the impulse restores the impulse's certain Bits. Building a certain physical Bit requires injection of Landauer's energy [15] for both erasure and memorizing the bit.

### 1.3.1. **The Virtual Observer. Certain Information Observer**

Since $\overline{p}(d\omega)$ increases in the growing impulse observations, the $S_B$ tends to decrease. Therefore, impulse observation minimizes uncertainty, from which automatically emerges the principle of Minimum Entropy:

$$\min_{n-1\leftarrow i} S_{Bi} = S_{Bo}.$$

At $\overline{p}(d\omega) > p(d\omega)$, $S_B < S$, this principle is applied to $S$ in form:

$$\min_{a^{\omega}\to a^{\varpi}} S \to S_{Bo}.$$

Observation generates $S_B$ and therefore the virtual observer. Removing uncertainty generates a certain information Observer. When the potential observer obtains a Bit through cutting the maximal entropy of the observing Bayes impulse, it automatically measures minimal $S_B$ of the diffusion process, approaching probability 1.

*The interactive impulse observations reduce the uncertainty minimizing in the growing probability of the impulse observation. The impulse probabilistic description generalizes a potential random interaction, multiple interactive impulse actions of the observing random process, unifying the Bit's common Information origin.*

Moreover, it is proved in [14-16] that the impulse observation leads to the Max-Min principle, which increases probability reduction in each following uncertainty.

Below we specify how a Virtual Observer, by applying the observation-cutting impulses that decrease uncertainty $S_B$, rises up to become an Information Observer.

Comments 1.1. *Transition from the Observing Markov Chain to the Diffusion Process*

Kolmogorov's 0-1 jumping probabilities, interacting with a stochastic process, emerging from the random field, sequentially stops and starts it, generating a Markov chain during an interactive random time.

A continuous random time Markov chain can converge to diffusion [35] if observation is a jumping motion transiting an average sequence of each three time states of the chain along the Markov chain. The generator for a Markov chain jump moves the chain if each observation averages three states. In the considered impulse observation, the correlating three impulses sequentially average each three probabilistic impulses covering three states. Thus, the observation of a random Markov chain provides both triple correlations of the chain states under Kolmogorov 0-1 probabilities, and their transition along the chain. The average of each three states' probabilities provides formula (2.1) applied for transitional probabilities of the Markov chain.

Under such observations, the Markov diffusion evolves to a Lévy process [2].

The Lévy process independent random increments model multiple observing impulses of the random bits. The Lévy-Khintchine decomposition [2:113-117] characterizes any Lévy process in terms of three components (the Lévy triplet) one of which is the Lévy measure.

This measure describes the distribution of the Lévy process potential jumps.

Until any two time intervals of the process' independent random increments do not overlap, the jump absent and Lévy measure is zero.

The overlap is a merge of the yes-no impulses, when Lévy measure is nonzero, finite.

The merge of the yes-no impulses indicates zero random bit.

When Levy measure becomes finite, the jumps have started.

The jumps initiate a microprocess within the observing Levy process, which releases the entropy-uncertainty hidden in the cutting correlation, finally generating a non-random real bit.

At the merge, the probability of the impulse carrying a bit is zero.

It means there is no reality, defined by the interaction, when the microprocess starts.

With zero probability of the microprocess start, it is unpredictable. The information process starts when the microprocess(Sec.II) generates real qubit and/or bit. Each specific triad through an observing microprocess is virtually forming related physical bits.

The considered path from emerging probabilistic impulse observation up to zero impulse identifies virtual (uncertain) process covering hidden information (as uncertain reality), which precedes the starting information process (beginning a certain reality).

This Markov chain model has been applied to chemical reactions transiting to a diffusion model [35], which is widely accepted for modeling different chemical processes with various kinetics. •

Linear Brownian movement, described in Markov diffusion, is an example of stochastic process.

The existence of trajectories of stochastic process satisfies limitation on stochastic movement [18:44].

### 1.3.2. **Evaluation of the Cutting Process EF Fractions by an Impulse Control**

Let us define control on the space $KC(\Delta,U)$ of a piece-wise continuous step functions $u_t$ at $t\in\Delta$:

$$u_{-}\overset{def}{=}\lim_{t\to\tau_k-o}u(t,\tilde{x}_{\tau_k}),\ u_{+}\overset{def}{=}\lim_{t\to\tau_k+o}u(t,\tilde{x}_{\tau_k}) \qquad (2.1)$$

which are differentiable on the set

$$\Delta^o=\Delta\setminus\{\tau_k\}_{k=1}^m,\ k=1,...,m, \qquad (2.1a)$$

and applied on diffusion process $\tilde{x}_t$ from moment $\tau_{k-o}$ to $\tau_k$, and then from moment $\tau_k$ to $\tau_{k+o}$, implementing the process's transformations $\tilde{x}_t(\tau_{k-o}) \to \varsigma_t(\tau_k) \to \tilde{x}_t(\tau_{k+o})$. The $n$-dimensional process holds $m$ such transformations.

In the vicinity of moment $\tau_k$, between the jump of control $u_-$ and the jump of control $u_+$, we consider a control *impulse*

$$\delta u_\pm(\tau_k) = u_-(\tau_{k-o}) + u_+(\tau_{k+o}). \quad (2.2)$$

The following statement evaluates the EF information contributions at such transformations.

*Proposition 2.1:*

The Entropy Functional (1.1.10) at the switching moments $t = \tau_k$ of control (2.2) takes values

$$\Delta S[\tilde{x}_t(\delta u_\pm(\tau_k)] = 1/2, \quad (2.3)$$

and at locality of $t = \tau_k$: at $\tau_{k-o} \to \tau_k$ and $\tau_k \to \tau_{k+o}$, produced by each of the impulse control's step functions in (2.1), is estimated by

$$\Delta S[\tilde{x}_t(u_-(\tau_k)] = 1/4,\ u_- = u_-(\tau_k),\ \tau_{k-o} \to \tau_k \quad (2.3a)$$

and

$$\Delta S[\tilde{x}_t(u_+(\tau_k)] = 1/4, u_+ = u_+(\tau_k), \tau_k \to \tau_{k+o}. \quad (2.3b)$$

*Proof.* The jump of control function $u_-$ in (2.1) from moment $\tau_{k-o}$ to $\tau_k$, acting on the diffusion process, might cut off this process after moment $\tau_{k-o} \to \tau_k$. The cutoff diffusion process has the same drift vector and the diffusion matrix as the initial diffusion process. The additive functional for this cutoff has the form [9]:

$$\varphi_s^{t-} = \begin{cases} 0, t \le \tau_{k-o} \\ \infty, t > \tau_k \end{cases}. \quad (2.4)$$

The jump of the control function $u_+$ (2.1) from $\tau_k$ to $\tau_{k+o}$ might cut off the diffusion process *after* moment $\tau_k \to \tau_{k+o}$ with the related additive functional

$$\varphi_s^{t+} = \begin{cases} \infty, t > \tau_k \\ 0, t \le \tau_{k+o} \end{cases}. \quad (2.5)$$

For the control impulse (2.2), the additive functional at a vicinity of $t = \tau_k$ acquires the form of an *impulse function*

$$\varphi_s^{t-} + \varphi_s^{t+} = \delta \varphi_s^{\mp}, \quad (2.6)$$

which summarizes (2.3) and (2.4).

Entropy functional (1.1.10) following from (2.4-2.5) takes values

$$\Delta S[\tilde{x}_t(u_-(t \le \tau_{k-o}; t > \tau_k))] = E[\varphi_s^{t-}] = \begin{cases} 0, t \le \tau_{k-o} \\ \infty, t > \tau_k \end{cases}, \quad (2.7a)$$

$$\Delta S[\tilde{x}_t(u_+(t > \tau_k; t \le \tau_{k+o}))] = E[\varphi_s^{t+}] = \begin{cases} \infty, t > \tau_k \\ 0, t \le \tau_{k+o} \end{cases}, \quad (2.7b)$$

from 0 to $\infty$ acquiring *absolute maximum* at $t > \tau_k$, and back from $\infty$ to 0, acquiring *absolute minimum* at $\tau_{k-o}$ and $\tau_{k+o}$.

The multiplicative functionals [6], related to (2.4-2.5), are:

$$p_s^{t-} = \begin{cases} 0, t \le \tau_{k-o} \\ 1, t > \tau_k \end{cases}, \; p_s^{t+} = \begin{cases} 1, t > \tau_k \\ 0, t \le \tau_{k+o} \end{cases}. \tag{2.8}$$

The control impulse (2.2) provides an impulse probability density in the form of a multiplicative functional

$$\delta p_s^{\mp} = p_s^{t-} p_s^{t+}, \tag{2.9}$$

where $\delta p_s^{\mp}$ holds $\delta[\tau_k]$-function, which determines probabilities

$\tilde{P}_{s,x}(d\omega) = 0$ at $t \le \tau_{k-o}, t \le \tau_{k+o}$, and $\tilde{P}_{s,x}(d\omega) = P_{s,x}(d\omega)$ at $t > \tau_k$. (2.9a)

For the cutoff diffusion process, transitional probability (at $t \le \tau_{k-o}$ and $t \le \tau_{k+o}$) turns to zero, and states $\tilde{x}(\tau_k - o), \tilde{x}(\tau_k + o)$ become independent, while their mutual time correlations *are dissolved*:

$$r_{\tau_k - o, \tau_k + o} = E[\tilde{x}(\tau_k - o), \tilde{x}(\tau_k + o)] \to 0. \tag{2.10}$$

Entropy increment $\Delta S[\tilde{x}_t(\delta u_{\pm}(\tau_k))]$ of the additive functional $\delta\varphi_s^{\mp}$ (2.5), produced within or at a the border of control impulse (2.2), defines the equality

$$E[\varphi_s^{t-} + \varphi_s^{t+}] = E[\delta\varphi_s^{\mp}] = \int_{\tau_{k-o}}^{\tau_{k+o}} \delta\varphi_s^{\mp}(\omega) P_{\delta}(d\omega), \tag{2.11}$$

where $P_{\delta}(d\omega)$ is a probability evaluation of impulse $\delta\varphi_s^{\mp}$.

The integral of symmetric $\delta$-function $\delta\varphi_s^{\mp}$ between the above time intervals on the border is

$E[\delta\varphi_s^{\mp}] = 1/2 P_{\delta}(\tau_k)$ at $\tau_k = \tau_{k-o}$, or $\tau_k = \tau_{k+o}$. (2.12)

The impulse, produced by deterministic controls (2.2) for each process dimension, is random with probability

$$P_{\delta c}(\tau_k) = 1, k = 1, ..., m \tag{2.13}$$

at each $\tau_k$-locality.

This probability holds a jump-diffusion transition probability in (2.12) (according to [19]), which is conserved during the jump. For each jump, condition (2.4) leads to $a^u(t, \tilde{x}_t) \to \infty, or \sigma(t, \tilde{x}) \to 0$, while both satisfy integral (1.1.9).

*Therefore, each jump increases Markov speed up to infinity within a finite impulse with fixed* $\tau_{k-o}$ *and* $\tau_{k+o}$.

From (2.11)-(2.13) it follows estimation of the EF increment under impulse control (2.2) applying at $t = \tau_k$ in form

$$\Delta S[\tilde{x}_t(\delta u_{\pm}(\tau_k))] = E[\delta\varphi_s^{\mp}] = 1/2 \tag{2.14}$$

which proves (2.3), while delta impulse at $\delta\varphi_s^{\mp} \to \infty$ brings absolute maximum to (2.14) within each $k$ cutoff impulse.

Symmetrical entropy contributions (2.6) at a vicinity of $t = \tau_k$:

$$E[\varphi_s^{t-}] = \Delta S[\tilde{x}_t(u_-(t \le \tau_{k-o}; t > \tau_k))] \tag{2.15a}$$

$$E[\varphi_s^{t+}] = \Delta S[\tilde{x}_t(u_+(t > \tau_k; t \le \tau_{k+o}))] \tag{2.15b}$$

estimate relations

$$\Delta S[\tilde{x}_t(u_-(t \le \tau_{k-o}; t > \tau_k))] = 1/4, u_- = u_-(\tau_k), \tau_k \to \tau_{k-o}, \tag{2.16a}$$

$$\Delta S[\tilde{x}_t(u_+(t > \tau_k; t \le \tau_{k+o}))] = 1/4, u_+ = u_+(t > \tau_k), \tau_k \to \tau_{k+o}, \tag{2.16b}$$

which prove (2.3a,b).

Entropy functional (1.1.10), defined through Radon-Nikodym's probability density measure (1.1.3), holds all properties of the considered cutoff-controllable process, where both $P_{s,x}$ and $\tilde{P}_{s,x}$ are defined. Thus, cutting correlations (2.10) extracts the entropy of hidden process Information which directly measures each $\delta -$cutoff:

$$\Delta I_k[\tilde{x}_t(\delta u(\tau_k))] = \Delta S[\tilde{x}_t(\delta u_{\pm}(\tau_k))] = 1/2. \qquad (2.17)$$

Commonly-known Information measures cannot provide such measurements. According to the definition of the Entropy Functional (1.1.2), it is measured in natural $\ln$, where each Nat equals $\log_2 e \cong 1.44 bits$. Therefore, the EF definition and measurement does not use Shannon's entropy measure. •

*1.3.2.1. Corollaries*

From Proposition 1.2 it follows that:

a. Stepwise control function $u_- = u_-(\tau_k)$, implementing transformation $\tilde{x}_t(\tau_{k-o}) \rightarrow \varsigma_t(\tau_k)$, converts the EF from its minimum at $t \le \tau_{k-o}$ (2.16a) to maximum at $\tau_{k-o} \rightarrow \tau_k$ (2.17);

b. Stepwise control function $u_+ = u_+(\tau_k)$, implementing transformation $\varsigma_t(\tau_k) \rightarrow \tilde{x}_t(\tau_{k+o})$, converts the EF from its maximum at $\tau_{k-o} \rightarrow \tau_k$ (2.17) to minimum at $\tau_k \rightarrow \tau_{k+o}$ (2.16b);

c. Impulse control function $\delta u_{\tau_k}^{\mp}$, implementing transformations $\tilde{x}_t(\tau_{k-o}) \rightarrow \varsigma_t(\tau_k) \rightarrow \tilde{x}_t(\tau_{k+o})$, switches the EF from its minimum to maximum and back from maximum to minimum, while the maximum of the Entropy Functional at a vicinity of $t = \tau_k$ allows the impulse control to deliver *maximal amount* of information (2.17) from these transformations;

d. The cutting correlation between the process's cutoff points (2.10) cuts *functional connections* at these discrete points, which border the Feller kernel measure [11];

e. The relation of that measure to the additive functional in form (1.1.7) allows the EF (1.1.5) to evaluate the *kernel's information*. The jump action (2.2) on a Markov process, associated with "killing its drift," selects the Feller measure of the kernel [20, 21], while the cutoff of EF *provides an Information measure* of the Feller kernel (2.17);

f. Stepwise control $u_- = u_-(\tau_k)$, transferring the EF from $\tau_{k-o} \rightarrow \tau_k$, maximizes (by moment $\tau_k$) the minimal Information increment (brought at $t \rightarrow \tau_{k-o}$), implementing condition

$$\max_{\tau_k} \min_{\tau_{k-o}} \Delta I_k[\tilde{x}_t(\delta u(\tau_k))]; \qquad (2.17a)$$

g. Stepwise control $u_+ = u_+(\tau_k)$, transferring the EF from $\tau_k \rightarrow \tau_{k+o}$, kills the additive functional at stopping moment $\tau_{k+o}$, minimizing the maximal Information increment by the end of this transformation, implementing condition

$$\min_{\tau_{k+o}} \max_{\tau_k} \Delta I_k[\tilde{x}_t(\delta u(\tau_k))]. \qquad (2.17b)$$

Such transformation associated with killing the Markov process at the rate of increment of the related additive functional $d\varphi_s^{ti} / \varphi_s^{ti}$ for each single dimension *i* [6].

Control $u_+ = u_+(\tau_k)$ transfers the rate of a killed Markov process to a process with probability (2.13), which is conserved during the jump, and starts a maximal probable (non-random) process with the eigenvalue of diffusion operator [22]. That process balances the killing at the same rate [23].

The step-wise controls, acting on the dimensions of the multidimensional diffusion process, sequentially stop and start the process, changing the probabilities 0-1 of each impulse. This allows measuring the Information of the Information functional for each dimension of the multidimensional process. The cutting element of the correlation matrix at these moments provides independence of the cutting off fractions, leading to orthogonality of their correlation matrix. •

*The control action on the additive functional with drift function* $a^u = a(x,t,u)$ *is an equivalent of Markov transitional probabilities (1.1.2) transforming random drift function* $a^{\omega} = a(\omega)$ *to drift function* $a^{\bar{\varpi}} = a(\bar{\varpi})$ (Sec.1.1). *Both are equivalents to changing* the EF *measure under the No-Yes actions. The above transformation carries the jump of Markov probabilities densities* $p(\omega)$ *in Sec.1.1, which in the Proposition 2.1 runs* $\delta p_s^{\mp}$ *under the controls.*

*Each observing impulse, acting on the Markov probabilities, virtually cuts the EF until the real control impulse converts the cutting EF entropy to an Information unit. Moreover, as it follows from the Proposition, each jump increases the Markov speed, allowing the observing maximin-minimax impulses to run faster than the initial Markov movement. Compared to any non-impulse observations, that speeds up getting the Information unit from the impulse observation.*

### 1.3.3. **The Impulse Action on the Entropy Integral**

*1.3.3.1. The Entropy Integral Functional for a Single Dimensional Markov Process*

The EF integrant in (1.1.10) is *partially observable* by measuring only the covariation function on the process's trajectories.

For a single dimensional Markov process with drift function $a = c\tilde{x}(t)$ at given nonrandom function $c = c(t)$ and diffusion $\sigma = \sigma(t)$, the Entropy Functional (1.1.10) acquires the form

$$S[\tilde{x}_t / \varsigma_t] = 1/2\int_s^T E[c^2(t)\tilde{x}^2(t)\sigma^{-2}(t)]dt, \qquad (3.1.1)$$

from which, at $\sigma(t)$ and nonrandom function $c(t)$, we get

$$S[\tilde{x}_t / \varsigma_t] = 1/2\int_s^T [c^2(t)\sigma^{-2}(t)E_{s,x}[x^2(t)]dt = 1/2\int_s^T c^2[2b(t)]^{-1}r_s dt \;. \qquad (3.1.2)$$

For the Markov diffusion process, the following relations hold true:

$$2b(t) = \sigma(t)^2 = dr/dt = \dot{r}_t, E_{s,x}[x^2(t)] = r_s \,. \qquad (3.1.3)$$

This allows *identifying* the EF on the observed Markov process $\tilde{x}_t = \tilde{x}(t)$ by measuring the above correlation functions, applying the positive function $u(t) = c^2(t)$, and *representing the* functional (1.1.10) through a regular integral

$$S[\tilde{x}_t / \varsigma_t] = 1/2\int_s^T u(t)A(t,s)dt \qquad (3.1.4)$$

with non-random functions

$A(s,t) = [2b(t)]^{-1}r_s = \dot{r}_t^{-1}r_s$ (3.1.4a) and $u(t) = c^2(t)$ (3.1.4b)

The *n*-dimensional form of the functional integrant (3.1.4a, b) follows directly from the related *n*-dimensional covariations (3.1.3), the dispersion matrix, and applying the *n*-dimensional function $u(t)$.

At given nonrandom function $u(t)$, a regular integral (3.1.4) measures the Entropy Functional of a Markov process at the probability transformation (1.1.2) with additive functional (1.1.7), where the integrant averages the additive functional.

*Proposition 3.1.*

Integral (3.1.4), satisfying variation condition (1.15) at linear function $c^2(t) = u \bullet t = c^2 \bullet t$, forms

$$S[\tilde{x}_t / \varsigma_t] = 1/2\int_s^T u(t)o(t)dt\,, \qquad (3.1.5)$$

where the extreme of function (3.1.5a) holds minimum

$$A(t, s_k^{+o}) = o(s)b_k(s_k^{+o}) / b_k(t) = o(t)\,, \qquad (3.1.5a)$$

which decreases with growing time $t = s_k^{+o} + o(t)$, at $t \to T$ and fixed both $b_k(s_k^{+o})$ and the starting

$$o(s) = A(s, s)\,. \qquad (3.1.5b)$$

Since satisfaction of this variation condition includes transitive transformation of a current distribution to that of Feller kernel, $b_k(t)$ is the transition dispersion at this transformation, which grows with the time of the transformation. •

*Proposition 3.2.*

The Entropy Integral (3.1.5) under impulse control $c^2(t, \tau_k) = \delta u_t(t - \tau_k)$ takes the following information values:

(a) at a switching impulse middle locality $t = \tau_k$ :

$$S[\tilde{x}_t / \varsigma_t]_{t=\tau_k} = 1/2 \text{ Nats}, \qquad (3.1.6)$$

(b) at a switching impulse left locality $t = \tau_k^{-o}$ :

$$S[\tilde{x}_t / \varsigma_t]_{t=\tau_k^{-o}} = 1/4 Nats, \qquad (3.1.6a)$$

(c) at a switching impulse right locality $t = \tau_k^{+o}$ :

$$S[\tilde{x}_t / \varsigma_t]_{t=\tau_k^{+}} = 1/4 Nats. \qquad (3.1.6b)$$

*Proof.* Applying delta-function $c^2(t, \tau_k) = \delta u_t(t - \tau_k)$ to integral

$$\Delta S[\tilde{x}_t / \varsigma_t]|_{\tau_k^{-o}}^{\tau_k^{+o}} = 1/2\int_{\tau_k^{-}}^{\tau_k^{+}} \delta u_t(t - \tau_k)o(t)dt, \tau_k^{-o} < \tau_k < \tau_k^{+o}\,, \qquad (3.1.7)$$

determines functions [24:678-681]:

$$\Delta S[\tilde{x}_t / \varsigma_t]|_{t=\tau_k^{-o}}^{t=\tau_k^{+o}} = \begin{cases} 0, t < \tau_k^{-o} \\ 1/4o(\tau_k^{-o}), t = \tau_k^{-o} \\ 1/4o(\tau_k^{+o}), t = \tau_k^{+o} \\ 1/2o(\tau_k), t = \tau_k, \tau_k^{-o} < \tau_k < \tau_k^{+o} \end{cases} \qquad (3.1.8)$$

Or such a cutoff impulse $o(\tau_k)$ brings amount of the EF on its middle locality $\Delta S[\tilde{x}_t / \varsigma_t]_{t=\tau_k} = 1/2o(\tau_k) = 1/2$ Nats, and on borders of interval $o(\tau_k)$, the EF amounts are $\Delta S[\tilde{x}_t / \varsigma_t]_{t=\tau_k^{-o}} = 1/4o(\tau_k^{-o}) Nats$ and $\Delta S[\tilde{x}_t / \varsigma_t]_{t=\tau_k^{+o}} = 1/4o(\tau_k^{+o}) Nats$ accordingly. •

These results concur with (2.2.3, 2.2.3a,b).

The cutoff delivers an entropy of $3/4 Nats$, transferring $1/4 Nats$ to the right border.

The sum of the impulse localities of the cutting interval $o(\tau_k)$ :

$$\sum_{t=\tau_k^{-o}}^{t=\tau_k^{+o}} \Delta S[\tilde{x}_t / \varsigma_t]_t = 1/4o(\tau_k^{-o}) + 1/2o(\tau_k) + 1/4o(\tau_k^{+o}) = o_k \qquad (3.1.9)$$

evaluates the constant-invariant $1 Nat$ fraction of the EF cutoff on this interval, which the interval encloses.

The invariant cutting fractions follow from the variation condition (1.1.7) imposed on (3.1.4), which leads to (3.1.9).

The impulse 0-1 probability has minimax probability and the related minimax entropy measure: $\Delta S_{P=0} \to -\infty, \Delta S_{P=0} = 0$ that is additive at each virtual–probabilistic EF cut which integrates the virtual observer with entropy measure $S_B$.

## 1.4. INFORMATION PATH FUNCTIONAL

The *Information Path Functional* (IPF) defines the distributed actions of a multidimensional delta-function on the Entropy Functional (1.1.10) through the additive functional for all dimensions:

$$I_{pm} = \delta_m\{S[\tilde{x}_t / \varsigma_t]|\} = 1/2E\{\int_s^T \delta_m[a(t,\tilde{x})^T (2b(t,\tilde{x}))^{-1} a(t,\tilde{x})dt)]\}. \qquad (4.4.1)$$

Relation (4.4.1) sums the discrete information measures $\Delta I_k[\tilde{x}_t(\delta u(\tau_k))]$ along the path of the cutting process intervals (3.1.9). In a limit it leads to

$$I_p = \lim_{m\to\infty} \sum_{k=1}^{m} \Delta I_k[\tilde{x}_t(\delta u(\tau_k))]. \qquad (4.4.2)$$

Formal definition (4.4.1) allows the IPF representation by Furies integral [24] leading to the frequency analysis with Furies series. The IPF is the sum of *extracted* information which approaches theoretical EF measure (1.1.10):

$$I_p = \lim_{m\to\infty} I_{mo} |_s^T = \lim_{m\to\infty} S_{mo} |_s^T \to S[\tilde{x}_t / \varsigma_t]_s^T, \qquad (4.4.3)$$

if all finite time intervals $t_1 - s = o_1, t_2 - t_1 = o_2, ..., t_{k-1} - t_k = o_k, ..., t_m - t_{m-1} = o_m$, at $t_m = T$ satisfy condition

$$(T - s) = \lim_{m\to\infty} \sum_{t=s,m}^{t=T} o_m(t). \qquad (4.4.4)$$

According to (3.1.9), each cutting interval $o(\tau_k)$ encloses an invariant entropy measure which converts to the equivalent information measure $\Delta I_k[\tilde{x}_t(\delta u(\tau_k))]$. The $I_p$ (4.4.3) limits the initially undefined upper time $T$ of the EF integral (1.1.10), which brings direct connection of $I_p$ and $T - s$, where each $o(\tau_k)$ limits the invariant discreet measure (3.1.9). Therefore, at infinite sequence of the integrated time intervals, this sequence limits only the zero discreet measure:

$$\lim_{m\to\infty} o(t_m) \to 0, \qquad (4.4.5)$$

and the sum of such a sequence is finite [24:130, 4.8].

A *random ensemble of the sequence, moving with time (4.4.5), would have infinite velocity at any finite size of its space distribution.*

The additive functional (1.1.3), Entropy Functional (3.14), and relation (4.4.4) *directly connect (1.1.10) with the time $T$ of the process.*

Realization (4.4.1), (4.4.4), (4.4.5) requires applying the impulse controls at each instant $(\tilde{x}, s), (\tilde{x}, s + o(s)), ...$ along the process trajectories with the conditional math expectation (1.1.10).

However, for any *finite* number $m$ of these instants, the *integral* process information, composed from the discreet information measured for the fractions of the process, is not complete.

*The $I_p$ properties:*

1. The IPF measures the information of the cutting process's interstate connections hidden by the state's correlations, which are not covered by the traditional Shannon Information measure.
2. Since each cutting value $\Delta S_k[\tilde{x}_t(\delta u(\tau_k))] = \Delta I_k[\tilde{x}_t(\delta u(\tau_k))]$ maximizes the cutting information on the path intervals, $I_p$ measures total (integral) maximal information on this path.

The cutting control provides equal maxmin-minimax information contributions

$$\max_{\tau_k}\min_{\tau_{k-o}}\Delta I_k[\tilde{x}_t(\delta u(\tau_k))]=\min_{\tau_{k+1}}\max_{\tau_k}\Delta I_k[\tilde{x}_t(\delta u(\tau_k))] \quad (4.4.6)$$

on each path $t_{k-1}\to(\tau_{k-o}\to\tau_k\to\tau_{k+o})\to t_k$ from cutting $t_{k-1}$ to following cut $t_k$ (*Corollaries a-c).*

3. If each $k$ -cutoff "kills" the $m$ dimensional process's correlation at moment $\tau_{k+o}$, then at $m=n$ relations (4.4.1-4.4.6) require infinite process dimensions.

4. At $m=n\to\infty, o_k=t_k-t_{k-1}\to\tau_k$, the process time

$$(T-s)=\lim_{n\to\infty}\sum_{t=s,k=1}^{t=T,k=n}\tau_k \quad (4.4.7)$$

approaches the summary of the discrete time intervals cutting all correlations on the path.

5. Sequential cuts transform the IPF discrete information contributions from each maximum through minimum to next maximal information contributions

$$\max_{\tau_k}\Delta I_k[\tilde{x}_t(\delta u(\tau_k))]\to\min_{\tau_{k+o}}\Delta I_k[\tilde{x}_t(\delta u(\tau_k))]\to\max_{\tau_k+1}\Delta I_k[\tilde{x}_t(\delta u(\tau_{k+1}))], \quad (4.4.8)$$

where each next maximum decreases at the cutoff moments

$$\max_{\tau_k+1}\Delta I_k[\tilde{x}_t(\delta u(\tau_{k+1}))]<\max_{\tau_k}\Delta I_k[\tilde{x}_t(\delta u(\tau_k))]. \quad (4.4.9)$$

Each Dirac delta-function preserves its cutting information (3.1.8).

The information contribution by the final interval $o_m$ at its inner ending moment $\tau_{m+o}$, according to (4.4.2-4.4.4), satisfies

$$\min_{\tau_{m+o}}\Delta I_m[\tilde{x}_t(\delta u(\tau_m))]\to 0, \quad (4.4.10)$$

which limits sum (4.4.3) at $m=n\to\infty$.

6. Since the EF functional $S[\tilde{x}_t/\varsigma_t]_s^T$ limits the growth of $S_{mo}|_s^T$ in (4.4.3), it limits the IPF in (4.4.8), (4.4.9). Hence, the IPF approaches the EF functional during time (4.4.7) at the unlimited increase of the process dimensions.

7. Because upper time $T$ in both the EF integral and IPF functional is limited by (4.4.3), (4.4.4) and (4.4.10), the cutting entropy integral converges in the Path Functional, and both of them are restricted at the unlimited dimension number.

8. For any of these limitations, the EF measure, taken along the process trajectories for time $(T-s)$, limits the maximum of total process information, while IPF extracts the maximum of the process Hidden Information during the same time and brings more information than the traditional Shannon Information measure for multiple states of the process.

9. Information density of cutting impulse

$$I_{ko_k}=I_k[\tilde{x}_t(\delta u(\tau_k))]/\tau_k \quad (4.4.11)$$

grows to absolute maximum at $\tau_k\to 0$:

$$I_{ko_n}\to\infty. \quad (4.4.12)$$

The time transition to each subsequent EF Nat decreases. Each such Nat integrates all preceding Nats concentrating the integral information in the final IPF Nat, which the Feller kernel absorbs.

The final finite integral information impulse approaches the Kronicker 0-1 impulse generated during finite $\tau_{k=n}$, which is the preserving measure (3.3.6).

The invariant probability measure applies to the impulse probes on an observable random process, which holds opposite Yes-No probabilities (as the unit of probability impulse step-function [25] preserves the max-min). •

All integrated information encloses the Feller kernel whose time and energy evaluate the results [20].

Minimal physical time interval limits the light time interval $\delta t_{\tau} \cong 1.33\times 10^{-15}\sec$ defined by the light wavelength $\delta l_m \cong 4\times 10^{-7} m$. This allows us to estimate maximal information density (4.4.14) for 1 Bit:

$$I_{ko_k} \cong \ln 2 / 1.33\times 10^{-15} \cong 5.2116\times 10^{+15} Nat / s. \quad (4.4.13)$$

Or, for each invariant impulse $1Nat$, the maximal density is estimated to be

$$I_{ko_{k1}} \cong 1/1.33\times 10^{-15} \cong 7.5188\times 10^{+15} Nat / s. \quad (4.4.14)$$

The variety of the impulse $\downarrow\uparrow$ physical interactions unites the impulse information model, which the EF-IPF integrate.

## 1.5. The EF-IPF Measures and Their Comparison with Other Entropy Measures

The notion of Information *originates* in the probabilistic Entropy's hidden correlation whose cut-erasure produces a physical Information Bit without the necessity of any physical particles.

The EF presents a potential (virtual) Information Functional of the Markov process until the applied impulse control, carrying the contributions of impulse cutoff entropy, transforming it to an Informational Path Functional (IPF) [26-28].

A Markov random process becomes a source of each Information contribution. The entropy increment of cutting random states delivers Information hidden in the correlation of these states. The cutoff virtually generates transitional probabilities densities $p_s^{t-}(0,1) \rightarrow p_s^{t+}(1,0)$ in (1.1) for current moments $t^-, t^+$ of the process.

These transitional probabilities allow us to *omit* a complete implementation of the initial formalism (Sec.1.1).

The impulses of the Bayes probabilities provide the entropy measure for the uncertainty of each impulse (Sec.1.3.2), which the EF integrates along the process.

The real cutoff converts this entropy to elementary Information, a Bit. The multiple real No-Yes actions convert the EF to the equivalent IPF measure. The IPF integrates the impulse's information in the Information process.

The EF functional entropy measured on trajectories is not covered by the traditional Shannon measure of entropy.

The cutting element of the process's functional correlation matrix at the cutoff moments provides independence of the process cutting off fractions. That leads to the orthogonal correlation matrix for these cutoff fractions.

Cutting the probability of a random ensemble is symmetrical.

The EF connects the observing *a priori* and *a posteriori* probabilities with the observing increment of correlations.

Let us have the ratio of *a posteriori* $P_t(\omega)$ to *a priori* probabilities $P_s(\omega_o)$: $P_t(\omega)/P_s(\omega_o) = p_{s,t}(\omega)$ for elementary events $\omega_o = (s, \tilde{x})$ preceding the current observation of events $\omega = (t, \tilde{x})$ with their probability density satisfying the relation for entropy

$$S_{s,t} = -\int_s^{\tau} \ln(p_{s,t}(\omega)) P_s(\omega_o) d\omega = -\int_s^{\tau} \ln(P_t(\omega)/P_s(\omega_o)) P_s(\omega_o) d\omega = 1/2 \int_s^{\tau} u^2(\bullet) \dot{r}_t / r_s dt, s < t < \tau \quad (5.1)$$

or

$$-\int_s^{\tau} \ln(P_t(\omega)/P_s(\omega_o)) P_s(\omega_o) d\omega = 1/2 \int_s^{\tau} u^2(\bullet) \dot{r}_t / \mathrm{r}_s . \quad (5.2)$$

Here $u^2(\bullet)$ is an initially undefined function which is supposed to equalize these integrals.

Let's take $u^2(t) = \delta(t-\tau)$ as a cutting action equivalent to the delta-function $d\omega = \delta(\omega - \omega_o) d\omega_o$.

Applying that function to the right integral together with the delta-function to the left integral (5.2) leads on the right to a ratio of the increment of posterior correlation to a prior correlation $\dot{r}_t / r_s \cong \delta r_t / r_s \delta t$ (where $\delta t$ is the time interval of the increment of the correlation at the cutoff action). This ratio directly evaluates the logarithmic relations for *a priori* and *a posteriori* probabilities for current random events $\omega_o, \omega$ along trajectory $\tilde{x}(s,t)$:

$\dot{r}_t / r_s = -2P_s(\omega_o)\ln(P_t(\omega)/P_s(\omega_o))$ (5.3)

Vice versa, these probabilities can identify the ratio of these correlations.

*Example.* At $P_t(\omega)/P_s(\omega_o) = 1/4$ it leads to $\dot{r}_t / r_s = -2\times 3.6889 P_s(\omega_o)$ with the negative derivative of *a posteriori* correlation, decreasing at the cut.

The connection of the probability density with additive functional's functions drift and diffusion (Eq.1.1.4) allows using these functions for solving Ito Eqs. [29] in form

$$d\,\mathrm{x}_u = a(t,\mathrm{x}_u)dt + \sigma(t,\mathrm{x}_u)d\xi_{\mathrm{x}}, \sigma(t,\mathrm{x}_u)\sigma^T(t,\mathrm{x}_u) = 2b(t,\mathrm{x}_u)\ . \quad (5.3a)$$

The resulting connections through (5.3a) leads to sequence

$$p(\omega) \to \varphi_s^t(\omega) \to [a(t,\mathrm{x}_u), b(t,\mathrm{x}_u)] \to \mathrm{x}_u\ , \quad (5.4)$$

which determines Markov state $\mathrm{x}_u$ which can control other Markov states along this process instead of $p(\omega)$.

The Markov process becomes self-controlling through the identified self-observing *a priori-a posteriori* probabilities. •

In the Markov diffusion process, each local time interval $\delta t$ is spent in the vicinity of each random $\tilde{x}$ decrease, and the distance from the origin of the random path is reduced. The Levy Walk is such a Markov process [18, Ref. 18:370]. •

Let us compare the EF with the definitions of Boltzmann entropy:

$$S_B = k_B \log W \quad (5.5)$$

where $k_B$ is the Boltzmann constant, and W is the number of accessible microstates of a system having a fixed energy, volume, and number of particles.

And let's compare the EF with Boltzmann H-function called H-entropy:

$$H(t) = \int f(v,t)\log[f(v,t)]\mathrm{dv} \quad (5.6)$$

which is defined for a distribution of velocities $v$: $f(v,t)$ integrating in volume $\mathrm{v}$ of microstates.

The Boltzmann entropy (5. 5) acquires the form of Gibbs entropy

$$S_G = k_B \sum_i p_i \log p_i \quad (5.7)$$

when $p_i$ are the probabilities of finding the system with fixed energy and volume, or fixed energy and number of particles in equilibrium. Both (5.5, 5.7) are time-independent, while (5.6) depends on time and volume of the distributed microstates in an equilibrium.

Entropy (5.7) is directly connected with Shannon entropy for any given probability distribution $p_i$ of states in local equilibrium [30]:

$$H_B = -K \sum_i p_i \log p_i\ . \quad (5.8)$$

It is seen that formulas (5.5, 5.7, 5.8) do not integrate the random process entropy, while the Boltzmann $H_B$ entropy measure integrates only the deterministic speed of the process microstates (molecules) in equilibrium.

None of these directly measures the time of the measurement.

Integral [31] applies for Hamiltonian systems, while the EF has more broad applications including non-Hamiltonian systems.

*The EF integral entropy measure, which averages both speed (drift) and diffusion of the observing Markov process, differs not only from* (5.6) *but from (5.5, 5.7, 5.8) by providing also the integral time of the measured process's non-equilibrium entropy.*

*The IPF directly measures information along a path of interacting impulses, concurrently cutting information.*

Comments 2.1.

Differential entropy defined on set of random variable $X$ with probability density $p(\upsilon)$, as the Shannon form of conditional entropy:

$$H_\upsilon = E_X[-\ln p(\upsilon)] = -\int_X p(\upsilon)\ln p(\upsilon)d\upsilon, \qquad (5.9)$$

is not equal to the EF, and is not invariant under the change of variables and the process.
Thus, it cannot be an entropy measure for arbitrary continuous process, which is unlimited by special requirements [32].
Extending Shannon's discrete entropy to continuous probability distributions requires the preservation of its invariant measure on any measured process. For these reasons, Shannon entropy (type $H_\upsilon$) with its density $p(\upsilon)$ is not suitable for continuous processes and is not sufficient and applicable for the considered random processes according to [33].
The EF allows measuring the integral entropy of the process with randomly correlated states.
The EF measures entropy additional to Shannon's entropy of a finite probabilistic sequence of the process-selected states. This reveals entropy hidden in the correlations that bind the selected states in the random process. Measuring EF along the trajectory of process $\tilde{x}_t$ requires a sequence of transformations changing the drift function of the additive functional, instantaneously converting the probability of the Markov process to the probability of a Brownian process and vice versa:

$$P_{s,x} \to \tilde{P}_{s,x}, \tilde{P}_{s,x} \to P_{s,x}. \qquad (5.10)$$

Such non-stop transformations bring an instant conditional probability's Brownian measure to the EF, implementing the direct and correct entropy measurement of the random process through Radon-Nikodym's probability density measure [3,4] (Sec.1.3.2).
Instant action on the functional drifts requires finding a function that immediately executes such a transformation. This requirement satisfies Dirac's delta-function. Its discrete form is represented through Heaviside's step-up and step-down functions, cutting the process on small intervals. Concurrently cutting each process dimension instantaneously implements these transformations for $n$-dimensional Markov processes under the manifold of cutoff impulses. •

Comments 2.2.

The Boltzmann H-function measures entropy of a physical process, which here models the Markov diffusion of a flow of particles exposed to random displacements at collisions with other particles and molecules.
In probability theory, the diffusion process is a solution to the Ito Eq. (5.3a). The stochastic diffusion included in "Ito's Jump Diffusion" has many physical applications. (In Sec. 2, we apply it to an impulse random jump.)
The EF functional integrates these probabilities along the trajectories of the Markov diffusion process, allowing the EF-IPF to integrate both probabilistic and physical properties of this process.
The physical Markov diffusion involves diffusion kinetics studied in [34].
The initial results of proposed IPF with the conditions of discrete interacting actions and time-space intervals are in [27].•

## 1.6. Imposing the Minimax Law. Invariant Impulse Logic. The Measure of Probabilistic and Information Causations

Within the Markov diffusion process (Sec.3.2), the impulse step-down cut generates maximal information, while the step-up action delivers minimal information from the impulse cut to the next impulse step-down.
Within each Markov impulse, the cutting action maximizes the cutting entropy of the observing impulse, while the reaction, carrying the minimum of the cutting maximum, minimizes the following Markov impulse entropy.
The impulse's maximal cutting No action minimizes the absolute entropy that conveys Yes action (increasing its probability), which leads to a maxmin of relational entropy between the impulse actions which transfer the probabilities.
Extracting the maximum of minimal impulse information and transferring minimal entropy between impulses imposes the maxmin-minimax principle of converting process entropy to Information.
*As soon as the initial impulse 0-1 actions are involved, the minimax principle is imposed.*

The variation problem, formulating for this principle and its solution [12], brings the invariant entropy increment of each discrete impulse. This preserves its probability measure and synchronizes an adjoint local time measure for n-process dimensions in an absolute time scale.

In physical terms, the sequence of opposite interactive actions models reversible micro-fluctuations produced within an observable irreversible macroprocess (like push-pull actions of a piston moving gas in a cylinder). A simpler example: When a rubber ball hits the ground, the energy of this interaction partially dissipates, increasing the interaction's total entropy The ball's subsequent reverse movement holds less entropy, leading to the max-min entropy of the bouncing ball. If a small amount of energy is added, compensating for the interactive dissipation, the bouncing will continue. The maximin-minimax principle does not contradict the Second Law of Thermodynamics.

Observing intervals between impulses are an imaginary-potential Information, since no real double controls apply within these intervals. The minimized increments of the cutting Entropy Functional between the cutoff intervals allow the prediction of each following cutoff with maximal conditional probability. Under this principle, the observing sequence of the functional *a priori-a posteriori* probabilities grows, providing Bayesian entropy that *measures* the probabilistic causality. When the growing posterior probabilities of the process approach maximum, *probabilistic* causality is transformed to *physical* causality. The observer logic depends on the sending sequence of probing impulses requested in observation. The multiple symbols $\downarrow\uparrow$ hold the inner probabilistic logic that integrates the EF-IPF measures.

Each triad in the field, which measures independent sets of the events, starts an independent observation by the defining initial conditions of the observing process. The EF-IPF of observing processes integrates the logic of the multidimensional observations. The sum of cutting Information contributions, extracted from the EF, approaches its theoretical measure (1.1.10) which evaluates the upper limit of the sum (4.4.6) of the observing logic.

Comments 2.3.

A.N. Kolmororov [97] proposed a fundamental solution for evolution of probability distribution at the Markov character of action, which in modern terms is called "time δ-correlated acceleration" [98]. In this approximation of processes, the correlation time of random action is much smaller than the reaction time. Or in the Markov diffusion, the correlated acceleration exists which is asymmetrical. It agrees with maxmin principle on a δ-cutting correlation of the impulse with action and reaction, where the time of cutting action is smaller than that for time reaction. That leads to increasing correlation along the impulses of observing process. •

Comments 2.4:

Results [36] experimentally confirm the biological existence of the maxmin principle for Yes-No (0-1) or (1-0) impulse•.

## II. A **MICROPROCESS WITHIN THE PATH**

## 2. EMERGENCE OF THE SPACE-TME OBSERVER, CONSTRAINTS, AND A MICROPROCESS

### 2.1. **The evaluation of impulses in the interactive observing process**

#### 2.1.1. *Discrete control action on the Entropy Functional*

Let us find a class of step-down $u_-^t = u_-(\tau_k^{-o})$ and step-up $u_+^t = u_+(\tau_k^{+o})$ functions acting on discrete interval $o(\tau_k) = \tau_k^{+o} - \tau_k^{-o}$ ,which will preserve the Markov diffusion process' additive and multiplicative functions within the cutting process of each impulse.

(These functions are controls implementing the interactive cutting actions required to reveal the Bit discrete impulse.)

Lemma 1.1.

1. Opposite discrete functions $u_-^t$ and $u_+^t$ in form

$$u_-(\tau_k^{-o}) = \downarrow_{\tau_k^{-o}} \bar{u}_-, u_+(\tau_k^{+o}) = \uparrow_{\tau_k^{+o}} \bar{u}_+ \qquad (1.1)$$

satisfy conditions of additivity

$$[u_+^t - u_-^t] = U_a \tag{1.1A}$$

and multiplicativity

$$[u_+^t \times u_-^t] = U_m \tag{1.1B}$$

at real

$$\mathrm{U}_a = U_m = c^2, \tag{1.1C}$$

where $U_a$ is measurable function of additive events, $U_m$ is measurable function of multiplicative events.

Example 1. The orthogonal functions $u_-^t = \bar{u}_- \times \xrightarrow{\tau_k^{-o}}$ and $\uparrow_{\tau_k^{+o}} \bar{u}_+$, where instance-jump $\downarrow_{\tau_k^{-o}}$ has time interval $\bar{u}_-$ and instance jump $\uparrow_{\tau_k^{+o}} \bar{u}_+$ has high $\bar{u}_+$, represent complex functions which satisfy conditions (1.1A), (1.1B) in forms

$$u_+ - u_- = (j-1) - (j+1) = -2,\ u_+ \times u_- = (j-1) \times (j+1) = (j^2 - 1) = -2,$$

and preserve negative real $c^2 < 0$ in (1.1C).

Example 2. The opposite complex functions

$$u_t(-u_\pm^{t1}) = u_t(u_\pm^{t2}), u_\pm^{t2} = [u_+ = (j+1), u_- = (j-1)], \tag{1.2b}$$

represent orthogonal functions $u_+^t = \bar{u}_+ \times \xrightarrow{\tau_k^{-o}}$ and $u_- = \uparrow_{\tau_k^{+o}} \bar{u}_-$ where instance jump $\downarrow_{\tau_k^{-o}}$ has time interval $\bar{u}_+$ and instance jump $\uparrow_{\tau_k^{+o}}$ has high $\bar{u}_-$ .

The complex functions (1.2b) satisfy (1.1A), (1.1B) in form

$$u_+ - u_- = (j+1) - (j-1) = 2, u_+ \times u_- = (j+1) \times (j-1) = (j^2 - 1) = -2$$

changing positive $c^2$ on negative one, and therefore not satisfying (1.1C), violating this Markov property.

Example 3.If function (1.2a), (1.2b) apply on borders of impulse interval $o(\tau_k) = (\tau_k^{-o}, \tau_k^{+o})$ in forms

$$u_-^{t1} = u_-(\tau_k^{-o}), u_+^{t1} = u_+(\tau_k^{-o}) \text{ and } u_-^{t2} = u_-(\tau_k^{+o}), u_+^{t2} = u_+(\tau_k^{+o}), \tag{1.2d}$$

$$\text{at } u_-^{t1} u_+^{t1} = c^2(\tau_k^{-o}),\ u_-^{t2} u_+^{t2} = c^2(\tau_k^{+o}),\ t = \tau_k^{+o}, \tag{1.2e}$$

then $c^2(\tau_k^{-o}), c^2(\tau_k^{+o})$ are discrete functions (1.1a), (1.2e) switching on interval $\Delta_\tau = \tau_k^{+o} - \tau_k^{-o}$ in form $c^2(\tau_k^{-o}) = \downarrow_{\tau_k^{-o}} c^2, c^2(\tau_k^{+o}) = \uparrow_{\tau_k^{+o}} \pm c^2$.

It follows that only at beginning of this time interval at $t = \tau_k^{-o}$, all Markov properties (1.1A,B,C) satisfy, while by the end of this time interval at at $t = \tau_k^{+o}$ the process satisfies only (1.1A,B). •

Example 4. At equal absolute values of actions $|u_+^t| = |u_-^t|$, imaginary functions

$$u_+^t = j\sqrt{2}, u_-^t = -j\sqrt{2} \tag{1.2c}$$

satisfy only multiplicative part $U_m = u_+^t \times u_-^t = -j^2 2 = 2$ at $c^2 > 0$ while the impulse additive measure holds $\mathrm{U}_a = 0$. •

Corollary 1.1.

1. The step-functions $u_-(\tau_k^{-o})=-1_{\tau_k^{-o}}\bar{u}_-, u_+(\tau_k^{+o})=+1_{\tau_k^{+o}}\bar{u}_+$ forming impulse $\delta u_t=[u_-(\tau_k^{-o})-u_+(\tau_k^{+o})]$ on interval $\Delta_\tau=\tau_k^{+o}-\tau_k^{-o}$ holds discrete function

$$\delta^o u_{t=\Delta_\tau}=[u_-(\tau_k^{-o})-u_+(\tau_k^{+o})]/(\tau_k^{+o}-\tau_k^{-o}) \quad (1.3a)$$

2. Discrete functions on $\Delta=(s_k^{+o},\tau_k^{-o})$:

$$u_+(s_k^{+o})=+1_{s_k^{+o}}\bar{u}_+, u_-(\tau_k^{-o})=-1_{\tau k^{+o}}\bar{u}_- \quad (1.3b)$$

are multiplicative:

$(u_-(\tau_k^{-o})-u_+(s_k^{+o}))\times(u_-(\tau_k^{-o})-u_+(s_k^{+o}))=[u_-(\tau_k^{-o})-u_+(s_k^{+o})]^2$,

which for functions $\bar{u}_+=j\bar{u}, \bar{u}_-=-j\bar{u}, \bar{u}\neq 0$ holds

$$[u_-(\tau_k^{-o})-u_+(s_k^{+o})]^2=-(j\bar{u})^2[-1_{\tau_k^{-o}}-1_{s_k^{+o}}]^2>0. \bullet \quad (1.3d)$$

Let us find discrete analog of the EF integral increments under *discrete* function (1.3a) in a form of delta-function on $\delta_o$:

$$\delta[u_-(\tau_k^{-o}),u_+(\tau_k^{+o})], u_-(\tau_k^{-o})=-1_{\tau_k^{-o}}\bar{u}_-, u_+(\tau_k^{+o})=+1_{\tau_k^{+o}}\bar{u}_+. \quad (1.3e)$$

Proposition 1.2.

1. Applying discrete delta-function (1.3e) to the EF integral in form (1.3.1.5) leads to

$$\Delta S[\tilde{x}_t/\varsigma_t]|_{t=\tau_k^{-o}}^{t=\tau_k^{+o}}=\begin{cases}0, t<\tau_k^{-o}\\ 1/4u_-(\tau_k^{-o})o(\tau_k^{-o})/\tau_k^{-o}, t=\tau_k^{-o}, 1/4\downarrow 1_{\tau_k^{-o}}\bar{u}_{ko}\\ 1/2(u_-(\tau_k^{-o})-u_+(\tau_k^{+o}))o(\tau_k)/(\tau_k^{+o}-\tau_k^{-o}), t=\tau_k, \tau_k^{-o}<\tau_k<\tau_k^{+o}, 1/2(\downarrow 1_{\tau_k^{-o}}-\uparrow 1_{\tau_k^{+o}})\bar{u}_{km}\\ 1/4u_+(\tau_k^{+o})o(\tau_k^{+o})/\tau_k^{+o}, t=\tau_k^{+o}, 1/4\uparrow 1_{\tau_k^{+o}}\bar{u}_{k1}\end{cases} \quad (1.4)$$

where

$$\bar{u}_{ko}=\bar{u}_-\times o(\tau_k^{-o})/\tau_k^{-o}, \bar{u}_{km}=(\bar{u}_+-\bar{u}_-)\times o(\tau_k)/(\tau_k^{+o}-\tau_k^{-o}), \bar{u}_{k1}=\bar{u}_+\times o(\tau_k^{+o})/\tau_k^{+o},$$
$$\bar{u}_{km}=1/2(\bar{u}_+-\bar{u}_-)=0.75, o(\tau_k)=\tau_k^{+o}-\tau_k^{-o}, o(\tau_k^{-o})/\tau_k^{-o}=0.5, o(\tau_k^{+o})/\tau_k^{+o}=0.1875, \quad (1.5)$$

and $|\bar{u}_-\times\bar{u}_+|=|1/2\times 2|=|\bar{u}_k|=|1|_k$ is multiplicative measure of impulse $(\downarrow 1_{\tau_k^{-o}}-\uparrow 1_{\tau_k^{+o}})\bar{u}_k$.

Measuring middle interval in (1.4), (1.5) by single impulse information unit $\bar{u}_k=|1|_k$, determines finite size of the impulse unit parameters $\bar{u}_{ko},\bar{u}_k,\bar{u}_{k1}$ in (1.5), which estimate value $\bar{u}_{km}$ on the unit border:

$$\bar{u}_{ko}=0.25=1/3\bar{u}_{km}, \bar{u}_{k1}=2\times 0.1875=0.375=0.5\bar{u}_{km}\ .\bullet \quad (1.6)$$

Proofs follow from Proposition 1.3 below. •

Let us introduce an entropy unit impulse $\bar{u}_s=|1|_s$ with moments $(s_k^{-o},s_k^{o},s_k^{+o})$ prior to impulse $\bar{u}_k=|1|_k$, which measures interval of impulse entropy $\bar{u}_{sm}$.

Then we will find increment of entropy $\Delta S[\tilde{x}_t / \varsigma_t]|_{s_k^+}^{\tau_k^{-o}}$ on a border of impulse $\bar{u}_k$ at prior $\Delta_{\tau s+} = \delta_{sk\pm} = (s_k^{+0} - \delta_k^{\tau-})$ and posterior $\Delta_{\tau s-} = \delta_{sk\mp} = (\delta_k^{\tau-} - \delta_k^{\tau+})$ moments under the impulse functions with unit $\bar{u}_s = |1|_s$:

$$\delta^o u_{\tau=(s_k^{+o}-\delta_k^{\tau-})} = (u_+(s_k^{+o}) - u_-(\delta_k^{\tau-}))(s_k^{+o} - \delta_k^{\tau-})^{-1} = \uparrow 1_{s_k^{+o}} \bar{u} - \downarrow 1_{\delta_k^{\tau-}} \bar{u} = [\uparrow 1_{s_k^{+o}} - \downarrow 1_{\delta_k^{\tau-}}]\bar{u}\,, \quad (1.7)$$

$$\delta^o u_{\tau=(\delta_k^{\tau-}-\delta_k^{\tau+})} = (u_-(\delta_k^{\tau-}) - u_+(\delta_k^{\tau+}))(\delta_k^{\tau-} - \delta_k^{\tau+})^{-1} = \downarrow 1_{\delta_k^{\tau-}} \bar{u} - \uparrow 1_{\delta_k^{\tau+}} \bar{u} = [\downarrow 1_{\delta_k^{\tau-}} - \uparrow 1_{\delta_k^{\tau+}}]\bar{u}\,, \quad (1.8)$$

$$\delta^o u_{\tau=(\delta_k^{\tau+}-\tau_k^{-o})} = (u_+(\delta_k^{\tau+}) - u_-(\tau_k^{-o}))(\delta_k^{\tau+} - \tau_k^{-o})^{-1} = \uparrow 1_{\delta_k^{\tau+}} \bar{u} - \downarrow 1_{\tau_k^{-o}} \bar{u} = [\uparrow 1_{\delta_k^{\tau+}} - \downarrow 1_{\tau_k^{-o}}]\bar{u}\,. \quad (1.9)$$

Here $\bar{u}$ evaluates each impulse interval, which according to the optimal principle is an invariant.
Since the EF is additive functional, applying functions (1.7)-(1.9) leads to additive discrete sum of its' increments:

$$\Delta S[\tilde{x}_t / \varsigma_t]|_{s_k^+}^{\tau_k^{-o}} = \Delta S[\tilde{x}_t / \varsigma_t]|_{s_k^+}^{\delta_k^{\tau-}} + \Delta S[\tilde{x}_t / \varsigma_t]|_{\delta_k^{\tau-}}^{\delta_k^{\tau+}} + \Delta S[\tilde{x}_t / \varsigma_t]|_{\delta_k^{\tau+}}^{\tau_k^{-o}} \quad (1.10)$$

along time interval

$$\Delta_{\tau sk\pm} = s_k^{+o} - \delta_k^{\tau-} + \delta_k^{\tau-} - \delta_k^{\tau+} + \delta_k^{\tau+} - \tau_k^{-o} = s_k^{+o} - \tau_k^{-o} = \Delta_{\tau s}\,. \quad (1.10a)$$

<u>Proposition 1.3.</u>

A. The increments of the entropy functional (1.10) collected on intervals (1.10a), under functions (1.7)-(1.9), bring the following entropy contributions:

$$\Delta S[\tilde{x}_t / \varsigma_t]|_{s_k^+}^{\delta_k^{\tau-}} = 1/2(u_+(s_k^{+o}) - u_-(\delta_k^{\tau-}))o(s_k^{+o} - \delta_k^{\tau-}))(s_k^{+o} - \delta_k^{\tau-})^{-1} = 1/2[\uparrow 1_{s_k^{+o}} - \downarrow 1_{\delta_k^{\tau-}}]\bar{u}_{ks} \quad (1.11)$$

at interval

$$\bar{u}_{ks} = \bar{u}(o(s_k^{+o} - \delta_k^{\tau-})(s_k^{+o} - \delta_k^{\tau-})^{-1}; \quad (1.11a)$$

$$\Delta S[\tilde{x}_t / \varsigma_t]|_{\delta_k^{\tau-}}^{\delta_k^{\tau+}} = 1/2(u_-(\delta_k^{\tau-}) - u_+(\delta_k^{\tau+}))o(\delta_k^{\tau-} - \delta_k^{\tau+}))(\delta_k^{\tau-} - \delta_k^{\tau+})^{-1} = 1/2[\downarrow 1_{\delta_k^{\tau-}} - \uparrow 1_{\delta_k^{\tau+}}]\bar{u}_{k\delta s}, \quad (1.12)$$

at interval

$$\bar{u}_{k\delta s} = \bar{u} \times (o(\delta_k^{\tau-} - \delta_k^{\tau+}))(\delta_k^{\tau-} - \delta_k^{\tau+})^{-1} \quad (1.12a)$$

and

$$\Delta S[\tilde{x}_t / \varsigma_t]|_{\delta_k^{\tau+}}^{\tau_k^{-o}} = 1/2(u_+(\delta_k^{\tau+}) - u_-(\tau_k^{-o}))o(\delta_k^{\tau+} - \tau_k^{-o})(\delta_k^{\tau+} - \tau_k^{-o})^{-1} = 1/2[\uparrow 1_{\delta_k^{\tau+}} - \downarrow 1_{\tau_k^{-o}}]\bar{u}_{k\delta}\,, \quad (1.13)$$

under function

$$[\uparrow 1_{\delta_k^{\tau+}} - \downarrow 1_{\tau_k^{-o}}]\bar{u}_{k\delta} = [\uparrow 1_{\delta_k^{\tau+}} + \uparrow 1_{\tau_k^{-o}}]\bar{u}_{k\delta}\,. \quad (1.13a)$$

Here each impulse interval acquires a specific entropy measure at known $\bar{u}_{k\delta}, \bar{u}_{k\delta s}, \bar{u}_{ks}$.

Let's find them. The $\bar{u}_{k\delta}$ at on the impulse invariant interval $\bar{u}$ acquires form

$$\bar{u}_{k\delta} = \bar{u} \times (o(\delta_k^{\tau+} - \tau_k^{-o}))(\delta_k^{\tau+} - \tau_k^{-o})^{-1} = \bar{u} \times o(\delta_k^{\tau+})(\delta_k^{\tau+} - \tau_k^{-o})^{-1} + \bar{u} \times o(\tau_k^{-o})(\tau_k^{-o})^{-1}(\delta_k^{\tau+} - \tau_k^{-o})^{-1}\tau_k^{-o} \quad (1.14)$$

Relation (1.14) leads to the impulse interval

$$\bar{u}_{k\delta} = \bar{u}_{k\delta o} + \bar{u}_{k\delta 1} \quad (1.14a)$$

with its parts

$$\bar{u}_{k\delta o} = \bar{u} \times (o(\delta_k^{\tau+}))(\delta_k^{\tau+} - \tau_k^{-o})^{-1})\,,\ \bar{u}_{k\delta 1} = \bar{u}_{ko1} \times \bar{u}_{ko2}, \quad (1.14b)$$

$$\bar{u}_{ko1}=\bar{u}\times(o(\tau_k^{-o}))(\tau_k^{-o})^{-1},\ \bar{u}_{ko2}=\bar{u}^{-1}\times\tau_k^{-o}(\delta_k^{\tau+}-\tau_k^{-o})^{-1}\ . \qquad (1.14c)$$

B. Intervals $\bar{u}_{ko1}$ and $\bar{u}_{ko2}$ are multiplicative parts of impulse step-up interval $\bar{u}_{k\delta 1}$, which satisfies relations

$$\bar{u}_{k\delta o}=\bar{u}_{k\delta 1}=1/2\bar{u}_{k\delta},\ \bar{u}_{k\delta 1}=\bar{u}_{ko1}\ , \qquad (1.15)$$

where invariant impulse $|\bar{u}_{k\delta}|=|1|_s$, acting on time interval $\delta_k^{\tau+}=2\tau_k^{-o}$, measures

$$\bar{u}_{k\delta}=\bar{u}_{ks} \text{ at} |\bar{u}_{k\delta 1}|=1/2\bar{u}_{sm}, \qquad (1.15a)$$

and relative time intervals of $\bar{u}_{ko}$ and $\bar{u}_{k1}$ accordingly are

$$o(\tau_k^{-o})(\tau_k^{-o})^{-1}=0.5,\ 1/4\bar{u}_{km}=o(\tau_k^{+o})(\tau_k^{+o})^{-1}=0.187. \qquad (1.15b)$$

Step-controls of impulse $\bar{u}_{k\delta}$ apply on two equal time intervals:

$$(\delta_k^{\tau+}-\tau_k^{-o})=\delta_k^{\tau+}/2 \ (1.16a) \qquad \text{and} \qquad \tau_k^{-o}=\delta_k^{\tau+}/2. \qquad (1.16b)$$

On first (1.16a), its step-up part $[\uparrow 1_{\delta_k^{\tau+}}]$ captures entropy increment

$$\Delta S[\tilde{x}_t/\varsigma_t]|_{\delta_k^{\tau+}}^{\tau_k^{-o}}=1/2[\uparrow 1_{\delta_k^{\tau+}}]\bar{u}_{-}=1/8[\uparrow 1_{\delta_k^{\tau+}}], \qquad (1.16)$$

on second (1.16b), its step-down multiplicative part in (1.14b) at $\bar{u}_{ko2}=\bar{u}^{-1}$ transfers entropy (1.16) to starting impulse action $[\downarrow 1_{\tau_k^{-o}}]$ which cut is within impulse (1.4) at $\bar{u}_{ko1}=1/2\bar{u}_{ko}$;

where $\bar{u}_{k\delta 1}$ in (1.14b) multiplies

$$\bar{u}[\uparrow 1_{\delta_k^{\tau+}}]\delta_k^{\tau+}/2\times\bar{u}^{-1}[\downarrow 1_{\tau_k^{-o}}]\tau_k^{-o}. \qquad (1.16c)$$

Both equal time intervals in (1.16b) are on the impulse border where the opposite inverse entropy increments are orthogonal.

C. The applied *extreme* solution (Proposition I.2.1.), decreasing the time impulse intervals, brings minimal increment (1.10a) and

(a)-persistence continuation a sequence of the process impulses;

(b)- balance condition for the entropy contributions;

(c)-each impulse invariant unit $\bar{u}_k=|1|_k$, supplied by entropy unit $\bar{u}_s=|1|_s$, *triples* information *increasing information density* in each following information unit. •

Proofs.

The additive sum of entropy increments under invariant impulses (1.7-1.9) satisfies balance condition:

$$\Delta S[\tilde{x}_t/\varsigma_t]|_{s_k^+}^{\tau_k^{-o}}=\Delta S[\tilde{x}_t/\varsigma_t]|_{s_k^+}^{\delta_k^{\tau-}}+\Delta S[\tilde{x}_t/\varsigma_t]|_{\delta_k^{\tau-}}^{\delta_k^{\tau+}}+\Delta S[\tilde{x}_t/\varsigma_t]|_{\delta_k^{\tau+}}^{\tau_k^{-o}}=$$
$$1/2[\uparrow 1_{s_k^{+o}}-\downarrow 1_{\delta_k^{\tau-}}]\bar{u}_{ks}+1/2[\uparrow 1_{\delta_k^{\tau-}}-\uparrow 1_{\delta_k^{\tau+}}]\bar{u}_{k\delta s}+1/2\uparrow 1_{\delta_k^{\tau+}}\bar{u}_{k\delta o}-1/2\downarrow 1_{|\tau|_k^{-o}}\bar{u}_{k\delta 1}=0\ , \qquad (1.17)$$

where action $1/2\downarrow 1_{|\tau|_k^{-o}}\bar{u}_{k\delta 1}\Rightarrow 1/4\downarrow 1_{|\tau|_k^{-o}}\bar{u}_{ko}$ transfers entropy increment $\Delta S[\tilde{x}_t/\varsigma_t](\tau_k^{-o})=1/4\downarrow 1_{|\tau|_k^{-o}}\bar{u}_{ko}$ on discrete locality $|\tau|_k^{-o}$ by step-down action $\downarrow 1_{|\tau|_k^{-o}}\bar{u}_{k\delta 1}$.

Fulfillment the relations

$$[\uparrow 1_{s_k^{+o}} \bar{u}_{ks} - \downarrow 1_{\delta_k^{\tau-}} \bar{u}_{ks} + \uparrow 1_{\delta_k^{\tau-}} \bar{u}_{k\delta s} - \uparrow 1_{\delta_k^{\tau+}} \bar{u}_{k\delta s} + \uparrow 1_{\delta_k^{\tau+}} \bar{u}_{k\delta o} - \downarrow 1_{|\tau|_k^{-o}} \bar{u}_{k\delta 1}] = 0$$

$$[\uparrow 1_{s_k^{+o}} \bar{u}_{ks} + \uparrow 1_{\delta_k^{\tau-}} [\bar{u}_{k\delta s} - \bar{u}_{ks}] + \uparrow 1_{\delta_k^{\tau+}} [\bar{u}_{k\delta o} - \bar{u}_{k\delta s}] = \downarrow 1_{|\tau|_k^{-o}} \bar{u}_{k\delta 1}, \downarrow 1_{|\tau|_k^{-o}} \bar{u}_{k\delta 1} = -1/2 \downarrow 1_{|\tau|_k^{-o}} \bar{u}_{ko}$$

leads to sum of the impulse intervals:

$\bar{u}_{ks} - \bar{u}_{ks} + \bar{u}_{k\delta s} + \bar{u}_{k\delta s} - \bar{u}_{k\delta s} + \bar{u}_{k\delta o} - \bar{u}_{k\delta 1} = 0$, and $\bar{u}_{k\delta 1} = -1/2\bar{u}_{ko}$,

or to

$$\bar{u}_{k\delta o} = \bar{u}_{k\delta 1} \ . \qquad (1.17a)$$

Impulse $[\uparrow 1_{\delta_k^{\tau+}} \bar{u}_{k\delta o} - \downarrow 1_{|\tau|_k^{-o}} \bar{u}_{k\delta 1}] = [\uparrow 1_{\delta_k^{\tau+}} + \uparrow 1_{|\tau|_k^{-o}}]\bar{u}_{k\delta}$ contains intervals $\bar{u}_{k\delta} = \bar{u}_{k\delta o} + \bar{u}_{k\delta 1}$,

where from (1.9), (1.13a) it follows $\bar{u}_{k\delta} = \bar{u}$ , and (1.17a) leads to

$$\bar{u}_{k\delta o} = \bar{u}_{k\delta 1} = 1/2\bar{u} \ . \qquad (1.17b)$$

Interval

$$\bar{u}_{k\delta} = \bar{u} \times [(o(\delta_k^{\tau+}))(\delta_k^{\tau+} - \tau_k^{-o})^{-1}) + (o(\tau_k^{-o}))(\tau_k^{-o})^{-1}(\delta_k^{\tau+} - \tau_k^{-o})^{-1}\tau_k^{-o}] \qquad (1.17c)$$

consists of $\bar{u}_{k\delta}$ components:

$$\bar{u}_{k\delta o} = \bar{u} \times (o(\delta_k^{\tau+}))(\delta_k^{\tau+} - \tau_k^{-o})^{-1}) \text{ and } \bar{u}_{k\delta 1} = \bar{u}_{ko1} \times \bar{u}_{ko2} / \bar{u} \ , \qquad (1.17d)$$

where

$$\bar{u}_{ko1} = \bar{u} \times (o(\tau_k^{-o}))(\tau_k^{-o})^{-1}, \ \bar{u}_{ko2} = \bar{u}^{-1} \times \tau_k^{-o}(\delta_k^{\tau+} - \tau_k^{-o})^{-1}.$$

Intervals $\bar{u}_{ko1}$ and $[\bar{u}_{ko2} / \bar{u}]$ are multiplicative parts of impulse interval $\bar{u}_{k\delta 1}$ covered by starting interval $|\tau|_k^{-o}$.

From (1.17b) and relations (1.17d) it follows

$$\bar{u}_{k\delta o} = \bar{u} \times (o(\delta_k^{\tau+}))(\delta_k^{\tau+} - \tau_k^{-o})^{-1}) = 1/2\bar{u} \ ,$$

$$(o(\delta_k^{\tau+}))(\delta_k^{\tau+} - \tau_k^{-o})^{-1}) = 1/2 \qquad (1.18)$$

and

$$\bar{u}_{ko1} = \bar{u} \times (o(\tau_k^{-o}))(\tau_k^{-o})^{-1} = 1/2\bar{u} \ . \qquad (1.18a)$$

That leads to

$$(o(\tau_k^{-o}))(\tau_k^{-o})^{-1} = 1/2 \ , \qquad (1.18b)$$

and from (1.18) to relation

$$(\delta_k^{\tau+} - \tau_k^{-o})^{-1} = (\tau_k^{-o})^{-1}, \delta_k^{\tau+} - \tau_k^{-o} = \tau_k^{-o}, \ \tau_k^{-o}(\delta_k^{\tau+} - \tau_k^{-o})^{-1} = 1, \qquad (1.18c)$$

then to

$$\tau_k^{-o} = 1/2\delta_k^{\tau+}. \qquad (1.18d)$$

From (1.18c) it follows

$$\bar{u}_{ko2} = \bar{u}^{-1}. \qquad (1.18f)$$

Applying the sequence of Eqs (1.7-1.9), at fixed invariant $\bar{u}$ , leads to

$$\bar{u} = u_+(s_k^{+o}) - u_-(\tau_k^{-o}) \ , \qquad (1.19)$$

$$\bar{u} = u_+(\delta_k^{\tau+}) - u_-(\tau_k^{-o}) \qquad (1.19a)$$

at $u_+(\delta_k^{\tau+}) - u_-(\tau_k^{-o}) = \bar{u}_{kb}$ ,

which brings invariant $|\bar{u}_s|=|1|_s$ to both impulses (1.19) and (1.19a).
Relation
$u_+(s_k^{+o})+u_-(\tau_k^{-o})=2[u_-(\delta_k^{\tau-})+(u_+(\delta_k^{\tau+})]=0$
following from the sequence of Eqs (1.7-1.9) leads to

$$u_-(\delta_k^{\tau-})=-u_+(\delta_k^{\tau+}), \qquad (1.19b)$$

reversing and mutual neutralizing these actions on related moments $\delta_k^{\tau-}\cong\delta_k^{\tau+}$ .

Impulse interval $\bar{u}_{k\delta}$, with $\bar{u}_{k\delta o}$ and $\bar{u}_{k\delta 1}$, starts interval of applying step-down action $o(\tau_k^{-o})(\tau_k^{-o})^{-1}=0.5$ in (1.4) at $\bar{u}_{k\delta 1}=\bar{u}_{k\delta o}=1/2\bar{u}_{k\delta}$.

Invariant impulse $|\bar{u}_s|=|1|_s$,consisting of two step-actions$[\uparrow 1_{\delta_k^{\tau+}}\downarrow 1_{|\tau|_k^{-o}}]\bar{u}_{k\delta}$, measures intervals
$\bar{u}_{kb}=\bar{u}_{sm}=\bar{u}_s$
at

$$\bar{u}_{k\delta 1}=1/2\bar{u}_{sm}. \qquad (1.19c)$$

At conditions (1.18c, d), limiting time-jump in (1.13a), step-actions of impulse $\bar{u}_{k\delta}$ apply on two equal time intervals following from (1.19c). On the first interval
$(\delta_k^{\tau+}-\tau_k^{-o})=\delta_k^{\tau+}/2$
step-up part of $\bar{u}_{k\delta}$ -action $[\uparrow 1_{\delta_k^{\tau+}}]$ captures entropy increment

$$\Delta S[\tilde{x}_t/\varsigma_t]|_{\delta_k^{\tau+}}^{\tau_k^{-o}}=1/2[\uparrow 1_{\delta_k^{\tau+}}]\bar{u}_-=1/8[\uparrow 1_{\delta_k^{\tau+}}]. \qquad (1.20)$$

On the second interval $\tau_k^{-o}=\delta_k^{\tau+}/2$, the captured entropy (1.20) through the step-down multiplicative part (1.17c,d) delivers to the cutting action $\bar{u}_{ko}=\bar{u}_-\times o(\tau_k^{-o})(\tau_k^{-o})^{-1}$ equal contributions

$$\Delta S[\tilde{x}_t/\varsigma_t]|_{\delta_k^{\tau+}}^{\tau_k^{-o}}=1/4[\downarrow 1_{\tau_k^{-o}}]\bar{u}_-=1/8[\downarrow 1_{\tau_k^{-o}}]. \qquad (1.20a)$$

The control action $[\downarrow 1_{\tau_k^{-o}}]$ at $\bar{u}_-=0.5$ cuts external entropy of correlation in impulse (1.4) at $\bar{u}_{ko1}=1/2\bar{u}_{ko}$.

<u>Comment 1.1</u>. Action $[\uparrow 1_{\delta_k^{\tau+}}]$ cuts the captured entropy from impulse $\bar{u}_s=|1|_s$, while multiplicative step-down part (1.17b) transforms the captured entropy to the cutting action in (1.4) at $\bar{u}_{ko2}=\bar{u}^{-1}$. •

At the end of $k$ impulse, control action $\bar{u}_+$ transforms the entropy (1.20) on interval $\bar{u}_{kio}=\bar{u}_-\times(o(\tau_k^{+o})/\tau_k^{+o})$ to information

$$\Delta I[\tilde{x}_t/\varsigma_t]|_{\delta_{k+}^{\tau+}}^{\tau_k^{+o}}=1/4[\uparrow 1_{\tau_k^{-o}}]\bar{u}_{kio}\bar{u}_+=1/4\times(-2\bar{u}_{kio})[\uparrow 1_{\tau_k^{-o}}] \qquad (1.21)$$

and supplies it to $k+1$ impulse.
(If between these impulses, the entropy increments on the process trajectory are absent (cut)).

That leads to balance equation for information contributions of $k$-impulse:
$$\Delta I[\tilde{x}_t/\varsigma_t]|_{\delta_k^{\tau+}}^{\tau_k^{-o}} + \Delta I[\tilde{x}_t/\varsigma_t]|_{\tau_k^{-o}}^{\tau_k} + \Delta I[\tilde{x}_t/\varsigma_t]|_{\tau_k}^{\tau_k^{+o}} = \Delta I[\tilde{x}_t/\varsigma_t]|_{\delta_{k+}^{\tau+}}^{\tau_k^{+o}}, \quad (1.21a)$$
where interval $\bar{u}_{kio}$ holds information contribution $\Delta I[\tilde{x}_t/\varsigma_t]|_{\tau_k}^{\tau_k^{+o}} = 1/4\bar{u}_{km}$ satisfied (1.4) at $\bar{u}_+ = -2$, which measures $\bar{u}_{km} = 0.75$ (1.5). That brings relations
$$0.125 + 0.75 + \bar{u}_{kio} = -2\bar{u}_{kio}, 0.125 + 0.75 + 3\bar{u}_{kio} = 0, \bar{u}_{k1} = 3\bar{u}_{kio} = 0.375 = \bar{u}_- \times o(\tau_k^{+o})/\tau_k^{+o}) \quad (1.22)$$
and
$$o(\tau_k^{+o})/\tau_k^{+o} = 0.1875, \quad (1.22a)$$
$$\bar{u}_{ko} + \bar{u}_{km} + \bar{u}_{k1} = 1.25 = 5/3\bar{u}_{km} \quad (1.22b)$$
from which and (1.21a) it follows
$$\Delta I[\tilde{x}_t/\varsigma_t]|_{\tau_k}^{\tau_k^{+o}} = 3\Delta I[\tilde{x}_t/\varsigma_t]|_{\delta_k^{\tau+}}^{\tau_k^{-o}}. \quad (1.23)$$
Ratio $\bar{u}_{k1}/2\bar{u}_{kio} = 3/2$ at $2\bar{u}_{kio} = 0.25$ evaluates part of $k$ impulse information transferred to $k+1$ impulse.
Relations (1.17b,d), (1.18b,d,f), (1.19c), and (1.22a) *prove* the Proposition parts A-B (including (1.16b)).•

Since $\bar{u}_- = 0.5$ is cutting interval of impulse $\bar{u}_k$, it allows evaluate the additive sum of the discrete cutoff entropy contributions (1.4) during entire impulse $(\downarrow 1_{\tau_k^{-o}} - \uparrow 1_{\tau_k^{+o}}) = \delta_k$ using $\bar{u}_- = \bar{u}_k$:
$$\Delta S[\tilde{x}_t/\varsigma_t]|_{\tau_k^{-o}}^{\tau_k^{+o}} = 1/4\bar{u}_k/2 + 1/2\bar{u}_k + 1/4 \times 3/2\bar{u}_k = \bar{u}_k. \quad (1.24)$$
That determines the impulse cutoff information measure
$$\Delta S[\tilde{x}_t/\varsigma_t]_{\delta_k} = \Delta I[\tilde{x}_t/\varsigma_t]_{\delta_k} = (\downarrow 1_{\tau_k^{-o}} - \uparrow 1_{\tau_k^{+o}})\bar{u}_k = |1|\bar{u}_k, \bar{u}_k = |1|_k \; Nat \quad (1.24a)$$
equals to $\cong 1.44$ Bit, which the cutting entropy functional of this random process generates.

The above single unit impulse $\bar{u}_k = |1|_k$ measures the relative information intervals
$$\bar{u}_{ko} = 1/3\bar{u}_{km}, \; \bar{u}_{km} = 1, \; \bar{u}_{kio} = 1/3\bar{u}_{km} = \bar{u}_{ko}, \text{ and } \tau_k^{+o}/\tau_k^{-o} = 3. \quad (1.24b)$$
From relations
$\bar{u}_{ko1} = 1/2\bar{u}_{sm}$ and $\bar{u}_{ko1} = 1/2\bar{u}_{ko} = 1/6\bar{u}_{km}$ it follows
$$\bar{u}_{km} = 3\bar{u}_{sm}, \quad (1.25)$$
which shows that impulse unit $\bar{u}_k = |1|_k$ triples information supplied by entropy unit $\bar{u}_s = |1|_s$ or interval $\bar{u}_k$ compresses three intervals $\bar{u}_s$.

At satisfaction of the extremal principle, each impulse holds invariant interval size $|\bar{u}_k| = |1|_k$ proportional to middle impulse interval $o(\tau)$ with information $\bar{u}_{km}$ which measures $o(\tau)$, and vice versa, time $o(\tau)$ measures this information.

Condition of decreasing $t - s_k^{+o} = o(t) \to 0$ with growing $t \to T$ and squeezing sequence $s_k^{+o} \to \tau_{m-1}^{+o}, k = 1,2....m$ leads to persistence continuation of the impulse sequence with transforming previous impulse entropy to information of the following impulse: $\bar{u}_s = |1|_s \to \bar{u}_k = |1|_k$.

The sequence of growing and compressed information increases at

$\bar{u}_{k+1} = |3\bar{u}_k| = |1|_{k+1}$. (1.25a)

The persistence continuation of the impulse sequence links intervals between sequential impulses ( $\bar{u}_{ks}$, $\bar{u}_{k\delta s}, \bar{u}_{k\delta o}$) whose imaginary (virtual) function $[\uparrow 1_{s_k^{+o}} - \downarrow 1_{\delta_k^{\tau-}} + \uparrow 1_{\delta_k^{\tau+}}]u$ prognosis entropies increments (1.11), (1.12), (1.10).

Information contributions at each cutting interval $\delta_{k-1}, \delta_k$, $k, k+1, \dots, m$: $\Delta I[\tilde{x}_t / \varsigma_t]_{\delta_{k-1}}, \Delta I[\tilde{x}_t / \varsigma_t]_{\delta_k}, \dots$ determine time distance interval $\tau_k^{-o} - \tau_{k-1}^{+o} = o_s(\tau_k)$, when each entropy increment

$\Delta S[\tilde{x}_t / \varsigma_t]|_{\tau_{k-1}^{+o}}^{t \to \tau_k^{-o}} = 1/2uo(\tau_k) = \bar{u}_s \times o_s(\tau_k)$ supplies each $\Delta I[\tilde{x}_t / \varsigma_t]_{\delta_k}$ satisfying

$\bar{u} \times o(\tau_k) = \Delta I[\tilde{x}_t / \varsigma_t]_{\delta_k}$ at $\bar{u} \times o(\tau_k) = \bar{u}_k(\tau_k^{+o} - \tau_k^{-o})$.

Hence, impulse interval

$$\bar{u}_k = \Delta I[\tilde{x}_t / \varsigma_t]_{\delta_k} / (\tau_k^{+o} - \tau_k^{-o}) \quad (1.26)$$

measures density of information at each $\delta_k = \tau_k^{+o} - \tau_k^{-o}$, which is sequentially increases in each following Bit. Relations (1.25a,b), (1.26) *prove* part C of Proposition 1.3. •

Such a Bit includes three parts:

-the first delivers multiplicative action (1.16c) by capturing entropy of random process;

-the second delivers the impulse step-down cut of the process entropy;

-the third is information, which delivers the impulse step-up control and then transfers to nearest impulse.

That keeps information connection between the impulses and provides persistence continuation of the impulse sequence during the process time $T$.

Corollaries 1.2.

A. The additive sum of discrete functions (1.4) during the impulse intervals determines the impulse information measure equals to Bit, generated from the cutting entropy functional of random process.

The step-down function generates $1/8 + 0.75 = 0.875 Nat$ from which it spends $1/8$ $Nat$ for cutting correlation while getting $0.75$ $Nat$ from the cut. Step-up function holds $1/8$ $Nat$ while $0.675$ $Nat$ it gets from cutting $0.75$ $Nat$, from which $0.5Nat$ it transfers to next impulse leaving $0.125$ $Nat$ within $k$ impulse. The impulse has $1/8 + 0.75 + 1/8 = 1Nat$ of total $1.25Nat$ from which $1/8$ $Nat$ is the captured entropy increment from a previous impulse. The impulse actually generates $0.75Nat \cong 1Bit$, while the step-up action, using $1/8Nat$, transfers $2/8Nat$ information to next $k$ impulse, capturing $1/8Nat$ from the entropy impulse between $k$ and $k+1$ information impulses (on interval $o_s(\tau_k)$).

B. From total maximum $0.875$ Nat, the impulse cuts minimum of that maximum $0.75$ $Nat$ implementing minimax principle, which validates variation condition (I.1.7) and results (I.2.17a,b).

By transferring overall $0.375Nat$ to next $k+1$ impulse, that $k$ impulse supplies it with its maximum of $1/3 \times 0.75Nat$ from the cutting information, thereafter implementing principle maximum of minimal cut.

C. Thus, each cutting Bit is active *information unit* delivering information from previous impulse and supplying information to following impulse.

It includes: the cutting step-down control's information delivered through capturing the external entropy of the random process; the cutoff information, which the above control cuts from the random process; the information delivered by the impulse step-up control, which, being transferred to the nearest impulse, keeps the information connection between the impulses providing persistence continuation of the impulse sequence.
D. The amount of information that each second Bit of the cutoff sequence condenses grows in three times, which sequentially increases the Bit information density. At invariant increments of impulse (1.4), every $\overline{u}_k$ compresses three previous intervals $\overline{u}_{k-1}$ thereafter sequentially increase both density of interval $\overline{u}_k$ and density of these increments for each $k+1$ impulse.•

2.3. THE EMERGING MICROPROCESS

As the Bayes a posteriori probabilities grow along the observations, neighbor impulses ↓↑ and ↓↑ may merge. A discrete merge generates interactive jump ↑ within the impulse border.
The merge location on the observing trajectory identifies the impulse' probability and entropy.
The merge meets causing action with reaction, superimposing cause and effect and their probabilities.
It could cover unpredictable events within the merge, as merge itself.
(J.A. Wheeler and R. Feynman, cited in this Introduction [6.7], show that when action and symmetric (adjunct) reaction merge, a microprocess rises from classical Maxwell field).
Mathematically the jump increases Markov drift (speed) up to infinity (Sec.1.1.3.2).
A starting jumping action ↑ interacting with opposites action ↓ of the bordered impulse initiates the impulse inner process $\tilde{x}_{otk} = \tilde{x}(t \in o(\tau_k)))$ called a microprocess.

Comments 2.3.
In a sub-Markov process [37, 38], potential kernel negative curvature exposes Markov drifts convergence, which could lead to the merge. •

**2.3.1. The conjugated entropy increments in the microprocess**

The microprocess is developing under step-function $u_{\pm}^{t1}$, $u_{\pm}^{t2}$ within the bordered impulse with the step-function $u_t(u_-^t, u_+^t) = c^2(t \in o(\tau_k))$ on a fixed impulse interval $o(\tau_k)$ within the discrete impulse (1. 4).
The impulse step-down $u_-^t = u_-(\tau_k^{-o})$ and step-up $u_+^t = u_+(\tau_k^{+o})$ functions, acting on the discrete interval $o(\tau_k) = \tau_k^{+o} - \tau_k^{-o}$ satisfying (1.1A-1.1C) and (1.2a-1.2d), generates the EF (2.1) increments:

$$\Delta S_- = \Delta S_-[u_-^t], \Delta S_+ = \Delta S_+[u_+^t], \quad (3.1)$$

which preserve the additive and multiplicative properties within the Markov process.
(But the merging actions may not simultaneously possess both of these Markov properties).
Here, step function $u_{\pm}^{t1}$ (1.1c) is the analog of $\overline{u}_{k\delta 1}$ (1.14b) which multiplies in (1.16c) at locality $\delta_k^{\tau+}/2$ of the beginning of impulse moment $\tau_k^{-o}$.
Opposite functions $u_{\pm}^{t1}(t^*)$ of jumps ↑↓, starting at beginning of the process with relative time

$$t^* = [\mp\pi/2 \times \delta t^{\pm *}/\mathrm{o}(\tau_k)], \delta t^{\pm *} \in (\delta t_{ok}^{\pm} \to 1/2\,\mathrm{o}(\tau_k)), \quad (3.2)$$

hold directions of opposite impulses

$$u_{\pm}^{t1}=[u_{+}=\uparrow_{t_o^{*-}}(j-1), u_{-}=\downarrow_{t_o^{*+}}(j+1)] \quad (3.3)$$

on interval $\delta_o[t_o^{*-}, t_o^{*+}]=\delta t^* < o(\tau)$ at a locality of the impulse initial time $\tau_k^{-o}$. (The multiplication (1.16c) applied to functions (3,3), leads to invariant impulse $\bar{u}_k = |1|_k$ by the end of the locality.)

Functions (3.3), holding $u=c^2<0$ (1.2d), brings imaginable $u$ and minimal time interval

$$o=(\delta_k^{\tau+}/2)^2=(\tau_k^{-o})^2. \quad (3.3a)$$

The microprocess increments at interval $o$ do not possess Markov properties (1.C) but holds (1.1B).

The jumps (3.3) initiate relative differential increments of entropy:

$$\frac{\delta S}{S}/\delta t^*=u_{\pm}^{t1},\ [u_{+}=\uparrow_{t_o^{*-}}(j-1), u_{-}=\downarrow_{t_o^{*+}}(j+1)], \quad (3.4)$$

which in a limit leads to differential Equations:

$$\dot{S}_{+}(t^*)=(j-1)S_{+}(t^*), \dot{S}_{-}(t^*)=(j+1)S_{-}(t^*). \quad (3.5)$$

The applied (3.3) with symbol $j$ of orthogonality to the microprocess entropy increments rotates them.

Solutions of (3.5) describe the microprocess with opposite conjugated entropies functions on relative time $t^*$:

$$S_{+}(t^*)=[exp(-t^*)(\mathrm{Cos}(t^*)-jSin(t^*))]|_{t_o^{*-}}^{1/2o(\tau_k)}, S_{-}(t^*)=[exp(t^*)(\mathrm{Cos}(t^*)+jSin(t^*))]|_{t_o^{*+}}^{1/2o(\tau_k)} \quad (3.6)$$

with initial conditions $S_{+}(t_o^{*-})$, $S_{-}(t_o^{*+})$ at moment

$$t_o^{*+}=t_o^{*-}=[\mp\pi/2\delta t_{ok}^{\pm}]. \quad (3.6a)$$

The wide of step-function $u_{\pm}^{t1}: \delta t_o^{\pm}/\mathrm{o}(\tau_k)=0.2+0.005=0.205$ relative to interval $\mathrm{o}(\tau_k)$ and the impulse beginning interval $\tau_k^{-o}/\mathrm{o}(\tau_k)=0.25$ relative to that interval determine the relative moment $\delta t_{ok}^{\pm}=\delta t_o^{\pm}/\tau_k^{-o}=\pm 0.82$ of starting this function.

From that, numerical solutions of (3.6) by the moment of time $\delta t_{ok}^{\pm}=\pm 0.82$ follow:

$$\begin{aligned}S_{+}(t_o^{+})&=[exp(-\pi/2\times 0.82)(\mathrm{Cos}(\pi/2\times 0.82))-jSin(\pi/2\times -0.82))]\approx 0.2758\times 1,\\ S_{-}(t_o^{-})&=[exp(\pi/2\times -0.82)(\mathrm{Cos}(-\pi/2\times -0.82)+jSin(-\pi/2\times -0.82))]\approx 0.2758\times 1\end{aligned}. \quad (3.7)$$

The numerical solutions by the moments of time

$$t^{*-}=-\pi/2\times 1/2\mathrm{o}(\tau_k)/\mathrm{o}(\tau_k)=-\pi/4 \text{ and } t^{*+}=\pi/2\times 1/2\mathrm{o}(\tau_k)/\mathrm{o}(\tau_k)=\pi/4 \quad (3.8)$$

are

$$\begin{aligned}S_{+}(t^{*-})&=S_{+}(t_o^{*-})\times\exp(-\pi/4)[\mathrm{Cos}(\pi/4)-jSin(\pi/4)],\\ S_{-}(t^{*+})&=S_{-}(t_o^{*+})\times\exp(-\pi/4)[\mathrm{Cos}(-\pi/4)+jSin(-\pi/4)]=S_{-}(t_o^{*+})\times\exp(-\pi/4)[\mathrm{Cos}(-\pi/4)-jSin(\pi/4)]\end{aligned}. \quad (3.9)$$

These vector-functions at opposite moments (3.6a) hold opposite signs of their angles $\mp\pi/4$ with values:

$$S_{+}(t^{*-})\cong 0.2758\times 0.455\cong +0.125, S_{-}(t^{*+})\cong 0.2758\times 0.455\cong -0.125. \quad (3.10)$$

Function $u_{\pm}^{t2}$ (1.2d), starting these opposite increments, turns them on angle $\varphi_{-}^2-\varphi_{+}^2=\pi/2$ that equalizes the increments and starts entangling both equal increments with their angles within interval $t=\tau_k\mp 0$:

$$S^2_-(t=\tau_k+0)=\delta S^1_-(t=\tau_k-0)\times\downarrow_{\tau_k+0}\pi/2=S^1_-(t=\tau_k-0)\times\exp(\pi/2\times t^{*+}_{\tau_k+0})[\mathrm{Cos}(\pi/2\times t^{*+}_{\tau_k+0})+\mathrm{jSin}(\pi/2\times t^{*+}_{\tau_k+0}), \quad (3.11)$$

$$S^2_+(t=\tau_k+0)=\delta S^1_+(t=\tau_k-0)\times\uparrow_{\tau_k+0}\pi/2=S^1_-(t=\tau_k-0)\times\exp(-\pi/2\times t^{*-}_{\tau_k+0})[\mathrm{Cos}(-\pi/2\times t^{*-}_{\tau_k+0})+\mathrm{jSin}(-\pi/2\times t^{*-}_{\tau_k+0})$$

at moments

$$t^{*\pm}_{\tau_k+0}=[\mp\pi\times 2\delta t^{\pm}_{1k}],\delta t^{\pm}_{1k}=\delta t^{\pm}_1/1/2\tau_k\cong 0.4375,\delta t^{\pm}_1=\pm(0.5-\delta t^{k1}_{\pm}),\delta t^{k1}_{\pm}=\tau^{-o}_k/\tau_k+\delta t^{ko}_{\pm}/\tau_k=0.25+0.03125=0.2895 \quad (3.12)$$

where $\delta t^{ko}_{\pm}/\tau_k\cong 32^{-1}$ evaluates dissimilarities between functions $u^{t2}_{\pm}=[u_+=(j+1),u_-=(j-1)]$ switching from moment $t=\tau_k-0$ to moment $t=\tau_k$. (The dissimilarities are following from (2.24)).

The resulting values at $t=\tau_k+0$ are

$$S^2_-(t=\tau_k+0)=0.125\exp(\pi/2\times 0.4375)\times 1\cong 0.25, S^2_+(t=\tau_k+0)=0.125\exp(\pi/2\times 0.4375)\times 1\cong 0.25 \quad (3.13)$$

which, being in the same direction, are summing at that locality:

$$S^o_{\mp}=2S^2_{\mp}[(\delta t^{ko}_{\pm}/\tau_k)]\cong\mp 0.5. \quad (3.14)$$

The entanglement, starting with entropy (3.13), continues to entropy (3.14) up to cutting all entangled entropy increments.

Thus, the entanglement starts at angle $(\pi/2)\times 0.4375<\pi/4$ takes relative time interval of the impulse $\delta t^{ko}_{\pm}/\tau_k\cong 0.03125$ to ends on angle $\pi/2$.

Since only at angle $\pi/2$ the space interval within impulse begins, it means that *the entanglement starts before the space is formed and ends with beginning the space.*

Here $\tau_k=1/2o(\tau),o(\tau)=1Nat$ and $\delta t^{ko}_{\pm}=0.03125\times 1/2o(\tau)=0.015625o(\tau)=\varepsilon_{ok}$. (3.14a)

*Moreover, the entanglement may even create the space during that time interval which is reversible.*

Comments 2.4. A potential path during creation of both entanglement and space could be a *wormhole*-a short cut in space-time predicted by general relativity. But real *space curvature does not exist at this time*. It may emerge only after entanglement by a moment of forming a Bit at the end of the impulse. Hence, space curvature may form at the *end of a microprocess (analog of a quantum process)* when the Bit, as the elementary unit of a macroprocess, emerges.

*Since the entanglement has no space measure, the entangled states can be everywhere in a space.* •

The $t=\tau_k\mp 0$ locality evaluates the $0_k$-vicinity of action of inverse opposite functions (3.9), whose signs imply the signs of increments in (3.14) and in the following formulas.

The subsequent step-up function jumping action $\downarrow$ changes increment (3.14) according to Eqs

$$S_{\mp}(\tau^{+o}_k)=S^o_{\mp}(\delta t^{ko}_{\pm}/\tau_k)\times\exp(t^{*+}_{\tau^{+o}_k}),t^{*+}_{\tau^{+o}_k}=[\pi/2\delta t^{*o}_k],\delta t^{*o}_k\in(\delta t^{*o}_{1k}\to\tau^{+o}_k/\tau_k), \quad (3.15)$$

at

$$\delta t^{*o}_{1k}=\delta t^{\pm}_{1k}/1/2\tau,\delta t^{\pm}_{1k}=\pm(0.5-\delta t^{k1}_{\pm}),\delta t^{k1}_{\pm}=\delta t^{ko}_{\pm}/\tau_k+\tau^{+o}_k/\tau_k=0.25+0.03125=0.2895,\delta t^{\pm}_{1k}=\delta t^{\pm}_1/1/2\tau_k\cong 0.4375$$

with resulting value

$$S_{\mp}(\tau^{+o}_k)=\mp 0.5\exp(\pi/2\times 0.4375)=\mp 0.5\times(\cong 2)\cong\mp 1, \quad (3.16)$$

which measures total entropy of the impulse

$$\bar{u}_k=|1|_k=1Nat. \quad (3.17)$$

Trajectories (3.10-3.16) describe anti-symmetric conjugated dynamics of the microprocess within the jumping impulse $\uparrow\downarrow$, which is reversible generating entangled entropy increments (3.16) up to the impulse cutting moment.

Comments 2.5. From relation (3.4) and Jacobi-Hamiltonian variation equation $\partial S / \partial t = -\tilde{H}$ it follows that the microprocess Hamiltonian gets form'

$\tilde{H}(t^*) = -u_{\pm}^{t2} S(t^*)$. (3.17a)

That Eq. admits the conjugated Hamiltonian with both real and imaginary parts:

$\tilde{H}(t^*) = -[(j+1)S + (j-1)S] = -[(\dot{S}_+(t^*) / S_+(t^*) + \dot{S}_-(t^*) / S_-(t^*)]S(t^*)$ . (3.17b)

At the entanglement, $S_+(t^*+) = S_-(t^*+), S = S_+(t^*+) + S_-(t^*+) = 2S_+(t^*+) = 2S_-(t^*+)$ and Hamiltonian acquires view

$\tilde{H}(t^*+) = -[(\dot{S}_+(t^*) / 2 + \dot{S}_-(t^*) / 2] = -\dot{S}(t^*+)$ (3.17c). •

Cutting the entangled joint entropy at moment $\tau_k^+ \cong 0_k + \tau_k^{o+}$ coverts it to equal information contribution

$S_{\mp}^o[\tau_k^+] = \Delta I[\tau_k^+] \cong 1.44$ bit (3.18)

which each $\overline{u}_k$ impulse produces.

An interactive action $\uparrow$ of this impulse with the microprocess jumping action $\downarrow$ delivers entropy on the $0_k$-vicinity of the cutting moment:

$S_c^*(\tau_k^+) = \exp 0_k = 1$ . (3.19)

The current impulse step-up action $[\uparrow_{\tau_k^{+o}} \overline{u}_+^o]$(in (3.6)) generates an information bit from the microprocess reversible entropy.

The jumping actions provide the minimal discrete displacement (3.3a,3.2), which rotates the entropy opposite increments. The interactive jump generates a pair of random interactive actions on the bordered impulse, which are equally probable, reversible within the probabilities of multiple random interactive actions. The curving shift initiates a microprocess within the bordered impulse running the superposition and entanglement of conjugates entropy fractions during time interval starting with the jump.

The entanglement starts before the space of the shift is formed and ends with beginning the space shift, being small part of impulse reversible time interval. •

2.3.2. **The rotating conjugated dynamics of the microprocess**

*Starting step functions* $u_{\pm}^{t1}$ initiates increments of the entropies on interval $o(\tau_k - 0)$ by moment $t = \tau_k - 0$:

$\delta S_+[u_+^{t1}] = \delta S_+^1(t = \tau_k - 0)) = \delta S_+^1(t = \tau_k^{-o}) \uparrow_{\tau_k^{+o}} (j-1), \delta S_-[u_-^{t1}] = \delta S_-^1(t = \tau_k - 0) = \delta S_-^1(t = \tau_k^{-o}) \downarrow_{\tau_k^{-o}} (j+1)$. (3.20)

Step functions $u_{\pm}^{t2}$ (1.2d) starting at $t = \tau_k - 0$ contribute the entropy increments on interval $o(\tau_k)$ by moment $t = \tau_k$:

$\delta S_+[u_+^{t2}] = \partial S_+^2(t = \tau_k) = \partial S_+^2(t = \tau_k - 0)) \uparrow_{\tau_k} (j+1), \delta S_-[u_-^{t2}] = \partial S_-^2(t = \tau_k) = \partial S_-^2(t = \tau_k - 0)) \downarrow_{\tau_k} (-j+1)$ .(3.21)

Complex function $u_+^{t1}$ turns on the multiplication of functions $\delta S_+^1(t = \tau_k^{-o})$ on angle $\varphi_+^1 = -\pi / 4$, and function $u_-^{t1}$ turns on the multiplication function $\partial S_-^1(t = \tau_k^{-o})$ on angle $\varphi_-^1 = \pi / 4$ by moment $t = \tau_k - 0$. This brings the entropy increments

$\delta S_+^1(t = \tau_k - 0)) = \delta S_+^1(t = \tau_k^{-o}) \times \uparrow_{\tau_k^{+o}} -\pi / 4, \delta S_-^1(t = \tau_k - 0)) = \delta S_-^1(t = \tau_k^{-o}) \times \downarrow_{\tau_k^{+o}} \pi / 4$. (3.22)

Analogously, step-functions $u_{\pm}^{t2}$, starting at $t = \tau_k - 0$, turn entropy increments (3.22) on angles $\varphi_-^2 = \pi / 4$ by moment $t = \tau_k$ and on angle $\varphi_+^2 = -\pi / 4$ the entropy increments by moment $t = \tau_k$:

$\delta S_-^2(t=\tau_k)=\delta S_-^2(t=\tau_k-0)\times\downarrow_{\tau_k}\pi/4, \delta S_+^2(t=\tau_k)=\delta S_+^2(t=\tau_k-0)\times\uparrow_{\tau_k}-\pi/4$ . (3.23)

The difference of angles between the functions in (3.22): $\varphi_+^1-\varphi_-^1=-\pi/2$ is overcoming on time interval $o(\tau_k-0)=\tau_k^{-o}+1/2o(\tau_k)$ .

After that, control $u_\pm^{t2}$, starting with opposite increments (3.23), turns them on angle $\varphi_-^2-\varphi_+^2=\pi/2$ equalizing entropy increments (3.23).

That launches *entanglement* of entropies increments and their angles *within* interval $o(\tau_k)$ (on a middle of the impulse $t=\tau_k$):

$\delta S_+^2(t=\tau_k)=\delta S_-^2(t=\tau_k)=\delta S_\mp^2$ . (3.24)

Control $\bar{u}_-=0.5$, turning the time-located vector-function at the impulse beginning:

$u_-^t=u_-(\tau_k^{-o}):\xleftarrow{\tau_k^{-o},\bar{u}_-=0.5,}\delta\varphi_1$ (3.25)

on angle

$\delta\varphi_1=\varphi_+^1-\varphi_-^1=\pi/2$, transforms it to space vector $u_+(\tau_k-0)=\uparrow_{\tau_k-o}\bar{u}_+=1$ during a jump from moment $t=\tau_k^{-o}$ to moment $t=\tau_k-0$ on interval $o(\tau_k-0)$ in (3.22).

Then, vector-function $\downarrow_{\tau_k}\bar{u}_-^o=2$, starting on time $t=\tau_k-0$ with space interval $\bar{u}_-^o=2,$ jumps to vector-function $\uparrow_{\tau_{k+0}}\bar{u}_+^o=2$, forming on time interval $o(\tau_k+0)=1/2o(\tau_k)+\tau_k^+$ the additive space-time impulse

$u_\mp=[\downarrow_{\tau_k+0}\bar{u}_-^o]+[\uparrow_{\tau_k^{+o}}\bar{u}_+^o]$. (3.26)

The first part of (3.26) equalizes (3.24) within *space-time* interval $\bar{u}_-\times1/2o(\tau_k)$, and then joins, summing them on $\bar{u}_-\times o(\tau_k+0)$, which finalizes the entanglement.

The last part of impulse (3.26) cuts-kills the entangled increments on interval $\bar{u}_+\times\tau_k^+$ at ending moment $\tau_k^+$ . Section 2.3.4 details the time –space relation and their measures.

Relations (3.1-3.26) lead to following specifics of the microprocess.

3.1a.Step-functions $u_\pm^{t1}$ initiate microprocess $\tilde{x}_{otk1}=\tilde{x}(t\in o(\tau_k-0))$ on beginning of the impulse discrete interval $o(\tau_k-0)$ with only additive increments (3.2). Opposite step functions $u_\pm^{t2}$ continue microprocess $\tilde{x}_{otk2}=\tilde{x}(t\in o(\tau_k+0))$ within interval $o(\tau_k+0)$ with both additive and multiplicative increments (3.3) preserving the process Markov properties.

3.1b.Space-time impulse (3.17) within interval $o(\tau_k+0)$ processes entanglement of increments (3.25) of microprocess $\tilde{x}_{otk2}=\tilde{x}(t\in o(\tau_k+0))$ summing these increments on $o(\tau_k)$ locality of $t=\tau_k$:

$S_\mp^o=2\delta S_\mp^2[(o(\tau_k)]$ . (3.27)

Then it kills entropies (3.27) at its ending moment $\tau_k^{o+}\to\tau_k^+$:

$S_\mp^o[\tau_k^+]=0$ . (3.27a)

The microprocess, producing entropy increment (3.27) within the impulse interval, is reversible before killing, which converts the increments in equal information contribution

$S_\mp^o[\tau_k^+]\Rightarrow\Delta I[\tau_k^+]$, (3.27b)

That amount depends on the entropy of merging impulses starting the microprocess.
The information emerging at the ending impulse time interval accomplishes injection of an energy with step-up control $[\uparrow_{\tau_k^{+o}} \bar{u}_+^o]$,.

The energy injection can be a result of the impulse's middle inter-action with environment.
From the impulse ending moment starts an irreversible information process.
3.1c. Transferring the initial time-located vector to equivalent space-vector $\uparrow_{\tau_k - o} \bar{u}_+$ transforms a *transition impulse*, starting within a jump of time $\tau_k^{-o}$ on interval of $\bar{u}_- = 0.5$ up to creating space interval with $\bar{u}_+ = 1$. The opposite space vector $\downarrow_{\tau_k} \bar{u}_-^o = 2$, acting on relative time interval $1/2o(\tau_k)/(\tau_k^{+o} - \tau_k^{-o}) = 0.5$, forms space-time function $\downarrow_{\tau_k} \bar{u}_-^1 : \bar{u}_-^1 = 2 \times 0.5 = 1$, which, as inverse equivalent of opposite function $\uparrow_{\tau_k - o} \bar{u}_+$, neutralizes it to zero. Both time duration of $\bar{u}_- = 0.5$ and $\bar{u}_+ = 1$ concentrate these functions in transition interval $\tau_k - (\tau_k - 0) = 0_k$.

Within the whole impulse, only step-down functions $[\downarrow_{\tau_k^{-o}} \bar{u}_-]$ on time interval $\bar{u}_- = 0.5$ and step-up function $[\uparrow_{\tau_k^{+o}} \bar{u}_+^1]$ on space-time interval $\bar{u}_+^1 = \bar{u}_+ \times \tau_k^+ = 2_{\tau_k^+}$ are left. That determines size of the discrete $1-0$ impulse by multiplicative measure $U_m = |0.5 \times 2| = |1|_k = \bar{u}_k$ generating an information bit.
Therefore, functions $u_+(\tau_k - 0) = \uparrow_{\tau_k - o} \bar{u}_+$ and $u_-(\tau_k) \downarrow_{\tau_k} \bar{u}_-^o$ are transitional during formation of that impulse and creation time-space microprocess $\tilde{x}_{otk} = \tilde{x}(t \in 1/2o(\tau_k), h_k \in 2_{\tau_k^+})$ with final entropy increment (3.27) and a virtual logic. The microprocess transits from the entropy at at ending moment $\tau_k^{0+} \to \tau_k^+$: -locality (3.27) to actual information (3.27b).

### 2.3.3. Probabilities functions of the microprocess

Amplitudes of the process probability functions at $S_\mp^*(\tau_k^{+o}) = |S_+^*| = |S_-^*| = 1$ are equal and independent:

$$p_{+a} = 0.3679, p_{-a} = 0.3679. \qquad (3.28)$$

That leads to

$$p_{+a} p_{-a} = p_{\pm a}^2 = 0.1353, S_{\mp a}^* = -\ln p_{a\pm}^2 = 2,$$

or at

$S_{\mp a}^* = 2$, to

$$p_{a\pm} = \exp(-2) = 0.1353, \qquad (3.28a)$$

where

$$S_{\mp a}^* = S_\mp^*(\tau_k^{+o}+) + S_c^*(\tau_k^+ +)$$

includes the interactive components at $\tau_k^{+o}+$ following $k$ impulse.
Functions $u_+ = (j-1), u_- = (j+1)$, satisfying (1.IA), fulfill the additive property at the impulse starting interval $o[t_o^\mp]$, running the anti-symmetric entropy fractions.
Opposite functions $u_+ = (1+j), u_- = (1-j)$, satisfying (1.IB) by the end of impulse at $\uparrow_{\tau_{k+}^{+o}} \bar{u}_\pm$, mount entanglement of these entropy fractions within the impulse' $|1/2 \times 2| = |\bar{u}_k| = |1|_k$ space interval $\bar{u}_\pm = \pm 2$.
The entangling fractions hold the equal impulse probabilities (3.28), which indicate appearance of both entangled anti-symmetric fractions simultaneously with starting space interval.

Interacting probability amplitudes $p_{+a}, p_{-a}$ of $p_{\pm a}$ satisfy multiplicative relation $p_{\pm a} = \sqrt{p_{+a} p_{-a}}$.
However sum of non-interacting probabilities: $p_{+} + p_{-} = \exp(-S_{+}^{*}) + \exp(-S_{-}^{*}) = p_{\pm} \neq \mathrm{p}_{a\pm}$ does not comply with it. The summary probability $p_{\pm am} = 0.7358$ of the non-interacting entropies components is unequal to probability $p_{\pm a}$ of interacting entropies. The interacting probabilities in transitional impulse $[\uparrow 1_{\tau|_k^-} \downarrow 1_{\tau|_k^+}]\bar{u}_k$ on $\tau_k$-locality violate their additive property, but preserve additive of the entropy increments.
The impulse microprocess on the ending interval preserves both additive and multiplicative properties only for the entropies.
The basic relations for the impulse's entropy and probability are equivalent for quantum mechanics (QM) probability amplitudes relations. However, the impulse cutting probabilities $p_{+}, p_{-}$ are the probability of random events in the hidden correlations, while probability amplitudes $p_{+a}, p_{-a}$ are attributes of the microprocess starting within the cutting impulse. That distinguishes the considered microprocess from the related QM equations, considered for physical particles.
The entropy of multiple impulses integrates the microprocess along the observing random distributions.
With minimal impulse entropy ½ Nat starting a virtual observer, each following impulse' initial entropy $S_{\pm}(t_o) = 0.25 Nat$ self-generates entropy $S_{\mp a}^{*} = 0.5 Nat$.
Thus, the virtual observer's time–space microprocess starts with probability $p_{a\pm} = \exp(-0.5) = 0.6015$.
Probability $p_{a\pm} = 0.1353$ is relational to the impulse initial conditions, which evaluates appearance of time–space actual impulse (satisfying (3.26)) that *decreases* its initial entropy on $S_{\mp a}^{*} = 2$ Nat.
The impulse's invariant measure, satisfying the minimax, preserves $p_{a\pm}$ along the time-space microprocess for multiple time-space impulses. Reaching probability of appearance the time-space impulse needs $m_p = 0.6015 / 0.1353 \cong 4.4457 \approx 5$ multiplications of invariant $p_{a\pm} = 0.1353$, which predicts a priori probability of the impulse's reactive action.
The space interval, beginning the displacement shift, starts within the interval of entanglement (3.15a) having probability

$$P_{\Delta}^{*}(\delta t_{\pm}^{k1}) = \exp(-| S_{\mp}^{*}(\delta t_{\pm}^{k1}) |, P_{\Delta}^{*}(\delta t_{\pm}^{k1}) = 0.821214 \quad (3.28b)$$

at

$$\delta t_{\pm}^{k1} = 0.2895, S_{\mp}^{*}(\delta t_{\pm}^{k1}) = \mp 0.125 \exp(\pi / 2 \times \delta t_{\pm}^{k1}) = \mp 0.1969415, \quad (3.28c)$$

and continues during the shift, extending to the space part of the impulse multiplicative measure after the displacement ends. Hence, each reversible microprocess within the impulse generates invariant increment of entropy, which enables sequentially minimize the starting uncertainty of the observation.
Assigning the entropy minimal uncertainty measure $h_{\alpha}^{o} = 1/137$-physical structural parameter of energy [30], which includes the Plank constant's equivalent of energy, leads to relation:

$$S_{\mp a}^{*} = 2h_{\alpha}^{o}, p_{\pm a} = \exp(-2h_{\alpha}^{o}) = 0.98555075021 \to 1, \quad (3.29)$$

This evaluates the probability of a real impulse's physical strength of coupling independently chosen entropy fractions.
The initially orthogonal non-interacting entropy fractions $S_{+a}^{*} = h_{\alpha}^{o}$ , $S_{-a}^{*} = h_{\alpha}^{o}$ at mutual interactive actions, satisfy multiplicative relation

$$S_{\mp a}^{*} = (h_{\alpha}^{o})^2 [\mathrm{Cos}^2(\bar{u}t) + Sin^2(\bar{u}t)] |_{t_o^{\mp}}^{t=1/2\tau} = (h_{\alpha}^{o})^2 = inv \quad (3.30)$$

which at $S^*_{\mp a} = (h^o_\alpha)^2 \to 0$ approaches $p^*_{\pm a} = \exp[-(h^o_\alpha)^2] \to 1$.

The impulse interaction adjoins the initial orthogonal geometrical sum of entropy fractions in linear sum $2h^o_\alpha$. Starting physical coupling with double structural $h^o_\alpha$ creates initial information triple with probability (3.29).

*The microprocess initiates the merge that starts with the jumping actions multiplication on the bordered impulse time according to (1.16b) succeeding displacement (3.3a) during the merge. Both follow from the EF extreme. Random microprocess' merge proves impossibility of deterministic prediction of QM.*

*The multiplication violates the Markov property (1.1B) leading to a complex control (1.2c), which starts the microprocess within the displacement and rotates the initial conjugated entropy increments.*

*The microprocess (2.3.3) emerges from multiple interactions starting with probabilities (3.28),inverse entropy* $S^*_{\mp a} = 2$*, and injection of the related random energy. With growing probabilities up to 1, this energy increases rising the equivalent entropy, which is leading to equal information Bit.*

*The energy aspect is in Sec.2.6, where the Jarzynski Equality (JE)*[39], *applied to the evolving microprocess, measures thermodynamic energy connecting the JE with this process's information measure.*

*Thus, this developing microprocess presents a Stochastic Quantum process with evolving thermodynamics and a path to Information Macrodynamics* [40].

*Results show that a window of interaction with an environment opens only on the impulse border twice: at the beginning between moments* $\delta^{\tau+}_k / 4$ and $\tau^{-o}_k$ *when the entropy flow with energy accesses impulse, and at the end of a gap when an entangled entropy with accesses of energy converts to equivalent information.*

2.3.4. **The relation between the curved time and equivalent space length within an impulse**

Let us have a two-dimensional rectangle impulse with plain $p$ measured in time length $[\tau]$ unit and orthogonal $h$ measured in space length $[l]$ unit, with the rectangle measure

$$M_i = p \times h . \qquad (3.31)$$

The problem: Having a measure of the plain part of the impulse $\mathrm{M}_p$ to *find* high $h$ at equal measures of both parts:

$$\mathrm{M}_p = \mathrm{M}_h \text{ and } \mathrm{M}_p + \mathrm{M}_h = M_i . \qquad (3.32)$$

From (3.42) it follows

$$\mathrm{M}_h = 1/2M_i = 1/2p \times h . \qquad (3.33)$$

Assuming the impulse has only equal plain parts $1/2p$, it measure $\mathrm{M}_p = (1/2p)^2$.

Then from $\mathrm{M}_p = (1/2p)^2 = \mathrm{M}_h = 1/2p \times h$ it follows

$$h / p = 1/2 . \qquad (3.34)$$

Let us find a length unit $[l]$of the curved time unit $[\tau]$ rotating on angle $\pi/2$ using relations

$$2\pi h[l]/4 = 1/2p[\tau] \quad (3.35a), \quad [\tau]/[l] = \pi h / p. \qquad (3.35)$$

Substitution (3.34) leads to ratio of the measured units:

$$[\tau]/[l] = \pi / 2 . \qquad (3.36)$$

Relation (3.36) sustains orthogonality of these units in a time-space coordinate system, but since initial relations (3.32) are linear, ratio (3.36) represents a linear connection of time-space units (3.35).

The impulse-jumps curve the time unit in (3.8). According to Proposition 1.3, the impulse' invariant entropy implies the multiplication, starting the rotation.
The microprocess, built in rotation movement curving the impulse time, adjoins the initial orthogonal axis of time and space coordinates (Fig.1a).The curving impulse illustrates Fig.1b.

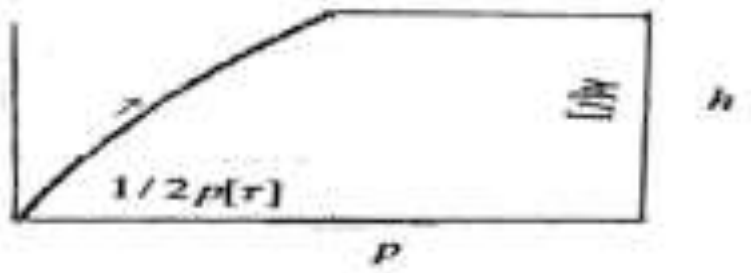

Fig.1(a)

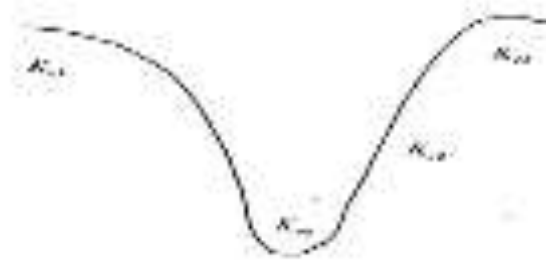

Fig.1 (b)

Fig.1(a). Illustration of origin the impulse space coordinate measure $h[l]$ at curving time coordinate measure $1/2p[\tau]$ in transitional movement.

Fig.1(b). Curving impulse with curvature $K_{e1}$ of the impulse step-down part, curvature $K_{eo}$ of the cutting part, curvature $K_{e2}$ of impulse transferred part, and curvature $K_{e3}$ of the final part cutting all impulse entropy.

The impulses, preserving the multiplicative and additive measures, have common ratio of $h/p=1/2$, whose curving part $p=1/2$ brings universal ratio (3.36), which concurs with Lemma 1.1, (1.2a).
At the above assumption, measure $\mathrm{M}_h$ does not exist until the impulse-jump curves its only time plain $1/2p$ at transition of the impulse. This transition is measured only in time. The following impulse transition is measured both in time $1/2p$ and space coordinate $h$.
According to (3.33), measure $\mathrm{M}_h$ emerges only on a half of that impulse' total measure $M_i$.
The transitional impulse could start on border of the virtual impulses ↓↑ , where the transition, curving time $\delta t_p = 1/2p$ under impulse-jump during $\delta t_p \to 0$ , leads to

$$M_p \to 0 \text{ at } M_h \Rightarrow M_i = p \times h \ . \qquad (3.37)$$

If a virtual impulse ↓↑ has equal opposite functions $u_-(t), u_+(t+\Delta)$, at $\bar{u}_+ = \bar{u}_-$, the additive condition for measure (1.2a): $\mathrm{U}_a(\Delta)=0$ is violated, and the impulse holds only multiplicative measure $\mathrm{U}_m(\Delta) \neq 0$ in relation (1.2C): $U_m(\Delta) = \mathrm{U}_{am}$ which is finite only at $\bar{u}_+ = \bar{u}_- \neq 0$.
If any of $\bar{u}_+ = 0$ , *or* $\bar{u}_- = 0$ , both multiplicative $\mathrm{U}_m(\Delta)=0$ and additive $\mathrm{U}_a(\Delta)=0$ disappear.
At $\bar{u}_- \neq 0,$ measure $\mathrm{U}_a(\Delta)$ is a finite and positive, specifically, at $\bar{u}_- = 1$ it leads to $\mathrm{U}_a(\Delta)=1$ preserving measure $\mathrm{U}_{amk} = |\mathrm{U}_a|_k$ . An impulse-jump at $o[t_o^{\mp}] \to \delta t_p \to 0$ curves a "needle pleat" space at the transition to the finite form of the impulse. The Bayes probabilities measure may overcome this transitive gap.
Since entropy (I.3.2.1) is proportional to the correlation time interval, whose impulse curvature $K_s = h[l]^{-1}$ is positive, this curved entropy is positive.
The curving needle cut changes the curvature sign converting this entropy to Information.

## 2.4. CURVATURE OF THE IMPULSE

An external step-control carries entropy which evaluates:

$$\delta_{ue}^{i} = 1/4(\mathrm{u}_{io} \text{-} \mathrm{u}_i) \, , \qquad (3.38)$$

where $\mathrm{u}_{io} = \ln 2 \cong 0.7 Nat$ is the total cutoff entropy of the impulse and $\mathrm{u}_i \cong 0.5 Nat$ is its cutting part.

The same entropy-information carries each impulse step-down and step-up control, while both controls carry $\delta^i_{ueo} \cong 0.1Nat$.

That evaluates information wide of each single impulse control's cut which the impulse carries:

$\delta^i_{ue} \cong 0.05Nat$. (3.38a)

To create information, the starting step-down part and the step-up part transfer entropy to the final killing part generating information. These three parts carry the entropy measures accordingly: $\delta^i_{ue1} \cong 0.025Nat, \delta^i_{ue2} \cong 0.02895Nat, \delta^i_{ue3} \cong 0.01847Nat$. (3.38b)

The first relation in (3.38b) allows estimate Euclid's curvature $K_{e1}$ of the impulse step-down part, related to currying entropy $0.25Nat$ and its increment $\delta K_{e1}$:

$K_{e1} = (r_{e1})^{-1}, r_{e1} = \sqrt{1+(0.025/0.25)^2} = \mp 1.0049875, K_{e1} \cong -0.995037, \delta K_{e1} \cong -0.004963$. (3.39)

The cutting part's curvature estimates relations

$K_{eo} = (r_{eo})^{-1}, r_{eo} = \mp\sqrt{1+(0.1/0.5)^2} = 1.0198, K_{eo} \cong -0.98058, \delta K_{eo} \cong -0.01942$. (3.39a)

The transferred part's curvature estimates relations

$K_{e2} = (r_{e2})^{-1}, r_{e2} = \sqrt{1+(0.02895/0.25)^2} \cong 1.0066825, K_{e2} \cong +0.993362, \delta K_{e2} \cong 0.006638$ (3.39b)

which is opposite to the step-down part.

The final part cutting all impulse entropy estimates curvatures

$K_{e3} = (r_{e2})^{-1}, r_{e3} = \sqrt{1+(0.01847/\ln 2)^2} \cong \pm 1.014931928, K_{e3} \cong +0.99261662, \delta K_{e3} \cong -0.00738338$.

Thus, the entropy impulse is curved with three different curvature values (Fig.1b).

These values estimate each impulse' curvature holding the invariant entropies.

The entropies emerge in minimax cutoff of the impulse carrying entropy $S_{ki} = 0.5$ and *a priori* probability $p_{a\pm} = \exp(-0.5) = 0.6015$ after multiple numbers $m_p$ of probing impulses observe this probability.

Since the rectangle impulse, cutting a time correlation, has measure $\mathrm{M} = |1|_M$, the curving impulse, cutting the curving correlation, determines measure

$r_{iM} = \mathrm{M} \times K_{ei}$. (3.40)

The rectangle impulse, not cutting time-correlations, possess Euclid's curvature $K_{iM} = 1$.

Accordingly, the impulse with both time and space measure $|M_{io}| = \pi$, which could appear in transitional impulse curvature of cutting part $K_{eo}$, determines correlation measures

$r_{icM} = \mathrm{M}_{io} \times K_{eo}$. (3.40a)

At appearance of the impulse with emerging space coordinate, the increment of the curved impulse correlations measure ratio of the measures for the curved correlation to one with only time correlation:

$r_{icM} / r_{iM} = \pi / |1| K_{ei} / K_{eio}$ (3.40b)

Counting (3.40b) leads to

$r_{icM} / r_{iM} \cong 3.08$.

Relative increment of correlation:

$\Delta r_{iM} / r_{iM} = (r_{iM} + r_{icM}) / r_{iM} = 1 + r_{icM} / r_{iM} \cong 4$

concurs with (2.12), which in limit:

$\lim_{\Delta r(\Delta t), \Delta t \to 0} [\Delta r_{iM} / r_{iM}) = \dot{r}_{icM} / r_{iM}$

brings the equivalent contribution to integral functional IPF (I.4.4.7), which is growing with increasing the above ratio.

Measure $|M_{io}| = |[\tau] \times [l]| = \pi$ satisfies relations

$[\tau] = \pi / \sqrt{2}, [l] = \sqrt{2}$ at $[\tau]/[l] = \pi / 2$ . (3.40c)

Shortening the cutting time intervals triples density (1.26) of each invariant curving correlation for the minimax impulse (1.25), preserving its measure (3.40).

Since any virtual cutting impulse preserves its virtual measure (3.40b), the related virtual time correlation is able to create the space during the entanglement that triple density measures.

For the invariant impulse that compresses the impulse curvature, the probability of both cutting time interval and emerging space coordinate increases.

The impulse measure $\mathrm{M}_{io} = r_{icM} \times (K_{eo})^{-1}$ .defines correlation multiplied on inverse impulse curvature. But since $\mathrm{M}_{io} = inv = \pi$ , it follows direct connection the correlation with curvature $r_{icM} = \pi K_{eo}$; with a growing correlation curvature increase, and vice versa. Growing the impulse density accompanied with the shortening of cutting time intervals increases the space interval for the invariant impulse measure, but changes the correlation only with the changing curvature. Since the increasing IPF with growing density accompanies the increasing curvature of rotating impulses, the correlations also grow.

After accumulating energy these information curvatures evaluate the impulse information gravity.

## 2.5. HOW THE OBSERVATION'S CUTTING JUMP ROTATES THE MICROPROCESS TIME AND CREATES SPACE INTERVAL

Each observation, processing the interactive impulses, cuts the correlation of random distributions.

The virtual impulse's curved cutting correlations evaluates the entropy measure of the curvature, which with growing probability eventually brings information-physical curvature to a real impulse.

The curved jump of the cutting correlation rotates the impulse time interval starting the impulse microprocess. The jump initiates a multiplicative impulse action $\uparrow_{\delta_k^+} \downarrow_{\tau_k^{-o}}$ on the edge of the starting instance $\tau_k^{-o}$. Applied to the opposite imaginary conjugated entropies, it rotates the entropies with enormous angular speed [41] up to the entanglement. The edge of interval $\tau_k^{-o}$ determines both the jump width-displacement and the curvature forming in the rotation.

The curved time interval $\delta t_{\pm}^{ko} / \tau_k \cong 0.03125$ relative to the impulse time, formed during *the entanglement, turns on beginning a space before the entanglement ends at angle* $\pi / 2$.

That illustrates Fig.1a.

Thus, the time and then space intervals emerge in the interacting impulse as a phase interval, whose probabilistic functions of frequencies enclose a fractional probability of the field available for the observation. The negative curvature of the curved impulse (Fig1b) attracts an observing positive curvature of an interacting impulse. The attraction in the interacting virtual impulses measures the entropy increment of the interacting curvatures as an analogy of a virtual gravitation.

A real impulse' negative curvature attracts energy from the random field necessary to create information, which causes gravitational attraction.
Hence, *the attracting gravitation starts with the creation of space at the entanglement.*
We detail it below.
The interactive impulse microprocess rotates in a transitive movement holding transitive action ↑.
This action, starting on angle of rotation |π/4|, initiates entanglement of the conjugated entropies.
The rotation movement, rotating action ↑ on additional angle, approaching |π/4|, conveys action ↓ that settles a transitional impulse, which finalizes the entanglement at angle| π/2|.
The transitional impulse holds temporal actions ↑↓ opposite to the primary impulse ↓↑ which intends to generate the conjugated entanglement, involved, for example in left and rights rotations ( $\mp$ ).
The transitional impulse, interacting with the opposite correlated entanglements $\mp$, reverses it on $\pm$.
The interacting movement along the impulse boundary ends with cutting the impulse correlation, which carries the potential erasure, becoming a real with delivering an external energy. Since the entropy' impulse is virtual, transition action within this impulse ↑↓ is also virtual and its interaction with the forming correlating entanglement is reversible, as well as the space and attractive entropy gravitation.

Comments 2.6.
The time-interactions are emerging actions of the initial probability field of interacting events at the beginning of the random microprocesses. From this field emerges the first time-correlation and then space coordinates nearby the middle of the impulse, making it possible to deliver the field's energy.
Within the probability field, the emerging initial time has a discrete probability measure, satisfying Kolmogorov law. •
Thus, the time-interactions hold a discrete sequence of impulses carrying entropy, from which emerges a space in the sequence: interactions-correlations–time-space. The sequence of the impulses replicates the frequencies of observation, creating a wave function. The information form of the Schrodinger equation was established in [17] and published in [42].

## 2.6. THE INTERACTING CURVATURES OF STEP-UP AND STEP-DOWN ACTIONS, AND MEMORIZING A BIT

Each impulse (Fig.1a) step-down action has negative curvature (3.39,3.39a ) corresponding attraction, step-up reaction has positive curvature (3.39b) corresponding repulsion, the middle part of the impulse having negative curvature transfers the attraction between these parts.
In the probing virtual observations, the rising Bayes probabilities increase the reality of the interactions bringing energy.
When an external process interacts with the entropy impulse, it injects energy capturing the entropy of the impulse's ending step-up action (Sec.2.3.4). The inter-action with another (internal) process generates its impulse's step-down reaction, modeling 0-1 bit (Fig.2A B).

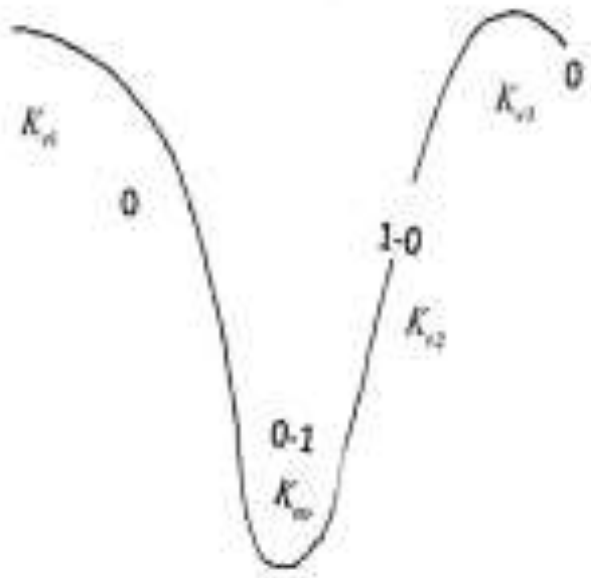

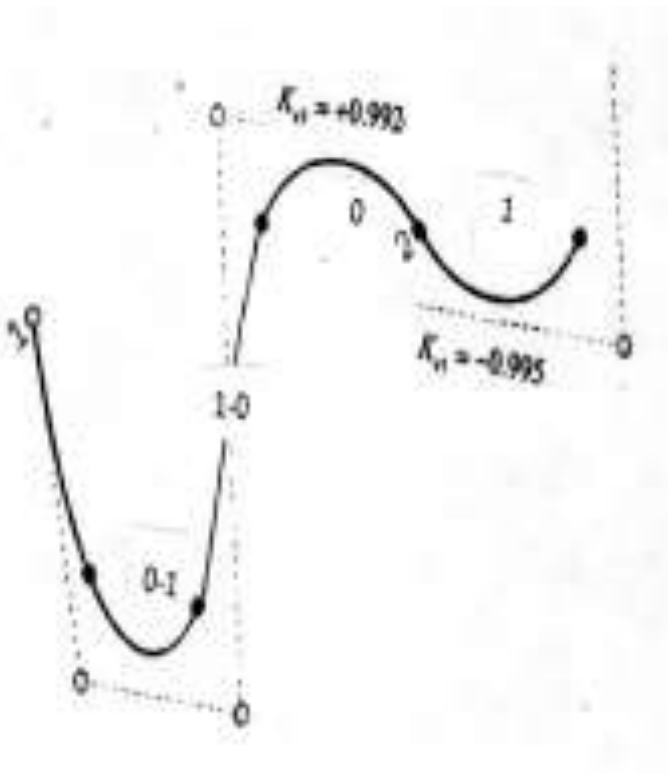

Figure 2A                                                                 Figure 2B

A virtual impulse (Fig.2A) starts step-down action with probability 0 of its potential cutting part; the impulse middle part has a transitional impulse with transitive logical 0-1; the step-up action changes it to 1-0 holding by the end interacting part 0, which, after the inter-active step-down cut, transforms the impulse entropy to information bit.

In Fig. 2B, the impulse Fig. 2A, starting from instance 1 with probability 0, transits at instance 2 during interaction to the interacting impulse with negative curvature $-K_{e1}$ of this impulse step-down action, which is opposite to curvature $+K_{e3}$ of ending the step-up action ($-K_{e1}$ is analogous to that at beginning the impulse Fig.2A). The opposite curved interaction provides a time–space difference (a barrier) between 0 and 1 actions, necessary for creating the Bit. When the interactive process provides Landauer's energy [43] with maximal probability (certainty) 1, the interactive impulse' step-down action ending state memorizes the Bit. Such certain interaction injects the energy overcoming the transitive gap including the barrier toward creation the Bit.

The step-up action of an external (natural) process' curvature $+K_{e3}$ is equivalent of potential entropy $e_o = 0.01847Nat$ which carries entropy $\ln 2$ of the impulse total entropy 1 Nat.

The interacting step-down part of internal process impulse' invariant entropy 1 Nat has potential entropy $1-\ln 2 = e_1$. Actually, this step-down opposite interacting action brings entropy $-0.25Nat$ with anti-symmetric impact $-0.025Nat$ which carries the impulse wide $e_w \cong -0.05Nat$ with total entropy $-0.3Nat$ that equivalent to $-e_1$.

Thus, during the impulse interaction, the initial energy-entropy $W_o = k_B\theta_o e_o$ changes to $W_1 = -k_B\theta_1 e_1$, since the interacting parts of the impulses have opposite-positive and negative curvatures accordingly; the first one repulses, the second attracts the energies. For erasing the Bit equivalent to the internal impulse minimal entropy $e_{10} = \ln 2$, the needed Landauer's energy is $W = k_B\theta\ln 2$.

If the interactive internal process accepts this Bit by memorizing (through erasure), the above Landauer energy should compensate the difference of these energies-entropy: $W_o - W_1 = W_B$ in balance form

$$k_B\theta_o e_o + k_B\theta_1 e_1 = k_B\theta\ln 2. \quad (3.41)$$

Assuming the interactive process supplies the energy $W_B$ at moment $t_1$ of appearance of the interacting Bit, we get $k_B\theta_1(t_1) = k_B\theta(t_1)$. That brings (3.41) to form

$$k_B\theta_o 0.01847 + k_B\theta(1-\ln 2) = k_B\theta\ln 2, \; \theta_o/\theta = (2\ln 2 - 1)/0.01847 = 20.91469199. \quad (3.42)$$

The opposite curved interaction decreases the ratio of above temperatures on increment $\ln 2/0.0187-(2\ln 2-1)/0.01847=16.61357983$, with ratio

$$(2\ln 2-1)/\ln 2 \cong 0.5573. \qquad (3.42a)$$

Natural impulse with maximal entropy density $e_{do}=1/0.01847=54.14185$ interacting with internal curved impulse transfers minimal entropy density $e_{d1}=\ln 2/0.01847=37.52827182$.

Ratio of these densities $k_d=e_{do}/e_{d1}=1.44269041$ equals

$$k_d=1/\ln 2 \qquad (3.43)$$

which identifies single impulse 1 measured by $k_d$ bits-or $1Nat$.

Hence, that ratio enables the production of the information impulse $1Nat=1.44bit$.

Here the interacting curvature, enclosing entropy density (3.43), lowers the initial energy and the related temperatures in the above ratio. From that follow

<u>Conditions creating a bit in interacting curved impulse</u>

1. The opposite curving impulses in the interactive transition require keeping entropy ratio 1/ln2.
2. The interacting process should possess the Landauer energy by the moment ending the interaction.
3. The interacting impulse should hold invariant measure $M=[1]$ of entropy 1 Nat whose the topological metric preserves the opposite curvatures.•

The last follows from the impulse' max-min mini-max law under its stepdown-stepup actions, which generate invariant [1]Nat's time-space measure' topological metric π(1/2circle) preserving opposite curvatures.

Results [44] prove that a physical process which holds the invariant entropy measure for each phase space volume (for example, minimal phase volume $v_{eo}\cong 1.242$ per a process dimension [41]), the above topological invariant characterizes and satisfies the Second Thermodynamic Law. It also shows that "decreasing entropy and negative entropy production arises in arbitrary coordinates", applied to self-organizing systems. Energy $W_B$ that delivers the external (natural) process will erase the entropy of both attracting and repulsive movements, covering energy of the both movements, which are ending at the impulse stopping states. The erased impulse total cutoff entropy is memorizes as equivalent information, encoding the impulse Bit in the impulse ending state.

The ending logic of natural step-up action captures its entropy, moving along the action positive curvature, transits to interacting step-down action' negative curvature, and by overcoming entropy-information gap [28] acquires the equal information that compensates for the movements logical cost.

Thus the attractive logic of an invariant impulse, converting its entropy to information within the impulse, performs the function of a *logical Maxwell's Demon* (MD) in the microprocess.

<u>Topological transitivity at the curving interactions</u>

The impulse of the external process holds its $1Nat$ transitive entropy until its ending curved part interacts, creating an information bit during the interaction.

Theoretically, when a cutting maximum of entropy reaches a minimum at the end of the impulse, the interaction can occur, converting the entropy to Information by getting energy from the interactive process.

<u>Comments 2.7</u>

Each cutting correlation on the EF time-space extremal produces Kronicker impulse, and the sequence of these correlations transits these impulses along a topology of the extremal.

The EF integrates entropy of each impulse, and the IPF integrates the impulses' information which every increment of energy coverts. That leads to the impulse consuming energy along the IPF extremal. This topological transition preserves both geometrical and entropy measures of the transiting Kronicker impulses.
Topological specific defines Euler invariant depending on the number of holes on the topological space, which determine the space dimension that only distinguishes the topological spaces.
Euler entropy [93] measures how many holes are in the broken extremal.
The EF-IPF integrates the manifold of cutting extremals covering the Kronicker impulses, and the rotating dynamics are sequentially joining the manifold in a single dimensional extremal (Sec.5.2).
When the observing Bayes probability approaches one, only a single curved Kronicker impulse holds the EF entropy measure whose the Euler's topological entropy is zero.
The considering topological transition runs through the curved Kronicker impulse, which measures its Euclid's radius curvature. This topological transition, connecting impulse topology with curvature, preserves both geometrical and entropy measures of the transiting Kronicker impulse.
The count curvature of that Kronicker impulse is asymmetrical (Fig.2).
Since the final Bayes entropy leaves only a single impulse before an influx of energy converts this entropy to the equivalent information, that Bayes entropy coincides with the impulse topological entropy. Before injection of energy on the entropy-information gap, only the Bayesian logic holds the single impulse. And that logic of the asymmetrical curvature measures asymmetrical entropy logic. That asymmetrical logic will be memorized as information logic through injection of the energy on the gap edge. Thus, the erasure of asymmetrical logic preempts forming a Bit. •

The invariant's topological transitivity has a duplication point (a transitive base) where one dense form changes to its conjugated form during orthogonal transition of hitting time.
During the transition, the invariant holds its measure (Fig.1b) preserving its total energy, while the densities of these energies are changing.
The topological transition separates (on the transitive base) both primary dense form and its conjugate dense form, while this transition turns the conjugated form to orthogonal.
At the transition-turning moment, a jump of the time curvature switches to a space curvature (Fig.1a) with potential rising space waves in a microprocess above. •
As a distinction from the traditional Maxwell's Demon, which uses an energy difference in temperature form [45], this approach reveals a Maxwell's Demon through the naturally created difference of the curvatures.
Forming a transitional impulse with the entangled qubits leads to possibility of memorizing them as a quantum bit. That requires first to provide the asymmetry of the entangled qubits, which starts the anti-symmetric impact through the main impulse step-down action $\downarrow$ interacting with opposite action $\uparrow$ of starting the transitional impulse. This primary anti-symmetric impact $-0.025\times 2 = -0.05 Nat$ starts curving both the main observing and transitional impulses with curvature $K_{e1} \simeq -0.995037$, enclosing $0.025 Nat$, while the starting step-up action of the transitional impulse generates curvature $K_{e2} \simeq +0.993362$ enclosing $e_o = 0.01847 Nat$. Difference $(0.025\text{-}0.01847) Nat$ estimates entropy measuring total asymmetry of main impulse $0.00653 Nat = S_{as}$.
Entangled qubits in the transitional impulse evaluate the entropy volume 0.0636 Nat, which spends its entropy on transfer the minimal entangle phase volume $v_{eo} \cong 1.242$ to the entropy-information gap; while the primary impact brings minimal entropy $0.05Nat$ starting the entangled curved correlation. Thus, the correlated curved entanglement can memorize $(0.05-0.0656)Nat$ in the equivalent information of two qubits. The middle part of the main impulse generates curvature $K_{e2}\simeq +0.993362$ which encloses entropy $0.02895Nat.$ The difference $0.02895-0.025=0.00395Nat$ adds asymmetry to the starting transitional entropy, while $0.02895-0.01845=0.0105Nat$ estimates the difference between the final asymmetry of the main impulse and the ended asymmetry of transitional impulse.

With starting entropy of the curved transitional impulse $0.05Nat$, the ending entropy of the transitional impulse asymmetry estimates(in Nat):

$0.0653\text{-}0.0105\text{-}0.00395=0.05085.$

Memorizing this asymmetry requires compensation from a source of equivalent energy. It could be supplied by opposite actions of the transitional step-down $\downarrow$ and main step-up interacting action $\uparrow$ ending the transitional impulse. That action will create the needed curvature at the end of the main impulse, adding $0.0653-0.05085=0.01445, 0.01445\text{-}0.0105=0.00395$ to entropy of transitional impulse curvature sum $0.05085$. Another part 0.0105 will bring the difference of entropy' curvature $0.02895-0.01845=0.0105$ with total 0.0653.

Thus, $0.05085Nat=s_{as}$ is entropy of asymmetry of entropy volume $s_{ev}=0.0636$ of transitional impulse, whereas $0.0653Nat=S_{as}$ is the entropy of the asymmetry of the main impulse.

This asymmetry generates the same entangled entropy volume that the step-action of the main impulse transfers for interacting with the external impulse. Thus, $s_{as}$ is the Information "demon cost" for the entangled correlation, which the curvature of the transitional impulse encloses.

The asymmetrical curvature of transitional impulse, holding the entangled volume, encloses the entangled correlation. Instead of direct evaluation of this correlation, allows memorizing the Information of two qubits in impulse measure 1 Nat.

That evaluation is closed to [46], obtained differently, and confirmed experimentally.

When the posteriori probability is closed to reality, the impulse positive curvature of step-up action, interacting with the merging impulse' negative curvatures of step-down action, transits a real interactive energy, which the opposite asymmetrical curvatures actions enfolds.

During curved interaction this primary virtual asymmetry compensates the asymmetrical curvature of a real external impulse, and that real asymmetry is memorized through the erasure by the supplied external Landauer's energy.

The ending action of the external impulse creates a classical bit with probability

$. P_k=\exp[-(0.0636)^2]=0.99596321.$

Since the entanglement in the transitional impulse creates an entropy volume $0.0636$, the potential memorized pair of qubits has the same probability.

Therefore, both the memorized classical bit and pair of qubits occur in a probabilistic process with high probability but less than 1.

The question is how to memorize the entropy enclosed in the correlated entanglement, which naturally holds this entropy and therefore has the same probability?

If a transitional impulse, created during interaction, has such a high probability, then its curvature holds the needed asymmetry, and it should be preserved for multiple encoding with the identified difference of the locations of both entangled qubits.

Information of the memorized qubits can be produced through interaction, which generates the qubits within a material-device (a conductor-transmitter) that preserves the curvature of the transitional impulse in a Black Box, by analogy with [47].
At such an invariant interaction, the multiple connected conductors memorize the qubits' code.
The needed memory of the transitional curved impulse encloses entropy $0.05085 Nat$.

The time intervals of the curved interaction
If the natural space action curves the internal interactive part, the joint interactive time-space curved action measures its interactive impact.
If the interaction at moment $t_o$ creates internal curvature $K_{e1} \simeq -0.995037$ enclosing $-0.025 Nat$ by moment $t_{o1} = 0.01845 Nat$, the interacting time-space interval measures the difference of these intervals $|t_{o1}| - t_o / |t_{o1}| - t_o| \cong 0.1449$ $0.0250 - 0.01847 = 0.00653 Nat.$
For that case, the internal curved inter-action attracts the energy of natural interactive action.
By the moment $t_1$ of appearance of the interacting Bit, ratio (3.43) selects part of the information impulse $i_{11} \cong (1.44 - \ln 2) \times 0.5573 \cong 0.2452 bit$ which the curve interaction deducts from the internal impulse' Bit. The anti-symmetric interaction involves middle part of the internal impulse with the asymmetry of curvature $K_{e2} \cong +0.993362$ which encloses entropy $0.02895 Nat.$
Difference $0.02895 - 0.025 = 0.00395 Nat$ adds asymmetry to the starting transitional entropy, while $0.02895 - 0.01845 = 0.0105$ estimates the difference between the final asymmetry of the main impulse and ended asymmetry of transitional impulse.
Taking into account the asymmetry information $i_{13} \cong 0.0105 \times 1.44 = 0.015 bit$, we get information

$$i_f \cong 0.2452 - 0.015 \cong 0.23 bit \qquad (3.44)$$

evaluating the total asymmetrical increment of the curved interaction. This is free information created in addition to the Bit, which measures the attracting action of the asymmetrical interaction. It amount simply evaluates~1/3/Bit. If No part of the interacting impulse emerges at $t_o$ and Yes part arises by $t_1$, then the invariant interacting impulse will spend $1 - \ln 2 + \ln 2 = 1$ Nat on creation bit ( $\ln 2 Nat$ ).
If inter-action of the natural process on the internal process delivers energy $W_B$ by moment $t_1$, this energy will erase the bit and memorize it according to the balance relations.
The interacting impulse spends ~1 Nat on creating and memorizing bit $\ln 2$ holding free information $(1 - \ln 2) \cong 0.3 Nat$. The curved topology of interacting impulses decreases the needed energy ratio, according to the balance relation above. Thus, the time interval $t_o - t_1$ creates the bit and performs the *Maxwell's Demon* (MD) function.

Comments 2.8. Interaction of two Kronicker impulses produces a curve analogical to potential of a neuron axon. The interaction predicts Hodgkin-Huxley equations [96].
The second interacting impulse on Fig.2B is analogous to the axon potential whose threshold models the between impulse barrier, and comparative ratios of the time intervals on Fig.2.B:
$(|t_{o1}| - t_o) / |(t_{o1}| + t_o|) \cong 0.1449$ is closed to the relative ratio for the axon: $(|T_{o1}| - T)_o / (|T_{o1}| + T_o) \cong 0.13$ . • .
Since the movement within the internal impulse ends at the impulse step-up stopping states, the thermodynamic process delivering this energy should stop in that state. Hence, the erased impulse cutoff entropy memorizes the equivalent information $1.44 bit$ in the impulse ending state. It includes $1.44 - 1.23 = 0.21$ where $0.21 \times 1.44 \cong 0.3 Nat$ is transferred to the next interacting impulse as the equivalent to is $-e_1$. In the ending observer's probing logic, such a curving interaction moving along the negative curvature of its last *a priori* step-up action, overcomes the gap by moving along the positive curvature of the *a posteriori* step-down. It acquires Information that compensates for the logical cost of the movement. Thus, the attractive free information logics of an invariant impulse, converting its entropy to Information within the impulse, performs the function of a *logical* (MD) in the microprocess. Coordination of an observer's external time-space scale with its internal time-space scale happens when an external step-down jump action interacts with the observer's inner thermodynamic time-space interval, which, in the curved interaction, measures the difference of the time (3.44).

Ratio $[\tau]/[l]=\pi/2$ leads to $\Delta l_{10}=2\Delta t_{10}/\pi, \Delta l_{10}\cong 0.00415 Nat$.

*Thus, the curvature of the rotating impulse encloses its time and space.*

The interacting jump injects energy, capturing the entropy of impulse's ending step-up action. This interaction models the 0-1 bit. The opposite curved interaction provides a time–space difference (an asymmetrical barrier) between 0 and 1 actions, necessary for creating the Bit.

The interactive impulse' step-down ending state memorizes the Bit when the observer interactive process provides Landauer's energy with maximal probability.

Within the microprocess (Sec.2.3.1) evolves a thermodynamic process, which generates the opposite jumps creating the transitional impulse above.

Both microprocess and thermodynamic process develop along the observing process.

Applying the Jarzynski equality (JE) of irreversible thermodynamic transition [39] to conversion energy in Information, and using results of its experimental verification [48], lead to the JE in form

$$e^{\Delta F/k_B\theta} - < e^{W/k_B\theta} >= \gamma. \quad (3.45)$$

Here $\Delta F$ is increment of free energy needed to produce energy $W, \gamma$ is parameter of the verification, which defines sum of probabilities $P_\Sigma \Rightarrow \gamma$ that an inverse trajectory is observed in the experiments.

At $\gamma=1$, the JE satisfies exactly:

$$e^{\Delta F/k_B\theta} - < e^{W/k_B\theta} >= 1, P_\Sigma = 1. \quad (3.45a)$$

A thermodynamic process, satisfying the JE for all its states sequence, evolves irreversibly in the microprocess enables generation qubits or Bit.

The quantity of information $I_{\delta t}$ at a curved cut has equivalent of free energy $\Delta F=\Delta F_{\delta t}$ to produce energy $W_t$ during a fixed time $\delta_t$, satisfying relation

$$\Delta F_{\delta t} = k_B \theta I_{\delta t}. \quad (3.45b)$$

Average exponential energy $<\exp W/k_b\theta>$, collected during natural interactions at interval $\delta T_t$, measures the EF (1.10) integral increment $\Delta S_{\delta t}$:

$$<\exp W/k_b\theta>=\exp \Delta S_{\delta t}. \quad (3.45c)$$

Since EF counts also curved observing probabilistic impulses, the average energy includes the curved impulses. Erasing that entropy by the natural source' high quality energy brings a non-random information $I_{\delta t}$ measures of the curved impulse during $\delta_t < \delta T_t$.

Influx of impulse energy $\Delta F_{\delta t}$ at $\delta_t$ enables converting entropy $\Delta S_{\delta t}$ to equivalent information $I_{\delta t}$.

Substituting $I_{\delta t}$ from(3.45b)and $\Delta S_{\delta t}$ from (3.45c) the JE(3.45a):

$$\exp I_{\delta t} - \exp \Delta S_{\delta t} = 1, \quad (3.45d)$$

brings the equivalence of JE in both formulas for transition the entropy to information at probability $P_\Sigma = 1$.

During the thermodynamic process, a free energy compensates the entropy dissipative energy.
By emerging real time on cutting intervals $\delta_t$ a certain logic with $I_{\delta t}$ appears with probability $P_{\Sigma} = 1$, which brings entropy $\Delta S_{\delta t} = 0$ (relates to $\ln(P_{\Sigma} = 1) \Rightarrow \Delta S_{\delta t} = 0$)).
Substitution that in the previous equation leads to $\exp I_{\delta t} - 1 = 1, I_{\delta t} = \ln 2$ which is equivalent to free energy $\Delta F_{\delta t} = k_B \theta \ln 2$.
Thermodynamic conversion of entropy to information Bit occurs during a probability approaching one when the entropy jumps from ln2 to zero, while related information jumps from zero to ln2.
While the first jump starts the microprocess with its thermodynamics, the second opposite jump occurs when the probability approaches to one.
*It confirms our initial statement about connection the entropy and information and condition of their conversion.* The equivalence of JE in both formulas for the transition to information requires invariant $I_{\delta t} = 1 Bit$, the unit of information per impulse to compensate for the Maxwell Demon energy at the time interval of transition.

Thus, to satisfy the MD, the information produced by each impulse time interval should be invariant, holding constant the unit (Bit, Nat) in $I_{\delta t}$.

It confirms that the impulse minimax extreme principle (EP) satisfies the JE for impulse information transition, or vice versa. Each impulse time interval enables encoding invariant unit of information.

Or, the EP follows from the JE in the physical process whose interactive time interval is an equivalent of the impulse information cutting from the correlation carrying the above energy.

The cutting correlation's time intervals hold the information equivalent of this energy, and any real time interval of interaction brings the entropy equivalent of energy $\Delta F_{\delta t}$ which compensates for the MD while producing information during the interaction.

In interactive random process, whose sequence of cuts satisfy the EP, each impulse encodes the cutting correlation, and all information of the process cutoff correlations encodes the information process fulfilling the minimax law which is independent on size and curvature of any impulse.

Moreover, with probability $P_{\Sigma} = 1$, the inverse conjugated trajectories of the interacting impulses in the microprocess are part of the natural interacting process.
That exactly satisfy the JE initial conditions [48, 49].
The evolving microprocess starts with probability (Sec.2.3.3) and relational entropy of inverse states $S^*_{\mp a} = 2$.
The 0-1 entropy units (potential bit) of the microprocess impulse connect the impulse inner correlation, while 01-0-1 entropy entities (a potential qubit) bind the microprocess' entanglement.
Such multiple microprocesses, which the observation generates, hold the statistical thermodynamic process where the JE automatically measures the energy of these impulse discrete units.
*Relations (3.45b,3.45c, 3.45d) had applied the JE for the first time in [50] for measuring energy within the impulse microprocess (quantum) connecting the JE with encoding this process' information measure at the cutting correlation. The random interactions on the path to the generation of Information naturally average the impulse microprocess' dissipative work in the JE thermodynamics.*
*The curved impulse thermodynamics on the rotating microprocess trajectories describe forming physical micro units encoding qubits, bits.*
The information process's last cutting impulse encodes the process's total Information integrated in its IPF.
Thus, the JE enables measuring energy of the entropy-information unit in both statistical thermodynamic microprocess and encoding thermodynamic macroprocess.
This approach is distinct from other JE applications by averaging the work in the JE during the evolving observations naturally, while others need multiple experiments and specific procedures of averaging their results.

Multiple interactions generate a code of the interacting process at the following conditions:

1. Each impulse holds an invariant probability-entropy measure, satisfying the Bit conditions.
2. The impulse interactive process which delivers the code is a part of a real physical process that maintains this invariant entropy-energy measure. That process memorizes the Bit and creates an Information process of multiple encoded Bits. By attracting free Information, they build the process's information dynamic structure.
(For example, water, cooling interacting drops of hot oils in the found ratio of temperatures, enables spending energy of its chemical components to encode the components chemical structures. Or the water kinetic energy will carry the multiple drops' bits as an arising information dynamic flow.
Such a physical -chemical process supplies the needed energy to generate the code.)
3. Building the multiple Bits code requires increasing the impulse information density three times with each following impulse acting on the interacting process (Sec 2.1.1).
To create a code of the Bits, each interactive impulse, producing a Bit, should follow three impulses measure π, i.e. frequency of interactive impulse should be f=1/3 π=~0.1061. •
The interval 3π provides the opportunity to join the impulses of three Bits in a triplet as an elementary macro unit. It combats the noise and redundancies from both the internal and external processes.
Natural encoding merges memory with the time of memorization, compensating the cutting cost by running time intervals of encoding.
The encoding process, preserving the invariant cutting information, connects its multiple bits or qubits in the invariant irreversible thermodynamics where each such discrete information unit's energy measures the JE.
Multiplication mass M on curvature $K_{e2}$ of the impulse equals to relative density Nat/ Bit=1.44 which determines M=1.44/ $K_{e2}$. At $K_{e2}$=0.993362, we get a relative mass M=1.452335645.
The opposite curved interaction lowers the potential energy, compared to other interactions for generating a bit. The multiple curving interactions create topological Bits code, which sequentially forms a moving spiral structure [16].

<u>How to find an invariant energy measure, which each Bit encloses starting Maxwell's Demon?</u>
Since its minimal energy is $W = k_B \theta \ln 2$, it's possible to find such temperature $\theta_1^o$ that is equal to inverse value of $k_B$. If the interacting process carries this temperature, then its minimal energy holds

$$W_1^o = \ln 2 \; at \; \theta_1^o = 1 / k_B, \qquad (3.46)$$

which equals the invariant time-space Nat measure of physical bit , or its entropy logic measure.
Let us evaluate $\theta_1^o$ at $k_B = 8617 \times 10^{-5} \mathrm{eV/K}$ and Kelvin temperature $K = 20/293 = 0.0682259386^{oC/K}$ equivalent to $20^{oC}$. Then $\theta_1^o = 588.19 \times 10^5 /\mathrm{eV}$.
If we assume that this primary natural energy brings the eV amount equivalent to quanta of light: $e_q = 1240\mathrm{eVnm}$, $1\mathrm{nm} = 10^{-9} m$, then we come to $\theta_1^o = 588.19 \times 10^5 \times 1.240 \times 10^3 / e_q \times 10^{-9} m \cong 72.9356^{oC/m} / e_q$.
Or each quant brings temperature' density $\theta_1^o = 72.9356^{oC/m}$, which is reasonably real.
With this $\theta_o^o$, the interacting impulse will bring energy $W_1^o = \ln 2$ to create its bit.
Following the balance relation, the external process at this $\theta_o^o$ holds temperature $\theta_o^o = 20.914469199 \theta_1^o = 1525,42^{oC/m}$ brought by a quant. This energy holds invariant impulse measure $|1|_M = 1Nat$ with metric π, or each such impulse has entropy density $1Nat/\pi$. The bit of the interacting impulse has minimal density energy equivalent to $\ln 2/\pi = 0.22$ at temperature $\theta_1^o$.
In cognitive dynamics [15], it allows spending energy ln2 for erasure of the observing bit and memorizes the equivalent cognitive quantity equal to Landauer's bit by the neuron information bits.

*With such energy, the information attraction-gravitation imitates free information* $0.23bit$ *enables attracting actions.* Therefore, the curving interaction dynamically encodes Bits in a *natural process*, developing the Information structure Fig.3 of the interacting Information process.
The growing curvature of the impulses during rotation increases the density of the Bit.
*The rotating thermodynamic process with minimal Landauer energy performs the natural memorizing of each natural Bit.*
The invariant Information-cutting process holds the invariant irreversible thermodynamics measured through piece-wise Hamiltonian and diffusion-kinetic matrix equations [26] $I_f = L_t X_t, L_t = 2b_t$ where $I_f$ is diffusion-kinetic flows, $X_t$ is thermodynamic forces, $b_t$ and is diffusion matrix. At $L_t \geq 2b_t$ kinetic flow transfers to diffusion $2b_t$ at $L_t \leq 2b_t$ the diffusion flow transfers to kinetics, where the transformation applies on a small $\varepsilon$-localities of the bordered impulse. These conditions are found in [34] where, at kinetic transition, the Hamiltonian includes increments of chemical potentials of interacting physical-chemical entities.
Hence, the JE with both connections to irreversible thermodynamic and kinetics describes increments of temperature, entropy, energy, diffusion, and physical-chemical components in the variety of thermodynamic processes within interacting impulses and their cooperative macrodynamics.
The Information macrodynamics describes all these through the equivalent Information parameters.

## 2.7. INTENTIONAL BINDING

From the EF extreme variation equations [26] follows the equation for entropy force on an extreme trajectory:

$$X_{o1}(\tau_k^1, t_k) = (2\mathbf{a}_o / t_k)^2 \exp(2\mathbf{a}_o) x(\tau_{k1}^o). \qquad (3.47)$$

Here at a fixed state $x(\tau_{k1}^o)$ and impulse invariant measure $\mathbf{a}_o$, the inverse ratio $(2\mathbf{a}_o / t_k)$ decreases each current time intervals $t_k$ between the impulses. And the force grows in square function of inverse time intervals. On the limit, at $t_k \to 0$, (3.57) it generates the infinite force:
$\lim_{t_k \to 0} X_{o1}(\tau_k^1, t_k) \to \infty$. This involves potentially merging the nearest impulses's interactive actions and starting the microprocess. This means that the microprocess follows from the EF variation principle. Even on a trajectory of probabilistic observations, a prior action could bypass the probabilistic impulse microprocess. Such observation pulls together the action and its result.
The related macroprocess, averaging such impulse's microprocess, overruns that impulse, bringing together its *a priori* and *a posteriori* actions.
This phenomenon, arising on the Information macrotrajectory, may appear in an Observer as Information brain processing, where it becomes known as *intentional binding* [51].

# III. **INFORMATION DYNAMIC PROCESSES DETERMINED BY THE EXTREME OF EF AND IPF FUNCTIONALS**

## 3.1. THE EF-IPF CONNECTION

Since the IPF Functional integrates a finite amount of Information and converges with the Entropy Functional, the EF covers both the contributions of the cutoff Information and the entropy increments between them.
Here is an explanation how the EF-IPF works for the $n$-dimensional Markov process approaching infinity. The IPF at $n \to \infty$ integrates an unlimited discrete sequence of the EF cutoff fractions.
It is a difficult mathematical task to integrate the discrete fractions and solve a classical variation problem for the IPF, in order to find the continuous extreme Iinformation Dynamics.

The EF describes the potential integral Information Functional of the Markov process until the high probability impulse is applied. Carrying the cutoff increments, it is transformed to a physical IPF.
The IPF maximal limit approaching EF at $o(t \to T) \to 0, n \to \infty$ avoids direct access to the Markov random process.
The extreme of this integral describes a dynamic process $x(t)$ which approximates a random $\tilde{x}_t$ transformation to $\varsigma_t$ that evaluates the dynamic transition to a Feller kernel.
Process $x(t)$ carries Information, collected by the maximal IPF at $n \to \infty$, and describes the IPF information dynamic macroprocess decreasing at each of its following time interval

$$\Delta_t = (t - s) \to o(t) . \quad (1.1)$$

The increment of the EF at the end of interval $o_m \to 0$ approaches zero, satisfying in limit:

$$\lim_{t_m = T} \Delta S_m[\tilde{x}_t(\tau_m \to t_m))] \to 0 . \quad (1.2)$$

The IPF extracts the finite amount of integral information on all cutoff intervals, approaching $S[\tilde{x}_t / \varsigma_t]$ value it cuts.
The sequential cuts during time $(T - s)$ break the process correlations and the EF functional connections, transforming the initial random process to a limited sequence of independent states.
The IPF extracts the extreme process which approaches the EF extreme trajectories, while the IPF collected information approaches its source, which is measured by the EF.
Only the impulse Yes-No actions, releasing information from the EF, make the IPF information feasible for the observer in the form of Bits. These Bits allow the impulse communication with a following observing process integrated by the EF.

## 3.2.ESTIMATION OF THE EXTREME PROCESS

Mathematical expectations of Markov drift in stochastic Eq. (Sec.1.1):

$$E[a] = \dot{\bar{x}}(t) = E[c\tilde{x}(t)] = cE[\tilde{x}(t)] = c\bar{x}(t) \quad (2.1)$$

approximates a regular differential Eq.

$$\dot{\bar{x}}(t) = c\bar{x}(t) , \quad (2.2)$$

whose common solution averages the random movement by a dynamic *macroprocess* $\bar{x}(t)$:

$$\bar{x}(t) = \bar{x}(s)\exp ct, \bar{x}(s) = E[\tilde{x}(s)] . \quad (2.3)$$

Within discrete $o(t) = \delta_o$, the opposite controls $u_+, u_-$, satisfying relations (Sec.2.1):

$$c^2 = |u_+ u_-| = c_+ c_- = \bar{u}^2, c_+ = u_+, c_- = u_-, \ |u_+ u_-| = \bar{u}^2$$

are opposite discrete conjugated complex:

$$u_+ = j\bar{u}, u_- = -j\bar{u} . \quad (2.4)$$

Conditions 2.1.1A,B of Sec.2.1 are fulfilled at

$$\bar{u} = -2j . \quad (2.4a)$$

when

$$u_+ u_- = j\bar{u}(-j\bar{u}) = \bar{u}^2, \ -j\bar{u} - (+j\bar{u}) = -2j\bar{u} , \ \bar{u}^2 = -2j\bar{u}, \bar{u} = -2j . \quad (2.4b)$$

The controls are real when

$$u_+ = j(-2j) = 2, u_- = -j(-2j) = -2 . \quad (2.5)$$

Relations (2.3), (2.4) satisfy two conjugated differential equations

$$\dot{x}_+(t) = j\bar{u}x_+(t), \dot{x}_-(t) = -j\bar{u}x_-(t) \quad (2.6)$$

describing a microprocess ( $x_+(t)$ , $x_-(t)$ ) under controls (2.4, 2.5) on time interval $\Delta_t = t - s$ , $\Delta_t \to o(t)$ .
Solutions of (2.6) take the forms

$$\ln x_+(t) = Cu_+t, \ln x_-(t) = Cu_-t, x_+(t) = C\exp(j\bar{u}t), x_-(t) = C\exp(-j\bar{u}t), \mathrm{C} = x_-(s^{+o}) = x_+(s^{+o}), \quad (2.7)$$

$$x_+(t) = x_+(s^{+o})(\mathrm{Cos}\,\bar{u}t + jSin\bar{u}t), x_-(t) = x_-(s^{+o})(\mathrm{Cos}(\bar{u}t - jSin\bar{u}t) \quad . \quad (2.7a)$$

The correlation function for the microprocess conjugated solution (2.7a) at $Cos^2(\bar{u}t) + \mathrm{Sin}^2(\bar{u}t) = 1$ holds

$$r(x_+(t), x_-(t)) = r_s = x_+(s^{+o}) \times x_-(s^{+o}) \quad . \quad (2.7b)$$

During this fixed correlation, the conjugated anti-symmetric entropies (2.3.6) interact, producing an entropy flow (2.3.16).
Correlation (2.7b) depends on the interaction on a border edge of the impulse at moments $\delta^{+o}/4, \delta^{+o}/2$.
Applying formula [1:27] for the correlation between moments $\delta^{+o}/4, \delta^{+o}/2$ leads to

$$r_s = \sqrt{(\delta^{+o}/4)(/\delta^{+o}/2)}\,, \; r_s = \sqrt{0.5} \quad . \quad (2.7c)$$

Within this correlation acts the impulse $[1_{\delta_k^{\tau+}/4}^{\delta_k^{\tau+}/2}]\bar{u}_k, \bar{u}_k = \mid \pm 1 \mid_k$ which includes both opposite conjugated controls instantly.
If a real control, cutting the influx of entropy at this correlation, does not compensate it, the states' correlation is not dissolved, and the states, carrying both opposite controls, will hold during the process's correlation.
These correlated states might be entangling states of a hidden microprocess within the entropy-information gap balancing the anti-symmetric actions.
Solutions (2.7a) describe a microprocess at $o(t) \to o(\tau_n^{-o})$ comparable to the impulse macroprocess (2.3) averaging (2.7a) at these intervals.
The microprocess becomes an inner part of the dynamics process, minimizing time distance (1.1.1), when the time intervals satisfy the optimal time (Prop.2.12) running between the impulse cutoff information at

$$\tau_k^{-o}/\tau_k^{+o} = 3, \delta_k = \tau_k^{+o} - \tau_k^{-o} = 2\tau_k^{-o}, \tau_k^{-o} = 3\delta_k/2 . \quad (2.8)$$

It implies that the microprocess (2.7a) imaginary time intervals between the impulses triples the cutoff's discrete intervals:

$$\Delta_k = 3\delta_k\,, \quad (2.8a)$$

Dynamic process $x(t)$ detewrmines solution of (2.3) under real control (2.5) starting at moment $t = t^e$ :

$$x_\pm(t^e) = x(s^+)\exp(u_\pm t^e)\,, \; t^e = s_k^{+o} b_k(t)/b_k(s_k^{+o}), \; u_\pm = \pm 2 \quad . \quad (2.9)$$

This process approximates an extreme of the Entropy Functional within each $\Delta_t = t - s$.
The solutions for $x(t)$ in form

$$dx(t)/x(t) = cdt, \ln x(t) = ct, t = s_k^{+o} b_k(t)/b_k(s_k^{+o}), \ln x(t) = cs_k^{+o} b_k(t)/b_k(s_k^{+o}) \quad (2.10)$$

starting at time $t = t^e$, integrate on minimal time distance $\Delta_t = t - s$ in the process:

$$x(t) = \exp[cs_k^{+o} b_k(t)/b_k(s_k^{+o})], x(s_k^{+o}) = \exp(cs_k^{+o}), c = \ln(x(s_k^{+o}))/s_k^{+o}, x(s_k^{+o}) = \bar{x}(s_k^{+o}),$$

$$x(t) = \exp[\ln(x(s_k^{+o}))b_k(t)/b_k(s_k^{+o})] = \exp[\ln(x(s_k^{+o}))t/s)], \ln x(t) = \ln(x(s_k^{+o}))t/s\,, \quad (2.10a)$$

which at $t \to T$ approaches

$$\ln x(T) = \lim_{t\to T}[\ln(x(s_k^{+o}))T/s], x(T) \to x(s_k^{+o}))T/s \,. \quad (2.10b)$$

Process $x(t)$ in (2.3), (2.10b) is the extreme solution of *macroprocess* $\overline{x}(t)$ which averages the solution of related stochastic Eq. under the optimal controls.
The solution integrates the cutoff increments from the EF for $n$-dimensional Markov process within intervals

$$\Delta_t = (t - s) \to o(t) . \qquad (2.11)$$

Thus, process $x(t)$ carries the EF increments, while the information dynamic macroprocess collects the maximal IPF cutoff contributions on each interval (2.11) at $o(t) \to 0$, $n \to \infty$.
Information, collected from the diffusion process by the IPF, approaches the EF entropy functional. The EF extremals formalize the minimax probabilistic trajectories evolving to the IPF extremals.
Finding extreme process $x(t)$ which the EF generates requires solution of the EF variation problem.

## 3.3. SOLVING THE VARIATION PROBLEM FOR THE ENTROPY FUNCTIONAL

Applying the integral functional form of entropy functional (1.1.10):

$$S = \int_s^T L(t, x, \dot{x}) dt = S[x_t] \;, \qquad (3.1)$$

allows formulating the variation problem, minimizing the Entropy Functional of the diffusion process:

$$\min_{u_t \in KC(\Delta, U)} \tilde{S}[\tilde{x}_t(u)] = S[x_t] \text{ on } Q \in KC(\Delta, R^n) . \qquad (3.1a)$$

Specifically, for integral (1.3.2.1), at (1.3.22), it leads to the variation problem

$$\operatorname{extr} S[\tilde{x}_t / \varsigma_t] = \underset{c^2(t)}{\operatorname{extr}} 1/2 \int_s^T c^2(t) A(t, s) dt \,, \; c^2(t) = \dot{x}(t) . \qquad (3.1b)$$

<u>Proposition 3.1.</u>
1. An *extremal solution* of variation problem (3.1a, 3.1) for the Entropy Functional (1.1.10) brings the following equations of extremals for vector $x$ and conjugates vector $X$ accordingly:

$$\dot{x} = a^u \,, \; a^u = a(u, t, x) \; (t, x) \in Q \;, \qquad (3.2)$$

$$\dot{X} = -\partial P / \partial x - \partial V / \partial x . \qquad (3.3)$$

where function

$$P = (a^u)^T \frac{\partial S}{\partial x} + b^T \frac{\partial^2 S}{\partial x^2} \;, \qquad (3.4)$$

is a potential of functional (1.1.10) that depends on the function of action $S(t, x)$ on extremals (3.2, 3.3); and $V(t, x)$ is integrant of additive functional (1.1.7), which defines probability function (1.1.3).
<u>Proof.</u> The Jacobi-Hamilton (JH) equations for function of action $S = S(t, x)$, defined on the extreme $x_t = x(t), (t, x) \in Q$ of functional (3.1), leads to

$$-\frac{\partial S}{\partial t} = H, H = \dot{x}^T X - L, \qquad (3.5)$$

where $X$ is a conjugate vector for $x$ and $H$ is a Hamiltonian for this functional.
(All derivations here and below have vector form).
From (3.1a) it follows

$$\frac{\partial S}{\partial t} = \frac{\partial \tilde{S}}{\partial t}, \frac{\partial \tilde{S}}{\partial x} = \frac{\partial S}{\partial x}, \qquad (3.6)$$

where for the JH we have

$$\frac{\partial S}{\partial x} = X, -\frac{\partial S}{\partial t} = H . \qquad (3.6a)$$

The Kolmogorov Eq. for functional (1.1.10) on a diffusion process, joining with Eq. (3.6a), holds the form

$$-\frac{\partial\tilde{S}}{\partial t}=(a^u)^T X+b\frac{\partial X}{\partial x}+1/2a^u(2b)^{-1}a^u=-\frac{\partial S}{\partial t}=H\,, \tag{3.7}$$

Here dynamic Hamiltonian $H=V+P$ includes differential function $V=d\varphi/ds$ of additive functional (1.1.7) and potential function (3.4). Function (3.4), at satisfaction (3.7), imposes constraint on transforming diffusion process to its extreme in form:

$$P(t,x)=(a^u)^T X+b^T\frac{\partial X}{\partial x}\,. \tag{3.8}$$

Applying Hamilton equations $\frac{\partial H}{\partial X}=\dot{x}$ and $\frac{\partial H}{\partial x}=-\dot{X}$ to (3.7) brings the extreme solutions for $x$ and $X$ in forms (3.2) and (3.3) accordingly.• More details in [12].

Proposition 3.2.

A *minimal solution* of variation problem (3.1a, 3.1) for the EF brings the equation connecting extreme $\dot{x}$ and $X_o$:

$$\dot{x}=2bX_o\,, \tag{3.9}$$

which satisfies condition

$$\min_{x(t)} P=P[x(\tau)]=0\,. \tag{3.10}$$

Condition (3.10) is a dynamic constraint, which is imposed on solutions (3.2), (3.3) at some set of the functional's field $Q\in KC(\Delta,R^n)$ holding the $\tau$-localities of cutting process $x(t)_{t=\tau}=x(\tau)$ :

$$Q^o\subset Q\,,\; Q^o=R^n\times\Delta^o,\Delta^o=[0,\tau],\tau=\{\tau_k\},k=1,...,m\,. \tag{3.11}$$

Hamiltonian

$$H_o=-\frac{\partial S_o}{\partial t} \tag{3.12}$$

defines the function of action $S_o(t,x)$ which on the extreme Eq.(3.9) satisfies condition

$$\min(-\partial\tilde{S}/\partial t)=-\partial\tilde{S}_o/\partial t\,. \tag{3.13}$$

Hamiltonian (3.12) and Eq. (3.9) determine a second order differential Eq. of extremals:

$$d^2x/dt^2=dx/dt[\dot{b}b^{-1}-2H_o]\,. \tag{3.14}$$

Proof. Using (3.4) and (3.6) allows us to find the equation for Lagrangian in (3.1) in form

$$L=-b\frac{\partial X}{\partial x}-1/2\dot{x}^T(2b)^{-1}\dot{x}\,. \tag{3.15}$$

On extremals (3.2, 3.3), the drift and diffusion in (1.1.10) are nonrandom.

After substitution the extreme Eqs (3.2, 3.9) to (3.7), integral functional $\tilde{S}$ (3.1) on the extremals holds:

$$\tilde{S}[x(t)]=\int_s^T 1/2(a^u)^T(2b)^{-1}a^u dt\,. \tag{3.15a}$$

Since both integrals are defined on the same extremals, they should satisfy variation conditions (3.1a), or

$$\tilde{S}[x(t)]=S_o[x(t)]\,. \tag{3.15b}$$

From (3.15) and (3.15a,b) follow Lagrangian

$L_o = 1/2(a^u)^T(2b)^{-1}a^u$, or $L_o = \dot{x}^T(2b)^{-1}\dot{x}$ . (3.16)

Both Lagrangian's expressions (3.15) and (3.16) coincide on the extremals.
Potential (3.7) on the extremals satisfies condition (3.10):

$$P_o = P[x(t)] = (a^u)^T(2b)^{-1}a^u + b^T\frac{\partial X_o}{\partial x} = 0 . \quad (3.17)$$

Whereas Hamiltonian (3.12), and the function of action $S_o(t,x)$ satisfies (3.13).
From (3.15b) it also follows

$$E\{\tilde{S}[x(t)]\} = \tilde{S}[x(t)] = S_o[x(t)] . \quad (3.17a)$$

Applying (3.16) to Lagrange's equation

$$\frac{\partial L_o}{\partial \dot{x}} = X_o , \quad (3.17b)$$

leads to the equations for vector

$$X_o = (2b)^{-1}\dot{x} \quad (3.17c)$$

and extremals Eq.(3.9).
Both Lagrangian and Hamiltonian here are the *Information forms* of JH solution for the EF.
Lagrangian (3.16) satisfies the maximum principle for the functional (3.1,3.1a) from which also follows (3.17a).
Functional (3.1) reaches its minimum on extremals (3.8), while it is a maximal on extremals (3.2,3.3) of (3.6). Hamiltonian (3.7), at satisfaction of (3.17), reaches minimum:

$$\min H = \min[V + P] = 1/2(a^u)^T(2b)^{-1}a^u = H_o \quad (3.18)$$

from which it follows (3.10) at

$$\min_{x(t)} P = P[x(\tau)] = 0 . \quad (3.19)$$

Function $(-\partial\tilde{S}(t,x)/\partial t) = H$ on extremals (3.2,3.3) reaches a *maximum* when constraint (3.10) is not imposed. Both the minimum and maximum are conditional with respect to the imposition of constraint.
Variation conditions (3.18) imposing constraint (3.10) selects Hamiltonian

$$H_o = -\frac{\partial S_o}{\partial t} = 1/2(a^u)^T(2b)^{-1}a^u \quad (3.20)$$

on the extremals (3.2,3.3) at discrete moments $(\tau_k)$ (3.11).
The variation principle identifies two Hamiltonians: $H$ -satisfying (3.6) with function of action $S(t,x)$, and $H_o$ (3.20), whose function action $S_o(t,x)$ reaches absolute minimum at moments $(\tau_k)$ (3.11) of imposing constraint $P_o = P_o[x(\tau)]$.
Substituting (3.2) and (3.17b) in both (3.16) and (3.20) leads to Lagrangian and Hamiltonian on the extremals:

$$L_o(x, X_o) = 1/2\dot{x}^T X_o = H_o . \quad (3.21)$$

Using $\dot{X}_o = -\partial H_o/\partial x$ brings $\dot{X}_o = -\partial H_o/\partial x = -1/2\dot{x}^T\partial X_o/\partial x$ .
Imposing constraint (3.10) brings

$$\partial X_o/\partial x = -b^{-1}\dot{x}^T X_o \text{ and } \partial H_o/\partial x = 1/2\dot{x}^T b^{-1}\dot{x}^T X_o = 2H_o X_o . \quad (3.22)$$

After substituting (3.17b) and (3.22) it leads to extremals (3.9).
From the Eq. for conjugate vector (3.3), Eqs. (3.7), (3.8), and (3.17c), constraint (3.10) acquires the form

$$\frac{\partial X_o}{\partial x} = -2X_o X_o^T \quad . \tag{3.23}$$

Differentiation of (3.9) brings a second order differential Eq. on the extremals:

$$\ddot{x} = 2b\dot{X}_o + 2\dot{b}X_o , \tag{3.24}$$

which after substituting (3.22) leads to (3.14). •
The above solutions simplify the proof of Theorem 3.1 in [12].

Comments.
The entropy force, as the gradient of entropy (3.6a), arises in a covariant coordinate of a unit of displacement, or at distance in each observation when the microprocess appears.
Until the space coordinate emerges, such a force does not exist.
With appearance of the logic Bit in a certain logic, the certain time-space emerges.
In such a time–space, time is a contra-variant vector and the force is a covariant vector in the rotating coordinate system. This coordinate system depends on its probabilistic choice during the time course of the observation. In the symmetric Lorentz group's transformation and Riemann geometry, the time and space are independent contra variant and covariant vectors accordingly, whose product is a scalar independent in this coordinate system.
Within the microprocess of a virtual observer, the time-space emerges as probabilistic.
In the certain logic of the Observer with the Hamiltonian Dynamics above, such a time-space is symmetrical. With emerging Information Bits, the information dynamics become irreversible and asymmetrical. Observation of the emerging multiple Information bits is accompanied by a piece–wise Hamiltonian dynamics, where a pieces of observation with the dynamics symmetry (reversible) alternate with pieces of asymmetrical (irreversible) information dynamics.
With observation of the Information Macrodynamics emerge each observer own real coordinate system.
A universal–single coordinate system may emerge for multiple Information observers in the Universe. •

### 3.4. THE INITIAL CONDITIONS FOR THE ENTROPY FUNCTIONAL AND ITS EXTREMALS

The initial conditions for the EF determine the ratio of primary *a priori- a posteriori* probabilities which begin the probabilistic observation:

$$p(\text{o}_s^p) = \frac{P_{s,x}^a}{P_{s,x}^p}(\text{o}_s^p) \quad . \tag{4.1}$$

Start of the observation evaluates the minimal probability and entropy [16]:

$$p(\text{o}_s^p) \cong 1.65 \times 10^{-4} \tag{4.2}$$

$$\Delta s_{ap}(\text{o}_s^p) = -\ln p(\text{o}_s^p) = 0.5 \times 10^{-4} \tag{4.3}$$

with minimal posterior probability $P_{poo} \approx 1 \times 10^{-4}$. (4.4)
That gives estimation of an average initial entropy of the observation:

$$S(\text{o}_s^p) = [-\ln p(\text{o}_s^p) \times P_{poo}] \cong 0.5 \times 10^{-8}\, Nat . \tag{4.5}$$

Based on physical coupling parameter $h_{\alpha}^{o} = 1/137$, physical observation theoretically starts with entropy: $S(\text{o}_{rs}^{p}) = 2/137 \cong 0.0146 Nat$, (4.6)
while entropy of the first real 'half-impulse' action starts at moment $t^{oe}$:
$S_{ko}(t^{oe}) = 0.358834 Nat$ (4.7)
with *a priori-a posteriori* probabilities $P_{ako} = 0.601, P_{pko} = 0.86$.
In this information approach, involving no material entities, the physical process begins with the real interactive action when the physical coupling may start with minimizing this entropy at beginning the Information process.
At the real cut with *a posteriori* probability $P_{po} \to 1$, the ratio of their *a priori-a posteriori* probabilities $P_{ao} / P_{po} \cong 0.8437$ determines the minimal entropy shift between interacting probabilities $P_{ao} \to P_{po}$ during a real cut: $\Delta s_{apo} = -\ln(0.8437) \cong 0.117 Nat$, which, after averaging at $P_{po} = 1$ leads to
$S(\text{o}_{r}^{p}) = 0.117 Nat$. (4.8)
The minimal entropy cost for covering the gap during its conversion to Information Observer is $s_{ev} \cong 0.0636 Nat$.
The theoretical start of observation (4.6) remains a potential until an gets Information from that entropy.
If we *define* the launch of a real Information Observer by minimal converting entropy (4.8), then opening real observation *defines* entropy of first real 'half-impulse actions (4.7). For a multi-dimensional process, there could be many.
Since the potential observing physical process with the logical structural entities is possible (virtual) when approaching entropy (4.6), this observation is *virtual* until a real observation generates an Information Observer. (The term 'virtual' here associates with physically possibility, until this physical objectivity becomes Information-physical reality for the Information Observer.)
Specifically, if the start of virtual observation associates with entropy binding primary *a priori-a posteriori* probability (4.5), then the virtual observer identifies the entropy of potential coupling structures (4.6).
Here, we have *identified* the beginning of both virtual and physical observations, and the virtual and real Information Observers, based on the actual quantitative parameters, which are independent for each particular observer.
However, specific limitations may be imposed after the Observer has formed [41].
The initial conditions for the EF *extremals* determine function $x_{\pm}(t^{e}) = x(s^{+})\exp(u_{\pm}t^{e})$ (from (2.9)), which at moment $t^{e} = s_{k}^{+o} b_{k}(t^{e}) / b_{k}(s_{k}^{+o})$, starts the virtual or real observations, depending on the required minimal entropy of related observations (4.5-4.8). It brings functions

$x_{\pm}(t^{e}) = x(s^{+})\exp(\pm 2t^{e})$ (4. 9a), $t^{e} = s_{k}^{+o} b_{k}(t^{e}) / b_{k}(s_{k}^{+o})$, (4.9b)

where (4.9b), at known dispersion function $b_{k}(t^{e}, s_{k}^{+o})$, identifies its time dependency $t^{e} = t^{e}(s_{k}^{+o}, b_{k})$, while (4.9a) identifies initial conditions for the EF extreme conjugated process:

$$x_{\pm}(t^{e}) = x(s^{+})\exp(\pm 2t^{e}(s_{k}^{+o}, b_{k})). \quad (4.9)$$

Applying the EF solutions (2.3.6) at opposite relative time $t_-^* = -t_+^*$ leads to the entropy functions:

$$S_\pm(t_\pm^*) = 1/2 S_+(t_+^*) \times S_-(t_-^*) = 1/2[\exp(-2t_+^*)(\mathrm{Cos}^2(t_+^*) + Sin^2(t_+^*) - 2Sin^2(t_+^*))] = 1/2[\exp(-2t_+^*)((+1 - 2(1/2 - Cos(2t_+^*))))] = 1/2\exp(-2t_+^*)Cos(2t_+^*) \quad . \quad (4.10)$$

This interactive entropy $S_\pm(t_\pm)$ becomes the minimal interactive threshold (4.8) at $t_+^* = t_*^e$, which starts the Information Observer. From (4.8) it follows:

$$S(t_+^* = t_*^e) = 1/2\exp(-2t_*^e)Cos(2t_*^e) = 0.117 \quad (4.11)$$

with relative time $t_*^e = \pm\pi / 2t^e$. The value (4.11) concurs with both (2.2.27) and (3.2.2).

Solution (4.11) will bring real $t^e$ which, after substitution in (4.9b), determines initial moment $s_k^{+o} = s_k^{+o}(t^e, b_k(t^e, s_k^{+o}))$ if dispersion functions $b_k(t^e, s_k^{+o})$ are known. Substituting moment $s_{k*}^{+o} = s_k^{+o}(t_*^e, b_k)$ relative to $t_*^e$ in relation

$$S(s_{k*}^{+o}) = 1/2\exp(-2s_{k*}^{+o})Cos(2s_{k*}^{+o}) \quad (4.12)$$

allows finding the unknown *a posteriori* entropy $S_\pm(s_k^{+o})$ starting a virtual observer at $s_k^{+o} = s_{k*}^{+o} / (\pi / 2)$.
To find the moment of time starting virtual observer at the maximal uncertainty measure (4.5) when dispersion functions are unknown, only joint pre-requirements (4.12) and (4.5) are available.
Eqs. (2.10) for initial conditions $\ln x(s_k^{+o}) = x(s_k^{+o})u_\pm s_k^{+o}$ after integration lead to
$1/2[\ln x(s_k^{+o})]^2 = u_\pm(s_k^{+o})^2$, $\ln x(s_k^{+o}) = \sqrt{2u_\pm}(s_k^{+o})$ and to

$$x_\pm(s_k^{+o}) = \exp(\pm\sqrt{2u_\pm})(s_k^{+o}), x_+(s_k^{+o}) = \exp(\pm\sqrt{2\times 2})s_k^{+o}, x_-(s_k^{+o}) = \exp(\pm\,\mathrm{j}\sqrt{2\times 2})s_k^{+o}, u_+ = 2, u_- = -2 .$$

It brings both real and complex initial conditions for starting extreme processes in the virtual observer:

$$x_+^i(s_k^{+o}) = \exp(\pm 2s_k^{+o}), x_-^i(s_k^{+o}) = \exp(\pm 2js_k^{+o}) = Cos(2s_k^{+o}) \pm jSin(2s_k^{+o}), \quad (4.13)$$

where first one describes virtual trajectory before the impulse generates the complex microprocess (2.3.6) in (4.13).
Finally, the trajectories of extreme processes (4. 9a) by moment $t_i^e$ on each $i$-dimension takes form

$$x_\pm^i(t_i^e) = x_\pm^i(s_k^{+o})[Cos(2s_k^{+o}) \pm jSin(2s_k^{+o})]\exp(\mp 2t_i^e) . \quad (4.14)$$

Let us numerically validate the results (4.11-4.14).
Solution of (4.11):

$$\ln 1/2 - 2t_*^e + \ln[Cos(2t_*^e)] = \ln 0.117, -0.693 + 2.1456 = 2t_*^e - \ln[Cos(2t_*^e)], 1.4526 = 2t_*^e + \ln[Cos(2t_*^e)]$$

leads to $2t_*^e \approx 1.45, t^e = 1.45 / \pi \approx 0.46$ as one possible result.
Applying condition (4.6) to (4.12):

$$S(s_{k*}^{+o}) = 1/2\exp(-2s_{k*}^{+o})Cos(2s_{k*}^{+o}) = 2/137 \quad (4.14a)$$

leads to solution

$$-0.693 + 4.22683 = 2s_{k*}^{+o} - \ln[\mathrm{Cos}(2s_{k*}^{+o})], 3.534 = 2s_{k*}^{+o} - \ln[\mathrm{Cos}(2s_{k*}^{+o})]$$

with the result $s_{k*}^{+o} \approx 1.767$, $s_k^{+o} \approx 1.12$.

Applying the condition of beginning virtual observation (4.5) at relative time $o_{s*}^p$ to relation

$$S(o_{s*}^p) = 1/2\exp(2o_{s*}^p)Cos(2o_{s*}^p) = 0.5\times 10^{-8} \quad (4.14b)$$

leads to $-0.693 + 12.8 = 2o_{s*}^p - \ln[Cos(2o_{s*}^p)]$ with solutions $2o_{s*}^p \approx 12$ and $o_s^p \approx 3.85$.

Applying the condition of starting Information observation (4.7) with Eq. (4.10), at relative time $t_*^{oe}$, leads to solution

$2t_*^{oe} \approx 0.33, t^{oe} = 2t_*^{oe} / \pi \approx 0.1$.

These time moments are counting from the real observer (back to starting observation) after the observing process overcomes the threshold (4.8).

This means that $o_s^p \approx 3.85$ evaluates the time interval of virtual observation, while the virtual observer starts on time interval $s_k^{+o} \approx 1.12$, and the real observer starts on $t^e \approx 0.46$.

Whereas observing the first real 'half-impulse' action takes *part* of this time: $t^{oe} \approx 0.1$.

States $x_+^i(s_k^{+o}) = \exp(\pm 2\times 1.12)$ hold probability $P_{ako} = 0.601$ and have multiple correlations $r_\pm^x(s_k^+) = x_+^i(s_k^{+o}) x_-^i(s_k^{+o})$,

starting virtual observer (4.12, 4.12a). Conjugates processes (4.14) interact through correlation

$$r_\pm^x(t_i^e) = x_+^i(t_i^e)\times x_-^i(t_i^e) = r_\pm^x(s_k^+)[Cos(2s_k^{+o})^2 + jSin(2s_k^{+o})^2]\exp(-2t_i^e)\exp(+2t_i^e) = 1 \quad (4.14c)$$

reaching the Information Observer's threshold (4.8) with relative probability $P_{ao} / P_{po} \cong 0.8437$ and two conjugated entropies

$$S_\pm(t^e) = 1/2\exp(\pm\pi / 2\times t^e)Cos(\pm\pi / 2\times t^e), \quad (4.15)$$

following from (4.11). The entangled entropies move Eq.(4.15).

The starting extreme process (4.14) evaluates two pairs of real states for this conjugated process:

$$\begin{aligned} x_\pm^r(t_i^e) &= 9.39\times 0.999\times 0.3985 = 3.738, x_\mp^r(t_i^e) = 0.1064\times 0.999\times 2.509 = 0.2666 \\ x_\mp^{r1}(t_i^e) &= 0.1064\times 0.999\times 0.3985 = 0.042, x_\pm^{r1}(t_i^e) = 9.39\times 0.999\times 2.509 = 23.536 \end{aligned}. \quad (4.15a)$$

Imaginary initial conditions evaluate four options:

$$x_+^{im1}(t^e) = \exp(2\times 1.12)[\pm j\,\mathrm{Sin}(2\times 1.12)]\times\exp(-0.92) = j3.064\times 0.039\times 0.3985 \cong \pm j0.0475 \quad (4.15b)$$

$$x_-^{im2}(t^e) = \exp(-2\times 1.12)[\mp j\,\mathrm{Sin}(2\times 1.12)]\times\exp(0.92) = \mp j0.323\times 0.039\times 2.509 \cong \mp j0.0316. \quad (4.15c)$$

Information force, acting by the end of microprocess on an edge of two nearest impulses, describes Eqs.:

$$X_\delta = \frac{\partial\Delta S_\delta}{\partial x}.$$

The force specifies relation $\partial x = 1/2r^{-1/2}\partial r$ from (4.14c), which at $x(t^e) = r_x^{1/2}$ at $r = r_x$, and $\Delta S_\delta \to \Delta I_\delta$ leads to

$\frac{\partial\Delta S_\delta}{\partial r} \to \frac{\partial\Delta I_\delta}{\partial r}, \frac{\partial\Delta I_\delta}{\partial x} = \frac{\partial\Delta I_\delta}{\partial\ln r} = -1/8r^{-1}$. Then at $X_{\delta M} = \frac{\partial\Delta S_\delta}{\partial r}$ we obtain

$$X_\delta = \frac{\partial\Delta S_\delta}{1/2r^{-1/2}\partial r} = -1/4r^{-1/2}X_{\delta M} = -1/4r^{-1/2}. \quad (4.16)$$

This is a finite Information force, which keeps the states of the impulse process bounded (Sec.1.1), counterbalancing the destructive force that arises at dissolving correlations at $r \to 0$.

Hence, it follows that this elementary ("weak") Information force, which binds the impulse's opposite states, defines the same variation equation as the macroprocess Information force (3.17c) which binds multiple microprocesses.

The "strong" information force (3.17c) determines the sum of correlations integrated along the macroprocess up to the process's current moment.

In the topological transitivity at the curving interactions (Sec.2.2.5), at the transition turning moment, a jump of the *time curvature* switches to a *space curvature*. Here the weak force (4.16) emerges, starting the space curvature by analogy with the gravitational force. Thus, both weak and strong Information forces arise along the observing Information process.

3.5. APPLYING EQUATION OF EXTREMALS $\dot{x} = a^u$ TO A DYNAMIC MODEL'S TRADITIONAL FORM:

$$\dot{x} = Ax + u, u = Av, \dot{x} = A(x+v), \quad (5.1)$$

where $v$ is a control $u$ reduced to state vector $x$.
Solving the initial variation problem (VP) for this model allows finding optimal control $v$ and identifying matrix $A$ under this control's action.

<u>Proposition 5.1.</u>
The reduced control is formed by feedback function of macrostates $x(\tau) = \{x(\tau_k)\}, k = 1,...,m$ in the form:

$$v(\tau) = -2x(\tau). \quad (5.2)$$

Or using (5.1), it applies to $u = u(x(\tau)) = u(\tau)$ as function of speed of the macroprocess in (5.1):

$$u(\tau) = -2Ax(\tau) = -2\dot{x}(\tau), \quad (5.3)$$

at the localities of moments $\tau = (\tau_k)$ (3.3.11), when matrix $A$ determines equations

$$A(\tau) = -b(\tau) r_v^{-1}(\tau), r_v = E[(x+v)(x+v)^T], b = 1/2\dot{r}, r = E[\tilde{x}\tilde{x}^T] \quad (5.4)$$

and $A$ identifies the correlation function with its derivative, or directly dispersion matrix $b$ from (1.2.1):

$$|A(\tau)| = b(\tau)(2\int_{\tau-o}^{\tau} b(t)dt)^{-1}, \ \tau - o = (\tau_k - o), k = 1...,m. \quad \bullet \quad (5.5)$$

<u>Proof.</u> Using variation Eq. for the conjugate vector (3.3) allows writing the constraint (3.10) in the form

$$\frac{\partial X}{\partial x}(\tau) = -2XX^T(\tau), \quad (5.6)$$

for model (5.1).
It leads to Eqs. for the conjugate vectors in this model:

$$X = (2b)^{-1}A(x+v), X^T = (x+v)^T A^T (2b)^{-1}, \frac{\partial X}{\partial x} = (2b)^{-1}A, b \neq 0. \quad (5.7)$$

After substituting (5.7) to (5.6), it acquires form

$$(2b)^{-1}A = -2E[(2b)^{-1}A(x+v)(x+v)^T A^T (2b)^{-1}], \quad (5.8)$$

from which, at a nonrandom $A$ and $E[b] = b$, the identification equations (5.4) follow.
Completion of both (5.6), (5.7) performs the control's action.
Using (5.4) leads to (5.8) which under control (5.3) takes the form

$$A(\tau)E[(x(\tau)+v(\tau))(x(\tau)+v(\tau))^T] = -E[\dot{x}(\tau)x(\tau)^T], \text{ at } \dot{r} = 2E[\dot{x}(\tau)x(\tau)^T].$$

This relation after substituting (5.1) leads to

$A(\tau)E[(x(\tau)+v(\tau))(x(\tau)+v(\tau))^T] = -A(\tau)E[(x(\tau)+v(\tau))x(\tau)^T]$ and then to

$$E[(x(\tau)+v(\tau))(x(\tau)+v(\tau))^T + (x(\tau)+v(\tau))x(\tau)^T] = 0,$$

which is fulfilled at applying control (5.2).
Since $x(\tau)$ is a discrete set of states, satisfying (3.11), (3.13), the control has a discrete form.

Each stepwise control (5.2), with its inverse value of doubling controlled state $x(\tau)$, applied to both Eqs. (5.7), implements (5.6). Eq. (5.6), following from variation conditions (3.1a), fulfills this condition.
This control, applied to the additive functional (1.1.7), imposes a constraint (3.8, 3.10) which limits transformation of random process segments to the process extremals.
By applying the control step-down and step-up actions to satisfy conditions (3.7) and (3.10), the control sequentially starts and terminates the constraint, while extracting the cutoff hidden information on the $x(\tau)$-localities. By performing the transformation, this control initiates the identification of matrix $A(\tau)$ in (5.4, 5.5) during its time interval $\tau - o, \tau, \tau + o$ (Sec.1.2.3), solving simultaneously the identification problem [52].• Obtaining this control here *specifies* some results of Theorems 4.1 [12].

<u>Corollary.5.1</u>.
The control, which turns the constraint (3.8,3.10) on, creates the Hamilton dynamic model with complex conjugated eigenvalues of matrix $A$. After the constraint's termination, the control transforms this matrix to its *real* form (on the diffusion process' boundary point [4]), which identifies the diffusion matrix in (5.4). Thus, within each extremal segment, the dynamics are reversible. Irreversibility rises at each constraint termination between the segments. The EF-IPF Lagrangian integrates both the impulses and constraint Information on its time space-intervals.•

<u>Proposition</u> <u>5.2.</u>
Let us consider controllable dynamics under feedback control (5.3) described by operator $A^v(t,\tau)$ with eigenfunctions $\lambda_i^v(t_i,\tau_k)_{i,k=1}^{n,m}$ , whose matrix equation:

$$\dot{x}(t) = A^v x(t), \tag{5.9}$$

includes the feedback control (5.3).
The drift vector for both models (5.1) and (5.9) has same form:

$$a^u(t,\tau) = A(t,\tau)(x(t,\tau)+v(\tau)), A^v(t,\tau)x(t,\tau) = A(t,\tau)(x(t,\tau)+v(\tau)). \tag{5.9a}$$

Then the following holds true:
(1) Matrix $A^v(t,\tau)$ under control $v(\tau_k^o) = -2x(\tau_k^o)$, applied during time interval $t_k = \tau_k^1 - \tau_k^o$, in form $A^v(t_k,\tau_k^o)$, depends on initial matrix $A(\tau_k^o)$ at the moment $\tau_k^o$ according to Eq

$$A^v(t_k,\tau_k^o) = -A(\tau_k^o)\exp(A(\tau_k^o)t_k)[2-\exp(A(\tau_k^o)t_k)]^{-1}. \tag{5.9b}$$

(2) The identification Eq.(5.4) at $\tau_k^1 = \tau$ brings

$$A^v(\tau_k^1) = -A(\tau_k^1) = b(\tau_k^1)r_v^{-1}(\tau_k^1), b(\tau_k^1) = 1/2\dot{r}(\tau_k^1), \tag{5.9c}$$

whose covariation function $r_v(\tau_k^o)$, starting at moment $\tau_k^o$ , by the end of this time interval $\tau_k^1$, acquires form

$$r_v(\tau_k^1) = [2-\exp(A(\tau_k^o)\tau_k^1)]r(\tau_k^o)[2-\exp(A^T(\tau_k^o)\tau_k^1)]. \tag{5.9d}$$

(3a) Applying control $v(\tau_k^o) = -2x(\tau_k^o)$ at moment $\tau_k^1 = \tau_k^o + o$ following $\tau_k^o$ to (5.9a, right)) changes initial matrix sign:

$$A^v(\tau_k^1) = A^v(\tau_k^o + o) = -A(\tau_k^o). \tag{5.9e}$$

(3b) When this control, applied at the moment $\tau_k^1$, ends the dynamic process on extremals in following moment $\tau_k^1+o$, at $x(\tau_k^1+o)\to 0$, function $a^u=A^v x(t)$ in (5.9a) turns to

$$a^u(x(\tau_k^1+o))\to 0; \qquad (5.9f)$$

which brings (5.9a) to dynamic form $a^u=A(\tau_k^1+o)v(\tau_k^1+o)\to 0$ which at $A(\tau_k^1+o)\neq 0$ requires turning the control off.

(3c) At fulfilment of (5.9f), the Markov process includes only its diffusion component, which identifies dynamic matrix $A(\tau_k^1+o)$.

This matrix, transforming in following moment $\tau_{k+1}^1: A(\tau_k^1+o)\to A(\tau_{k+1}^1)$, identifies correlation matrix $r(\tau_{k+1}^1)$ in form

$A(\tau_{k+1}^1)=1/2\dot r(\tau_{k+1}^1)r^{-1}(\tau_{k+1}^1)$ at $r(\tau_{k+1}^1)=r^v(\tau_{k+1}^1)_{v(\tau_k^1+o)\to 0}$.

(3d) Dispersion matrix $b(\tau_k^o,t_k)=1/2\dot r_v(\tau_k^o,t_k)$ on the extremals, identifies by the dynamic matrix (5.9b) in form

$$\partial r_v/\partial t_k=-A(\tau_k^o)\exp(A(\tau_k^o)t_k)r(\tau_k^o)[2-\exp(A^T(\tau_k^o)t_k)]+ [2-\exp(A(\tau_k^o)t_k)]r(\tau_k^o)[-A^T(\tau_k^o)\exp(A^T(\tau_k^o)t_k)] \qquad (5.10)$$

which for symmetric matrix $A(\tau_k^o)$ leads to relations $\partial r_v/\partial t_k=-2A(\tau_k^o)\exp(A(\tau_k^o)t_k)r(\tau_k^o), b(\tau_k^o,t_k)=-A(\tau_k^o)\exp(A(\tau_k^o)t_k)r(\tau_k^o)$,

At $\tau_k^1=\tau_k^o$, $t_k=0$, it brings

$$b(\tau_k^1=\tau_k^o)=-A(\tau_k^o)\exp(A(\tau_k^o)0_k)r(\tau_k^o)=-A(\tau_k^o)r(\tau_k^o),\ b(\tau_k^1)\times b(\tau_k^o)^{-1}=\exp(A(\tau_k^o)t_k). \qquad (5.10a)$$

Last ratio in (5.10a) for a single dimension, at $A(\tau_k^o)=\alpha_1(\tau_k^o), t_k=\tau_k^1-\tau_k^o$, leads to

$$b(\tau_k^1)/b(\tau_k^o)=\exp(\alpha_1(\tau_k^o)-\alpha_1(\tau_k^o)) \qquad (5.10b)$$

which after applying relation $b(\tau_k^1)/b(\tau_k^o)=\tau_k^1/\tau_k^o$, leads to

$$\tau_k^1/\tau_k^o=\exp(\alpha_1(\tau_k^1)-\alpha_1(\tau_{k1}^o)), \qquad (5.10c)$$

connecting interval $t_k=\tau_k^1-\tau_k^o$ with eigenvalue $\Delta\alpha_1=\alpha_1(\tau_k^1)-\alpha_1(\tau_{k1}^o)$.

That time interval measures information speed $\Delta\alpha_1$ and vice versa.

(3e) Equation for conjugated vector (5. 7) on each extremal segments follows from relation:

$$X_o(t_k)=2b(t_k)^{-1}\dot x(t_k)=2A(\tau_k^o)\exp(A(\tau_k^o)t_k)x(\tau_k^o)A^T(\tau_k^o)\exp(A^T(\tau_k^o)t_k). \qquad (5.11)$$

(3f) Entropy increment $\Delta S_{io}$ on optimal trajectory, measured on the cutting localities $\tau$ (3.11) at

$$E[\frac{\partial\tilde S}{\partial t}(\tau)]=1/4Tr[A(\tau)]=H(\tau),\ A(\tau)=-1/2\sum_{i=1}^n \dot r_i(\tau)r_i^{-1}(\tau),(r_i)=r, \qquad (5.11a)$$

determines the time interval ratio $\tau_i/\tau_{i-1}$ of the nearest segments along the EF increment $\Delta S_{io}$ transforming to IPF( $I_{x_t}^p$ ):

$$\Delta S_{io}=I_{x_t}^p=-1/8\int_s^T Tr[\dot r r^{-1}]dt=-1/8Tr[\ln(r(T)/\ln r(s)],(s=\tau_o,\tau_1,...,\tau_n=T). \bullet \qquad (5.12)$$

Proof (1). Control $v(\tau_k^o) = -2x(\tau_k^o)$, imposing the constraint at moment $\tau_k^o$ on both (5. 6) and (5.9) and terminating it at $\tau_k^1$ during time interval $t_k = \tau_k^1 - \tau_k^o$, brings solutions of (5.1) by the end of this interval:

$$x(\tau_k^1) = x(\tau_k^o)[2 - \exp(A(\tau_k^o)t_k)]. \quad (5.13)$$

Substituting this solution to $\dot{x}(\tau_k^1) = A^v(\tau_k^1)x(\tau_k^1)$ both to derivative on the left side and the right side leads to $-x(\tau_k^o)(A(\tau_k^o)t_k)\exp(A(\tau_k^o)t_k) = A^v(\tau_k^1)x(\tau_k^o)[2 - \exp(A(\tau_k^o)t_k)])]$, or to connection of both matrixes $A^v(\tau_k^{1`})$ and $A(\tau_k^1)$ (at the interval end) with matrix $A(\tau_k^o)$ (at the interval beginning):

$$A^v(t_k, \tau_k^o) = -A(\tau_k^o)\exp(A(\tau_k^o)t_k)[2 - \exp(A(\tau_k^o)t_k)]^{-1}. \quad (5.14)$$

That confirms (5.9b) and leads to $A^v(\tau_k^1) = -A(\tau_k^1)$ by moment $\tau_k^1$, from which follows (5.9e).

Other *Proofs* of the parts are straight forward. •

The identified functions drift and diffusion of the EF through the controllable dynamics automatically transforms the EF to IPF revealing the integrated information hidden in the observing process cutting correlations. The initial conditional for probability (1.1.2) determines the probability measure along the extremal trajectory:

$$p[x(t)] = p[x(s)]\exp(-S[x(t)]) \quad (5.14a)$$

where starting probability $p[x(s)]$ follows from (4.12) with numerical values (4.14a),(4.15,4.15a,b).

The described method and procedure identify parameters of Markov diffusion process, which define both the structure EF and differential Eqs. of the Observer's dynamics. The results were published in [19-23].

### 3.5.1. **Finding the invariant relations**

Using(5.3) in form $u(\tau) = -2Ax(\tau) = -2\dot{x}(\tau)$, and $c^2 = |u_+u_-| = c_+c_- = \bar{u}^2, c_+ = u_+, c_- = u_-$ from (2.4,2.10) leads to

$$c^2 = \dot{x}(\tau) = -2Ax(\tau), \alpha_{ko}(\tau - s) = \mathrm{u}_\pm(\tau - s) \quad (5.15)$$

where $\alpha_{ko}$ is the eigenvalue of matrix $A(\tau_k^o) = \|\alpha_{ko}\|$ starting Information speed on invariant interval $t_{ko} = (\tau - s)$ of imposing constraint (3.10,318).

Applying constraint (3.10) to (5.15) brings

$$\alpha_{ko}(\tau - s) = \mathrm{u}_\pm(\tau - s) = inv = \mathbf{a}_o, \quad (5.15a)$$

where invariant $\mathbf{a}_o = \mathbf{a}_o(\gamma_k)$ depends on ratio of imaginary to real eigenvalues of matrix (5.5): $\gamma_k = \beta_{ko} / \alpha_{ko}$.

Since $\mathrm{u}_\pm = 2$ is a real act of the interacting Markov process, it determines the real eigenvalue $\alpha_{ko} = 2$ at $k = n = 1$.

From (5.15a), and (2.2.29) real eigenvalues $\alpha_{ko} = \pm 2$ of matrix $A(\tau_k^o)$ determines $\mathbf{a}_o / |\alpha_{ko}| = (\tau - s)$.

For the optimal model with information invariant $\mathbf{a}_o(\gamma_k \to 0.5) = \ln 2$, it leads to $\mathbf{a}_o / |\alpha_{ko}| = \ln 2 / 2 = 0.346 = \tau - s$, which evaluates $\delta_k = \tau_k^{+o} - \tau_k^{-o} \cong 0.35$ [5].

Invariant $\mathbf{a}_o = \ln 2 \cong 0.7$ measures information at each impulse real cut for a multi-dimensional observing process.

Correlation matrix (5.9d), measured by the optimal model's invariant, takes the form

$$r_v(\tau_k^1)=[2-\exp\mathbf{a}_o\ )]r(\tau_k^o)[2-\exp(\mathbf{a}_o)]=r(\tau_k^o)\times\left[1.5^2\right], \quad (5.16)$$

where vector $x(\tau_k^1)=x(\tau_k^o)\times[1.5]$ applies to relation (5.13) for the optimal model. Conjugate vector

$$X_o(\tau_k^1)=2A(\tau_k^o)\exp(\mathbf{a}_o)x(\tau_k^o)A^T(\tau_k^o)\exp(\mathbf{a}_o) \quad (5.16a)$$

for a single dimension holds

$$X_{o1}(\tau_k^1)=2\alpha_1(\tau_{k1}^o)^2\exp(2\mathbf{a}_o)x(\tau_{k1}^o), \quad (5.16b)$$

or at $\alpha_1(\tau_{k1}^o)=2\mathbf{a}_o/t_k, t_k\to 0,\ t_k\to 0$, it approaches $X_{o1}(\tau_k^1)=(2\mathbf{a}_o/t_k)^2\exp(2\mathbf{a}_o)x(\tau_{k1}^o)\to\infty$.

It means at decreasing time intervals $t_k$ between the impulse's generated information invariants $\mathbf{a}_o$ with the information force (5.16b) grows infinitely.

Growing in square function of time intervals, the force leads to potentially overrunning the impulse with a minimal $t_k$. It leads to the possibility of pulling together the real action and its result-reaction for the minimal time impulse (Sec.2.7).

The EF-IPF *estimate* invariant measure $\mathbf{a}_o(\gamma_k)$, counting increments of both the segments and inter-segments:

$$\tilde{S}_{\tau m}^i=\sum_{k=1}^m(\mathbf{a}_o(\gamma_k)+\mathbf{a}_o^2(\gamma_k)), \tilde{S}_\tau=\sum_{i=1}^n\tilde{S}_{\tau m}^i,, \quad (5.17)$$

where $m$ is number of the segments, $n$ is the model dimension (assuming each segment has a single $\tau_k$-locality).

To *predict* each $\tau_k$-locality, where information generates, only the invariant measure $\mathbf{a}_o(\gamma_k)$ is needed. The sum of process's invariants

$$\tilde{S}_{\tau m}^{io}=\sum_{k=1}^m\mathbf{a}_o(\gamma_k),\ \tilde{S}_\tau^o=\sum_{i=1}^n\tilde{S}_{\tau m}^{io} \quad (5.18)$$

estimates the EF entropy with process's maximal probability (5.14a) expressed through $\mathbf{a}_o=\mathbf{a}_o(\gamma_k)$ [6].

This entropy allows the encoding of the observing process using Shannon's formula for an average optimal code-word length:

$$l_c\geq\tilde{S}_\tau^o/\ln D, \quad (5.19)$$

where $D$ is the number of letters of the code's alphabet, which encodes $\tilde{S}_\tau^o$ (5.18).

An elementary code-word to encode the optimal process' segment is

$$l_{cs}\geq\mathbf{a}_o(\gamma_k)/\log_2 D_o, \quad (5.20)$$

where $D_o$ is code alphabet which implements the states invariant connections on the extremal segment.

At $\mathbf{a}_o(\gamma_k\to 0.5)\cong 0.7$, $D_o$=2, it follows $l_{cs}\geq 1$, or (5.20) encodes a bit per encoding the alphabet letter.

With the values of $x_\pm^r(t^e), x_-^{im1}(t^e), x_-^{im2}(t^e)$ in (4.15, 4.15a,b,c) start Hamiltonian process and correlation (5.16). The correlation identifies the eigenvalues of matrix (5.9.b) and initial segment's state $x(\tau_{k1o}^o)=x(\tau_k^o)$ in (5.13) during the process forming Information units.

The connection to each following segment on the extremal trajectory determines the segment state $x(\tau_k^o)$ (5.13). Each $x(\tau_k^o)$, starting the macroprocess segment, enfolds the impulse microprocess.

Moment $t^e$ identifies starting dispersion and correlation on the optimal trajectories that determine start of dynamic matrix $A(t,\tau_k)$, Hamiltonian, and both EF-IPF on the trajectory segments.

Optimal control (5.3) starts with each segment initial states bringing the feedback transforming to controllable dynamics.

The details of Information micro-macrodynamics (ID), based on the *invariant* $\mathbf{a}_o(\gamma_k)$ *description*, are in [26], where scale parameter $\gamma^{\alpha}_{k,k+1} = \alpha_k / \alpha_{k+1}$ of the dynamics depends on the frequency spectrum of observations detecting through the identified $\gamma_k$. The observer self-scaled observation initiates its time-space which builds a distributed Information network [53].
*The above dynamic equations finalize both math description of the micro-macro-processes and validate them numerically.*

## 4. ARISING THE OBSERVER COLLECTIVE INFORMATION

### 4.1. **Rising a cooperative attraction**

Cooperative rotation and ordering start with entangling entropy increments in the entropy volumes.
Then, the sequential impulse' entropy increments with their volumes involve in collective rotating movement.

Since each following impulse may start only after the previous impulse cutting time $\delta^t_{ei}$ (3.3.3b) will triple. Time interval between the impulse cutoff actions:

$$\Delta_t = 3\delta^t_{ei} \cong 1.2\times10^{-15}\sec \qquad (1.1)$$

imposes limitation on adjoining a pair of the impulse entropies in collective movement in imaginary time course $\Delta_t$.
This limitation is applicable to virtual observer with its virtual impulse, which can initiate a virtual collective movement of the adjoining entangled volumes. That requires an entropic force

$$X_e = \Delta s_o / \Delta l_o, \qquad (1.2)$$

where $\Delta s_o \cong 0.25 Nat$ is a minimal entropy increment between the impulse, and $\Delta l_o \cong \Delta_{lo}$ (3.2.14) is a distance between nearest impulses with that entropy. At these relations, the entropic force measures

$$X_e \cong 0.25/14.4\times10^{-5} \cong 0.1736\times10^4 [Nat/m]. \qquad (1.3)$$

Speed of rotating moment $\delta M_e / \delta t$ defined by force (1.3) and velocity (3.2.12a) in relation

$$\delta M_e / \delta t = X_e w_o \qquad (1.4)$$

characterizes intensively of the rotation which evaluates

$$\delta M_e / \delta t = 0.344\times10^6 [Nat/m\sec]. \qquad (1.4a)$$

Since both the rotating moment's speed and entropy force proceed between the impulses, relation (1.4a) describes intensively of attracting rotation intended on capturing next impulse's entropy increment in joint distributed rotating movement. It measures cooperative connection of the impulses entropy increments prior forming next information units.
Thus virtual collective movement (between the entangle increments entropy increments) may emerge before a following impulse kills the entropy volumes.
Perhaps, that movement engages a cooperative transition of the entangled entropy volumes to the cutting transition *gap*. It makes possible a collective entanglement that not requires spending energy.
Study [54] "demonstrates that the spreading of entanglement is much faster than the energy diffusion in this nonintegrable system". Such a cooperative virtual distribution rotates the involving entangled groups- an ensemble jointly preparing them to the following killing-cut producing information units.
Therefore, the distributed rotation primary involves the entropies of the Bayesian linked probabilities, resulting from virtual interactive probes of different frequencies. The entangled entropies' volumes, in a cooperative rotation, engage in the transition movement up to cutting them on information units. The rotation continues connecting the forming information units.

4.2. **Forming a triple consolidation of information units during the cooperative rotation**

Imaginary entropy in each virtual impulse predefines information of a certain impulse starting a movement generating a single information unit with invariant information $\mathbf{a}_{io}$ satisfying the minimax.

Speed of cutting correlation is an initial source of real information speed, while imaginary information speed arises in the microprocess at $s$-locality on the minimax trajectory of observing process.

Hence both participate in forming $\mathbf{a}_{io}$. Speed $\alpha_{io}$ of generation $\mathbf{a}_{io}$ starts the single real information unit after killing entropy volume moving on time interval $t_{io}=(\tau-s)$ to form invariant $\mathbf{a}_{io}=\alpha_{io}t_{io}$.

Imaginary $\beta_i$ and real $\alpha_i$ speeds are components of the dynamic process with complex eigenvalues which begin the movement:

$$\operatorname{Re}\lambda_{io}=\alpha_{io}, \operatorname{Im}\lambda_i=\beta_{io}\ \lambda_{io}=\alpha_{io}\pm j\beta_{io} \qquad (2.1)$$

at a known ratio $\gamma_{io}=\beta_{io}/\alpha_{io}$ of the starting imaginary and real components.

This movement begins with speed $c_{ev}$ transiting entropy speed $\beta_{io}$ to $\alpha_{io}$ while forming $\mathbf{a}_{io}$:

$$\beta_{io}=c_{ev}\ . \qquad (2.2)$$

When the unit gets complete information $\mathbf{a}_{io}$, the imaginary speed will turn to zero by the end of time interval $t_{io}$. That requirement connects $\mathbf{a}_{io}$ and $\gamma_{io}$ by Eq. [2]:

$$2\sin(\gamma_{io}\alpha_{io})+\gamma_{io}\cos(\gamma_{io}\alpha_{io})-\gamma_{io}\exp(\alpha_{io})=0\,. \qquad (2.3)$$

From that, at invariant $\mathbf{a}_{io}\cong\ln 2$ follows ratio

$$\gamma_{io}=\beta_{io}/\alpha_{io}\to(0.4142-0.5) \qquad (2.4)$$

Here invariant $\mathbf{a}_{io}(\gamma_{io})$ determines the invariant values of $\gamma_{io}$ which impose limitation on $\beta_{io}$ by minimal speed of transferring entropy volume in (3.3.6):

$$\beta_{io}=c_{ev}=0.587\times10^{15}\,Nat/\sec\,. \qquad (2.5)$$

Then, from minimal $(\gamma_{io})^{-1}$ in (2.4) and (3.3.9) it follow starting information speed $\alpha_{io}(t_{io})$:

$$\alpha_{io}=(\gamma_{io})^{-1}c_{iv}\cong 2.4143\times0.587\times10^{15}\,Nat/\sec\cong1.41\times10^{15}\,Nat/\sec\,, \qquad (2.6)$$

which determines minimal time interval of completion information unit $\mathbf{a}_{io}$:

$$t_{io}^{o}\cong0.491\times10^{-15}\sec\ . \qquad (2.6a)$$

Estimation (2.6a) is close to (3.3.9), where the difference $\delta_i^t=0.0535\times10^{-15}\sec$ approximates time interval of the information impulse' wide forming during total time (2.6a).

Moving $\beta_{io}$ during a time interval $t_{ko}$ determines increment of imaginary contribution $\mathbf{b}_o^{'}=\beta_{io}t_{ko}$.

Condition of turning this contribution to zero by the end of time interval $t_{ko}$ connects $\mathbf{b}_o^{'}$ to $\gamma_{io}$ by Eq.

$$2\cos(\gamma_{io}\mathbf{b}_o^{'})-\gamma_{io}\sin(\gamma_{io}\mathbf{b}_o^{'})-\exp(\mathbf{b}_o^{'})=0\ . \qquad (2.7)$$

Solution of (2.7): $\mathbf{b}_o^{'}=\beta_{io}t_{ko}=\pi/6$ at (2.3) evaluates $t_{ko}:t_{ko}=\pi/6/0.596\times10^{15}=0.8785\times10^{-15}\sec$, which determines minimal interval of turning imaginary speed (2.6) to zero.

It’s seen that imaginary (virtual) time interval is wider than real (2.6a).

Here $t_{ko}$ is close to the constraint interval between impulses (1.1.1).
Contribution $\mathbf{b}_o^{'}(\gamma_{io})$ at $\gamma_{io}$ (2.4) is invariant following from (2.7).
Ratio of the invariants

$$\beta_{io}t_{ko} / \alpha_{io}t_{io} = \gamma_{io}t_{ko} / t_{io} = \gamma_{io}^{ko}, \gamma_{io}^{ko} \cong 0.915 \qquad (2.7a)$$

evaluates how imaginary entropy increment $\delta_{Eio} = \beta_{io}\delta_{tio}$ at a fixed time $\delta_{tio}$, during the transitive movement enables *attracting real information* at forming the single unit.
That allows evaluates *attractiveness* a real speed by the imaginary speed during the transitive movement

$$\delta_{Iio} = \alpha_{io}\delta_{tio}. \qquad (2.8)$$

According to (2.6), (2.7a) and (2.8) , at a minimal coefficient of attraction:

$$\delta_{Eio} = 0.4142\delta_{Iio}, \qquad (2.8a)$$

a unit of entropy $\delta_{Eio} = 1$ may attract $0.4142\ \delta_{Iio}$ units of information.

Entropy volume $s_{ve} = 0.0636Nat$, moving with minimal speed $c_{ev}$ (3.3.6) at coefficient of attraction in (2.8a) may attract potential-information (entropy) $i_v \cong 0.02634Nat$ .from a nearest impulse on minimal time interval $\Delta_t$ .
The information speed of attraction evaluates.

$$c_{ivo} = i_v / \Delta_t,\ c_{ivo} \cong 0.0548 \times 10^{15} Nat / \sec. \qquad (2.9)$$

That might increase minimal transition speed $c_{ev}$ up to

$$c_{ev} + c_{ivo} \cong 0.65 \times 10^{15} Nat / \sec. \qquad (2.9a)$$

With transition speed (2.9a), the rotating entropy volume moves to its cut on the gap, while engaging other entangled entropy volumes to a joint (collective) rotation with the speed of rotating moment (1.4).
Impulse, carrying $\cong 0.25Nat$, cuts random process with entropy volume which evaluates entropy (3.3.3c):

$$\delta_e \cong 0.4452Nat. \qquad (2.9b)$$

The control, cutting the volume, should compensate for the interacting entropy increments $0.117Nat$. That requires increase information of the applied control up to $0.25 + 0.117 = 0.367Nat$.
This control, cutting the phase volume, could bring information

$$s_{evo} = 0.367 \times 1.272 = 0.466824Nat. \qquad (2.10)$$

Additional potential information following from (2.8a) conveys the entropy of transferring volume $s_{ve} = 0.0636Nat$. That may decrease the amount of attracting potential information (entropy) $s_v = i_v \cong 0.02634Nat$ .
The difference of the entropy volumes is

$$\delta s_{ve} = s_{ve} - s_v = 0.0374. \qquad (2.10a)$$

Sum $\delta s_{ve} + s_{evo} \cong 0.5Nat$ (2.10b)
coincides with that, which determines the amount of information delivered by the impulse cutting the random process. This information compensates for entropy of the virtual probing that delivers the entangled entropy volume.
The potential cut of this volume brings total entropy (2.10b), which the real impulse converts to equivalent information through memorizing.
Assuming the virtual impulse spends part $\delta s_{ve}$ of the entropy volume on transition to cutting gap, the real impulse overcomes entropy threshold $s_{evo}$ by the real cut, producing information $0.5Nat$, which includes information compensation for the virtual entropy volume

$\delta s_{ve} = \delta s_{iv}$. (2.10c)

In multi-dimensional virtual process, correlations grow similarly in each dimension under manifold of impulse observation. The correlations, accumulated sequentially in time, increase with growing number of currently observed process' dimensions, entropy volumes, and collective attractions.

Each impulse cuts the increasing entropy volume, leading to rising density of the cutting entropy even at the invariant impulse size. The current density, which measures the ratio of the impulse volume to the cutting impulse wide, increases the impulse speed.

Killing the distinct volumes densities converts them in the Bits distinguished by information density. Between these different Bits, an information gradient of the attracting force rises, minimizing the difference, which prompts a collective memory. That connects elementary information Bits in units of information process starting the IPF.

The interacting impulse with minimal time interval $\delta_{io} \cong 0.8785 \times 10^{-15}\sec + 0.491 \times 10^{-15}\sec \cong 1.3695 \times 10^{-15}\sec$ covers $\Delta\delta_{io} = 0.1695 \times 10^{-15}\sec$ of a nearest interacting impulse. A bit emerges during relative time $\delta_{io}^{*} \cong 0.3629$ of the impulse under the attracting imaginary entropy within the impulse.

**4.3. Conditions of forming an optimal triplet with a stable cooperative information structure. The emerging information macroprocess**

Information of each previous impulse starts attracting next cutting information in the rotating movement during the impulse imaginary time interval with entropy force, moment's speed (1.4, 1.4a), and potential attracting information

$$i_v \cong 0.02634 Nat \quad (3.1)$$

Information ratio in (2.8a) allows evaluate the attracting information brought by each cutting impulse:

$$i_{vo} = 0.5 \times 0.4142 = 0.207 Nat. \quad (3.1a)$$

Sum of the values in (3.1) and (3.1a) adds information for the attraction:

$$i_{vf} = i_v + i_{vo} = 0.23334 Nat. \quad (3.1b)$$

A part of information delivered with the impulse $i_{vf}$ evaluates its *free* information enables attracting next cutting information until this information will spend on adjoining the following cutting information. Each step-down cut requests information (of ~1/3 of the impulse):

$$1/3(\ln 2 + 0.05) Nat \cong 0.24766 Nat \approx 0.25 Nat \quad . \quad (3.\ 2)$$

That, concurs with (3.1b) measuring the cooperative attraction, which could deliver the free information (3.1). Elementary information unit Bit is an impulse, which encloses the equivalent of entropy $\ln 2$ Nat, whose cut converts this entropy to information.

The nearest impulses' step-down and step-up actions, while cutting entropy of the final probe, memorizes it delivering an information equivalent cost of energy (Sec.2.6). The impulse step-up action limits the cutting information to equivalent Nat that satisfies the minimum of maximal cutting entropy volume.

The Bit encloses the probing impulses logic, conserving through the information cost of energy as a logical equivalent of Maxwell Demon's energy spent on this conversion, and also generates the free information of attraction with its logic.

Thus, the information Bit is a self-participator in both converting entropy to equivalent information, which memorizes logic of its entropy prehistory, and in extending a posterior logic during persistence attraction, involving the multiple impulses' Bits sequence in the information process.

The self-participating inter-action holds the difference between the entropy of last virtual step-up and real step-down cut actions within the ending impulse actions. That could emerge in natural or artificial processes. The real interactive impulse carries both real microprocess, attracting next Bit, and the information cost of getting the Bit.

That distinguishes the real impulses from the virtual impulses in the observing process.

The information equivalent of the impulse wide $\delta_{ue}^{i} \cong 0.05 Nat$ limits its size and the extension minimal time interval $\delta_{te} \approx 1.6\times10^{-14}\sec$ (concurring with $\Delta\delta_{io} = 0.1695\times10^{-15}\sec$).

Potential information speed of attraction within the impulse:

$$c_{ia} = 1/3(\ln 2 + 0.05)/\delta_{te} \cong 0.1548\times10^{14} Nat/\sec, \quad (3.3)$$

is less than both maximal speed $\alpha_{io}$ (2.6), starting a single real information from imaginary entropy, and the speed between the nearest impulses on time interval $\Delta_t$:

$$c_{ika} \cong 0.0516\times10^{14} Nat/\sec. \quad (3.3a)$$

Maximal speed (3.3a) conveys a flow of the formed information Bits in the information process, which carries the enclosed entropy, logic, energy, and memory, while attaching free information (3.1) between the impulses. A single Bit, moving with speed (3.3a), spends its free information of $\sim 1/3 Nat$ to attract next Bit on time interval $t_{ika}$:

$$t_{ika} = 1/3\ln 2/c_{ika} \cong 1/3\ln 2/0.0516\times10^{-14} = 4.477\times10^{-14}\sec. \quad (3.3b)$$

Minimal distance $\Delta_{lo} \approx 14.4\times10^{-15} m$ to a next Bit limits a maximal dynamic spatial speed of information attraction:

$$c_{lo} \approx 14.4\times10^{-15} m/4.477\times10^{-14}\sec = 3.216\times10^{-1} m/\sec. \quad (3.3c)$$

In this process, the impulses pair adjoins in a doublet which encloses bound free information spent on the attraction. The cutoff attracting bits start collecting each three of them in a primary basic triplet unit at equal information speeds, whose resonates frequencies when coheres join the triplet units of information process. The triple resonance concurs with Efimov's Scenario [55-57], early proposed in Borromean Universal three-body relation, including Brunnian knot [58] and Borromean ring [59]. Ancient Borromean Rings represent symbols for strength in unity. Information triplet, satisfying the minimax, is forming during the cooperative rotation of information units, applied to each eigenvector of information dynamics, as well as to the unit groups: doublets and triplets.

### 4.4. **The space-time trajectory solving the minimax variation problem**

Since the entropy functional is defined on Markov diffusion, which includes both micro and macroprocessses, the EF extremals of the VP describe the Hamiltonian dynamic, time-space movement rotating the opposite directional-complimentary conjugated trajectories $+\uparrow SP_o$ and $-\downarrow SP_o$.

The trajectories form spirals located on conic surfaces Figs.3, 3a.

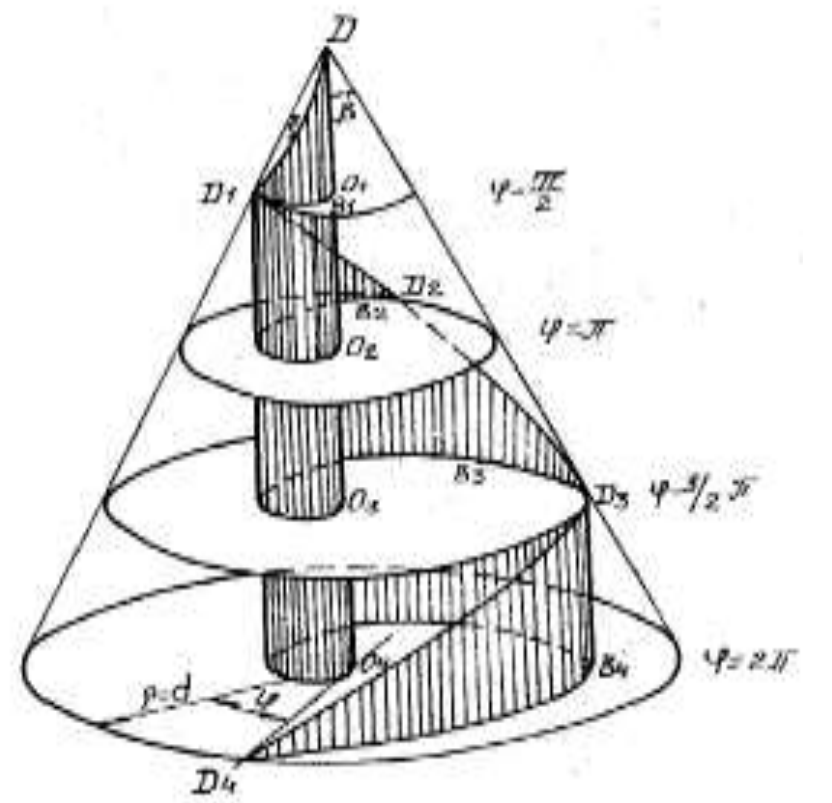

Fig. 3. Forming a space -time spiral trajectory with radius $\rho = b\sin(\varphi\sin\beta)$ on the conic surface at the points D, D1, D2, D3, D4 with the spatial discrete interval DD1= $\mu$, which corresponds to the angle $\varphi = \pi k/2$, $k = 1,2,...$ of the radius vector's $\rho(\varphi,\mu)$ projection of on the cone's base (O1, O2, O3, O4) with the vertex angle $\beta = \psi^o$.

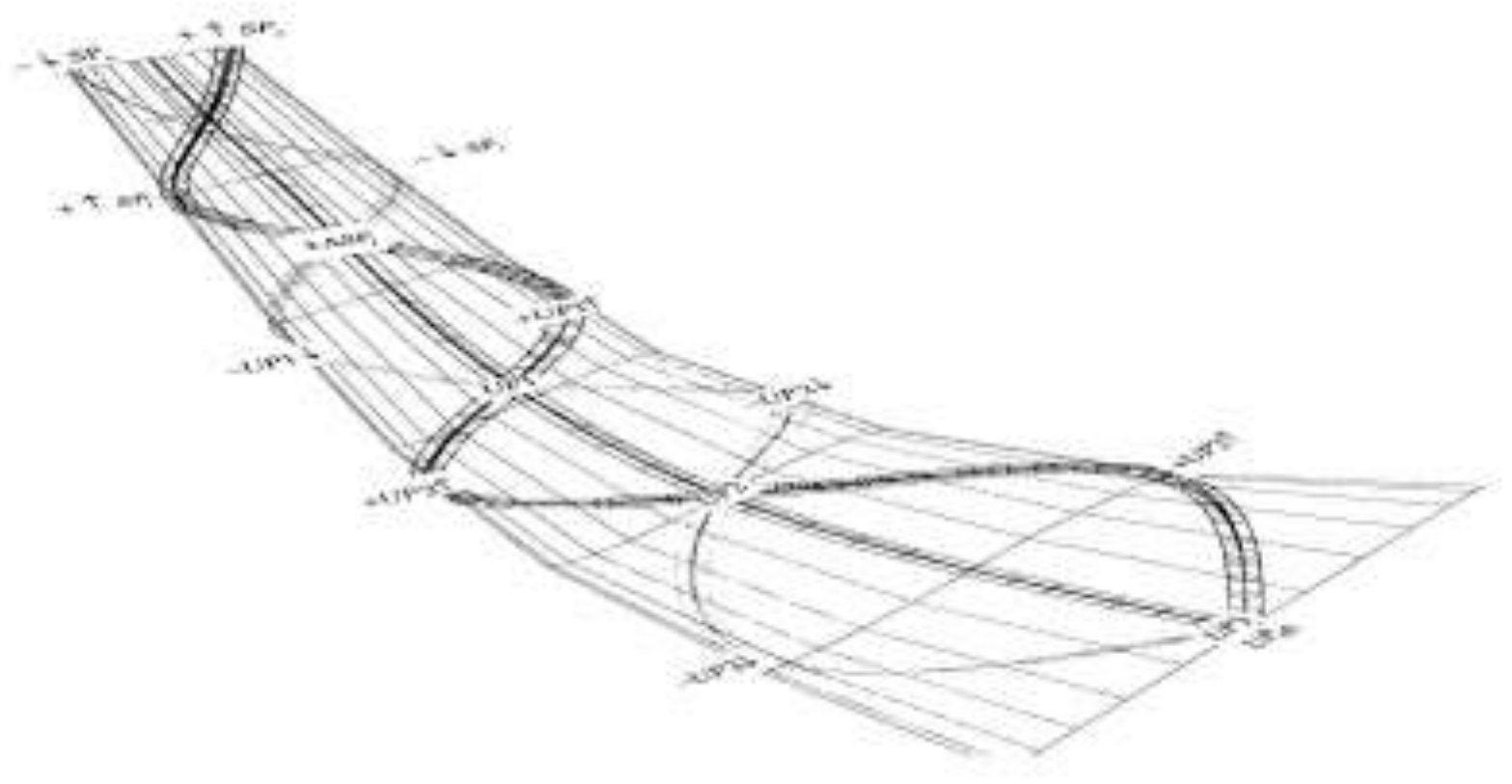

Fig.3a. Time-space opposite directional-complimentary conjugated trajectories $+\uparrow SP_o$ and $-\downarrow SP_o$ of Hamiltonian process (Sec. 3), forming the spirals located on the surfaces (Fig.3). Trajectory on the join bridges $\pm\Delta SP_i$ binds the contributions of process unit $\pm UP_i$ through the impulse joint No-Yes actions, which model a line of switching inter-actions (the middle line between the spirals).

In an observing process, the trajectories entangle its opposite *segments carrying potential qubits with growing probabilities.* When the probability approaches the gap, the trajectory holds a high probable pair of qubits entropy, which the step-down action on the pair first kills its entropy, creating information qubits, and then joins them in bit. Such jumping step-down action, ending the microprocess, is analogous of the jumping step-up actions starting the microprocess. This means, the microprocess proceeds between the merging No and Yes probabilistic actions of observation. It implies that each of such action covers a sub-markov opposite jumping actions. Or, the Yes-No probabilistic actions actually are not separated until the opposite jumping actions of growing probabilities displays border of each impulse with No-Yes actions within the Markov process.

The virtual qubits measure the impulse virtual observation ending in the entangled entropies.

Until the impulses are separated, the microprocess exists within a probability of "eroding a fuzzy border" of the impulse, and the microprocess disappears automatically when the between borders probability approaches zero. Or, the impulse ending border rises with following probability approaching 1, when classical physical bit emerges.
Fig.3a illustrates how a pair of qubits from the opposite segments (at end of the conic tube) binds a bit under the step-down actions within the gap. If the entangled qubits hold their probabilities, there joining in the bit not occurs.
The illustration also indicates and explains how the conjugated entropies entangle $\pm$ or $\mp$ complementary parts of the trajectory units $\pm UP$ or $\mp UP$, which are assembling by mutual attractions along the observing trajectory(Fig.1). Such opposite triple qubits can assemble a triplet bit, spending on binding $(3\times \ln 2/2-1)Nat \cong 0.0397 Nat$ to assemble each 1 Nat. That Bit would have free information $(1-\ln 2-0.0397)\cong 0.267 Nat$ which will be enough for spending it on attraction and binding another two qubits for structuring new triplet.
When the real triple qubits assemble a triplet, it joins virtual observation with multiple trajectories of microprocess in the information process. The trajectories assemble opposite *segments of information process* $\pm SP_i, i=1,...,n$ dimensions compiling them up to maximal $n$. Each $\pm SP_i$ segment averages the microprocesses entangling the attracting qubits. The impulse step-up and step-down actions selects segments $-SP_i$, $\pm SP_i$ ending on the interacting trajectories bridge $\pm \Delta SP_i$ which binds each $i$ pair segment in a spiral structure (Fig.3a).
Each opposite directional segment's enables attracting half of each unit $\pm UP_i$. of the process.
The impulse, joining No-Yes action, connects opposite units $-UP_i$ and $+UP_i$ in unit $UP_i$ -a Bit through bridge $\pm \Delta SP_i$. Dynamics on the bridge borders describes trajectory of switching actions located on middle between the opposite spirals (Fig.3a). Each pair of opposite sections $+SP_i$ and $-SP_i$ forms local circle $\uparrow \text{o} \downarrow$ with sequentially reverse direction of their movement, while total direction along conjugated trajectories $+\uparrow SP_o$ and $-\downarrow SP_o$ preserves. The cyclic process temporary exists during the assembling.
The persistent attraction assembles the Bits in information process, whose information integrates and measures information path functional (IPF) on the macroprocess trajectories.
Transfer from the VP extremal's maximum to minimum limits the dynamic constraint (Sec.4.3).
Each $UP_i$ integrates into $UP_{i+1}$ along the space-time trajectory, and when $-SP_i$ transfers to $+SP_i$, the local EF maximum transforms to the minimum on the bridge.
The qubits doublets, integrated in triplets, form elementary structural units of the emerging information process.
The observation process builds the conjugated dynamics of microprocess which disappears after the information qubits emerge, composing triple units in the macroprocess segments.
In multidimensional observations, the recursive step-down-step-up actions between segments feed the following segment with microlevel entropy, currently converting in information, and connecting each previous one.
The conjugated trajectories describe the EF extremals, while the emerging bits on the bridge between the segments integrates the IPF enclosing integral information in its final bit.
Each bit, memorized in the conjugated interactive bridge, divides the trajectory on reversible process segment, excluding the bit' bridge and irreversible bridge between the reversible segments.
Thus, the observer irreversible dynamic trajectory includes the reversible segments ending with each bridge, where each irreversible bit emerges from the current observation.
It brings irreversibility to the composing segments' conjugated Hamiltonian dynamics.

The EF-IPF linear and nonlinear equations describe information flows initiated by gradients-the information forces starting physical equivalents of the thermodynamics flows and forces satisfying Irreversible Thermodynamics [60,61].

### 4.4.1. The information mechanism of assembling information units in triplets

While the ending microprocess binds each pair $\pm UP_i$ in information unit $UP_i$ on the bridge, each three *information units* $UP_i$ assemble new formed triplet's units $UP_{oi}$ through their ending minimal information speeds having opposite directions.

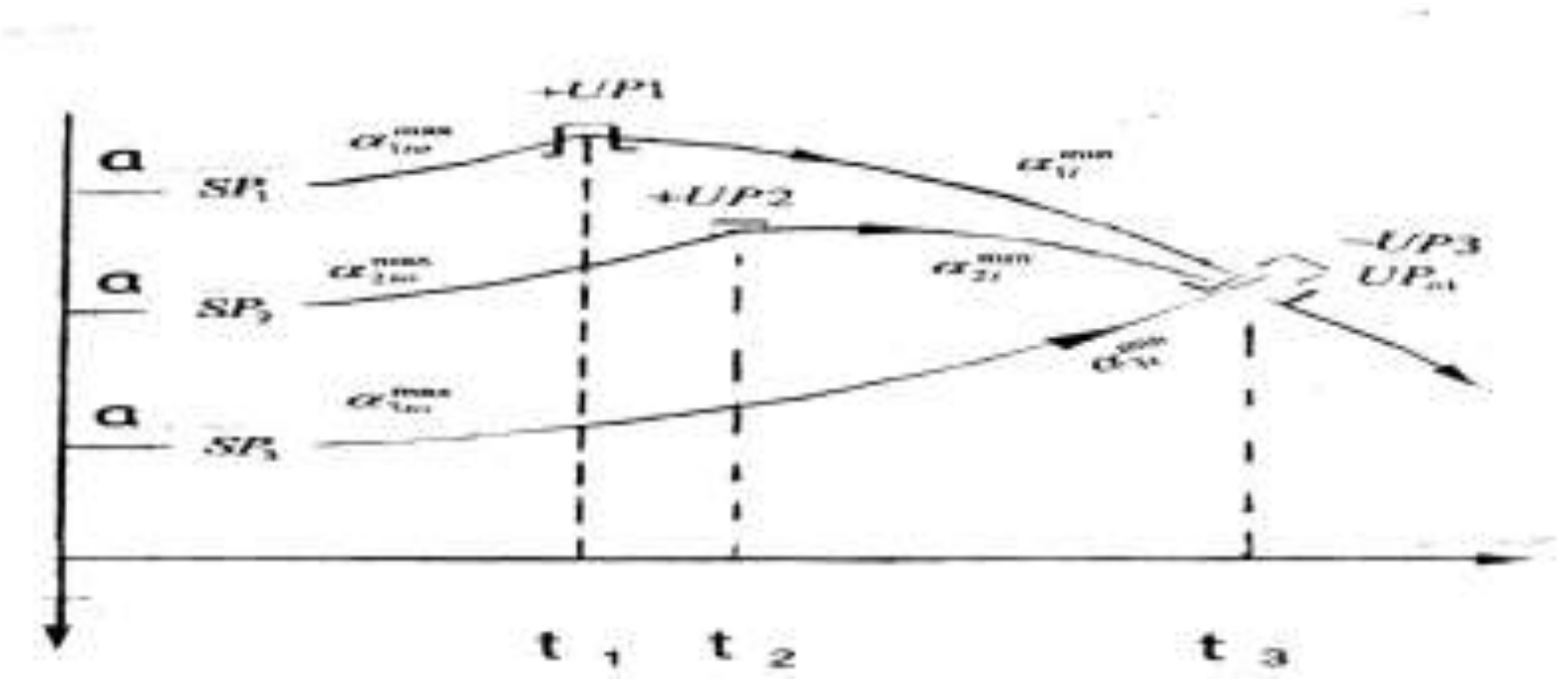

Fig.4. Illustration of assembling a triple qubit units $+UP1, +UP2, -UP3$ and adjoining them to the composed triplet unit $UP_{o1}$ at changing information speeds on the space-time trajectory from $\alpha_{1to}^{\max}$, $\alpha_{2to}^{\max}, -\alpha_{3to}^{\max}$ to $\alpha_{1t}^{\min}$, $\alpha_{2t}^{\min}, -\alpha_{3t}^{\min}$ accordingly; **a** is dynamic information invariant of an impulse.

Fig.4 illustrates simplified dynamics of assembling tree units $UP_i$ $i=1,2,3$ and adjoining them in triplet $UP_{o1}$ along the segments of space–time trajectory $\pm SP_i$ .The assembling proceeds at changing the opposite information speeds on the trajectory between the segments from maximal $|\mp\alpha_{ito}|^{\max}$ to $|\pm\alpha_{it}|^{\min}, i=1,2,3$, while the *bridge o*f unit $UP3$ connects these units with $UP_{o1}$.(Fig.4 shows a single symmetrical part of the conjugated dynamics).

Suppose segment $-SP_1\downarrow$ starts moving unit $-UP1\uparrow$ with maximal speed $|-\alpha_{1t}|$ , and segment $+SP_1\uparrow$ starts moving unit $+UP1\uparrow$ with speed $|+\alpha_{1t}|$. The rotating movement of both units involves a local circle rotating these segments' speed in right direction $\uparrow \mathrm{o} \downarrow$.The next segment $-SP_2\downarrow$ starts moving $-UP2\downarrow$ with maximal speed $|-\alpha_{2t}|<|-\alpha_{1t}|$ and $+SP_2\uparrow$ starts moving $+UP2\uparrow$ with speed $|+\alpha_{2t}|<|+\alpha_{1t}|$. Each of these speed forms second local circle $\downarrow \mathrm{o} \uparrow$ rotating in left direction with absolute speed less than those in the previous circle. Third segment $-SP_3\downarrow$ starts moving unit $-UP3\downarrow$ with maximal speed $|-\alpha_{3t}|<|-\alpha_{2t}|$, and $+SP_3\uparrow$ starts moving unit $+UP3\uparrow$ with maximal speed $|+\alpha_{3t}|<|+\alpha_{2t}|$.

Third circle rotates in right direction $\uparrow \mathrm{o} \downarrow$ with absolute values of the speeds satisfying relation

$$|\alpha_{3t}|<|-\alpha_{2t}|<|\alpha_{1t}|. \qquad (4.1)$$

The opposite directional speeds within each circle attract each $-UP_i$ to $+UP_i$, minimizing the ending speeds down to $|\alpha_{it}|=\alpha_{it}^{\min}$ .

When these pairs approach, the starting attracting force (1.3) (Sec.4.1) binds them in related units $UP_i$ with sequence of speeds

$$\alpha_{1t}^{\min} \to \alpha_{2t}^{\min} \to \alpha_{3t}^{\min} . \qquad (4.2)$$

Current units $UP1, UP2$ have assembled at growing time-space intervals $t_{o1} < t_{o2} < t_{o3}$, at $t_{o3} \geq t_{o1} + t_{o2}$, which automatically integrate unit $UP3$, and then condenses the units forming the triple knot $UP_{o1}$ that memorizes the triple.

The IPF is sequentially summing information of units $UP_i$, $i = 1,2,3$ and memorizing only current $UP_{o1}$, while the previous $UP_i$ are erased as their information integrates and memorizes $UP_{o1}$.

When the condensed information of $UP_{o1}$ integrates sequence $UP1 \to UP2 \to UP3$ in a primary triplet, it automatically implements the IPF with minimization of total time of building the triplet.

When speeds $+\alpha_{13t}$ and $-\alpha_{13t}$ of each opposite segments move close to speeds $+\alpha_{3t}, -\alpha_{3t}$ by the end of interval $t_3$, according to (4.1, 4.2), it makes possible connecting the conjugated information

$$a_{+23} = +\alpha_{23t3} t_3 \text{ with } a_{-23} = -\alpha_{23t3} t_3 . \qquad (4.3)$$

which is joining each of two complementary units in a third time-space loop.

The triple knot generates free information with the loop of the joining triplet's processes, which attracts the following units self-forming new triplet assembling the joint knot.

Such a triple self-supporting cyclic process requires an initial flow of entropy converted to information.

The bridge between the segments convert entropy of observing impulse to information, whose free information through the loops in the cycles, assembles each triplet.

At forming $UP3$, new triplet Bit may appear if the $UP1$ ending speed, minimized by the opposite speed of $UP2$, equalizes it with third segment minimal speed $\alpha_{3t}^{\min}$; and the $UP2$ ending speed, being also minimized in the opposite movement, equalizes it with third segment minimal speed $\alpha_{3t}^{\min}$ by the moment of forming $UP3$.

Adjoining the two with the third allows forming $UP3$ *during* formation of $UP1$ and $UP2$, which are joining with a minimal information speed equals to ending speeds of the two moving segments (Fig.4).

These speeds minimize attracting information $\mathbf{a} = 1/3bit \approx 0.23Nat$ from each $UP1, UP2, UP3$ whose sum $3\mathbf{a} \cong \mathbf{a}_o$ can join all three in new triplet Bit with information $\mathbf{a}_o$.

(More precise calculation brings free information $\mathbf{a}^* = (1 - \ln 2 + 0.0397)/1.44 \cong 0.24bit$ which includes both attracting and binding information $\mathbf{a}_b = 0.0397/1.44 \cong 0.02757bit$ in forming the triplet). That confirms relation (3.2).

Actually, the attracting information $\mathbf{a}$ of unit $UP1$ decreases speed of its starting movement $\alpha_{13t} = \alpha_{1t} - \Delta\alpha_{1t} \to \alpha_{1t}^{\min}$ on such increment $\Delta\alpha_{1t}$ which can equalize the units speed with $\alpha_{3t}^{\min}$ forming $UP3$.

Attracting information $\mathbf{a}$ of $UP2$ decreases speed of its starting movement

$$-\alpha_{23t} = -\alpha_{2t} - \Delta\alpha_{2t} \to -\alpha_{2t}^{\min} \qquad (4.4)$$

on such increment $\Delta\alpha_{2t}$ which allows also equalizing this unit speed with $\alpha_{3t}^{\min}$ .
Here the speed signs hold the directions of rotation in each local circle.
The attracting movement connects these speeds in the triple movement arranging in direction

$$+\alpha_{1t}^{\min} \Rightarrow -\alpha_{2t}^{\min} \Leftarrow +\alpha_{3t}^{\min} \qquad (4.\ 5)$$

when ending speed $-\alpha_{2t}^{\min}$ emanated from $UP2$ joins equal minimal speeds $+\alpha_{1t}^{\min}, +\alpha_{3t}^{\min}$ forming the end of triple knot which is assembling the triplet unit $UP_{o1}$.
The knot binding information memorizes the accumulated information at moving joint speed

$$\alpha_{1t}^{\min} = |\alpha_{2t}^{\min}| = \alpha_{3t}^{\min} = \alpha_{uo1}\ . \qquad (4.\ 5a)$$

An example of assembling a triplet space structure shows Fig.5.

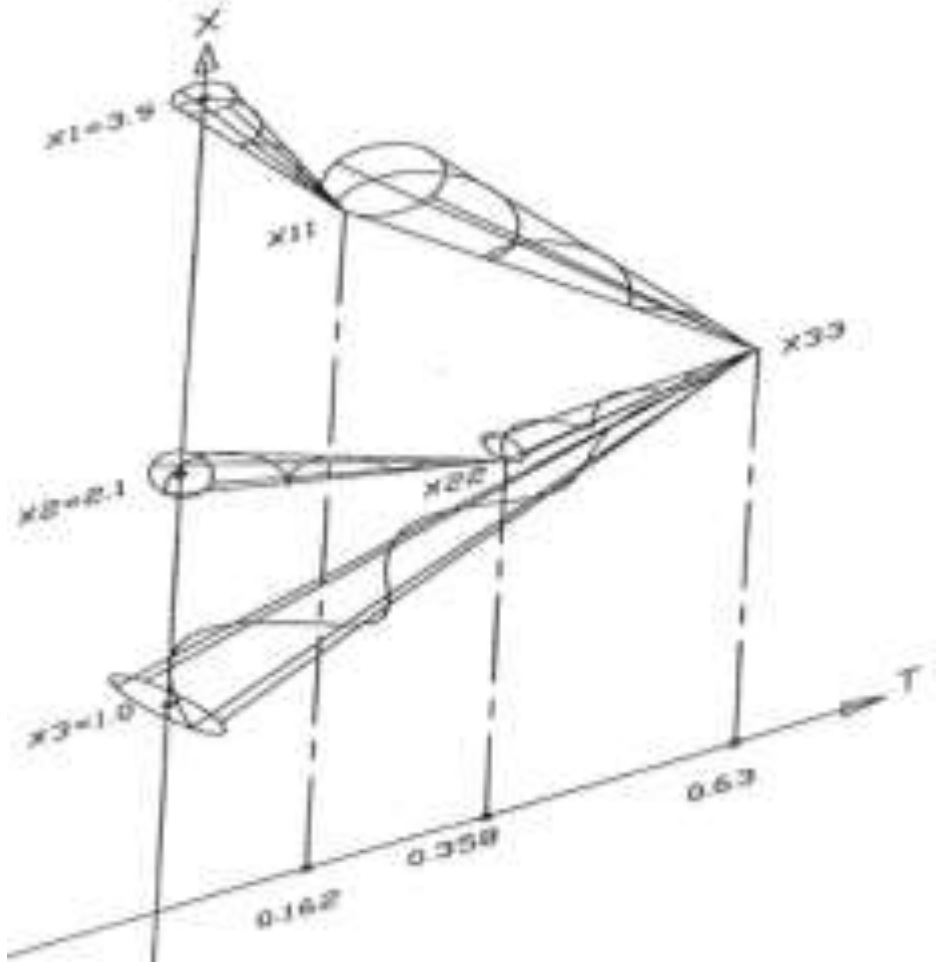

Fig. 5. Forming a triplet's space-time structure.

Let us identify the time of formation of this space-time structure during cooperative dynamics.
Assume forming $UP1$ Bit on segment $SP_1$ needs its moving time interval $t_1$ and time interval $\Delta t_{13}$ to attract $UP3$. Forming second $UP2$ Bit on segment $SP_2$ needs its moving time interval $t_2$ and time interval $\Delta t_{23}$ to attract $UP3$. Forming $UP3$ Bit takes time interval $t_3$ of moving third segment $SP_3$ .
(The attracting time intervals belong to segments free information).
Since all three segments move decreases information speeds, it is possible to reach equality

$$t_{13} = t_3 = t_{23} + t_{12} \qquad (4.\ 6)$$

where $t_{12}$ is time interval between $UP1$ and $UP2$.

Satisfaction of (4. 6) allows adjoining all three segments during time interval $t_3$ of forming $UP3$.

Since intervals $t_1, t_2, t_3$ are connected by the same free information invariant $\mathbf{a} = 1/3 bit \approx 0.23 Nat$,
it brings joint relation

$$\alpha_{1to}^{\max} t_1 = \alpha_{2to}^{\max} t_2 = \alpha_{3to}^{\max} t_3 = \mathbf{a}_o \cong 3\mathbf{a}\ . \qquad (4.7)$$

The simulated dynamics of the units' speeds determine information delivered at each of these intervals:
$\mathbf{a}^{13} = \alpha_{13t} t_{13} \cong 0.232$ , $\mathbf{a}^{23} = \alpha_{23t} t_{23} \cong 0.1797$ and $\mathbf{a}^{33} = \alpha_{3t} t_3 \cong 0.268$. (4.7a)

At $t_{13} = t_3$ and $\alpha_{3to} t_3 \cong \mathbf{a}_o$, $\alpha_{13t} t_{13} \cong 1/3\mathbf{a}_o \cong \mathbf{a}$, we get $\alpha_{3to} / \alpha_{13t} \cong 3$. (4.7b)

To satisfy (4.7) at $\mathbf{a}^{13} = \mathbf{a}^{33} = \mathbf{a}$, the information spent on attraction and assembling triplet unit $UP_{o1}$, should also be $\mathbf{a}$. Therefore, if information spent on attraction $UP3$ is $\alpha_{23t} t_{23}$, the difference

$$\Delta\mathbf{a}^{23} = \mathbf{a} - \alpha_{23t} t_{23} \cong 0.23 - 0.1797 \cong 0.05 \qquad (4.7c)$$

spends on assembling $UP_{o1}$ using the delivered information $2\mathbf{a} = \mathbf{a}^{13} + \mathbf{a}^{33}$.
Following relations

$$\Delta\mathbf{a}^{23} = |\alpha_{23t}|\,\delta t_{23}, |\alpha_{23t}| \cong 1/3\alpha_{3ot}, \alpha_{3ot} = \mathbf{a}_o / t_3, \qquad (4.8a)$$

determine the time interval on assembling $UP_{o1}$:

$$\delta t_{23} \cong 3 t_3 \Delta\mathbf{a}^{23} / \mathbf{a}_o, \qquad (4.8b)$$

which evaluates

$$\delta t_{23} \cong 0.214 t_3. \qquad (4.8c)$$

Assembling information $\mathbf{a}_o \cong 3\mathbf{a}$ in the knot with speed $-\alpha_{23t}$ determines $-UP_{o1}$ sign.

The assembling loop, connecting speeds (4.5a), builds the triplet knot $-UP_{o1}$, which binds primary information of $UP1, UP2, UP3$ .The knot free information will spend on the information attraction of a second forming triplet $+UP_{o1}$. The attracting free information from triplet $-UP_{o1}$ or $+UP_{o1}$ forms secondary information unit $UP_{o1}$. The time of building $UP_{o1}$ minimizes the time of sequential connections of units $UP_i$ , $i = 1,2,3$ in the information process. The $n$-dimensional process trajectory locates multiple conjugated pairs $+\uparrow SP_{oi}$ and $-\downarrow SP_{oi}$, which could assemble each triple unit $UP_i$ , $i = 1,2,3$ in cooperative $\mp UP_{oi}$ Bit, where the sign of each unit depends on the sign each second segment $SP_{2i}$ which binds each of these units in the triple segment knot.

This attracting movement assembles their three Bits in new formed $\mp UP_{i0}$ Bit forming new circular loops on higher (second) level which connects the equal speeds of each triple (Fig.6).

Particularly, the attractive motion of rotating triples units $-1_o(+UP_{o1}, -UP_{o2}, +UP_{o3})$ can cooperate next one to form triplets unit $-UP_{4o}$ depending on sign of segment $-SP_{2i}$.

The sign of $+SP_{2i}$ at cooperating opposite triple $+2_o(-UP_{o5}, +UP_{o6}, -UP_{o7})$ participates in forming triplets unit $+UP_{5o}$, other three rotating triple $-3_o(+UP_{o8}, -UP_{o9}, +UP_{o10})$ cooperates in unit $-UP_{6o}$ that closes each of three rotating circles.

Cooperative motion of units $+4_{1o}[-UP_{4o}, +UP_{5o}, -UP_{6o}]$ assembles them in new formed triplet $+UP_{I0}$ Bit composing third levels of triplets. If above units $-1_o, +2_o$ $-3_o$ have equal attracting speeds by the moment of cooperation, they may join in composite unit $+4_{10}$ during building this triple.

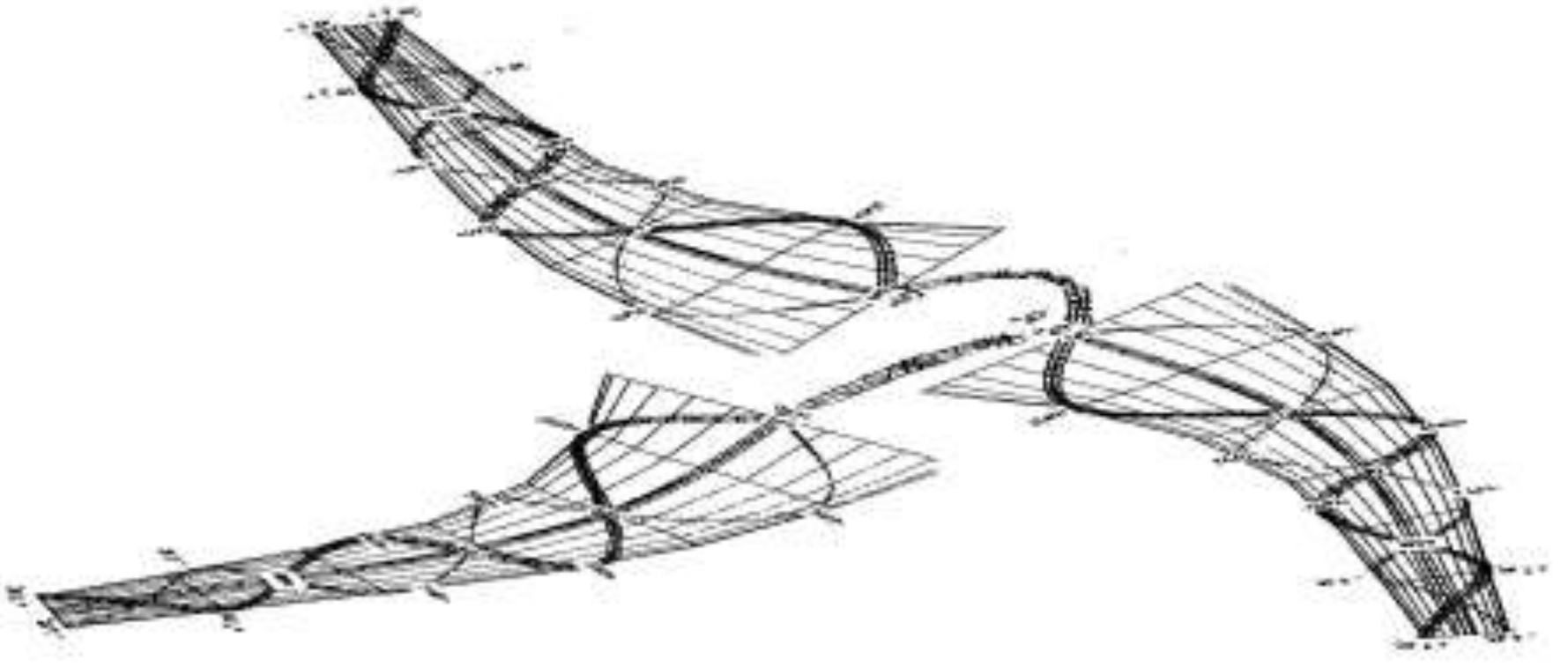

Fig.6. Assembling three formed units $\pm UP_{i0}$ at a higher triplet's level connecting the units' equal speeds (Fig.3). The attracting circles, analogous to that on Fig.5, are not shown here.

Particularly, the attractive motion of rotating triples units $-1_o(+UP_{o1}, -UP_{o2}, +UP_{o3})$ can cooperate next one to form triplets unit $-UP_{4o}$ depending on $-SP_{2i}$. The sign of $+SP_{2i}$ at cooperating opposite triple $+2_o(-UP_{o5}, +UP_{o6}, -UP_{o7})$ participates in forming triplets unit $+UP_{5o}$; other three rotating triples $-3_o(+UP_{o8}, -UP_{o9}, +UP_{o10})$ cooperate in unit $-UP_{6o}$. That closes each of three rotating circles. Cooperative motion of units $+4_{1o}(-UP_{4o}, +UP_{5o}, -UP_{6o})$ assembles them in new formed triplet $+UP_{I0}$ Bit composing third levels of triplets. If above units $-1_o, +2_o$ $-3_o$ have equal attracting speeds by the moment of cooperation, they may join in composite unit $+4_{10}$ during building this triple.

Information of units $UP_{o1}, UP_{o2}$, currently formed at the growing time-space intervals $t_{io1} < t_{o2} < t_{io3}$, at $t_{io3} \geq t_{io1} + t_{io2}$, automatically integrate and memorize $UP_{oit}$ which along with $UP_{o3}$ forming the triple knot $UP_{4ot} = UP_{oit}$. The sequential built triplet knots memorize only current $UP_{oit}$, while the previous units' information has erased. The cooperative dynamics automatically implement the IPF minimax, minimizing total time of building each composite information unit. The minimax leads to sequential decreasing ending information speed of each $UP_{oi}$, and therefore to decreasing starting information speed of a next cooperative unit. The minimax composes the ordered connection of binding speeds at forming triplets, which memorizes the bound information units, sequentially structures the observer's information dynamics, binding and connecting each $UP_{oi}$ in new cooperating triple units.

Building the units of dynamic information structure from multiple $\pm SP_{ij}$ segments of $n$ -dimensional process trajectory requires assembling each $UP_{ij}$ from two segments, taking from conjugated segments on the trajectory of one dimension, with other opposite directional segment of the conjugated pair which belongs to other dimension.

Building a cooperative forming triplet $UP_{oij}$ involves three such complimentary parts from three different process' dimensions illustrated on Figs.3a,6. Building more involves more process dimensions.

In the minimax attracting movement of $\pm UP_{ij}$, three of the time–space segments spend only the time-space interval of a third segment, joining three segments while both the first and second segments attract third segment during their moving segments' speeds intervals accordingly. The opposite moving segments on the trajectory cooperate the symmetrical-complimentary $\pm UP_{oij}$ triplets, enclosing half of each complementary Bits, which through self-joining enable creating complete Bit whose attracting information can assemble other $UP_{oij}$ triplets. The minimax movement decreases ending information speed of each complementary units of $\pm UP_{oij}$. The units enclosed in the knot increase information density of each following Bit. Each such triple unit contains four Bits whose forth Bit encloses the triplet information in a triple knot and provides free information for subsequent assembling. The attracting space-time movement selects, orders, and assembles the rotating segments' cooperating speeds along the time-space dynamic trajectories.

*Indeed*. Each primary $\pm UP_{o1}$ from different conjugated pairs emanates three time –space segments $\pm SP_{o1}, \pm SP_{o2}, \mp SP_{o3}$ selected on the minimax trajectory. The dynamic symmetry of opposite moving segments on the trajectory actually generates half of each complementary $\pm UP_{o1}$ unit's $\pm BitI4, I = 0,1,....m$ which could self-joins, creating complete Bit $|BitI4|$ of $UP_{o1}$, whose attracting free information can assemble other triplets units $\pm UP_{Io1}$ and so on.

The triplets' knots create *new class* of information Bits that distinct from the first class information Bits assembling units, which were generated via virtual probes with entropies of cutting random process.

Each forming Bit of the following class grows density of the enclosed information, its geometrical density, and curvature.

Forming the closing loop in processing a complimentary triplet allows self-formation of such joint triplet.

The triple knot generates free information for a next consecutive attraction, increasing number of cooperating triplets.

The joint triplet with free information produces cooperative assembling which brings new information in each triple knot.

Each dynamic triple cooperation builds basic mechanism equalizing information speeds of moving impulses which carry coherent speeds, whose frequencies resonate rotating in a cycle that assembles the triplet.

### 4.5. The conditions for building space-time information network, observer geometry, and restriction on its parameters

Multiple moving triplets, sequentially equalizing their speeds-frequencies in resonance, assemble nested layers of bound triplet units-nodes. The attracting nodes' logic self-organizes information networks (IN) in logical information structure of the nested hierarchy of the nodes. The self-building proceeds the time-space information dynamics, which self-assemble space-time information nested structure of *information network* (IN) (Fig.7), enclosing the growing number of triplets.

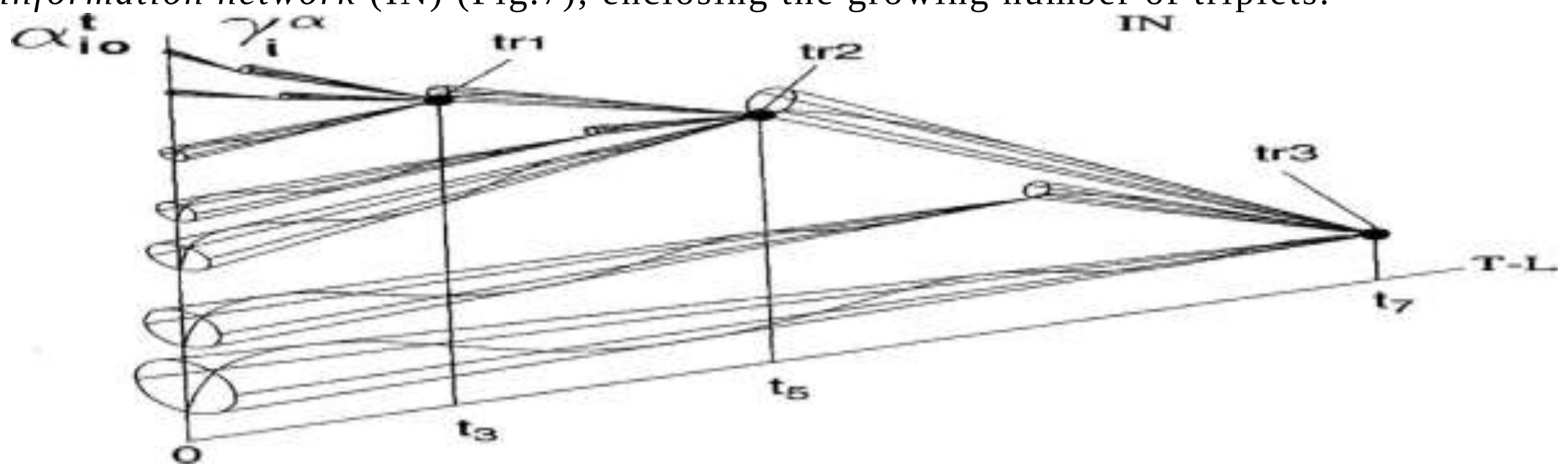

Fig.7. The IN information geometrical structure of hierarchy of the spiral space-time dynamics of triplet nodes (tr1, tr2, tr3,…); $\{\alpha_{io}\}$ is a ranged string of the initial eigenvalues, cooperating on (t1, t2, t3) locations of T-L time–space, $\{\gamma^\alpha_i\}$ is parameter measuring ratio of the IN nodes space-time locations.

The attracting process of assembling triplet's knots forms a rotating loop shown on Fig.7 at forming each following cooperative triplet tr1,tr2, . The scales of curves on Figs. 5,6 distinct from the interacting knots' curves on Fig. 7, since these units, at reaching equal speeds, resonates, which increases size of the curves on Figs. 5,6 while building each triple.
Each triplet accumulates three Bit's information logic enfolded in the knot's information Bit.
Such self-forming structure automatically implements the IPF integration of the process information in a last IN node.
The *triplet information dynamics* start with information speed of its first segment's eigenvalue (2.6):

$$\alpha_{io} \cong 1.41\times 10^{15} Nat/\sec \,, \qquad (5.1)$$

where $\alpha_{io}=\alpha_{1o}$ is potential information speed predicted by the moving entangled entropy volume.
Each self-forming triplet joins two trajectory segments with positive eigenvalues by reversing their unstable eigenvalues and attracting a third segment with negative eigenvalues. This segment's rotating trajectory moves it to the two opposite rotating eigenvectors and cooperates all three information segments in a triplet's knot (Figs 3,4,5,7).
Prior joining, the segment entropy-information, satisfying the minimax, ends with its minimum which evaluates relations (2.9b) and (2.10).
The optimal triplet requires minimal time interval spending on equalization each its segment's eigenvalues with the following segment's eigenvalues.
At forming triplet, the cooperation of two segments may bind attracting information $2i_{vf}=2(i_v+i_{vo})=0.4668Nat$ of both segments, which equals to that requires to overcome entropy threshold $\delta_e \cong 0.4452Nat$ (3.2.9b) for joining the third segment. Third segment encloses information of the two triplet segments, including information that binds them.
Its ending free information attracts next information segment of following triplet, which builds the attracting information units in the subsequent information dynamics.
The ratios of starting information speeds $\gamma_1^{\alpha}=\alpha_{io}/\alpha_{i+1o}$ and $\gamma_2^{\alpha}=\alpha_{i+1o}/\alpha_{i+2o}$ on triplet's cooperating segments, satisfying (2.3),(2.4), determine information dynamic invariant $\mathbf{a}(\gamma)$ according to formulas [26]:

$$\gamma_1^{\alpha}=\frac{\exp(\mathbf{a}(\gamma)\gamma_2^{\alpha})-0.5\exp(\mathbf{a}(\gamma))}{\exp(\mathbf{a}(\gamma)\gamma_2^{\alpha}/\gamma_1^{\alpha})-0.5\exp(\mathbf{a}(\gamma))},\ \gamma_2^{\alpha}=1+\frac{\gamma_1^{\alpha}-1}{\gamma_1^{\alpha}-2\mathbf{a}(\gamma)(\gamma_1^{\alpha}-1)} \qquad (5.2)$$

where multiplication $\gamma_1^{\alpha}\times\gamma_2^{\alpha}=\gamma_{13}^{\alpha}$ holds segment's eigenvalue ratio $\gamma_{13}^{\alpha}=\alpha_{io}/\alpha_{i+3o}$.
It's seen that $\gamma_1^{\alpha}$ is basic triplet scale factor which depending on invariant $\mathbf{a}(\gamma)$ holds invariant measure. Invariants $\mathbf{a}_i(\gamma)=\alpha_i^t t_i$ and $\mathbf{a}_{io}(\gamma)$, at complex eigenvalue dynamics: $\lambda_i^t=\lambda_{io}^t\exp(\lambda_{io}^t t_i)[2-\exp(\lambda_{io}^t t_i)]^{-1}$, connect Eq.

$$\mathbf{a}=\mathbf{a}_o\exp(-\mathbf{a}_o)(1+\gamma^2)^{1/2}[4-4\exp(-\mathbf{a}_o)\cos\gamma\mathbf{a}_o+\exp(-2\mathbf{a}_o)]^{-1/2}] \ . \qquad (5.3)$$

In the dynamics of real eigenvalue: $\alpha_i^t=\alpha_{io}^t\exp(\alpha_{io}^t t_i)[2-\exp(\alpha_{io}^t t_i))]^{-1}$, invariants $\mathbf{a}_i(\gamma_i)$, $\mathbf{a}_{io}(\gamma_i)$ connect Eq:

$$\mathbf{a}_i(\gamma_i)=\mathbf{a}_{io}(\gamma_i)\exp\mathbf{a}_{io}(\gamma_i)(2-\exp\mathbf{a}_{io}(\gamma_i))^{-1} \ . \qquad (5.4)$$

Optimal ratio $\gamma_{io}=0.4142$ satisies the minimax with $\mathbf{a}_{io}(\gamma_{io}=0.4142)\cong 0.73$ and $\mathbf{a}_i(\gamma_{io})\cong 0.23$.
Each triplet structure identifies its starting invariants $\gamma_i$, $\mathbf{a}_{io}(\gamma_i)$ and then both $\gamma_1^{\alpha},\gamma_2^{\alpha}$ .
At the known starting eigenvalue $\alpha_{io}=c_{iv}\cong 2.4143\times 0.596\times 10^{15}Nat/\sec\cong 1.44\times 10^{15}Nat/\sec$ and $\alpha_{io}=\alpha_{1o}$, the next such speeds in the triplet are $\alpha_{1o}/\gamma_1^{\alpha}=\alpha_{2o}, \gamma_1^{\alpha}(\gamma_{io})=2.236$, at. $\alpha_{2o}\cong 0.644\times 10^{15}Nat/\sec$.

Starting information speed of a first cooperating segment's eigenvalue $c_{ika}=\alpha_{1o}, c_{ika}\cong 0.0516\times 10^{14} Nat/\sec$ leads to second segment's eigenvalue information speed $\alpha_{2o}\cong 0.02345\times 10^{14} Nat/\sec = 0.2345\times 10^{13} Nat/\sec$.

The known $\alpha_{2o}, \gamma_{32}^{\alpha}$ settle the third starting eigenvalue's speed:

$$\alpha_{3o}\cong \alpha_{2o}/\gamma_2^{\alpha} = 0.02345/1.6\times 10^{14} Nat/\sec = 0.1465\times 10^{13} Nat/\sec. \quad (5.4a)$$

Invariants $\gamma_1^{\alpha}, \gamma_2^{\alpha}$ limits each triplet's time intervals and related intervals of rotating space dynamic movement at

$$\alpha_{io}t_{io}=\alpha_{i+1o}t_{i+1o}=\alpha_{i+2o}t_{i+2o}=\mathbf{a}_{io}(\gamma_i),\ \gamma_1^{\alpha}=\alpha_{io}/\alpha_{i+1o}=t_{i+1o}/t_{io} \text{ and } \gamma_2^{\alpha}=\alpha_{i+1o}/\alpha_{i+2o}=t_{i+2o}/t_{i+1o}. \quad (5.5)$$

Particular observations leads to specific ratios of the triplet's initial eigenvalues $\alpha_{1o}/\alpha_{2o}=\gamma_1^{\alpha}, \alpha_{2o}/\alpha_{3o}=\gamma_2^{\alpha}$, satisfying the invariant relations (5.2-5.4).

Each iinteracting impulse with information measure $\mathbf{a}_{io}=\ln 2 Nat$ has multiplicative information measure (Sec.2):

$$U_m=(\mathbf{a}_{io})^2, \quad (5.6)$$

which binds the following double connection in a triplet and provides invariant information measure of each interaction with total information

$$(\mathbf{a}_{io}(\gamma_{io}))^2+\mathbf{a}_i(\gamma_{io})\cong 0.7\cong \mathbf{a}_{io}(\gamma_{io}). \quad (5.7)$$

This invariant approaches the information delivered from each previous impulse.
Forming a stable triplet, which enables the attraction, limits the maximal ratio

$$4.8\geq \gamma_1^{\alpha}\geq 3.45. \quad (5.8)$$

That ration determines a boundary of the triplet scale factor $\gamma_1^{\alpha}(\gamma_i)$. Approaching $\gamma_1^{\alpha}=\gamma_2^{\alpha}\to 1$ leads to repeating the triplet's eigenvalues that limits related theoretical admissible invariant $|\gamma_i|\in(0.0-1.0)$.

Approaching information locality of $\gamma_i=1-o$ at $\mathbf{a}_o(\gamma_i=1-o)$ indicates a jump of an event information, moving to the event information $\mathbf{a}_o(\gamma_i)$ that rises the time ratio of the following to preceding intervals.

For example, when $\gamma_i$ approaches 1, $\mathbf{a}_o(\gamma_i)$ is changing from $\mathbf{a}_o(\gamma_i=1-o)=0.56867$ to $\mathbf{a}_o(\gamma_i)=0$, and the above time's ratio of reaches limit $\tau_{i+1}/\tau_i=1.8254$.

This changes sign of the eigenvalues' ratio: $\alpha_{io}/\alpha_{it}\cong -1.9956964$ (following from (5.5) at $\gamma_i=1$) and leads to $\mathbf{a}_o(\gamma_i=1)=0$ which brings information contributions for regular control $\mathbf{a}(\gamma_i=1)=0$ and impulse control $\mathbf{a}_o^2(\gamma_i)=0$. This jump of time or related eigenvalues is a *dynamic indicator* of breaking up the dynamic constraint (Sec.4.3). It leads to cutting off the model's dynamics from the initial random process with a possibility of getting more uncertainty. The appearance of an event, carrying $\gamma_i\to 1$, leads to *decoupling* the events chain and rising chaotic diffusion dynamics. Whereas the moment of this event's occurrence *predicts* measuring a current event's information $\mathbf{a}_o(\gamma_i)\neq 0$ and using it to compute $\gamma_i$ applying (5.2-5.4). That allows predicting the decoupling, losing stability, and chaotic dynamics.

Practically admissible maximal $\gamma_{ia}\to 0.8$ leads to a *minimal stable* triplet with $\gamma_1^{\alpha}=\gamma_2^{\alpha}\to 1.65$ which limits the acceptable $\gamma_i\to(0-0.8)$. That, for $\gamma_{io}\to 0$ determines $\mathbf{a}_{io}(\gamma_{io})$=1.1 bit and $\mathbf{a}_i(\gamma_{io})$= 0.34 bits. The triplet dynamics with $\mathbf{a}_i(\gamma\to 0)\cong 0.23$ hold $\gamma_1^{\alpha}\cong 2.460, \gamma_2^{\alpha}\cong 1.817$ and $\gamma_{13}^{\alpha}\cong 4.6$.

Optimal $\gamma_{io} = 0.4142$ brings $\gamma_1^{\alpha} \cong 2.21, \gamma_2^{\alpha} \cong 1.76, \gamma_{13}^{\alpha} \cong 3.89$, and $\gamma_i = 0.8$ brings $\gamma_1^{\alpha} \cong 1.96, \gamma_2^{\alpha} \cong 1.68, \gamma_{13}^{\alpha} \cong 3.3.$

Sequence of the model eigenvalues (... $\alpha_{i-1,o}^t$ , $\alpha_{io}^t$ , $\alpha_{i+1,o}^t$ ), satisfying the triplet's formation, is *limited* by boundaries for $\gamma \in (0 \to 1)$, which forms a *geometrical progression* with

$(\alpha_{io}^t)^2 = \alpha_{i-1o}^t \alpha_{i+1o}^t, i = 2,3,...n$, (5.9)

representing the geometric "gold section" at ratio

$$G = \frac{\alpha_{i+1,o}^t}{\alpha_{io}^t} \cong 0.618. \quad (5.10)$$

At $\gamma \to 1$ the sequence $\alpha_{io}^t, \alpha_{i+1,o}^t, \alpha_{i+2,o}^t, ...$ forms the Fibonacci series, where the ratio $\alpha_{i+1,o}^t / \alpha_{i+2,o}^t = \gamma_2^{\alpha}$ determines the "divine proportion" $PHI \cong 1.618$, satisfying

$PHI \cong G + 1$; (5.11)

and the eigenvalues' sequence loses its ability to cooperate.

*Indeed*. At $\gamma \to 0$, solutions of (5.2-5.4) determine

$\mathbf{a}_o(\gamma_i \to 1) = 0.231, \gamma_1^{\alpha} \cong 2.46, \gamma_2^{\alpha} \cong 4.47, \gamma_2^{\alpha} / \gamma_1^{\alpha} = \gamma_{23}^{\alpha} \cong 1.82.$

Ratio $(\gamma_1^{\alpha})^{-1} = \alpha_{io}^t / \alpha_{i-1,o}^t \cong (2.46)^{-1} \cong 0.618$ for the above eigenvalues forms a "golden section" (5.10) at $G = (\gamma_1^{\alpha})^{-1} = \alpha_{io}^t / \alpha_{i-1,o}^t \cong (2.46)^{-1} \cong 0.618$ *and* the "divine proportion" $PHI \cong 1.618$.

Above relations hold true for each primary pair of the triplets' eigenvalues sequence, while the third eigenvalue has ratio

$\alpha_{i+1,o}^t / \alpha_{io}^t = (\gamma_{23}^{\alpha})^{-1} \cong 0.549$. (5.11a)

Solution of the equation for invariants (5.2-5.4) at $\mathbf{a}(\gamma_i \to 1) = 0$) brings $\gamma_1^{\alpha} = \gamma_2^{\alpha} = 1$ with repeating $\alpha_{i-1,o}^t = \alpha_{io}^t = \alpha_{i+1,o}^t = \alpha_{i+2,o}^t =, ...$

The eigenvalues' sequence loses ability to cooperate, disintegrating to the equal and independent eigenvalues. The network, built through the attracting resonance of the units free information' eigenvalues, has limited stability and therefore each IN encloses a finite structure. That's why the observing process might self-build only multiple limited IN.

**4.5.1.The restrictions on the space-time rotating trajectory growing with increasing dimensions** $1,...,i,...,n$.

Each rotating movement presents $n$-three-dimensional parametrical equations of helix curve located on a conic surface (Figs.3, 8).

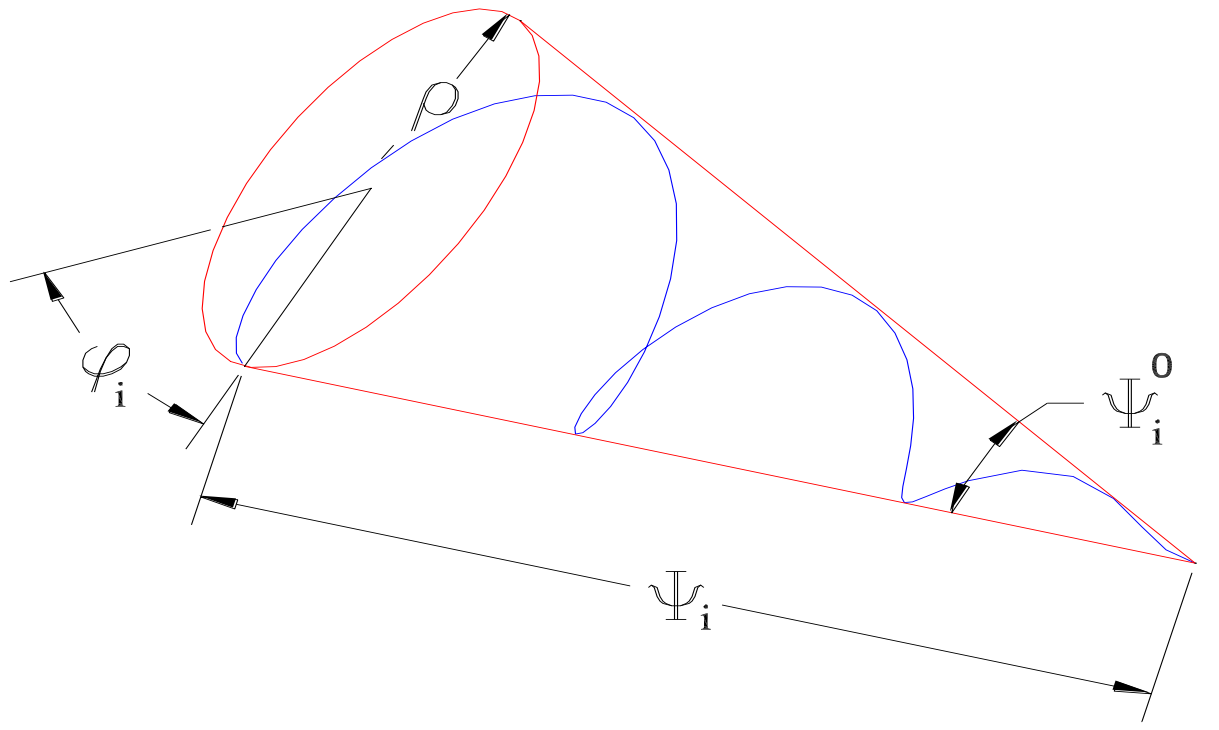

Fig. 8. The cone's parameters (the indications are in the text).

A projection of the space trajectory' radius-vector $\bar{r}(\rho, \varphi, \psi^o)$ on $i$-cone's base is spiral trajectory with radius

$$\rho_i = b_i \sin(\varphi_i \sin \psi_i^o) . \qquad (5.12)$$

At angle $\varphi = \pi k / 2$, $k = 1,2,\ldots$, the trajectory transfers from one cone to another cone trajectory on the cone space points $l_i$ which satisfy the extreme condition for the equation (5.12) at angles (Fig.8):

$$\varphi_i(l_i) \sin \psi_i^o = \pi / 2 , \qquad (5.13)$$

where $\psi_i^o$ is angle on the cone vertex, and the base' radius is

$$\rho_i = b_i \sin(\pi / 2) = b_i . \qquad (5.14)$$

The angle at the cone vertex takes values

$$\sin \psi_i^o = (2k)^{-1}, k = 1, 2, ., m, m+1, \ldots \qquad (5.15)$$

where at $k = 1$, $\psi_i^o = \pi / 6$ .

The angle on the cone space points $l_i$ takes values

$$\varphi_i(l_i(\tau_i)) = k\pi . \qquad (5.15a)$$

The minimax imposes optimal condition on these angles:

$$\varphi_i = \pm 6\pi, \psi_i^o = \pm 0.08343. \qquad (5.15b)$$

Projection of moving vector $l(\bar{r}) = l(\rho, \varphi, \psi^o)$ on the cone base satisfies Eq

$$dl = [(\frac{d\rho}{d\varphi})^2 \sin^{-2} \psi^o + \rho^2]^{1/2} d\varphi. \qquad (5.16)$$

The spiral space angle $\psi$ depends on angle $\psi_i^o$ (5.15) according to Eq

$$tg\psi = \frac{(1 - \sin\psi^o \cos\psi^o + \sin^2\psi^o)}{(1 \pm \sin\psi^o \cos\psi^o + \sin^2\psi^o)} \qquad (5. 17)$$

which for $\psi_i^o = \pi/6$ brings $\psi = 0.70311$. This angle, for the right directional spirals, at small angle $\psi^o$, satisfies realtion

$$\psi = \pi / 4 - \psi^o . \qquad (5.17a)$$

For the spirals with opposite directions, this angle is $\psi^1 = \pi / 4$ .

A relative increment of information volumes $\Delta V_{m,m+1}$ (Fig.8) between the volumes $V_m$ and $V_{m+1}$ of two sequential triplets' $m$ and $(m+1)$, where $\Delta V^*_{m,m+1} = (V_{m+1} / V_m - 1)$ depends on these triplets scale factor $\gamma^{\alpha}_{m,m+1}$:

$$\Delta V^*_{m,m+1} = (\gamma^{\alpha}_{m,m+1})^3 - 1 . \qquad (5.18)$$

While the triplet initial volume determines realtion

$$V_c = 2\pi c^3 / 3(k\pi)^2 tg\psi^o \qquad (5.19)$$

which depends on angle $\psi_i^o$ , initial space speed $c_{io}$

$$c_{io} \approx 14.4 \times 10^{-15} m / 4.477 \times 10^{-14} \sec = 3.216 \times 10^{-1} m / \sec , \qquad (5.20)$$

and parameter $k$ (5.15) of merging the $m$-volumes in $V_{m+1}$, starting with volume (5.19) at $k = 1$.

Velocity of rotation attraction $\omega_i[l_i(t_{ika})] = 0.1646 \times 10^{-14} radian/\sec$ determines space angle $\psi' = \pi/4$ at moment $t_{ika}$ (3.3b), where $\omega_i$ relates to formulas (3.2.12a), (3.3c)). The information dynamics at moment $t_{ika}$ determine Eqs (5.12-5.19) and above rotating angles, space interval $l_i(\tau_i)$ for the rotating volumes, and the eigenvalue ratios (speed) depending on triplet parameter $\gamma_1^{\alpha}, \gamma_2^{\alpha}$ in (5.2).

The triplet joint three eigenvalues form a first speed on its cone vertex, which delivers information to next triple units that join in next triplet in the proceeding rotating movement, generating an observer time course and space intervals.

Transfer from one cone's trajectory to another one locates on the cone's base, where each location satisfies extreme condition for entropy–information.

The sequential transfer requires rotating each spiral on the space angle up to adjoin a next optimal trajectory and relocate it in cooperation (Fig.6).

The space-time trajectories, rotating on the cones and cooperating in the triplet, shapes its geometrical structure (Figs.5, 7) evolving during each triplet formation with growing parameter $k$.

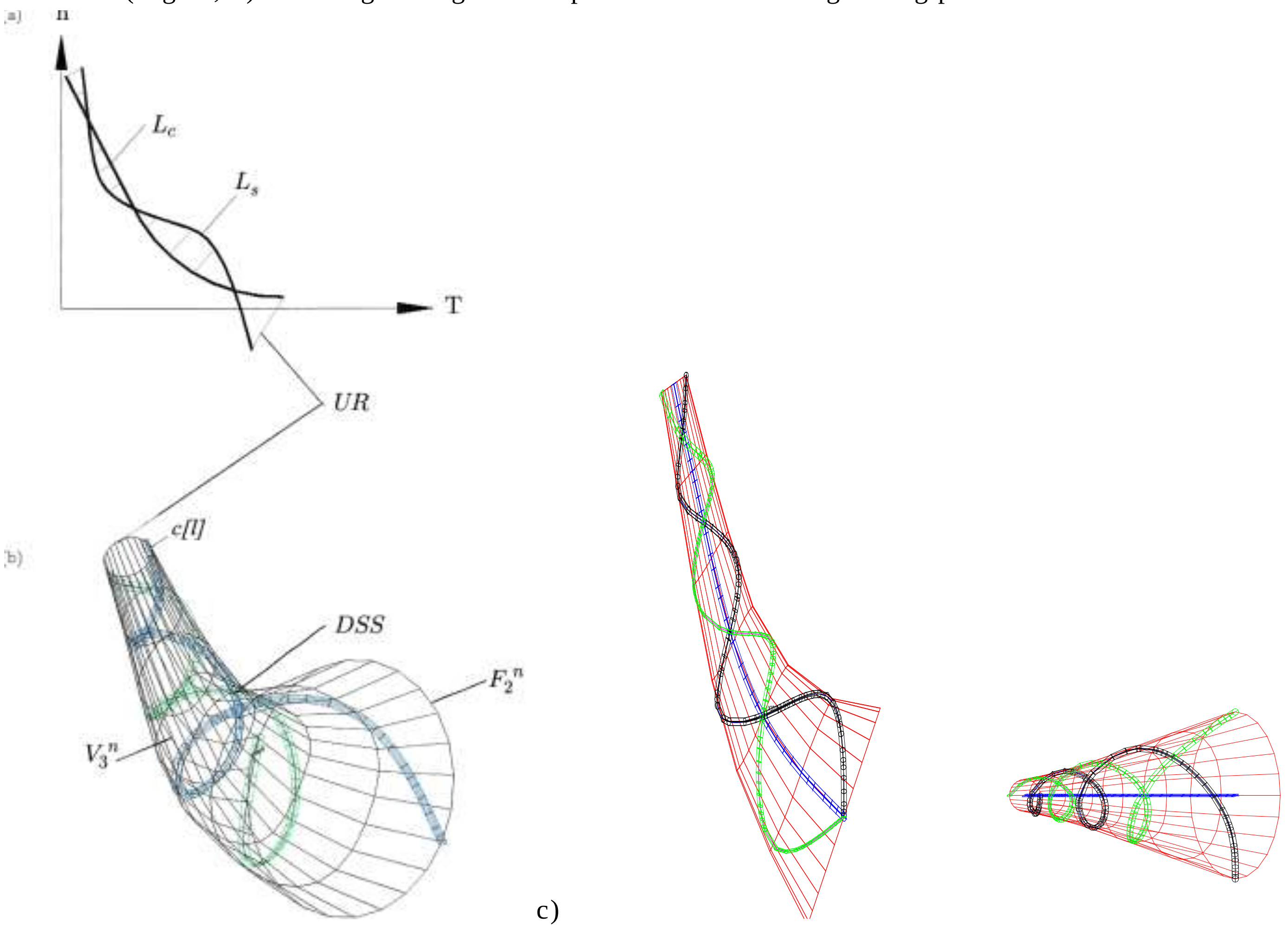

Fig. 9. Simulation of the double spiral cone's structure (*DSS*) with the cells (c[l]), arising along the switching control line *Lc* (a); with a surface $F_2^n$ of uncertainty zone (*UR*) (b), surrounding the *Lc*-hyperbola in the form of the *Ls*-line, which in the space geometry enfolds a volume $V_3^n$ (b,c).

Both information_dynamics and its space structure evolve concurrently, producing each other (Fig.9).

Each IN triplet accumulates three Bit's enfolded in its knot, which forms the IN node.The nested nodes enclose information logic enfolded in an ending IN knot. This ending triplet in every network contains the maximum amount of free information. The INs can be self-connected through the attraction of their ended triplet'*s* information logic. The attracting information logic of multiple moving INs sequentially equalizes their ending speeds-frequencies in resonance assembling the joint INs logic. Each forming IN emerges with logic of assembling triplets, whose knot memorizes and encodes the triplet code of the triple logic. The code of multiple IN holds geometrical double spiral structure (DSS), Fig.9 enfolding each triple informational knot. Each IN ending knot encloses the cell which condenses its local DSS code. The double structure of the conjugated segment builds Hamiltonian dynamics becoming irreversible at composing the triplet knot on a bridge between segments (Fig. 3a).

Since each bit of the knot code holds energy, the multiple IN knot-code physically organizes their local codes in coding information structure of information Observer.

The Egs of rotating time-space trajectory on the cones (Figs.3,7) and the space volume determine observer geometry, generated by the information dynamics (ID). The IN scale parameter $\{\gamma_i^{\alpha}\}$ identifies the rotating velocity and cooperating volumes of each knot, transferred to next triplet.

The multiple IN geometry structures the Observer's information geometry by a manifold of the cellular DSS (Fig.10).

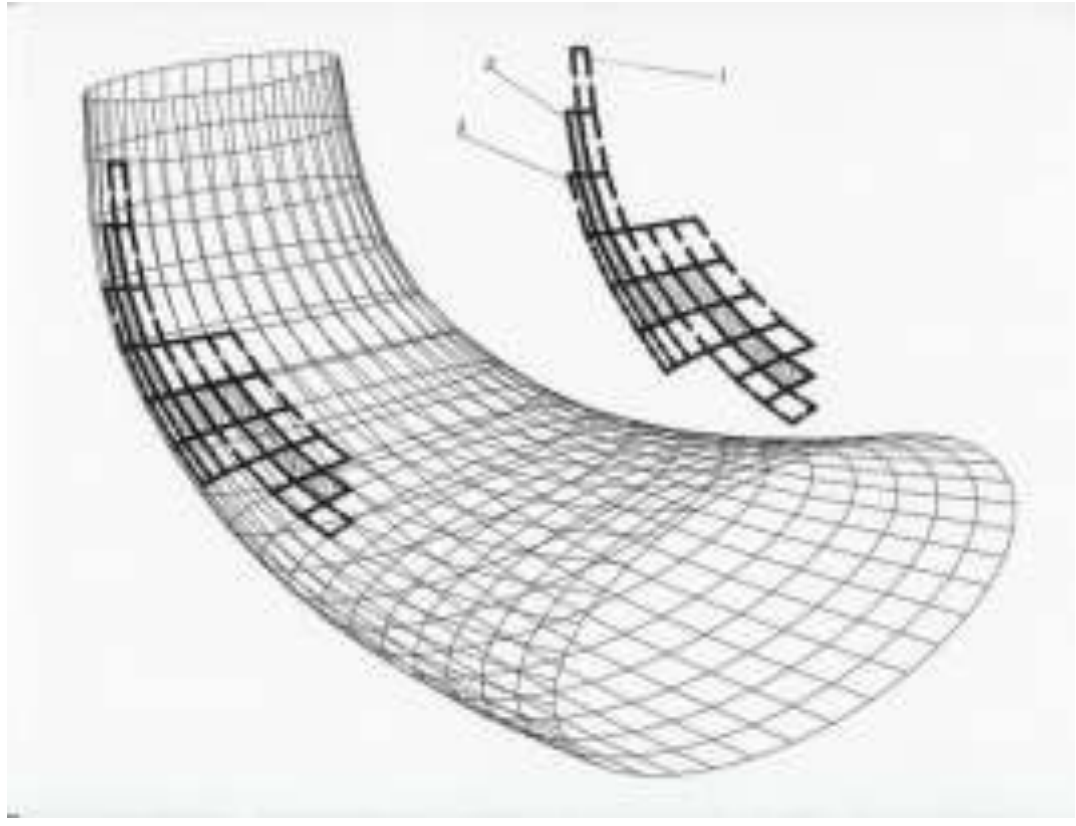

Fig. 10. Structure of the cellular geometry, formed by the cells of the DSS triplet's code, with a portion of the surface cells (1-2-3), illustrating the observer space formation.

## V. THE OBSERVER INFORMATION COGNITION AND INTELLIGENCE

5.1 EMERGING THE OBSERVER'S COGNITION AND INTELLIGENCE

### 5.1.1The Observer logic

The Observer logic emerges on the path from the collected observing interactive probabilistic logic, the curving impulses interactive certain logic, the rotating triplet logic, and the nested Information logics of attracting triplets [62].

The rotating speeds of each three logical Bits generate frequencies with the attractive logic of free Information.

Equalizing the speeds of the attracting free Information synchronizes the frequencies of the triplet's logic in a resonance which assembles a logical loop. That loop assembles the triple logic of the attracting logical Bits. The assembled logic forms a non-chaotic logical attractor [63].

The attractor assembles triplets' logic in a logical knot. The attractor-knot free information attracts new forming triplet logic in resonance loop which assemble other logical attractor.
These logically organized triplets' attractors-knots build a logical chain. The chain resonances assemble the attractors-knots in nested logical nodes. The nodes logically organize the nested hierarchy of Information network (IN).
The IN nodes self-organize the *hierarchical* frequencies of the nested attractors-knots. The ending node of the IN encloses all the *attracting* logic of its triplets' node hierarchy. The IN ending node holds a frequency which could attract in resonance other Observer INs.
The logical attractors-knots of the multiple INs triplets' logic form a logical chain which encloses the hierarchy of the assembling local loops with their hierarchical frequencies, up to the Observer's highest level IN.
This hierarchical logic of the Observer consists of the mutual attracting free logic which self-organizes the assembling logical rotating loops in a chain enclosing all observing time-space logical structures.
The logically organized Observer's highest level IN *structures* the Observer Logic.

### 5.1.2 The Observer cognition

The observing time-space *logical* structure composes a chain of the multiple cooperative logical loops called the Observer *cognitive logic*, or cognition.
The Observer's cognition assembles the logical chain through the multiple resonances of nested logical loops forming the IN-triplet hierarchy.
Such a multi-level logical structure possesses observations of virtual probabilistic causality, real Information causality, and complexity, all of which measure the cognitive intentional actions during each cooperative logical loop.
Each nested coherent loop accepts only such units that each IN node logic recognizes.
Finally, the IN finally memorizes each recognized unit in its Information node.
The cognitive rotating movement, upon forming each IN node and level, processes a *temporary loop* (Fig. 6) which might disappear after the newly formed IN triplet unit is memorized.
The Observer cognition emerges as an evolving intentional ability to request, integrate, and predict the needed Observer Information that builds the Observer's growing networks.
The free logic resonances self-organize a logical chain of *cognitive functions*, which are distributed along hierarchy of the assembling logical units: triplets, IN nested nodes, and the IN ending nodes.
These local cognitive functions self-organize the Observer's cognition, up to the cognitive functions at the upper level's synchronizing frequency.
Along the IN hierarchy runs the distributed resonance frequencies spreading a chain of the nested loops.
The chain rotates with minimal energy of the synchronized frequencies.
These cognitive thermodynamics process the chain's resonance logic.

### 5.1.3 Arising Observer intelligence

Since each assembling logical unit possesses the free logic, its topological opposite curved interaction (Sec. 2.6) with an external Yes-No impulse brings the asymmetry of interaction. The interactive Yes action can access the external Landauer's energy which starts erasure entropy and memorizes the resulting Information up to encoding a Bit.
The Bit's Free Information "No-action" stops delivering this energy. Such logical switching actions encode the logical triplet assembling in the attractor-knot with its Free Information.
The nested triplet's logic knots are encoding in cognitive logical codes on the IN hierarchal levels.

The multiple local Bits with the level's Free Information self-organize cognitive logic at all Observer's IN hierarchical levels. In that hierarchy, the loop logic, assembling in each the knot lower level, will be released after this knot logic is memorizing and encoding. The Observer hierarchical codes holds the energy of memorizing and encoding. This code hierarchy physically organizes the multiple IN, self-encoding their local codes in the Information structure of Information Observer.
We call this coding structure the *Observer intelligence*.
The logical switching actions perform *intelligence functions*, which generate each local code.
These functions are distributed hierarchically along the assembling units' local codes.
It is the Observer intelligence cooperative code which self-organizes these local functions.
The question is: What initiates the switching to memorize the Bit and its encoding?
Evidently, it requires opening the access of external energy to each local logical unit at every hierarchical level. That requires a coordinating connection of the Observer's inner and external times, which schedule the switching along the Observer hierarchy.
As it had shown (Sec. 2.6), such coordination takes place exactly at the moment ending interval of the free logic at each unit level. The switching time interval $\Delta t_{o1}$ runs units with specific time intervals $\delta T_{cm}$ and related frequencies of switching $f_{cm} = 1/\delta T_{cm}$. The $\delta T_{cm}$ rate changes from $\delta T_{co} = 12$ (for an elementary objective Observer with IN two triplet units) up to $\delta T_{cs} = 69242.359$ for a subjective Observer (with 9 triplets IN). If the unit runs one hour, then the objective Observer opens to get external energy with frequency $f_{co} = 1/12$, or each 12 hours.
The subjective Observer gets external energy with the frequency $f_{cs} = 1/69242.359$, or ~1/30 min, which is equivalent to one opening for two seconds, or 30 times per minute.
This means that each Observer has its own time clock with its time course which commands the hierarchical switching. The clock interactive switches command the encoding of each cognitive function to an intelligence function, starting the Observer intelligence. (The details of coordination of the times and frequencies of switching are in [41]).
Within the sequential segments of the observing dynamics (Fig. 3a), each switch links two segments through a bridge between them, while the third segment ends with its bridge. This bridge holds the triple segments' logical attractor, which is memorizing and encoding at the scheduled time. Each third bridge shapes a knot of the IN node hierarchical level.
At forming each node and level, the cognitive rotating process holds the coherent loop of harmonized speeds-Information frequencies at different levels. These frequencies determine the clock time units at different levels. At the unit level with a longer interval $\delta T_{cm}$, its frequencies are lower, and vice versa up to the highest frequencies of the subjective and an intelligence Observer.
The higher frequency grows the Information density of the encoding Information Bits.
The subjective Observer, self-assembling a hierarchy of logical structures, possesses a hierarchy of the frequencies and clock courses. The *harmonized speeds-Information frequencies* automatically setup the switching times and the frequencies when the cognitive loop at each level is self-established.
Finally, the coherent cognitive dynamics, assembling the cognitive functions of the units, self-organize the hierarchy of intelligence functions, encoding the Bits in the hierarchical codes. Or, the local cognitive units, involved in the resonance chain movement, self-organize themselves in local cognitive function which self-forms the Observer's cognition. The hierarchical cognition schedules the switching clock of intelligence functions, encoding the Observer hierarchy in the Observer cooperative code.

Since each assembled unit encodes the triplet code, the Observer cooperative code integrates the structural units of the triplet code at each level. These local codes have increasing densities of encoded impulses according to their hierarchal locations. The cooperative code, which the clock synchronizes, has a rhythmical sequence of time intervals scheduling access the external energy to each Observer logical structural unit as it needs. The clock time course assigns the frequency through the repeating time intervals, which determine each local resonance frequency of the assembling structural unit.
These frequencies-local rhythms identify the moments of the ending intervals of Free Information at each unit level, or the interacting cognitive and intelligence local actions. Each stable Observer logically conserves its switching time intervals. Therefore, the Observer code enhances multiple rhythms of the local structural units. That’s why the rhythms of an external melody, resonating with the Observer code rhythms, supports the cognitive functions, intelligence actions, and generation of both the cooperative Observer’s logic and the code encoding this logic [64].
Recently experiments have confirmed the influence of music on neural encoding [65, 66].

### 5.1.4 Arising structure of Observer code

The curving interactive movement, starting the curving impulses, rotates the trajectories of the observing process, forming spirals on a cone surface (Fig. 3).
During probabilistic observations, these spiral trajectories hold a random periodic sequence. With the emerging space and the space-time conjugated entropy increments, the rotating trajectories shape the conjugated space-time spirals (Fig. 3a). The emerging Information process continues rotating its trajectories in double spirals, assembling them according to Fig. 6.
The path from each interacting entropy impulse along the observing process’s entropy integrates the EF.
The minimax principle, preserving each interacting impulse measure, leads to the minimax variation principle (VP) for the EF. The VP described the conjugating trajectories of the observing process as the EF extremals converging to the Information processes’ conjugated Information trajectories as the IPF extremals. These EF-IPF trajectories satisfy the Hamilton equations for the Observer dynamic process. The trajectories integrate the observing process logic including the logic enclosed in the code of each ending IN node.
The multiple IN ending triple codes integrate the double space spiral coding structure (DSS) (Fig. 9).
The interactions of opposite directional conjugated spirals form bridges between the spirals (Fig. 3a).
Each spiral segment integrates its observing logic in the bridge. The bridges locate the switching interactions which will memorize this logic according to the time schedule.
The attracting Free Information of the triple Bits connects the memorized Bits within the triplets’ logic.
The triplet’s knot encodes the logic of the triple segments in the bridge code according to the time schedule. The bridges, located along the observing trajectory, connect the encoding knots of the INs nested nodes, and then, in the ending IN triplet code.
The time schedule sequentially builds a hierarchy of the memorized triplets and their encoding. That schedules sequential transfers from the logical to the memorized, and the encoding bridges along the observing trajectory. The hierarchical cognition schedules the switching clock of intelligence functions, encoding the hierarchy in the triplet codes. Encoding starts when the hierarchical switches open access to the external energy. Then, sequentially emerges the triple code and the intelligence cooperative coding structure DSS.
The bridges localize the coding units where external interactions may change the DSS coding structure.

Each triplet, memorized in the conjugated interactive bridge, divides the trajectory on a reversible proces' segment which does not include the bridge and irreversible bridge between the reversible segments. Thus, the Observer's irreversible dynamic trajectory includes the reversible segments, ending with each memorized bridge. Each irreversible triplet emerges from the encoding of the observation logic in the trajectory segment ending with the bridge. The conjugated trajectories integrate the EF extremals, while the emerging encoding Bits on the bridge integrate the IPF. The IPF encloses the integral Information in its final encoding Bit.

Therefore, the Observer's integral Information identifies the IPF final code, which the observation predicts through the VP minimax optimal Information law. This IPF code has increasing densities, which triple with each following Bit (Sec. 2.12).

The prediction, based on the EF-IPF integration of the observing process with its both probabilistic and Information logic may project *artificially designed Observer cognition*.

The EF logic predicts the conversion of cognition to the IPF Information, memory, and to the Observer cooperative encoding in an artificially designed Observer intelligence. It starts with the persisting Information speeds-frequencies of the attracting observing impulses on the EF-IPF trajectories (Sec.2.1.2).

The attracting Information sequentially equalizes the Information speeds-frequencies in the attracting resonances which assemble the cooperative (cognitive) logic.

The evolving logic self-organizes each hierarchical level's specific time-space Information logical structural unit that assembles the triplets, builds the IN, and the domains.

The assembling Information memorizes the self-organized logic.

The assembling resonance frequencies identify the clock, commanding the logical switching, which coordinates cognitive and intelligence actions. The cognitive functions perform the impulse switches delivering Landauer's energy for memorizing each unit's logic.

The intelligence functions encode the triplets' Bits logic.

The Observer EF-IPF integrates the observing process in the multiple Information process, coordinates and unifies the observing Information in the IPF code.

In the *artificial designed Observer*, where the EF integrates cognitive logic and IPF integrates its encoding Information, the automatic conversion of the EF in the IPF implements the triplet bridges which encode its cognitive logic. The IPF frequencies initiate the Observer clock time course which the EF prognosis in the observation. The time course sustains the EF-IPF integration in an optimal observing process of Information dynamics, and maintains the optimal Observer DSS double spiral structure which finally encloses the predicting Observer code. (Each persisting time interval encloses the entropy of the interval impulse, Sec. 2).

The Observer triplet code memorizes the Observer cooperative Information structure which encloses multiple rhythms of the local structural units. The DSS coding structure memorizes total collected Observer quantity and quality Information, which determine the Observer cooperative complexity [5].

The Observer assembled Information self-organizes the functions of cognition commanding the encoding intelligence on all hierarchical levels. These functions self-connect the local codes in the Observer code, which encodes all these structures in the space-time Information structure of Information Observer.

The EF-IPF observing process and Information dynamics artificially design the DSS.

The DSS measures the total Information IQ of this Observer. The DSS code integrates each Observer's IQ. The difference of these IQs measures the distinctness of their intelligence. The maximal

Information, obtained in the observation, allows designing the DSS with the maximal achievable IQ measure in the optimal AI Observer.

The space-time Information structure, enclosing the encoded EF-IPF, integrates the observed Information in the analytically designed AI Information Observer (Fig. 10).

The observing Information of a particular Observer is limited by the constraints of each observation[28].

The constraints also limit the conversion of the observing process to the Information process.

The thresholds between the evolving stages of the observation limit the stages' evolution, which can stop at any stage. All these limit the integral cognitive Information and the following intellective actions, which also limits the amount of Free Information that reduces ability to make the intelligent IN's connections.

### 5.2.1 Essentials of the Information Cognition

The Observer logical structure possesses both virtual probabilistic and real Information causality and complexity [64].

A virtual Observer, forming the rotational space-time displacement of the impulse's opposite actions during virtual observation, starts accumulating virtual Information by temporally memorizing it in probabilistic logic, initiating cognitive movement. The rotating cognitive movement connects the impulse microprocess with the Bits in the macroprocess.

The microprocess disappears automatically when the probability of the impulse borders approaches zero (Sec.4.4), and certain logic arises. It means the microprocess has *no going past*, as well as the microprocess cognitive movement.

The forming macroprocess composes the triple macrounits through Free Information, which assembles each evolving IN. The INs ending triplets integrate multiple nested IN's Information logic in Information domains with the evolving growth of the quality Information.

The Observer's cognitive dynamic movement models the Observer's hierarchical rotation mechanism, which enables transferring the Observer through the evolution stages by overcoming the stage thresholds [62].

The mechanism rotating movement characterizes potential of power $P_{in}(i)$ which measures the current $(i)$ rotating moment $M(i)$ multiplied on angular speed $\omega(i)$:

$$P_{in}(i) = M(i) \times \omega(i). \qquad (5.1.1)$$

This power compensates the resonance movement along each loop of the Observer cognition. The loop rotates the thermodynamic process (Sec. 2.6.6) with minimal energy of cognitive thermodynamics.

The cognitive movement, upon forming each node and level, processes a temporary loop (Fig. 6) which might disappear after the newly formed IN triplet is memorized. That memorizes the loop logic. After the memorized Bit emerges during the observation, the rotation movement develops Information form of double helix dynamic movement (Figs.3, 3a).

The Observer's cognition assembles the multiple resonance logical loops, forming in the IN-triple logic hierarchical levels, which accepts only the Information units which each cognitive loop recognizes.

The rotating process in the *coherent* loop *harmonizes speeds-Information frequencies* at different hierarchical levels. The cognitive functions model the correlated interactions and feedbacks between the IN levels, which the highest domain level's feedback controls. Both cognitive processes and cognitive functions emerge from the evolving interactive observations with their emerging properties.

The discrete impulses provide a discrete Information language for the cognitive logic.

The rotating double spirals (DSS) compose the evolving Information logic by running the macrodynamic process segment's logic. The DSS knots memorize the sequence of process' reversible segments and encode the cognitive process in the Observer time course. In the rotating DSS, the cognition merges with the natural memorizing of each Bit on all evolution levels.

The cognition emerges in two forms: a virtual rotating movement processing temporal probabilistic logic following certain Information process's logic rotating in a double helix structure.

The DSS concurrently organizes the observing Information Bits in the IN nodes.

On the observing trajectory, the sequential knots memorize the Information causality and logic and structure of the IN node hierarchy.

These processes start with the elementary virtual Observer and the emerging Bit at the microlevel, which holds the irreversible prehistory and participates in the evolving Information Observer.

The starting cognitive thermodynamic has no actual physical cost.

Results [67] confirm that cognition arises at the quantum level as a kind of "entanglement in time" in the process of measurement, where cognitive variables are represented in such a way that they don't really have values (only potentialities) until you measure them and memorize, even "without the need to invoke neurophysiologic variables," while "perfect knowledge of [a cognitive variable] necessitates uncertainty for the others."

Moreover, this analysis shows that both cognition and intelligence have an Information nature.

Recent work in cognitive maps [68] confirms "large-scale internal representations of navigable spaces," showing how cognitive maps are encoded, anchored to environmental landmarks and used to plan routes.

Similar neural mechanisms might be used to form 'maps' of *nonphysical spaces*, and "applied to nonspatial domains to provide the building blocks for many core elements of human thought."

**5.2.2 The self-forming hierarchical distributed logical structure of cognition**

Multiple moving INs, sequentially equalizing the nodes speeds-frequencies of attracting Information logic in a resonance, assembles total Observer logic. The mutual attracting free Information logic, sequentially interacting, self-organizes the cooperative logical rotating spiral loops in a chain which encloses the observing logic.

Each curved impulse invariant time-space measure $\pi$ enfolds the Information measure $1Nat$ which includes a Bit, Free Information, and the Information needed to encode the impulse Bit.

This Free Information attracts the impulse with intensity 1/3 Bit per impulse.

The IPF integrates the Bits with Free Information connecting the sequence in information trajectory.

The minimax principle, applied along the IPF extreme trajectory, maximizes the Information enclosed in each current impulse, squeezing its time interval. The growing attracting Free Information along this trajectory minimizes the time interval between the nearest impulses proportionally to 1/3Bit. For each third impulse, that interval of Information distance becomes proportional to 1 Bit, preserving the impulse's invariant Information and the time-space measure $\pi$. By the end of the IPF integration, all integrated Information is concentrated in a final impulse, whose Information density approaches the maximal limit. Since Free Information encloses Information logic, the multiple Bits with triple growing density increase the process's Information logic.

The IPF integral Information with its logic is condensed in the last integrated impulse time-space interval volume. For multiple Information impulses, each third curved impulse having invariant measure $\pi$ appears in the Information process with time frequency $f_i = k_i, k_i = 3,5,7,9,...$, which indicates the entrance of the triplets and their specific sequences.

The invariant impulse time measure $\tau_i = \pi / \sqrt{2}$, the flat surface space measure $l_i = \sqrt{2}$, and the orthogonal to surface space measure $h_i = \pi$ determine geometrical volume $v_i^S = \tau_i \times l_i \times h_i = \pi^2$ of each $i-$ three-dimensional impulse.
Information time-space density $D_i^I = k_i Nat / v_i^s$, concentrating $k_i$ Nat for each third impulse, increases with growing $k_i$ while the time-space geometrical volume $v_i^S$ holds the invariant density.
The Information density measure $D_i^I = k_i Nat / \pi^2$ grows only with each $k_i$.
Let us evaluate the relative Information density of each triplet's Bit encoding Information in its relative time-space interval $u_k = 1, 1/3, 1/9, \ldots 1/k_i, \ldots,$
Since each triplet's Bit encodes 3 Bits in its starting relative time-space interval $u_1 = 1$, its Information density is $N_d^1 = 3$.The following triplet also encodes 3 Bits in interval $u_2 = 1/3$, but 3 Bits are already encoded from the previous triplet's Bits. Thus, the Information density of such two triplets equals $N_d^2 = 3^2$ and so on. Hence, for the $m$-th triplet density is $N_d^{mn} = 3^m$.
The density of the process's dimension $.n$ with $m = n/2$ triplets is $N_d^{m=n/2} = 3^{n/2}$.
Thus, each current impulse time-space geometry encloses Information, its density and frequency, concentrating Information logic, and the Information of all previous impulses along the Integrated Information Path (IPF).
The EF extreme trajectories, starting from the multi-dimensional observing process, the EF-IPF transformation converts to the multi-dimensional orthogonal processes (Secs. 1-2) whose curved impulses hold the above Information measures.
The EF-IPF space-time extremal trajectories rotate the forming spirals located on conic surfaces (Fig. 3), which starts from virtual (entropy) process and continues as the Information process.
Since the cutting entropy in the impulse observation converts to the Bit on this trajectory, the trajectory consists of the segments of Information process dynamics and the intervals between segments delivering each Bit to the following segment. On Fig. 3 each segment starts on the cone vertex-point D and ends on point D4, which connects to a vertex of the following cone. The observing Bit is delivered at each cone vertex. The segment includes the observing impulse with its logical Bit, intervals of free logic, and the correlation connecting the nearest segment, temporarily memorizing the segment logic.
The logical and Information dynamics describe the process of sequential logical interactions of the multiple impulses, rotating with Information speed determined by the impulse density $D_i^I$.
The dynamics on trajectory between the cone points D and D4 is reversible and symmetrical described by Hamiltonian equations (Sec. 4.3). The logical anti-symmetry brings the anti-symmetrical logical Bit prior to the interaction with an external impulse which starts delivering the external energy.
This Bit is supplied at each cone vertex, as well as the interactive impulses, with intervals currying the Bit, intervals memorizing and encoding the Bit.
After the external energy generates the physical multiple Bits, the physical Information process starts.
According to Sec. 2.2.6, the moment of appearance of the interactive logical Bit from beginning of the impulse is $t_{11} = 0.2452$, which defines interval $\Delta t_1 = 0.2452 / 1.44 \cong 0.17$ relative to invariant information measure of each impulse. $1.44 Bit = 1 Nat$.The time interval for memorizing the Bit $\Delta t_B$ identifies the Bit information measure $\ln 2$ which is the equivalent of the invariant impulse relative time part $\Delta t_B = \ln 2 / 1.44 = 0.481352$. For impulse relative measure $|| 1 ||_M$, the relative difference $1-(0.17+0.481352)=0.348648$ includes intervals of supplying external energy, erasing the asymmetric logical Bit, memorizing this Bit, interval of encoding the Bit, and interval of Free Information logic $\Delta t_{fo}=0.23/1.44\cong 0.1597$. The deduction brings interval $0.188948$ which includes the interval following the opposite asymmetrical interaction $0.01847$ (Sec. 2.2.6). The interval of interaction $0.01847$ disappears at the end of the encoding interval $\Delta t_{eno}=0.188948$. Therefore, the interval of encoding is $\Delta t_{en}=0.188948-0.01847=0.17039$ which approximates the time interval of the appearance of the asymmetrical logical Bit $t_1$. The encoding Bit releases the Free Information interval $0.01847$ emerging after other impulse is encoding. The external impulse, erasing the asymmetric logical Bit, spends interval $\Delta t_B$ and ends with the interval of encoding $\Delta t_{en}$ for each invariant impulse. Along the IPF trajectory, this moment follows interval $\Delta t_B$ of creation logical Bit, which ends forming the triple logic that binds the free logic knot. The interval of memorizing the physical Bit requires the same interval $\Delta t_B$ during which the entropy of logical Bit erasure.

The external energy, supplied in time interval $t_{\Sigma b}=0.481352+0.18948=0.67083$, includes both the erasure of the logical Bit and its encoding. Since the external impulse interactive part measures $0.025Nat$, it brings the total $t_{\Sigma bo}=0.67083+0.025=0.69583\cong\ln 2$ measuring the interval of supplying the external Bit. (It also numerically confirms the direct connection of the Information and its time measures in Secs. 1-2.).

Since each impulse's curved measure $\pi$ with its time interval $\Delta t_1$ appears in the Information process with frequency $f_{10}=\pi$, the relative frequency is $f_1=(0.17/\pi)\times\pi=0.17$.

The frequency of spectrum $\omega_1=2\pi f_1=1.068$ identifies the time of opening supply of the external energy, memorizing the Bit. The time interval of memorizing the Bit $\Delta t_B$ identifies the frequency $f_B$ of the appearance of that interval within the impulse with the frequency of spectrum $\omega_2=2\pi f_B=2\pi\ln 2/1.44=3.02<\pi$. The time interval of the impulse encoding $\Delta t_{en}$ determines the spectrum frequency $\omega_3=\omega_1$, estimating also the process frequency $f_1$.

Therefore, the frequency spectrum $\{\omega_1\omega_2\omega_1\}$ initiates the frequency of the appearance of the logical Bit, following the frequency of the memorized Bit and the frequency of encoding, which equals $\omega_1$.

Spectrum $\{\omega_1,\omega_2,\omega_1\}=\omega_o,\omega_o=(1.068,3.0.2,1.068)$ delivers the logical Bit, energy memorizing it, encoding the Bit, which includes the Free Information attracting the Bit along the trajectory

Or, these frequencies approximate spectrum $\omega_o=\{1,28277,1\}\times 1.068\approx\{1,28277,1\}$.

After supplying the external energy during the sum of the impulse' invariant intervals, the impulse becomes the segment of a physical Information process.

Therefore, physical dynamics describe the IPF extremal trajectory rotating on sequential cones (Fig. 3). Each cone vertex encodes the Bit memorized with the frequency $\omega_2$ delivered from a previous impulse-segment with the frequency of a logical Bit $\omega_1$.

Hence, each physical Information impulse carries spectrum $\{\omega_1\omega_2\omega_1\}=\omega_o$, while their sequential pair on the trajectory carries the impulses spectrum $\{\omega_1,\omega_2,[\omega_1=\omega_1],\omega_2,[\omega_1=...]\}=\omega_\Sigma$, where $[\omega_1=\omega_1]$ is the resonance frequency for two impulses whose distance is shortening by 1/3. That allows closely connecting the impulses in the resonance.

Along the trajectory, each of these pairs appears with the growing frequency $f_{io} = 1/k_i$, $k_i = 3,5,7,...$

Since the fixed time intervals $\Delta t_1$, $\Delta t_B$, $\Delta t_{en}$ are relative to the invariant impulse measure, they are repeating for each invariant impulse with the increasing Information density and growing frequency.

Thus, along the extreme trajectory, each third impulse will deliver triple frequency of spectrum $\{\omega_1, \omega_2, [\omega_1 = \omega_1]_{\Delta t_{10}}, \omega_2, [\omega_1 = \omega_1]_{\Delta t_{20}}, \omega_2, [\omega_1 = \omega_1]_{\Delta t_{30}}, \} = \omega_{\Sigma 10}$ with related time intervals $|\Delta t_{10}, \Delta t_{20}, \Delta t_{30}|$. .

These time intervals are sequentially proportional to the distance between each Bit in the ratio $1/k_i$, or the invariant time measures of these impulses $\tau_i = \pi / \sqrt{2}$ are shortening in the sequence $\Delta t_{10} = \pi / 3\sqrt{2}, \Delta t_{20} = \pi / 5\sqrt{2}, \Delta t_{30} = \pi / 7\sqrt{2}$ .

In the sequentially shortened distances between the impulses on the extreme trajectory, each such three impulses (with their Free Information) assemble in a triple of the resonance frequencies.

The triple resonance frequencies, in collective resonance, assemble the trajectory of the triplet segments. Sum of these interval assembling triplet equals $\sum_{ko=1,2,3} \Delta t_{ko} = \pi / 2.09 \cong 0.478\pi$ .

Adding the interval of binding the knot $\Delta t_{kn} = 0.025\pi$ we get $\sum_{ko=1,2,3} \Delta t_{ko} + \Delta t_{kn} = 0.503\pi \cong \pi / 2$ .

The trajectory of the three not shortening time segment turns on $3\tau_{io} = 2.12\pi$ or on a circle assembling each three segments in the triplet loop.

**5.2.3 Self-forming triplet logical structures and their self-cooperation in the IN hierarchical logic**

In the multi-dimensional observing process, a minimum of three logical Bits with free logics can appear, which, attracting each other, would cooperate in a logical triple.

Multiple probabilities of interacting impulses (in this multi-dimensional process) produce the numerous frequencies. Some of those, a minimum of three, can generate an attractive resonance, cooperating in a triple. This triple logic starts temporarily memorizing two sequential pair cross-correlations during their time of correlation. A locally asymmetric cross-correlation memorizes the asymmetrical logic during this correlation process.

Comments 5.1

As reported recently [69], such anti-symmetric cross-correlations have been observed. •

When this cooperating process is ending, the triple correlations temporarily memorize the triple logical Bits. According to Sec. 2.6, the minimal entropy of cross-correlation $\ln 2$ can be memorized at a cost of the equivalent minimal energy of the logical Bit.

This is the Information cost of memorizing the triple logical Bit, which includes the free logic.

The attracting free logic of the emerging three logical Bits starts the Bit's self-cooperation in the following sequence. The free logic of the emerging logical Bit holding the frequency $\omega_2$ attracts next logical Bits toward a resonance with the equal frequency of next Bit's free logic, assembling the two in joint resonance. This resonance process links these Bits in duplets.

The free logic from one Bit of the pair gets spent on binding the duplet.

The free logic from the duplet's Bit attracts the third Bit and binds all three in a knot Bit, creating the triplet logical structure. The knot Bit still has Free Information, and it is used to attract a different bound pair of emerging Bits, creating two bound triplets. This process continues creating nested layers of bound triplets, three triplets and more (Figs. 4-6).

Hence the triplet logical structure creates the resonance frequencies of the attracting logic, joining the triple Bits. The free logic attraction toward the triple resonance of their equal frequencies is the *core Information mechanism* structuring an elementary triplet.
The trajectory of the forming triplet describes the rotating segments of their cones (Fig. 5), whose vertexes join the triplet knot and start the base of the following cone. When the next rotating segment starts, the knot frequency joins the cone vertexes in resonance along the cone base. It connects the next triplet in the resonance, and so on, creating the nested layers of a logical space-time Information network (IN), where the knot hierarchy identifies the nested nodes of the IN hierarchy.
Each triplet unit generates three symbols from three segments of Information dynamics, and one when the segment attracting triple logic binds the three in the logical triplet knot.
These symbols can produce a triplet code, while the knot logic symbol binds the triple code, potentially encoding all triples. Encoding the knot will release its Free Information logic, which transfers this triple code to the next triplet node. Thus, the nodes logically organize themselves in the IN code.
The external energy, encoding the IN triple code, sustains the frequency spectrum mentioned above.
The IN emerging logical structure carries the triple code on each node's space-time hierarchy, and the last triplet in the network collects and encloses the entire network's Information code.
The network, built through the resonance, has limited stability. Therefore, each IN encloses a finite structure. That's why the observing process self-builds multiple limited INs through the Free Information of its ending nodes.
The final triplet in every network contains the maximum amount of the enclosed Free Information.
The limited networks develop self-connection through the attraction of their *ending triplets.*
Even after each IN potentially loses stability, evolving in chaos, it possesses the ability of self-restoration [41].
The multiple INs develop self-cooperation in a hierarchical domain, starting as each three ended triplets' Free Information cooperates in a knot which joins this IN's triples in resonance. This IN ending knot's Free Information resonates with the other three INs ending Free Information, forming a triplet structure analogous to the elemental triplet. This high-level triplet joins these three INs, structuring the next IN of the domain hierarchy. The hierarchical logical trajectory describes the space-time spiral structure (Figs. 7, 9), evolving in observations. This hierarchy enables generating sequential triple codes located on the rotating trajectory of the cone vertexes, which are spatially distributed at the different hierarchical levels of the multiple IN and the domain hierarchy.
Such a discrete space-time code (DSS) integrates the observing process in space-time Information geometry, self-organizing an Observer.
Still the question is: *What self-organizes the structuring of the Information units in the Observer geometrical space-time shape during their movement along the observing trajectory?*

### 5.2.4 The Observer wave function self-forming hierarchical distributed logical structure of cognition

The self-creating units of the hierarchy generate frequency delivering spectrum $\{\omega_1,\omega_2,[\omega_1=\omega_1]_{\Delta t_{10}},\omega_2,[\omega_1=\omega_1]_{\Delta t_{20}},\omega_2,[\omega_1=\omega_1]_{\Delta t_{30}},\}=\omega_{\Sigma 10}$ which is growing in the triple sequentially shortening intervals $|\Delta t_{10},\Delta t_{20},\Delta t_{30}|$ for each $i$ trajectory of the segment.
The answer on the previous question is specified in the following Propositions with the common initial conditions.
The space-time spiral trajectory of the EF extremal (Fig. 3) describes the sequence of multi-dimensional curving rotating segments, representing the interacting impulses of the observing process, which integrates the observing process's logic. Each segment's impulse with invariant entropy measure $1Nat$ moves along the trajectory, rotating the impulse with invariant geometrical measure $\pi$.

This curved impulse's three-dimensional measure includes the time coordinate measure $\tau_i = \pi / \sqrt{2}$, the flat surface's space coordinate measure $l_i = \sqrt{2}$, and space coordinate measure $h_i = \pi$ orthogonal to both of them. The Information measure $1Nat$ includes the impulse logical Bit $\ln 2$ and the free asymmetric logic measure $f_{li} = 1 - \ln 2 \cong 0.3Nat$ on each trajectory segment. The logic density per each third segment volume $v_i^S = \tau_i \times l_i \times h_i = \pi^2$ increases according to $D_i^I = k_i Nat / v_i^s, k_i = 3,5,7,...$

The asymmetric logic divides the sequential segments by barriers: logical, Information, and physical. Between segments, the barriers transfer the anti-symmetrical interval $\Delta t_1$ of interactive logic, following the interval $\Delta t_B$ of memorizing the Bit, and interval $\Delta t_{en}$ of encoding and releasing the Free Information. These three intervals identify the bridges along the space-time trajectory. Each sequential bridge on the trajectory repeats this triple with invariant frequency spectrum $\{\omega_1, \omega_2, \omega_1\} = \omega_o, \omega_o \cong (1.068, 3.0.2, 1.068)$. The ratio of the spectrum's adjacent parts to its middle part repeats with the frequency $f_{io} = 1 / k_i$, where $k_i$ indicates each third density logic per segment.

Thus, along the trajectory, the invariant triple frequencies of the spectrum alternate on the sequences of the bridges. Or, each bridge on the trajectory identifies the repeating frequency of this spectrum.

*Propositions* 5.1.

Along each of $i$-dimensional space-time segment rotates *three-dimensional space wave function* on the cone external shape, Fig. 3 with the following rotating speeds:

(a) around each spiral cross-section $\alpha_i^s = 1[square / radian]$, or $\alpha_i^{s_o} = \pi / radian$, and

(b) orthogonal to this rotation space speed $\alpha_i^h = 1[volume / radian]$, or $\alpha_i^{h_o} = \pi / radian$.

Accordingly, the related frequencies of each orthogonal rotation are $\omega_i^s = \alpha_i^s / 2\pi, \omega_i^{so} = 1/2$ and $\omega_i^h = \alpha_i^h / 2\pi, \omega_i^{ho} = 1/2$.

Each $i$-dimensional segment's cross-sectional rotation spins the rotation on the space interval $\pi$ of the segments invariant measure. The three-dimensional wave function distributes the space rotation along the trajectory's segments with the above invariant speeds, delivering the invariant spectrum $\{\omega_1, \omega_2, \omega_1\} = \omega_o, \omega_o \cong (1.068, 3.0.2, 1.068)$.

*Proof*. We use the equation of a wave $u = F(vt - x)$ depending on velocity of movement $v$ and distance $x$ of the movement along a trajectory. This equation we apply to a wave function $u(u_s, u_h)$ whose component $u_s$ describes the rotation of the wave running along its forming cross-section square $s_i^w = \tau_i \times l_i = \pi$, and component $u_h$ running along the rotating orthogonal space length $h_i = \pi$. The wave function $u_s = F(f_{ws})$ argument $f_{iws} = \alpha_i^s \rho_i^s - s_i^w$ describes a two-dimensional rotation of the trajectory rotating $i$-dimensional cone's radius $\rho_i^s$ of the cross-section with speed $\alpha_i^s$, where $\rho_i^s$ is the analog of the curved distance moving to reach cross-section's square $s_i^w = \pi$.

From the geometry of rotating movement, each $i-$segment is rotating along trajectory on the cone square-basis reaching angle $\varphi_i^s = k_{io} \pi_{io}$, where $k_{io} = 1,2,3,..$ is sequence of the cone basis for segment $1,2,3,..,i,...$ (Figs. 3, 8). This rotation reaches distance $s_i^w = \pi$ at $f_{iws} = \alpha_i^s \rho_i^s - s_i^w = 0$ when radius $\rho_i^s$ rotating the wave cross-section with speed $\alpha_i^s$ reaches the segment geometrical measure $\pi$. At $s_i^w = \pi$, $\rho_i^s = \pi$, it determines $\alpha_i^s = 1$.Since, the wave function argument $f_{iws}$ reaches $f_{iws} = 0$ by the end of each $i$-segment with period equal to impulse measure $\pi$, the wave function $u_s$ is periodical with period $\pi$.

The wave function $u_s$, moving along its cross-section with speed $\alpha_i^s = 1[square / radian]$, or $\alpha_i^{s_o} = \pi / radian$,holds the related frequency $\omega_i^s = \alpha_i^s / 2\pi$.

The wave function's $u_h = F(f_{wh})$ argument $f_{iwh}$ moves along the segment length $h_i^w$ with space-rotating speed $\alpha_i^h$ to reach the impulse volume $v_{ih}^t$ according to Eq. $f_{iwh} = \alpha_i^h h_i^w - v_i^w$. The movement reaches volume $v_i^w = \pi^2$ at $f_{iwh} = \alpha_i^h h_i^w - v_i^w = 0$, $h_i^w = h_i = \pi$ with speed $\alpha_i^h = \pi^2 / \pi = 1[volume / radian]$ and related frequency $\omega_i^{ho} = 1/2$. Thus, the wave carries the frequency $\omega_i^{ho} = 1/2$ along the spiral trajectory and the equal frequency $\omega_i^{so} = 1/2$ along its cross-section. Or, each $i-$rotation with frequency $\omega_i^s$ equal to frequency $\omega_i^h$ brings the space rotation during the cross-sectional rotation, or vice-versa.

Since the wave function argument $f_{iwh}$, decreasing along the spiral trajectory, reaches $f_{iwh} = 0$ by the end of each $i$-segment, with the period equals impulse measure $\pi$, the wave function $u_h$ is also periodical, with period $\pi$. The arguments of these orthogonal components of the wave function connect the relation $\arg(F) = f, f = f_{ws} \times f_{wh}$. •

*Therefore, the space-time trajectory, starting on the cone shape basis and moving on the cone's external shape, reaches the cone vertex when the projection of this trajectory reaches the base shape center* (Fig. 3, 8).

When each $i-$rotating segment reaches the cone vertex it develops a logical bridge between $i$ and the next $i+1$ segment on the moving space-time trajectory. The bridge holds the relative interval $0.00653$ of this logic (2.6.3.55 in Sec. 2.2.6). Depending on the time course on the moving trajectory, the logical bridge moves to a memory bridge following an encoding bridge. During the movement along the trajectory, the cone segments $i, i+1, i+2$ join a triple whose bridges develop sequence of knots: for a logical triple, a memorized triplet, following an encoding triplet.

Moreover, since these triplet units sequentially join in a node of the IN, forming according to the time course, the logical, the memorized, and the encoding structures, these knots become the nodes of the related INs.

One scenario illustrating the assembling trajectories of the space-time triplet is shown in Fig. 6,

The knots of the triplets, cooperating in the IN, are shown in Fig. 7. The bridges at moment $t_1, t_2$ (Fig. 4) form as the double segments assemble. At moment $t_3$ they join in the triplet's knot. •

*Proposition* 5.2

Let us consider $i, i+1, i+2$ three-dimensional segments along the multi-dimensional rotating segments on the extreme trajectory located on the related dimensions of the multi-dimensional trajectory with their location specific enclosed logic. Along the extreme trajectory, each segment of equal measure $\pi$ increases density which is proportional to the segment's shortening invariant intervals $|\Delta t_{10}, \Delta t_{20}, \Delta t_{30}|$ on their locations along the trajectory. From these locations, these segments deliver the related invariant

spectrum $\{\omega_1,\omega_2,\omega_1\}=\omega_o,\omega_o \cong (1.068,3.0.2,1.068)$ through the cross-section rotation, which is speeding the space rotation that distributes the spectrum along these three-dimensional space segments.

*Then*, the wave function's frequencies synchronize the triple segment's logic in a collective resonance. The sequentially forming triple barriers-knots are squeezing the initial observing multi-dimensional process' segments, first to three-dimensional rotation, and finally to a single dimensional Information process encoding the Bits of all multiple knots. •

*Proof*. The wave consecutive three-dimensional space movements picks segments $i$, $i+1$, $i+2$ sequentially from each of these segments' trajectory-specific locations in these dimensions, and simultaneously starts rotating each of them during interval $|\Delta t_{10},\Delta t_{20},\Delta t_{30}|$, placing these shortening intervals between segments $i$, $i+1$, $i+2$ accordingly. The densities increase proportionally to the shortening-squeezing of time interval measures along each of these trajectory dimensions.

The first of the wave three-dimensional rotation moves $i$ segment rotating during interval $\Delta t_{10}=1$ (equivalent to space interval $\pi$ with density proportional $k_i=1$). The second of wave three-dimensional rotation moves $i+1$ segment during interval $\Delta t_{20}=1/2\Delta t_{10}$ (equivalent to space interval $\pi$ with density proportional to $k_i=2$). The third of wave three-dimensional rotation moves segment $i+2$ during time interval $\Delta t_{30}=1/3$ (equivalent to space interval $\pi$ with density proportional $k_i=3$).

With growing Information density along the trajectory, these three-dimensional movements repeat the shortening these intervals for each triple segment with increasing frequency $f_i=k_i, k_i=3,5,7,...$

Since each of the segments deliver the equivalent spectrums, the equal frequencies of the sequential segments' spectrum $\{\omega_1,\omega_2,[\omega_1=\omega_1]_{\Delta t_{10}},\omega_2,[\omega_1=\omega_1]_{\Delta t_{20}},\omega_2,[\omega_1=\omega_1]_{\Delta t_{30}},\}=\omega_{\Sigma 10}$ are synchronized during the sequence of these time intervals.

According to the Proposition 5.1 initial condition, the invariant spectrum frequency $\omega_1$ repeats time interval $\Delta t_1$ of the logical anti-symmetrical interaction on a bridge separating $i-1$ and $i$ segments on the trajectory. The end of this interval indicates the beginning of the time interval $\Delta t_B$ on $i$ segment repeating with frequency $\omega_2$. During time $\Delta t_B$ the segment Bit is memorized. The end of $\Delta t_B$ indicates beginning of time interval $\Delta t_{en}$ of Free Information logic, which identifies beginning of bridge separated $i$ and $i+1$ segments. The Free Information attracts the separated segments.

The time intervals of the sequentially squeezing segments hold, first, the double synchronization during interval $\Delta t_{20}=1/2\Delta t_{10}$, and second, the double synchronization during interval $\Delta t_{30}=1/3\Delta t_{23}=\Delta t_{20}-\Delta t_{30}=1/2-1/3=1/6$. The sum $\Delta t_{33}=\Delta t_{20}+\Delta t_{23}+\Delta t_{30}=1/2+1/6+1/3=1$ is equal to the first interval $\Delta t_{10}$, during which all doublets are forming. Three segments finally deliver three memorized Bits with their three Free Information intervals, which sequentially attract the synchronizing doublets during the rotation movement.

The Information attraction on these time intervals adjoins the synchronized interval Information in a triple during the $i$ dimensional interval $\Delta t_{10}=1$. Forming the triplet completes Free Information which delivers each $i+2$ segment with triple frequency while holding the invariant spectrum.

The wave function's frequencies synchronize the triplet logic in collective resonance.

Free Information of the triplet joins the three memorized Bits in a triple bridge, where during an additional interval of Free Information $0.01847$ the Bits are encoded in the triplet knot.

The frequencies of the shortening time intervals distribute the orthogonal space rotations along the segments of the multiple dimensional observing trajectory which is moving in a *three-dimensional space wave function* for each of this trajectory's multiple dimensions. Each of the three dimensions' shortening time intervals, which the three-dimensional rotation moves, brings the triplet knot that joins the three dimensions to one.

Applying the sum of the shortening interval assembling triplet in knot $\cong \pi / 2$ to the segments located on three independent orthogonal dimensions $3\pi / 2$, leads to squeezing these dimensions to one on the triplet knot. Whereas, at forming the segments' triple during resonance, the trajectory is assembling each three segments in a cyclical loop.

The sequentially forming triple knots squeeze the initial observing multi-dimensional process first to three-dimensional rotation, and then to single-dimensional Information process encoding the Bits of all multiple knots. Squeezing dimensions accompanies sequentially memorizing and encoding the IN hierarchical levels knot. That shortens number of the IN cognitive levels, releasing cognitive logical loops memorized in the encoding knot.

Finally, the periodical wave function includes the sequence of repeating arguments along both orthogonal rotations:

$u_{sh} = u_s \times u_h, f_{ws} = \{f_{iws}\}, f_{wh}\{f_{iws}\}$, which performs the multiple three-dimensional movement with *three-dimensional space wave functions*.

The movement distributes the extreme trajectory segments on structuring the space located information networks which join the synchronizing triplets in assembling knots, and compose multiple structure of the information Observer.

The shape of the multiple wave functions describes the extreme multi-dimensional trajectory, formalizing the minimax observation process which models rotating segments on cones (Fig. 3), and continues on the spiral cone structure DSS analogous to Figs. 9, 9a. •

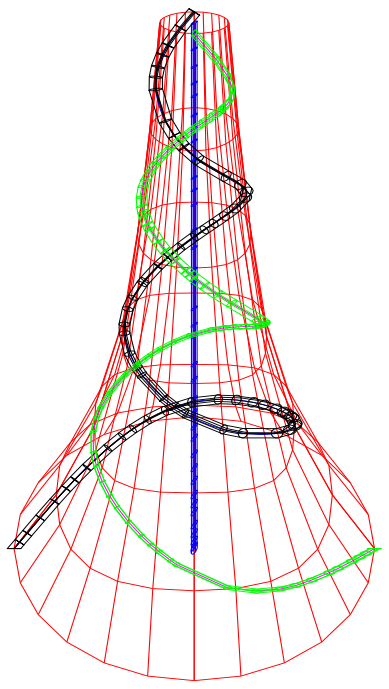

Fig.9a. Simulation of the 3-dimensional double spiral cone structure shortening the dimensions

*The spin-rotating trajectory of the invariant time space impulses builds geometry of information Observer.*

Comments 5.2

The conditions of sequential joining $n$ dimensional extremals in a common single dimensional extremals determine results [94].

That proves that such extremals should have Jacobean $J$ limited by condition

$$a \leq J < b \qquad (5.1.2)$$

where condition $J < a$ defines $k$-type of the extremal.

These extremals will be joint in sequence $N_o, N_1, N_2, ...N_k$ where $N_k$ is number of extremals of type $k > 1$ with two adjacent types, $N_1$ is number of extremals of type 1 with $N_1 + 1$ extremals of two adjacent types (i.e. one extremal with $J < a$), and $N_o$ is type of extremals with zero adjacent points (i.e., it is a single extremal already joining these points or closing the gap between these points).
From that follows that these extremals can join reducing their number sequentially from $n$ to 1.
However, in the case, when the sequentially joining extremals collect more observed information, that information could limit the IPF extremal' Jacobean by the condition (5.1.2) which hold maximal $J \ge a$ satisfying minimal $J < b$. Thus, the Jacobean-Hamiltonian $H = V + P$ (in Sec.3.3) should satisfy max min condition, which limits its potential function $P$ and the integrand $V$.
Since the both entropy functional and path integral satisfy the minimax conditions, each of their Jacobeans for the extremals, which are joining sequentially, is fixed and limited by a fixed minimax condition $a, b$ above. The order of the sequence, connecting the ordered sequence of the joining extremal' pair points, is distinct. Moreover, each joining extremals' point represents a border connecting the extremals, whose information is sequentially growing.
Finally, the multiple dimensional extremals will be joining in the above sequence in a single-dimensional extremal, while the sequentially collected information automatically limits the minimax principle. That confirms Proposition.5.2. •

<u>The wave function frequencies and properties</u>

1. The wave function with the above speeds and frequencies emerges in the observation process when a space interval appears within the impulse microprocess during a reversible time interval $\varepsilon_{ok} = 0.015625$ of the impulse invariant measure $\pi$ equivalent to $1Nat$. Before that, the observing trajectory has described the probabilistic time function whose probability $P_{\Delta}^{*} \cong 0.821214$ indicates the appearance of a space-time probabilistic wave. During the probabilistic time observation, entropy of the Bayes *a priori-a posteriori* probabilities measures the probabilistic symmetric logic of a sequence of these probabilities.
Thus, the wave function starts emerging in probabilistic observations as a probability wave in a probability field.
At the beginning of the microprocess, the probabilistic wave measures only the time of its propagation.
2. Within the microprocess, the asymmetrical logic emerges with the appearance of a free logic interval $\Delta t_{fo} \cong 0.1597 \cong 1/2\pi$ which, repeating with equal wave frequency $\omega_i^s$, indicates the beginning of the interactive rotating asymmetry on a primary bridge and segment. From that point, the observation logic on the trajectory becomes the asymmetric part of total free logic $f_{li} = 1 - \ln 2 \cong 0.3Nat$. The asymmetric *logical wave* emerges. Approaching $p_{\pm a} = \exp(-2h_{\alpha}^{o*1}) \cong 0.9866617771$, the asymmetric logic probability appears with the certainty-reality of the previously hidden asymmetrical Bit. Such logic temporary memorizes correlation with the probability which carries a logical Bit of the certain logic.
Such a certain logical Bit may carry energy in a real interactive process, described the Markov diffusion process. The path to creation of the certain Bit includes an increment of the probability $\Delta P_{ie} = 0.9855507502 - 0.981699525437 \cong 0.004$ starting injection of energy from an interacting impulse of the Markov process (Sec. 3.5). Thus, the certain free impulse logic carries the certain logical attraction.
The wave function in the microprocess is probabilistic until the certain logical Information Bit appears.
The certain asymmetrical logical Bit become physical Bit through erasure the entropy of this logic, which allows replacing the logic by memorizing its Bit.

3. The wave function starts on the observing process which the EF extreme trajectory prognosis, carrying the probabilistic wave which transforms the observing process to the certainty of real observation. The spinning movement of the space-time trajectory describes the invariant speed encircling the cross-section of its rotating impulses-segments, which spreads the invariant rotation space speed along the segment trajectories. The segment's invariant spectrum $\{\omega_1, \omega_2, [\omega_1 = \omega_1]_{\Delta t_{10}}, \omega_2, [\omega_1 = \omega_1]_{\Delta t_{20}}, \omega_2, [\omega_1 = \omega_1]_{\Delta t_{30}},\} = \omega_{\Sigma 10}$ repeats the triple frequencies of these three-time intervals between them. That shortens the distance of the equal spectrum frequencies and assembles them in a resonance creating joint logical structures-triplets up to the IN hierarchies and domains. The frequency absolute maximum indicates the finite end of its creation. A minimal energy of the resonance supports the forming logical loop. •

Distribution of the space-time hierarchy

1. The hierarchy of self-cooperating triplet units distributes the space rotation emerging along the EF segments of the time-space extreme trajectory, where each third impulse progressively increases the Information density measure of its Bit in the triple. The time-space hierarchy of the units starts emerging in observation of the symmetrical logic at the appearance of the space interval in the microprocess. This logic self-forms a hierarchy of the logical unit structures through the impulse's mutual attracting free logic which, sequentially attracting the moving unit's speeds, equalizes their frequencies in resonance that assembles the Observer logic along the hierarchy of units.

2. The hierarchy of the logical cooperating units becomes asymmetrical with appearance of certain logical Bit on the extreme trajectory. The repeating free logic interval indicates the wave frequency $\omega_1 = f_i^s = 1/2\pi$.

The EF rotating trajectory of three segments equalizes their Information speeds joining in the resonance frequency during the space rotation, which cooperates each third logical Bit's segment on the trajectory and logically composes each triplet structure in the unit space hierarchy.

3. The appearance of the asymmetrical logical Bit on the extreme trajectory indicates entrance the IPF Information measure $\ln 2$ on its path to forming a logical Bit. The path starts on the relative time interval $\Delta t_{fo} = 0.23/1.44 \cong 0.1597$ of the logical asymmetry, which identifies the segment bridge. During the triple impulses, the third time interval $\Delta t_{3r} = 3\Delta t_{1r} = 3/2\pi \cong 0.4775152$ indicates the end of the triple cooperative logic, starting to build the triplet knot. Forming the triplet knot requires a time interval, during which the triple free logic binds in the triplet Bit. The time interval of creating the Bit approaches $\Delta t_B = \ln 2/1.44 \cong 0.481352$. The difference $\Delta t_B - 3\Delta t_{3r} \cong 0.004$ evaluates the time of binding the triplet. Thus, the wave space interval delivers the logical Bit with the wave spectrum frequency $\omega_2 = 2\pi\Delta t_B = 2\pi \ln 2/1.44 = 3.02 < \pi$, while the triplet knot repeats with the spectrum frequency $\omega_{20} = 2\pi 3/2\pi = 3$.

4. Delivering external energy for memorizing the logical Bit identifies relative moment $t_1 = 0.2452/1.44 \cong 0.17$ ending the interval of the asymmetry. By this moment, the resonance frequencies of the asymmetrical logic have already been created.

Along the IPF path on the trajectory, this moment follows interval $\Delta t_B$ of creation the logical Bit, ending the emergence of the knot that binds the free logic. The interval of memorizing the physical Bit requires the same interval $\Delta t_B$ during which the entropy of logical Bit is erased. The necessary external impulse, erasing the asymmetric logical Bit, starts with interval $\Delta t_B$ and ends with the interval of encoding the Bit $\Delta t_{en} = 0.17$. The external energy, supplied on time interval $t_{\Sigma b} = 0.481352 + 0.19 = 0.671352$, includes both erasure of the logical Bit and its encoding. Whereas the interval of Information free logic $\Delta t_{fo} = 0.23/1.44 \cong 0.1597$ is left for attracting a new Bit (at interaction with an external impulse carrying energy, Sec.2.6). Since the external impulse interactive part is $0.025$, it brings total $t_{\Sigma bo} = 0.67083 + 0.025 = 0.69583 \cong \ln 2$ for the interval of the external Bit.
Therefore, the frequency spectrum, initiating the encoding, equals $\omega_1$ in sequence $\{\omega_1 \omega_2 \omega_1\}$. This triple sequence identifies the segments alternating on the trajectory with the repeating ratio of the bridge-middle part-starting next bridge, which measure the bridge relative interval $\Delta t_{en} = 0.17$.
Thus, the sequence of segments on the EF-IPF extreme trajectory carries its wave function's frequencies which self-structure the space-time unit of the logical Bits hierarchy that self-assembles the Observer logic. The logic controls memorizing and encoding physical Bits as well as the hierarchical structure of the space-time Information geometry of the units.
5. The segments-impulses on the EF spiral trajectory sequentially interact through the frequencies repeating on the bridge time-space locations connecting the segments in the trajectory. The segment sequences on the EF-IPF extreme trajectory (Figs. 3, 4) carry its wave function frequencies, self-structuring the unit logical Bit hierarchy that self-assembles the total Observer logic. This logic controls memorizing and encoding of both physical Bits and the hierarchical structure of their units. •
The Observer cognitive logic encloses both probabilistic and Information causalities distributed along all Observer hierarchies. The logical functions of the self-equalizing Free Information in the resonance perform the cognitive functions, which are distributed along the hierarchy of assembling units: triplets, IN nested nodes, and the IN ending nodes. These local functions self-organize the Observer cognition.
Assembling runs the resonance frequencies $[\omega_1 = \omega_1]$ spreading along this hierarchy. Each unit, ending high level hierarchical structure encloses all its Information logic, whereas the high unit' impulse invariant time-space interval, containing this Information, increases more Information density than the unit of lower level hierarchy. The resonance frequencies of spectrum $\{\omega_1, \omega_2, \omega_1\} = \omega_o$, holding the cognitive logic loop, self-creates the unit hierarchy.

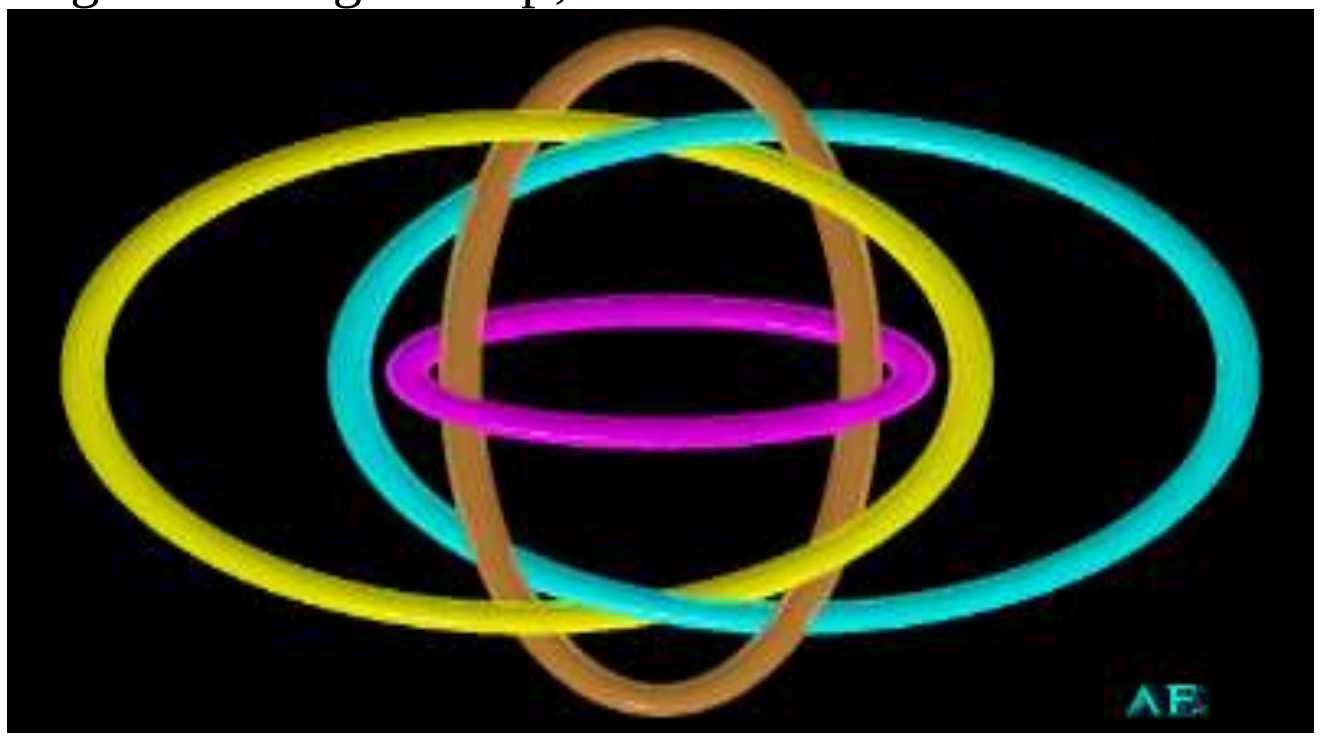

Fig.B. An illustrative example of a Brunnian link, potentially enclosing the cognitive loops from orthogonal hierarchical units. (That link with four loops discovers Brunn in 1892, source Alain Esculier's website).

### 5.2.5 Structure of the Observer logic and cognition

1.The Observer logical structure self–forms the attracting Free Information, which self-organizes the hierarchy of the logical triplet units assembling in the resonance frequencies. Each triplet logical structure is an analog of Borromini ring consisting of three topological circles linked by a Brunnian link-loop, Fig.B. (The forth loop represent a free logic)

The EF-IPF time-space trajectory distributes the hierarchical logic.

2. The Observer logical structure carries the wave along the trajectory segments, where each third segment delivers the triple logic of the Information spectrum $\{\omega_1, \omega_2, [\omega_1 = \omega_1]_{\Delta t_{10}}, \omega_2, [\omega_1 = \omega_1]_{\Delta t_{20}}, \omega_2, [\omega_1 = \omega_1]_{\Delta t_{30}}, \} = \omega_{\Sigma 10}$ with sequentially shortening intervals $|\Delta t_{10}, \Delta t_{20}, \Delta t_{30}|$ and the increasing segment Information density.

Two sequential segments synchronize resonance frequencies $[\omega_1 = \omega_1]_{\Delta t_{10}}$ and $[\omega_1 = \omega_1]_{\Delta t_{20}}$ while the triplet synchronizes resonance frequency $[\omega_1 = \omega_1]_{\Delta t_{30}}$. This triple logic holds one Bit in each Observer's triplet logical structure unit. The attracting free logic of sequential triplets conveys the resonance spectrum with progressively shortening time intervals and growing frequencies, which cooperate to form the logical units in IN nested hierarchy. The necessary spectrum with the increasing frequencies automatically carries each consecutive segment along the EF-IPF trajectory. The emanating wave function delivers the frequencies, cooperatively growing a hierarchy of the logical units. The self-built hierarchy of the logical structures self-integrates the observed logic which the structure encloses.

3. The hierarchy of distributed logical loops self-connects logical chain.

The logical chain width determines the invariant impulse's relative interval enclosing the assembled logical code. The growing density of consecutive impulses along the trajectory sequentially squeezes the absolute value of this interval, whose ratio preserves the invariant impulse. The absolute time-space sizes of the logical chain are squeezing through the multi-level distributed hierarchy.

4. The cognitive logical chain composes the coherent triplet loops which assemble the nested attractors-knots logically cooperating the IN nodes' hierarchy. This composite logical cognition synchronizes the triple rhythms along the EF-IPF trajectory, which schedule the access energy for encoding the observer intelligence. Thereafter the cognitive chain predicts the intelligence encoding.

However, during sequential encoding the hierarchy of logical knots, the cognitive logic of the knot, encoded at a current hierarchical level, dissolves, its predictive action vanishes.

Thus sequential hierarchical encoding successively removes the cognition which predicts this encoding.

5. The logical chain rotation, carrying the frequencies of the synchronized spectrum, requires a minimal energy to support the chain. This energy is equivalent to each Bit's logical code.

The integrated chain logic holds this code, and the physical encoding intelligence encloses the cognitive thermodynamics. •

Therefore, the wave function frequencies, initiating the self-forming Observer cognition, emerge along the EF-IPF extreme trajectory in the form of a probabilistic time wave in a probability field.

The probabilistic impulse observation starts the microprocess, where the entangled space rotation code develops the rotating space-time probability wave.

The emerging opposite asymmetrical topological interaction shapes the space-time wave function, becoming certain, as well as the Observer's cognitive logic, predicting the intelligence code.

*These results conclusively and numerically determine the structure and functions of cognition.*

### 5.2.6 Specifics of Information Intelligence and estimation its Information values

The causal probabilities, following from a Kolmogorov-Bayes probability link, start the Markov correlation connection with minimum of three probabilistic events. An Observer integrates the observing events in the Information networks, which accumulate the nested triple connections, depending on the IN Information invariant properties.

Each IN has an invariant Information geometrical structure and a maximal number of nodes-triple Bits, whose ability to cooperate more triplet nodes limits the possibility of the IN self-destruction by arising a chaotic movement.

The intelligence measures the *memorized ending node of the IN highest levels*, while the cognitive process at each triplet level preempts its memorization. This means each memorizing node encloses cognition. The Information measure of intelligence is *objective for each particular Observer*, while the IQ is an *empirical subjective* measure.
This theory shows that during the current observation, an Observer can build each IN with maximum 24-26 nodes with average $3^{26}$ Bits and enfold the maximum of 26 such INs.
Since each IN following level integrates Information from all the IN previous levels, measuring the relative Information quality, the built multi-levels INs hold Information quality relationships between the levels in the triplet forms.
Because the subsequent relationships have been enclosed by the cognitive rotating mechanism, they formalize a causal comparative Information quality meaning for the observing process events.
The *Observer Intelligence* has the ability to uncover causal relationships enclosed in the evaluated Observer $\mathrm{N}_{o1} = 3^{26} \times 26 bits$ networks Bits. That requires not only building each of $\mathrm{N}_{1I} = 26$ IN, but also to sequentially enfold them in a final node whose single Bit accumulates $\mathrm{N}_{oI}$ Bits:

$$\mathrm{N}_{oI} = (3^{26}) \times 26 = 2,541.865.828329 \times 26 = 66,088.511.536.554 \cong 6.61 \times 10^{9} \text{ Bits.} \quad (5.1)$$

However, since each IN node holds single triplet's Information, the final IN node's Bit keeps the triple causal Information relationship with density $\mathrm{D}_{oI} = \mathrm{N}_{oI} / bit$ -per Bit.
To support the IN node impulse feedback communication (Sec.6.3) with the requested attracting Information, this node requires Information density:

$$i_{md} \cong 1.8 \times 10^{14} Nat / \sec = 1.44 \times 1.8 \times 10^{14} bit / \sec, \quad (5.2)$$

where each such Bit accumulates $\mathrm{N}_{oI}$. Thus, the total Information density of the Observer final IN Bit:

$$i_{do} \cong 1.44 \times 1.8 \times 10^{14} \times (3^{26}) \times 26 bit / \sec \quad (5.3)$$

evaluates the intelligent Observer's Information density.
With this density, the intelligent Observer can obtain maximal Information from the EF through the impulse interaction with entropy random process during time observation $T$.
Let us evaluate the EF according to (Sec. 4.4):

$$I_e = 1/8 \ln[\mathrm{r}(T) / \mathrm{r}(\mathrm{t}_s) \approx 1/8 \ln(\mathrm{T}/\mathrm{t}_s), \mathrm{T} = m_N \mathrm{t}_s .$$

Here $m_N$ is a total number of the IN nodes needed to build the intelligent Observer, $\mathrm{t}_s$ is the time interval of the invariant impulse. At $m_N = 26 \times 26$ it allows estimate $I_e = 1/8 \ln 26^2 = 11.729 Nat$.
Therefore, the intelligent Observer needs $N_i \cong 12$ invariant impulses to build its total IN during the time interval of observation $T$.

Comments 5.3.

The human brain consists of about 86 billion neurons [70], which approximately in 14 times exceed $\mathrm{N}_{oI}$ (5.1), if each single Bit of the cognition commands each neuron. If each neuron builds own IN with about five-six triplets (with levels $3 + 2^4 = 11, or 3 + 2^5 = 13$), while the ending triplet Bit condenses this $\mathrm{N}_{oI}$, ability of the neuron building a net concurs with [70] and [71]. If this is true, then $\mathrm{N}_{oI}$ measures the Information memory of a human being. •

According to estimates [71, others], the maximal Information in the Universe approximates

$$I_U \cong 3 \times 10^{29} Nat = 4.328 \times 10^{29} bits, \quad (5.4)$$

from which each invariant intelligent Observer can get $I_{ob} \cong 6.61 \times 10^{9} bits$.

To obtain all $I_U$ Information, such intelligent Observers need $M_{ob} \cong 1.527\times10^{16}$ numbers of such invariant Observers.

Each IN triplet node may request $I_m \cong (3.45-2.45)bits$, which measures this IN level of quality Information that memorizes the node Bit. Such node's level accumulates average Information between $I_m bits$ and $\mathrm{N}_{om} = 3^{26} bits$, depending on each IN's levels quantity $\mathrm{N}_{1I}$..

Quantity $\mathrm{N}_{oI}$ (5.1) measures the invariant transformation to build the extreme IN node's structure during the observation, which transforms a probable observing process to an Information process in the emerging Information Observer with intelligence.

The initial probability field of random processes, evaluated by the Entropy Functional, contains the potential Information which an intelligent Observer can obtain through the invariant transformation.

The Information threshold $\mathrm{N}_{oI}$ limits the level of intelligence of the intelligent Observer, satisfying the minimax variation principle. The intelligent (human) Observer can overcome this threshold requiring highest Information up to $I_U$.

An Observer that conquers the threshold, possesses a superior intellect, which can control not only its own intellect, but other Observers. Multiple joint superior intellectual Observers can form a super-intellectual system (with $I_U$) controlling Universe, or would destroy themselves and others. However, in an intelligent machine, collecting the observing Information, the emerging invariant regularities of the minimax law limits the AI Observer actions.

### 5.3 Interacting intelligent observers through communication

An Information intelligent Observer emerges during the evolving observations, which have delivered invariant information, built Information IN' nodes' hierarchy, and double helix rotating structure with DSS intelligent code. The important issue is interaction of such Observers in a mutual communications, which preserve the invariant Information properties and benefit their information qualities.

Suppose an intelligent Observer sends a message, containing Information encoding its meaning.

Another intelligent Observer, receiving this Information, would be able to *read the message, recognize its meaning, select and accept* it if this Information *satisfies the Observer's needed Information quality,* which is being memorized through its DSS code. Next, we consider the fulfillment of these five issues.

#### 5.3.1 How the interacting intelligence Observers can understand meaning in each communication

Let an intelligent observer sends a message enclosing its information logic, quality, and bits encoding this information, which emanates from some intelligent observer's IN nodes.

Other intelligent observer, requesting the growing quality of needed information, sends specific qualities of free Information emanating from its IN nodes that need that quality. In the communication interaction, the receiving quality will add the needed quality compensating the need. (Each observer quality information classifies the node location in the IN hierarchy enfolding this information [62]).

The intelligent observer, receiving that information quality, identifies the nodes locations in the IN hierarchy, being equivalent to this quality. Each node location encloses logic of its triplet with a resonance cognitive loop. That quality may belong to ending node of the observer IN enclosing its cognitive loop. The ending node emanates the Observer sending free information quality enclosing the loop. The message quality associates with the node free logic attraction, carrying the related information frequency (Sec.5.2). If the cognitive loop of the sending free information quality accepts the receiving information quality, enclosing its frequency, then the receiving observer enables recognizing the message meaning. That implicates understanding of the message meaning encoding this quality.

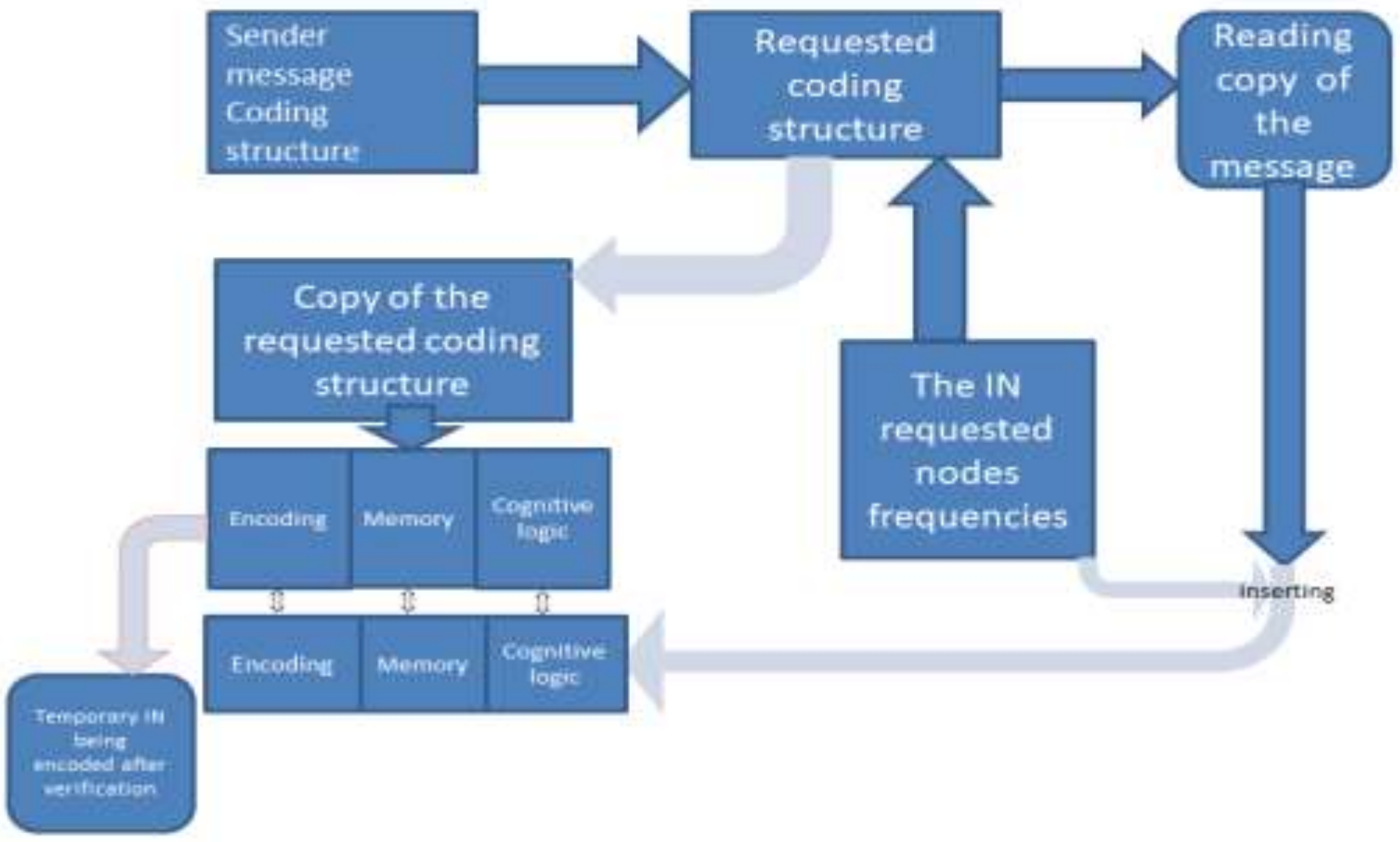

Fig.C. Schematic illustration of functional organization of comparative acceptance a message.

The schematic illustration shows

1. How the sender message coding structure is read by the requested Observer coding structure which initiates the IN requesting nodes frequencies;
2. The frequencies identifies the requested coding structure that allows comparing the reading message intelligence information with that in the Observer requested coding structure;
3. The comparison requests the same nodes frequencies which insert the Reading message copy' Bits into the copy of the requested coding structure (DSS Copy).
4. If the comparison verifies the receiving observer intelligence information, this observer accepts the message intelligence information. Assuming that this information identifies the observer meaning, this observer understands the message meaning.

Hence, the receiving Observer recognizes the message meaning if the cognitive loop of the temporal IN sending node free Information quality accepts the receiving Information quality, enclosing its frequency, while this frequency reveals the DSS's equivalent logics. Revealing the message logic, its memorizing Bit, and encoding, takes place during the message information copy's insertion and moving in the Observer Copy DSS.

Starting the message copying begins the time of the movement, which starts this DSS enclosed logic, up to revealing the Bits and encoding in the IN being temporal. Those IN nodes hold the message quality needed by the Observer request. Thus, the observer request with the specific free information initiates recognition of the needed information. That process comprises the following steps.

1.The IN nodes of Observer-receiver request the needed quality through the nodes free information which, attracting the message bits, starts copying them message on an Observer double spiral helix structure. The DSS structure, while copying, integrates and reads the message information. While reading the message, the DSS is moving along its time-space information structure and allocates a temporal IN with the nodes related quality.
(The admissible number of the temporal IN nodes constrains the message information.)
2. The requested INs ending nodes quality information mirrors the information from the IN temporal node. The free information of the ending nodes impulses initiates involvement of the copy information frequencies in resonance.
3. If the copies frequencies cohere in the resonances, assembling the temporal logical loops, the requesting information impulses insert the mirror copy information logic in the IN requested nodes.
4. During the insertion of the mirror copy, transitive impulses of the requested nodes provide asymmetrical free logics with $\Delta t_1$ intervals. This indicates that the requesting nodes reveal and accept the mirror copy of message logics.
5. Each of these logics intervals allows access to the ending node the interval $\Delta t_B$ currying the physical bit from the receiving Observer DSS which had copied the message bit. The physical bit energy erases logic of the mirror copy, revealing its information bit and starting process of memorizing bits and decoding the message enclosed triples.
Access of the Observer bits, which initiates the acceptance of the mirror copy logics, indicates recognizing the message bits by the observer bits.
6. Decoding of each memorized Bit takes interval $\Delta t_{en}$ of its encoding. The decoding of ending IN node impulses reveals the IN hierarchy of the enclosed message information logic.
(Revealing the message logic, its memorizing bit, and decoding takes place during the time of the message information' integration on the Observer DSS. Starting the message copying begins the time of the integration which starts the DSS logic up to revealing the bits and encoding-decoding. The Observer DSS, emerging with copying the message, is main information apparatus allowing the Observer comparative acceptance of the message information).
7.The message information, located on the DSS, delivers the wave function frequencies of the recognized decoding impulses. The IN node' frequencies generates the cognitive loops' coherent frequencies *recognizing* the message logic delivered to the IN receiving nodes. The frequencies spectrum may update the needed information quality.
8. Decoding finalizes requesting IN nodes acceptance of the message comparative qualities. These qualities indicate ability to cohere and cooperate the message quality with the quality of the IN node, enclosing it in the Observer-receiver IN structure. The observer logic' coherence with the message logic allows memorizing the decoded message information.
9. Accepting the message quality, the intelligent Observer recognizes the message information and encodes its triple logic' digital bits (images) by the observer time-space codes, being self-reflective in understanding the meaning of the message. Since the acceptance of the message quality changes the existing observer logic encoded in the INs hierarchy, understanding the meaning of the message extends the level observer intelligent logic. Thus, the intelligent observer uncovers a meaning of communicating message in the self-reflecting process, using the common message information language, temporary memorized logic, the cognitive acceptance, and the logic of the memorized decoding.
Understanding the meaning of an observing process includes the coherence of its Information with the Observer current coding structure, which evolves all previous observations, interactions, and communications.

**5.3.2 How does a biological Observer accept receiving information?**

The acceptance Information formalism includes:

1. Multiple interactive observations forming long-term correlation, where information emerges from entropy of cutting correlations, which integrate the EF.

2. The IPF integral connectivity performs free information resulting from memorizing the bit within the pair-wise correlated impulse. By moment of memorizing bit, the correlation of the impulse is cutting through the gap while capturing energy for memorizing.

3. The free information connection with attracting action is week (until resonance couples the bits), compared with already connected bits in the formed the IPF chain.

4. The integral connection builds a shortest information path (satisfying the VP). Eliminating free information loses the bits connectivity and memory.

5. Asymmetric interactions preempt the Bit memory with created free information which performs encoding the memorized Bit. When the IPF composes the sequence of bits, it increases density of each following bit-its "strength" but not changing the bit measure.

6. The IPF composes the rotating double helix (DSS) structure which distributes and shapes the hierarchical Information Networks. The EF-IPF time-space trajectory encloses the observed logics. Multiple moving INs assemble the Observer logic whose attracting nodes unite the hierarchical logical loops in cognition commanding the Observer information intelligence.

The DSS encodes the Observer Information intelligence especially in a human brain.

These features of the information integration mathematically emerge from observation, but its natural (biological) existence was not proven. Experimental results of study [72] verify the following.

1."Global integration of information in the brain, forming the largest connected nucleus node network", is result of the "giant connection component G that connects the largest number of the nodes trough a shortest path".

2. In a range correlation structure, forming the network functional connectivity through pairwise interactions, 'the G component identifies the *week node* covering the nucleus accumbens shell. ("accumbens is a region of the hypothalamuses responsible for thermo-regulation and receiving nervous stimulation from chemoreceptors in the skin, covers the surface of internal organs, and hypothalamus itself", which include all sensors of observation).

3. "The G serves as "crucial bridge (a shortcut) influencing large number of nearest nodes' collective in the encoding network memory". The encoded nodes "predict long-range influence".That influence initiates capturing". "The capture effect is emerging from the long-range correlation structure". "The capture provides the strength of weak ties". (The "weak degree nodes are as influential as long as they are surrounded by high degree nodes in their spheres of influence"). The bridge bundles a shortcut nodes by "the densely knit".

4. The G accumbens shell is "strategically located in the memory network". That memory, binding two nodes, "approximates $1mm$ cub" in the brain network". (Without "driving formation of the memory network, the G component not appears"). "Inactivation of the nucleus G accumbens eliminates the formation of the memory network",

5. The G accumbens, "in addition to performing on-line processing"(of capturing), provide "action of selection and encodes the output of the selected action (positive or negative relative to expectation) into memory, which in turn will condition future selections." The G acts "as a downstream station working as a limbic-motor interface with a role in selecting behaviorally relevant actions".

6. “The interaction between the hippocampus and frontal neocortical regions networks indicates of the G operation as gating mechanism that couples these networks’ two key structures for the storage of new information, providing a mechanism for updating memories to guide future behaviors”.
7. Information exchange, “that excludes some non-causal correlation, is result in directed (asymmetric) interactions e.g. Granger Causality”.
This experimental study uses “Segregation removing the node from brain network for disruption a long–range correlation in functional connectivity”, considering “global integration of information in the brain resulting from complex interactions of segregated brain networks”. The experiments study ”synaptic long-term potentiation (LTP) resulting of long term-correlation involved during memory encoding as basic of learning and memory.”
The acceptance formalism also contains copying the integral information on the cognitive moving helix, temporarily memorizing it as triplet entropies in a virtual IN structure.
This converting mechanism includes compression of the observing image in a virtual impulse ending the virtual IN. A receiving impulse makes the mirror copy of the IN ending impulse.
This impulse, holding the asymmetric equivalent of the ending IN information impulse, scans the DSS helix along the observer INs. The scan movement ends when a negative curved step-up down action of the impulse, carrying the entropy equivalent of energy, will attract a positive curvature of the IN node Bit‘s step-down action. The forming Bit encloses the equivalent energy’s quality measured by its entropy value. When the IN Bit’s step-down action interacts with the moving image’s step-up action, it injects energy capturing the entropy of the impulse’s ending step-up action. This inter-action models 0-1 Bit (Fig. 2A, B). The opposite curved interaction provides a time-space difference (an asymmetrical barrier) between 0 and 1 actions, necessary for creating the Bit. The interactive impulse’s step-down ending state memorizes the Bit when the Observer interactive process provides Landauer’s energy with maximal probability (up to a certainty). Such energy drives the cognitive helix movement, having minimal entropy production to overcome the bridge to the intellectual action memorizing and encoding the Bit. This is the energy of the cognitive thermodynamic process (Sec. 2.6.6), spending minimal quantity equal to binding the triplet structure by Landauer‘s energy ln 2.
When a triplet cooperates, this energy can be spent on memorizing the joint triple Bit in the knot after the third Bit gets the asymmetrical structure needed for memorizing.
Such triplets carry information of a message to accept.
The erasure and then memorization of each observing Bit can run *equal neuron Information Bits.*
If the incoming Information coheres with the cognitive loop, the message can be accepted and memorized in the receiver’s IN.
The forming IN Bit encloses the equivalent energy’s quality measured by its entropy value.
Therefore, the cognitive thermodynamic process practically has no thermodynamic cost, which models a cognitive software with the minimal algorithmic complexity.

The important coordination of an Observer external time-space scale with its internal time-space scale happens when an external step-down jump action interacts with the Observer inner cognitive thermodynamics time-space interval. The curved interaction measures the difference of these intervals (Sec. 2.6.6).
Understanding the receiving Information includes classifying and selecting such Information that concurs with this Observer’s memory of other comparative images.
Thus, cognitive movement, beginning in a virtual observation holding its imaginary form, composes an entropy microprocess, until the memorized IN Bit transfers it to an Information macromovement.

That brings two forms for a cognitive helix process: imaginary reversible with temporal memory, and real-Information which moves irreversible cognitive thermodynamics and ends memorizing the incoming Information.

Explaining the mechanism modeling the message acceptance and understanding requires admitting, first, that the developed math-Information formalism is considered as software controlling a brain structure, that is, a hardware. Connecting them requires a converting mechanism, which copies an observation and starts action on intelligence hardware. Those perform different sensors bending neurons, which make a mirror virtual copy of the observing image-message (analogous to transition impulse (Sec. 2.6.6, Figs. 2A, B) on the cognitive moving helix.

For example, eyes scan a TV screen, integrating selected visual features in a reflected image, accompanied with the accumulating sound of the visual image. This primary cognition allows intelligent memory and encoding. The Observer may not need to memorize each currently observed virtual image, which is reflected temporally in some sequence. Accordingly, such multiple virtual copies are formed by temporary triplet units composing a temporal collective IN whose ending node encloses a virtual impulse entropy Bit. Such virtual IN with temporal memory forms in a reversible logical process without permanent memory, which comprises a part of the observation process (when Kolmogorov-Bayes probabilities link the triple events).

This converting mechanism includes a virtual compression of the observed image in a virtual impulse ending a virtual IN with the following coordination internal and external time course. Specifically, the time-space difference between particular 0 and 1 actions determines the clock coordinating the Observer external and internal time course. When the image Bits memorize the Observer IN's specific node, this node Information quality and its precise position allow the Observer to *recognize* this image Information among other distinctive Information qualities.

The node's positions already contain the Observer INs' memorized Information qualities.

Recognizing a collective Information image is associated with understanding it by that Observer-enclosed Information. Understanding implies that the Observer can classify and select such Information according to this Observer's memorized *meaning*, among other comparative images.

<u>The Information model of the Observer's understanding of a receiving information includes:</u>

1. Sensor conversion of the observing image through building the virtual IN of the message, as a virtual mirror copy of the image collective Information, which the IN compresses in the virtual impulse.
2. Copying on the moving cognitive helix, which scans the Observer IN's Information enclosed from all IN and domain levels.
3. Interaction of a sensor's neuron impulse, initiating yes-no actions, with the virtual impulse of the image through its yes- action, which injects an energy capturing the virtual entropy of the impulse's ending step-up action when the scanning helix cognitive movement contacts the Observer IN node that provides this energy.
4. Memorizing the Yes-No interactive Bit by the neuron interactive impulse's step-down No-action through the cognitive dynamic interactive process, providing Landauer energy for erasing the observing image. That builds a mirror's Bit's memory, which decodes the message-image.

In this neuron-message communication, the neuron Yes-action captures the virtual impulse's ending step-up action, connecting it with this neuron's No-action, which provides a step-down action memorizing the message through the cognitive dynamic energy. Thus, the neuron curving interaction connects virtual and real actions, which actually binds the cognitive software with the brain hardware structure.

5. The memorized Information Bit stops the scanning cognitive mechanism on such an IN level, where this Information is understood through its Observer's IN recognition. That ends the process of understanding a current message.
By scanning the meanings understood by the Observer, the reading semantics of the message can be recovered, and then encoded in a sending message. The virtual impulse of cognitive interaction provides a logical Maxwell's Demon (DM), while transformation to the memorized IN Information runs the physical DM. It presumes that neuron's Yes-action starts its impulse entropy microprocess until the neuron's No-action, interacting with the Observer macroprocess via the IN node Bit by the jumping No-action, memorizes the incoming image in the Observer IN structure. Thus, cognitive movement, beginning during virtual observation, holds its imaginary form, composing an entropy microprocess, until the memorized IN Bit transfers it to an Information macro movement.
That brings *two forms* for the cognitive helix process: *imaginary reversible without memory, and real-Information moving by the irreversible cognitive thermodynamics memorizing incoming Information.* The imaginary starts with the neuron Yes-action and ends with the neuron No-action at the ending state of the neuron impulse, while the real starts with the IN Bit Yes-action memorizing the accepted Bit, which processes the cognitive thermodynamics continuing to move the cognitive helix irreversibly.
The threshold between the imaginary and real cognition holds the memory and energy of the cognitive thermodynamics.
This is how the observing quantity and quality of interacting Information emerge in Observer as the memorized quality encoding the Observer cognition.

### 5.3.3 Analysis of some experimental brain studies

Study [73] provides explicit quantities for the energetic cost of processing sensory Information.
The findings in the blowfly visual sensory system revealed that for visual sensory data, the cost of one Bit of Information is around $5\times10^{-14}$Joules, or equivalently $10^4$ ATP molecules. That amount of Information was delivered to the blowfly retina's photoreceptors in the form of fluctuations of light intensity. This neural processing efficiency is still far from Landauer's limit and its Bit's minimum *ln2*, but it is still much more efficient than modern computers.
This limit evaluates the minimal cost of a neuron's Yes-action, which starts capturing the virtual observation upward to a real observing action.
"A number of studies conclusively demonstrate that the large monopolar cell (LMC), the second-order retinal neuron, is optimized to maximize Bit rate."
That unique single cell holds "photoreceptors and an LMC of the blowfly retina code light level in a single pixel of the compound eye. Six photoreceptors carrying the same signal converge on a single LMC and drive it via multiple parallel synapses. The signals are intracellular recordings of the graded changes of membrane potential induced by a randomly modulated light source. Analysis of these analog responses yielded the rate at which photoreceptors and LMCs transmit Information.
"The low capacity (55 Bits per second) synapse transmits at a much lower cost per Bit than the high capacity (1600 Bits per second) interneuron, the LMC capacity 1,500Bits per second."
That result limits the Yes-No neuron transmission rate in our cognitive model.
The brain neurons communicate [74] when presynaptic dopamine terminals demand neuronal activity for neurotransmission; in a response to depolarization [75], dopamine vesicles utilize a cascade of vesicular transporters to dynamically increase the vesicular pH gradient, thereby increasing dopamine vesicle content.

Recent study [76] shows that “midbrain dopamine neurons activity also encodes sensory prediction errors unrelated to reward. By signaling errors in both sensory and reward predictions, dopamine supports a form of reinforcement learning”.
In [77], neurogenesis provides fresh fields of interaction at the cellular (neuronal) level. Prior responses are generalized and the stage is set for future responses to assess probabilities and store temporary and intermediate data.
Study [78] found that the brain computes a Bayes probability distribution which generates a current observation, and this “belief distribution” representing the (log-transformed) posterior distributions encoded in the pattern of brain activity reinforcing learning and decision making”.
Cortical networks exhibit different modes of activity such as oscillations, synchrony and neural avalanches [79].
Study [80] shows that dopamine modulates the brain dynamics boosting the cognitive performance of large–scale cortical networks. Specifically, “enhanced dopaminergic signaling modulates the two potentially interrelated aspects of large-scale cortical dynamics during cognitive performance.” Thus “dopamine enhances Information-processing capacity in the human cortex during cognitive performance.” That confirms the communication of interacting Bits modeling the neurons.
The structure of thought arises from Hidden Markov cognitive models [81].
Experiments using mathematical “pyramid problems” support the pyramidal space structure of the Information Observer’s domain intelligence.
More supporting results follow from the author’s image of conscience [82]:

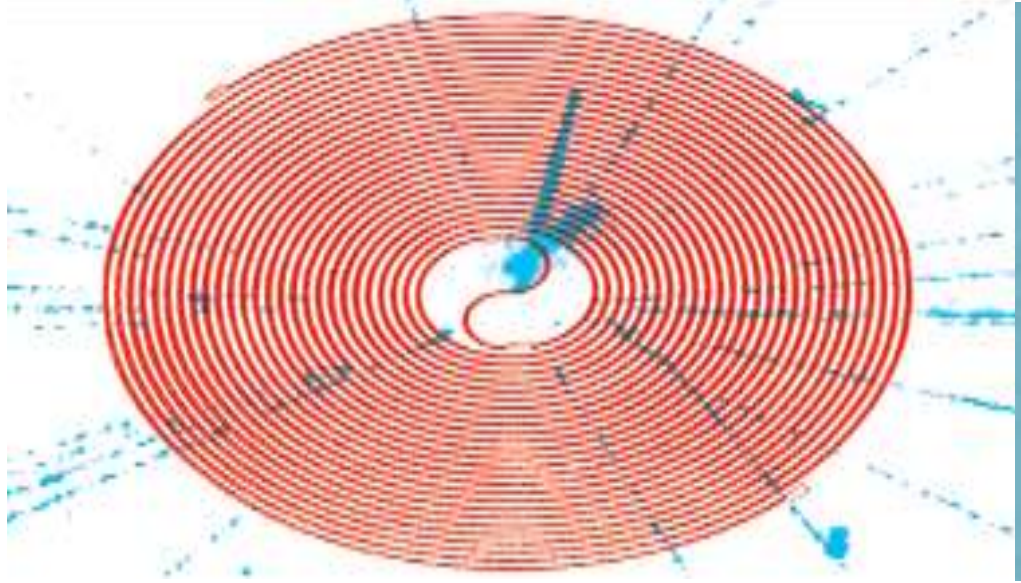
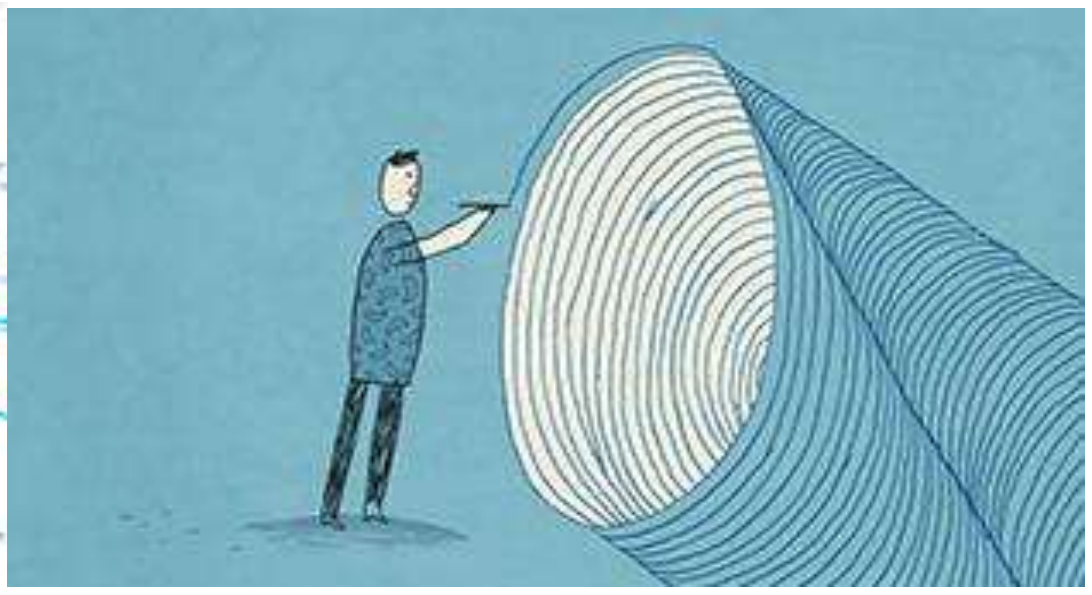

In this image, the first represents a plane projection of Fig.3, and the second replicates Fig.10.
The role of Thermodynamics and Information in brain activity and cognition is discussed in [83].
The idea of *predictive Information* (as the mutual Information between the past and the future of a time series) unifies connections between learning and complexity [84].
According to [85] “Intelligence measures general cognitive functioning capturing a wide variety of different cognitive functions. It has been hypothesized that the brain works to minimize the resources allocated toward higher cognitive functioning, and working memory performance is associated with the excitation–inhibition balance in the brain.”
"If consciousness is inherent to all mentation [86], it may be fundamental in nature..." That requires a physical account, a detailed description of the physics of *attention*, and self-reflection which is *interaction*.
Results [87] experimentally confirm the coherent dynamic cognition and conceptual retrieval of semantic cognition. According to experimental study [84], the brain’s energy supply regulates astrocytes which adjust blood flow, forming an anatomical bridge between the vasculature and neuronal synapses. That molecular mechanism confirms our Information model of supplying external energy for the Observer intelligence through a cognitive bridge, raising the intelligence.

## 5.4 Individuality of the Observer

The Observer's individuality includes the following specifics:

1.The particular probability field's *triad* observing a specific set of probabilistic events in the emerging Information Observer.

2.Depending on which pair of impulses on the observing random trajectory merges, appears the microprocess final specific qubits or Bit. Since the impulse position on the observing trajectory measures amount of the impulse random entropy, that entropy becomes the measure of forming information qubits or Bit. The asymmetry of the final Bit will conserve this specific.

Such Bits, starting the macroprocess, building triplets, INs, and forming the Observer cognition and intelligence, determine this Observer individuality.

3.Time of observation, measuring the quantity and density of Information in the delivered Bits.

4.Cooperation of the observing Information in a restricted number of IN-triplet nodes, and a limited number of the Observer's IN, which depends on the individual Observer's selective actions [41].

5.The selective actions defining the cooperative Information forces, which depend on the number of the IN nodes. The minimal cooperative force, forming the very first triplet, defines minimal selective Observer.

6.The individual ability for selection classifies Information Observers by levels of the IN hierarchy, timespace geometrical structures, and inner time scales whose feedback holds the *admissible* Information spectrum of observation.

The individual Observers INs determine its explicit ability of self-creation.

7.The produced specific self-generated Bit starts new generation of the personal Observers.

These specifics also classify the Observers also by level of cognition and intelligence.

8.The individual DSS code identifies the Observer's goal, which encloses the integral quality of Information events that the Observer accumulates during the observing process, starting from the probability field triad. These individual specifics, originating from both observing particular environment and evolving attributes of observation process, do not involve Observers' building mechanisms. The Information *mechanism* of building all Observers is *invariant*, which describes the invariant equations of Information dynamics following from the minimax variation principle.

## 5.5 How the intelligence code self-controls the observer's physical irreversible processes

The distributed intelligence coding actions at each hierarchical level control the needed external physical processes. The DSS encodes the triplet dynamics in an Information macrodynamic process which implements the Observer encoding logic.

The Observer requests the needed energy to implement own actions from such levels of his hierarchical structure, which enclose the requested code. The request follows the same steps that perform communication, except encoding the levels' Information macrodynamic in a related physical irreversible thermodynamic. After the request approval from the Observer cognition, the request interactive action attracts impulses with the needed external energy, bringing entropy gradient $\delta S / \delta x \cong dS / dx$ between the interactive actions' states $dx \cong \delta x$.The gradient provides an equivalent Information force $X = dS / dx$. The impulse correlation determines diffusion $b$.Acting on the diffusion, the force initiates a thermodynamic flow (speed) $I = bX$ of the needed external thermodynamic process. The thermodynamic forces and flows-speeds determine through the process Hamiltonian $H = X \times I$ the power to physically implement the requested actions.

The following observation of performance of these actions provides feedback to the Observer, self-controlling the performance.

**5.6 The observer self-controlling evolution**

1. The observing inter-action of the impulses with the field's energy, cutting the impulse entropy, develops its conversion to the emerging Bit of Information, which self-participates in evolving interactions which reduce the observable uncertainty. These interactions, holding probabilistic and then Information logic, evolve the stages and levels of the process.

2. Formal analysis of the stages and levels of the evolution, starting from multiple interacting impulses of observable random fields, show that each following level is enabled to self–generate the next level and self-form a nested timespace pyramidal evolution, a hierarchical network structure. The continued interaction delivers the new level's Information through each level's feedback with other levels along the hierarchy. Information attraction, which measures the quantity of Information requested by the IN stage, determines the evolution *potential* of this evolutionary stage. Information complexity of the evolution dynamics [88] measures the density of collective Information enfolded in the IN stage, which defines the Information value of the potential of cooperation for each stage.

3. The specific constraints imposed on each level, stage, and domain, limit their structural units.

When the Observer attempts to increase Information quality by overcoming-destroying its specific constraint, accidentally arising singularities [62] enable renovating the Observer's constraint location, bringing a new original (individual) level, or stage, and the domain quality distinctive from the evolution of Information dynamics within the constraints. Other non-cooperating singularities contribute the random field, which self-closes a current chain of the Observer's *individual* evolution. Interactive acquisition, bringing increasing quality Information, allows automatically self–overcoming some thresholds, decreasing the Observer *diversity*.

4. Acquisition Information through its interactive binding increases the tendency of growing Information quality in the evolving IN. That delivers growing Information forces, which enable overcoming a threshold and transfer to next stage of the growing quality. The evolving observation allows self-adjusting the constraints and threshold in a creative Observer.

5. The Observers, reaching a potential threshold on the timespace locality along the process trajectory, but unable to overcome it, will settle between the thresholds, and eventually disintegrate.

6. Evolution automatically selects the Observer remaining on the trajectory and eliminates others by memorizing the threshold through its encoding. Emanating from the cooperative code, the local intellectual functions control each stage of evolution. It establishes the evolution *hierarchy* of the evolving nested hierarchical structure of the IN levels, stages and domains.

The evolution *stability* depends on ability of memorizing Information of each evolving stage.

An Observer, unable to cross the threshold of the stage, stays stable within its stage.

That memorizes the *diversity of the selective and stable Observers.*

7. The evolution develops without any preexisting laws at each Observer's trajectory, which includes all its levels, stages and domains, and potential thresholds between them.

The Observer regularity rises in impulse observation from the self-created virtual up to real Observers, where each impulse is a max-min action transferred to the following through mini-max action. This arising variation principle imposes the Information form of the law, which encloses the following regularities. The process extreme trajectory, implementing that law's mathematical form, releases these regularities in the most general Information form. The Observer self-develops specific regularities in prolonging observation and self-evolution, which self-creates a law with extending regularities.

These abilities initiate the chain of virtual, logical, and Information causalities, which the extreme trajectory includes.

8. Using this code, Self-encoding Information units in the IN code-logic and the Observer's computation serve for common external and internal communications, allowing encoding different interactions in universal Information language and conduct cooperative operations both within and outside the domains and Observer. That unites the Observers.

9. The emergence of Observer time, space, and Information at multiple hierarchical levels follows the emerging evolution Information dynamics creating multiple evolving Observers with the Information mechanisms of cognition and intelligence.

*These results formalize the Observer's regularities in a comprehensive Information-physical theory, connecting the virtual quantum world with the physical classic and relativistic world.*

**5.7 Metaphysical aspects of the rising information observer**

According to [89], God initiates randomness creating an initial random process' field.

The field of random waves, interacting with an Observer, exposes discreet impulses acting as a random control (Sec.1.4). An analogy is a light wave which produces multiple quanta upon interaction (measurement). The linking Kolmogorov-Bayes probabilities' objective measures generate observing process of interacting random impulses where the Information Observer emerges.

Quantity $N_{oI}$ measures the invariant transformation built during the observation, which transforms a probable observing process to Information process of the emerging Information Observer (comparable to a human being). That shows that multi-interacting random waves finally can produce a real Information of the God-created Universe.

*Acceptance of the initial axiom means that Universe emanates interactions which are only reality.*

The question is how a natural intelligent Observer, a human being, can self-transform the God-initiated randomness in God's Information? We assume that such a transformer is a human faith.

According the Bible, Hebrews 11:1: "Now faith is the assurance of things hoped for, the conviction of things not seen." And following [90]: 'That definition of faith contains two aspects: intellectual assent and trust. Intellectual assent is recognizing the true and agreeing that it supports a person'…'Trust is actually relying on the fact that the something is true. '

Information cognitive model of intellectual Observer is recognizing Information through human Observer faith, agreeing with the Observer accepted Information, which the Observer had collected.

The Information Observer relies on the fact–Information obtained during the evolving observation path from random uncertainty to Information certainty with maximal probability approaching 1.

Since faith, arising intuitively in a person-Observer, encloses the described mathematical Information formalism, it connects faith and science, as we believe. Not each person-Observer can obtain the God sent Information. From that point of view, the cognitive movement, both virtual and real, modeling the faith, enables the human transformation of the God's created natural multi-interactions generating randomness to the human Information. The transformation starts with the virtual logical cognitive process up to cognitive Information and thermodynamics as a receptor of intellectual Information. Therefore, the describe math formalism could open a scientific path for a human Observer knowledge to God. But only some Observer enables overcoming the gap and many described restrictions.

According to contemporary astronomy, the Universe is limited by a boundary-edge or horizon.

Assuming that this boundary generates a huge amount of Information which produces the random field inside the Universe, I suppose that the generator of this Information is God located on the boundary horizon. The distributed intellectual multiple DSS code (Sec.1.8.3) might modeling some functions of the Generator.

The curved horizon topology may generate the resonance frequencies of spectrum $\{\omega_1, \omega_2, \omega_1\} = \omega_o$ holding the cognitive logic loop self-creating multiple Information in its encoding. Specifically, the first one enables creation of logic, the second generates Information Bits, and the third is capable to origination a code of intelligence in an Observer. (All these, according to ID, emerge within the Observer geometrical structure). Humans, and other possible Observers, select and accept only portions of the field of Information, following a path from uncertainty to certainty. These Observers are thus lawfully created by God and perform His Laws. Moreover, the Observers enable communicating with God through possibility of selecting Information from the random field.

Each Observer is created from God's Bits without pre-existing Law becoming Observer –Participator that selects new Bits from random field and self-evolves.

It follows that (1)-God creates all Observers inside Himself generating Information, (2)-God physically builds Himself though the Information-structured Observer, (3)-The Universe is physically built by God, (4)-Observer functions and destiny result from performing the emergent divine Information assignments, (5)-All Observers, including human beings, live inside God, continuing to build the physical world.

Thus, the Universe is God while all Observers are within God's Universe. Each Observer, evolving from God's Bits, becomes Observer–Participator that selects new Bits from random field and self-proceeds without pre-existing Law.

*The formalism of Information approach has many other philosophical, metaphysical, and sociological interpretations.*

**5.8. About information structure of artificial designed observer: the basic concepts**

The described path of emerging Information Observer and evolving the self-organizing stages reveal *functional Information structure of artificial designed Observer*, which will not only performs routine manipulations, but self-generates its brain performing both human and super human intellectual functions. Information structure *of artificial designed Observer* includes the following main functional components:

1. Probabilistic observation measured by discrete probabilistic impulses of the observing process.
2. Reduction of the process entropy under the observing probing impulse, and rising probabilistic logic.
3. Emergence time and space geometry in the impulse observing process.
4. The impulse transformation the observing process entropy to Information.
5. Integration of the process entropy and its transformation to the process Information.
6. Emergence a Bit and multiple Bits in Information micro- and macro-processes, and raising the process Information logic.
7. Self-forming triplet logical structures and their self-cooperation in an Information network' (IN) hierarchical logic.
8. Self-forming hierarchical distributed logical structure of cognition.
9. Self- cooperation of the multiple INs information structures in hierarchical domain.
10. The multilevel self-encoding the hierarchical cognitive logic in intelligence code.
11. The intelligence cooperative code enclosing the created Observer Information geometrical structure.
12. The intelligence code self-controlling the Observer evolution.
13. Understanding a meaning in each communication of the interacting intelligent Observers.

*The developed mathematical formalism allows the description of these functional regularities and the Information structure of the artificially designed Observer toward artificial brain.*

## 6. SUMMARY OF THE INFORMATION PATH OF ARISING INFORMATION OBSERVER

### 6.1.1. THE BASICS

1. Interactions are fundamental natural phenomena in the universe.
2. Each elementary interaction is an action and a reaction. It is an impulse which can be represented as Yes-No symbols modeling a binary 1-0 value, a Bit. That connects the phenomenon of Interaction with the phenomenon of Information, emerging in the impulse' observations.
3. The introduced notion of Information leads to an Information Observer evolving in interactive observations. It changes Information's relation to entropy, the origin of causality, logic, Information dynamics, micro-macroprocesses, complexity, Observer cognition, and intelligence logic, and many other essential concepts.

This study substantiates all of these and validates them analytically and numerically.

### 6.1.2. THE SPECIFICS

1. Interactions of different events (objects, particles) primarily indicate their occurrence in multiple traces revealed by observations.
2. Multiple interactions produce a manifold of random elementary 1-0 events, whose occurrence describes the axiomatic probabilities of the Kolmogorov 0-1 Law. The probabilities of these random events, emerging from the probability field, interact through random processes which are modeled as Markov chains of multiple Bits.

This field of axiomatic probability is the source of Information and physical events.

3. The probabilistic trace of multiple Bits objectively observes a formal act which changes the probabilities of a Markov chain to the transitional *a priori-a posteriori* probabilities of a Markov diffusion process, analogous to Bayesian probabilities.
4. The Bayesian probabilities of a Markov diffusion process model discrete 1-0 probabilistic impulses acting to observe the Markov chain. These observing impulses with Yes-No or No-Yes actions formally model the sequence of Kolmogorov 0-1 Law events. The field's axiomatic probabilities link the Kolmogorov's Law discrete probabilities with the Bayesian probabilities of the Markov diffusion process.
5. Particular probabilities of the field observe its specific set of events, which identify a potential Observer.
6. The Markov correlations hold the relative entropy measure of the uncertainty of random events between the impulse Yes-No probabilities, or uncertain multiple impulses of uncertain Bits.
7. The entropy contribution measures each observing impulse's interactive impact on the process being observed, which the observation changes. The impacts of observations, collected by EF-IPF integral measure, finally create the path to the Information Observer.
8. Certainty is produced by removing the entropy of the correlation or uncertainty, originating Information which emerges from a particular set of the observing probabilistic events. A specific Information Observer is created from the objective probabilities of the observations.

9 As shown in the previous chapters, this Information Observer emerges without any pre-existing physical law. •

### 6.1.3. SUMMARY OF THE HIERARCHICAL EVOLVING LEVELS IN THE PATH OF IMPULSE OBSERVATIONS

- The objective Yes-No probabilities measure the virtual probing impulses. Processing the interactions, they generate an idealized (virtual) probability measurement, from a finite uncertainty in the observable Markov process to the observing Bayes probability of the potential (virtual) observer, up to the certainty of the real Information Observer.

- The impulse of the interactive No-action cuts the maximum entropy, while its Yes-action transfers a minimum cut to the next impulse, thus creating multiple impulses on the maxmin-minimax principle, decreasing the uncertainty of the observing process.
- The reduced relational entropy along the trajectory of the observing process conveys the Bayes *a priori-a posteriori* probabilistic causality of the impulse. Correlation of the impulse temporarily memorizes the sequential probe's logic of the probabilistic causality.
- The correlations hold the hidden inner connections of the impulse's entropy which integrates the Entropy Functional (EF) along the observing process. The EF also integrates the time interval of the correlation connections along the observing process. The correlations connect events in random time course. In addition, this allows the integration of the probabilistic logical causality.
- The connection of the cutting impulses decreases the potential number of multiple virtual Observers, indicating a threshold which limits the number of observers not overcoming the threshold.
- As the Bayes *a posteriori* probability grows, neighboring impulses may merge, generating an interactive jump on each impulse border. A pair of random interactive actions on the bordering impulses becomes equally probable. The merge converges a causing action with a subsequent reaction, superimposing the cause and effect.

The emerging microprocess within the bordered impulse runs the superposition and the entanglement of conjugated entropy fractions. The fractions entangle during the time interval within the space is formed.

Since a beginning of the entanglement has no space measure, the entangled states can be everywhere in a space. That indicates non-locality of the entangle states. The microprocess ends as the entropy volume is entangled.

- The interaction curves the interacting impulse, which creates the inner impulse rotating movement along the observing process.
- The microprocess connects the entangled entropy with a Bit's formation through an entropy-Information gap. The gap holds a hidden real locality which the rotating potential momentum, growing with the increased entropy volume transition over the gap, can overcome. The real local gap reveals a physical Markov diffusion whose entropy erases an external energy impulse. The momentum acquires physical property near the gap end when the momentum curves a physical cut of the transferred entropy volume. The cutting bits conserve the causal logic in Information logic.
- Emerging during the interaction, energy kills the entropy volume within the gap, memorizing a logical Bit or two qubits. The $1Nat = 1.44bit$ of each impulse contains the $1bit \cong 0.7Nat$ and the Free Information of the cutting correlation $0.123bit$, enabling the attracting actions. The difference $1.44 - 1.23 = 0.21$ Bit with $0.21 \times 1.44 \cong 0.3Nat$ is transferred to the next interacting impulse as its entropy equivalent.
- The opposite curvature, enclosing the entropy of the interacting impulses, lowers the potential energy that converts entropy into a Bit of the interacting process.
- In a multi-dimensional observing process, the multiple cuts reveal multiple Bit units which the Hidden Information attraction binds in the collective dynamic movement of the Information macrodynamic process.

The macroprocess integrates both the entropy between impulses, the microprocesses, and the cutoff Information of real impulses, which sequentially convert the collected entropy in an Information physical process during the macro movement.

- Multiple interacting Bits self-organize the Information process in an Information structure, encoding Information causality, probabilistic logic, and complexity.
- The trajectory of the observing process carries the wave function (both probabilistic and certain), self-building the Information structure hierarchy.

- The Information Path Functional (IPF) integrates the Information process, enclosing the cutting correlations of the EF, creating the Bits which connect to the IPF along the extreme trajectory of the observing process. The IPF condenses all the integrated bits in the trajectory's final Bit.
- The Information Macrodynamics (IMD) are reversible within each EF-IPF extreme segment, whereas irreversibility rises at each border between the segments which encloses the memorized Information. The borders impose a dynamic constraint on the Hamiltonian of the irreversible IMD. The IMD Lagrangian integrates both the impulse's and constraint's Information on timespace intervals.
- The EF-IPF integrate the timespace intervals of the invariant impulses in an Information Geometry
- A flow of the moving cutoff Bits forms a unit of the Information macroprocess (UP), whose size limits the unit's starting maximal and ending minimal Information speeds, attracting a new UP through its Free Information.

Selected automatically during the minimax attracting macro movement, each UP joins two cutoff Bits with a third Bit, delivering Information for next cutting Bit.

- A minimum of three self-connected Bits assembles the optimal UP-basic triplet, whose Free Information requests and binds a new UP triplet that joins three in a knot that accumulates and memorizes the triplet's Information in the trajectory segments.
- During macro-movement, multiple UP triples adjoin the timespace hierarchical network (IN) whose Free Information's request produces new UP at higher level's knot-node and encodes it in triple code logic. Each UP has a unique position in the IN hierarchy, which defines the exact location of each code's logical structures. The IN node hierarchical level classifies the quality of the assembled Information, while the currently ending IN node integrates the Information enfolding all IN levels.
- Each Yes-No action, transformed in the UP's Bit logic and Information impulse, differentiates the impulse Information density and quality which identify the UP location on the IN hierarchical level.
- New Information for the IN delivers the requested node Information's interactive impulse impact on the needed external Information. Cutoff entropy of the observation converts to Information. The resulting new quality of Information concurrently builds the IN temporary hierarchy, whose high level enfolds the Information logic that requests new Information for the running observer's IN, extending the logic up to the IN code.
- The emergence of the current IN level indicates the Observer's Information surprise, measured through the IN feedback's interaction with both external observations and the internal IN's Information, delivering new self-renovating Information quality.
- The growing IPF Information, condensed in the integrated Bit with a finite impulse geometrical size, strengthens the Bit Information density, running up to finite maximal Information at infinite process dimension.
- The timespace Information geometry, emerging in observations, connected with the macro-movement in rotating timespace coordinate systems, shapes the Observer asymmetrical structure by confining its multiple INs. The time scale of the accumulation of Information determines the Observer's time of inner communications.
- Each Observer owns the time of inner communication, depending on the requested Information, time scale, and density of the accumulated Information.
- The Observer optimal multiple choices, evaluated through the minimax self-directed strategy, implement the cooperative forces emanating from the INs integrated nodes.
- The current Information cooperative force, initiated by Free Information, measures the Observer's selective actions, attracting new high-quality Information. Such quality delivers a high density-frequency of related observing Information through the selective mechanism. These actions engage acceleration of the Observer's Information processing, coordinated with the new selection, quick memorizing and encoding each node Information with its logic and space-time structure. All these implement the minimax strategy which minimizes spending Information and IN cooperative complexity.

- The self-built Information structure, under self-synchronized feedback, drives self-organization of the IN and evolution of the macrodynamics through its self-creation.
- The macro units logically self-organize Information network INs, encoding the units in the geometrical structures enclosing the triplet code.
- Multiple INs bind their ending triplets, enclosing Observer Information, cognition, and intelligence. The Observer cognition assembles common units through multiple attractions in resonances loops at the forming IN triplet hierarchy.
- The cognitive logic self-controls the process encoding the intelligence in a double helix coding structure (DSS). The clock time intervals open access to the external energy at each specific level of the IN multiple hierarchy, enabling the memorization and encoding of the hierarchy of these Bits.
- The maximal number of accepted triplet levels in multiple INs measures the Observer's maximum comparative Information intelligence. The intelligent Observer recognizes and encodes these digital images in message transmission.
- The intelligent Observers connect the Information transmission and communications. Such an Observer, being self-reflective through the DSS invariant helix code, enables reading and understanding the message meaning.
- Understanding implies that the Observer can classify and select such Information according to this Observer's memorized meaning, among other comparative images.
- The multiple code memorizes the IN assembled logical structure in the Observer's cooperative code. Since such code holds the energy of cognitive thermodynamics, it physically organizes the multiple INs with their local codes in the coding Information structure of Information Observer.
- The Double Spiral Structure triplet code, self-organizing all multiple local codes along the hierarchy, *encodes Observer distributed Intelligence*, which automatically includes the cognition integrating the observing processes.

The *Observer Intelligence* includes an ability to uncover causal relationships enclosed in evolving Observer networks, and self-extending the growing quality Information and the cognitive logic upon building the collective Observer intellect. The IN highest level ending node Information measures the Observer Intelligence.

- The intelligence of different Observers integrates the Information of their IN's node codes, which enclose a knowledge of the observations in the communicating observer's IN levels that enhance integrated knowledge.
- -Growing with its time interval, the Intelligence increases the Observer's lifespan.
- Observation processes with entropy-Information and micro-macroprocesses are Observer-dependent. The Information of each particular Observer is distinct. Each specific probability field triad generates an Information process creating its Observer.
- The invariant Information minimax law leads to common Information regularities for different Observers.

By observing even the same process, each Observer gets Information needed by its current IN during its optimal time-space Information dynamics. This creates specific (individual) Information processes of Individual Observer.

The constrained level identifies multiple individual Observers, each of which stops evolving.

- Integrating the process entropy in the Entropy Functional and its Bits in the Information Path Integral measures formalize the variation problem in the minimax law, determining all regularities of the processes. Solving the problem mathematically describes the micro-macro processes, the IN, and invariant conditions of Observer's self-organization and self-replication. These self-create law of evolving the multilevels processes and the Observer.
- These functional regularities create a united Information mechanism whose integral logic self-operates, transforming interacting uncertainties into physical reality (matter, human Information).

- The introduced path of interactions connects uncertainty of random process to certainty of observer information process which formalizes physical processes interacting with energies different qualities. The path integrates the multiple superimposing processes in the Observer.
- The united Information mechanism analytically synthesizes the AI enables modeling a brain processing.
- Both Information and Information processes emerge as phenomena of natural interactions.
- The Information equations developed here analytically finalize the main results, validate them numerically, and present Information models of many interactive physical processes.

**6.1.4. HOW THE OBSERVING PROBABILITY FIELD, CONSERVING ENERGY, CREATES PHYSICAL UNITS WITH THE CONDENSED QUALITIES ENERGY AND INFORMATION**

- The observing probability connects the uncertainty of random interactions to the certainty of the Information process. The connection includes physical processes interacting with energies of different qualities. The quality of energy is evaluated by the level of its order (disorder) or symmetry (asymmetry). This level measures the minimal entropy, ln2, which is equivalent to a Bit. The minimal entropy classifies the quality of energy (from the high-quality light energy to the low-quality energy of heat dissipation).
- The impulse interacting actions curve the impulse geometry whose curvature creates asymmetry of the impulses. Such interaction logically erases each previously rotating entangled entropy units of the entropy volume. Each process's high–quality energy compensates for entropy of lesser quality. That removes the causal entropy with symmetrical reversible logic, created by Bayes *a priori–a posteriori* probabilities, bringing asymmetrical Information logic equivalent to logical Bit. Such a Bit is naturally extracted or erased at a minimal cost of the Quality Energy through topological transitivity in a phase transition and compression.That involves a transitional impulse inside the virtual impulse, logically memorizing the entangled units by making their mirror copy. The asymmetry created qubits is encoded in a memorized Bit. Such operations perform the function of a logical Maxwell Demon. Asymmetrical logic covers physical Markov diffusion whose energy erases entropy impulse memorizing logical bit.
- Transferring entropy through interaction unifies physical and nonphysical processes. Such interactions naturally observe the probability's equivalent of entropy, transformed to Information. The Information creates the Information Observer.
- The energies of different qualities and quantities interact through the entropy-Information gap. By overcoming the gap, an Information Bit is produced. Such Bit measures the Information of a physical unit.
- The minimal energy ln2 creates the curvature of the Bit's geometry. The curved geometry enables binding. That creates the Bit's Free Information, enabling Information attraction and binding Information units. Free Information is a discrete Information form of a free energy (Gibbs-Landau).
- Each Bit binds and composes different units of Information with energy of high quality. As more Bits are composed, the quantity of this quality in the composite unit grows.
- An elementary triple encloses the minimal quantity of that quality. The enclosed triple quality binds the equal triple quantity.
- Since the IPF extremal shortens the time interval of each subsequent composed unit, the density of the enclosed quantities and quality increases. Each invariant external impulse brings energy ln2 during the time needed for both erasure of the reversible logical bit and memorization of the Information logic Bit. Each logical Bit is memorized by delivering Landauer's minimal energy during that time. Composing an Information triplet, the triple logical Bit unit memorizes and then encodes a knot of the forming Information Network's (IN) node. The *quality* of the IN measures the *number* of nodes in the IN. By enclosing all previous node Information, each hierarchical level of the IN is determined. Since each knot of this level measures an equal quality of the bound Information ln2, the *knot's Quality Energy and Information coincide* all along the IN hierarchy.

- Sec. 5.2.5 proves that the density of each impulse encloses an *equal measure of Quality Energy and Information.*
- The quantity of Quality Energy and Information identifies the anatomy of Information units: from qubits, to Bits, Free information, triplets, Information Networks (IN), and a final triplet which binds multiple INs. Physical units arise, ranging from the elementary structure of particles to various macro units: molecules, electro-chemical forms, cells, biological organisms, and humans. Each unit, bound by an invariant triple structure, preserves an invariant Information measure.
- Interacting with other triplets, a triplet of bound Bits is connected in a macroprocess. The physics of the Information macroprocess describes the irreversible thermodynamics of interacting particles. With the same measure of quality, but a growing amount of quantity, entropy increases, measuring the irreversibility of the macroprocess. The entropy difference ln2, classifying the disorder between the process impulses, have spent on forming the related units, is preserving along the macroprocess.
- As the composite Information units grow, entropy increases. The minimum of process quality is complete dissipation. To continue binding the composite Information units, the number of the interacting thermodynamic processes has to increase.
- As the number of IN nodes grows the quality of the enclosing physical process energy decreases. Each node number identifies the quality of a particular process having such energy quality. The growing hierarchy of binding triple structures requires a multi-dimensional structure of the physical macroprocess. In the limit, a maximal density of high–quality and high–quantity energy requires an infinite high-dimensional process.
- The physical structure's fundamental constant of bound Bits imposes an *Information* connection on the time and space.This constant identifies a *bridge between micro–and macroprocesses* emerging along the observing impulse interactions as they progress from maximal uncertainty to Information certainty. The connection concurrently forms a spatial structure of Information units during the time-space observation.
- The triplets which enclose composite units build an Information Network (IN). Each IN knot-node enables the memorization of the bound triplets. Such a bound memory causes an Information mass which holds the bound bits together. A physical Bit, memorized on the entropy-Information gap, binds other Bits in a physical macroprocess, being a source of physical mass. The evaluated physical mass measures the volume of a physical triplet and its Information invariant of Free Information.
- The IN builds the hierarchical path of interacting energy qualities by binding the growing levels of knot-nodes into a chain. The IN of the memorized knots encodes Information in a physical code.
- Multiple physical triple units of the macroprocess (UP) adjoin the IN hierarchical structure of growing nodes. Free Information produces new UP at a higher level node and encodes the triple spiral code logic (DSS). The unique position of each UP in the IN hierarchy defines the location of each code's logical structure.

- The hierarchical levels of the IN nodes classify the quality of assembled Information and energies. The ending IN node enfolds all IN levels.
- Each specific level of the IN hierarchy generates the specific clock time intervals at which access to the next quality measure of external energy is opened. This enables the memorization and encoding of the logical Bit hierarchy. The encoding logic encloses cognitive Information. The energy quantity (power) and quality of specific interaction limits the DSS code length through its final Bit's Information density. The total length of Information code limits the finite maximal dimension of the high–quality external energy which is delivered.
- Multiple INs enclose Observer Information, cognition and intelligence. By being self-reflective to its DSS, the intelligent Observer can read and understand the meaning of the message.
- Thus, the probability field of observing impulses enables the generation of various Information–physical units. These units satisfy the emergent Information minimax law, which dictates the allowable combinations of the invariant units being composed. For each allowable unit's combination, the fundamental constant and the emerging constraints provide the values of specific properties. The energy quality, evaluated by the energy entropy measure, limits the initial process observing probability and its entropy. That also limits the code length when the observation starts.

### 6.2. Analytical and Numerical Attributes Distinguishing Main Stages of the Evolutionary Regularities, Their Thresholds and Constraints

1. Starting virtual observation with minimal probability and maximal uncertainty identifies the following primary threshold. Minimal increasing probability approximates formula $\Delta p_N \to 2^{-N}$, where $N$ is the number of impulses starting the virtual observation (under Plank's physical uncertainty).

2. At a given accuracy $\varepsilon_k \in (0,1), i = 1,2,....n$, the number of impulses $m_o$ within each of $n$ process' dimensions estimates

$(1-\varepsilon_k)^3 / \varepsilon_k{}^2 = 1/2m_o S_{ki}$, (2.1)

where $N = n \times m_o$ measures the total number of each $m_o$ entropy $S_{ki}$ increment. . Minimal realistic accuracy $\varepsilon_k = 4.5\times10^{-4}$ estimates $m_o = 8800$ with relative probability increment $\Delta p_k \cong 4.5\times10^{-4}\exp(-1) \approx 1.65\times10^{-4}$, where $S_{ki} = 1/2\sqrt{1-\varepsilon_k}$ estimates $S_{ki}$, and $\Delta p_k$ measures the ratio of Bayesian *a priori* probability $P_{ao}$ to *a posteriori* $P_{po}$ starting with frequency $f_o = 8800Hz$.

3. The entropy of error $S_k = 2(1-\varepsilon_k)^3 / N\varepsilon_k{}^2$ at $\varepsilon_{kN} = 1/2^N$ and $N \to \infty$ leads to $S_{kN} = 2^{N+1}/N$, which estimates the *potential start of observation with a posteriori probability*

$P_{poo} = P_{po} / m_o \cong 0.977\times10^{-4}$. (2.1a)

4. If increasing correlation brings an impulse with entropy $S_{ki} = 0.5$, such an impulse temporarily holds the probabilities difference (closeness) consistent with accuracy $\varepsilon_{ko}$ of the starting correlation and minimal *a posteriori* probability $P_{poo} < P_{ao}$. The recursive action, overcoming a threshold of a maximal uncertainty with *minimal a priori* probability $P_{aoo} < P_{poo}$, automatically starts a virtual observation that connects the probing impulses in a potential virtual test.

5. Observing a random process under a Markov process's Bayesian probabilities reduces the difference (distance) between a random event $\xi_m$ and $\xi_n$ measured by $|\xi_m - \xi_n|$.

That may start and increase each posterior correlation, reducing conditional entropy measures at the following conditions beginning with the correlation and temporal memory.

According to [1:90], a coefficient correlation between above random events: $\mathrm{r}_{mn} \le c(|\xi_m - \xi_n|)$ reaches the required stability at the sufficient condition

$$\lim_n n^{-2} \sum_{k=o}^{n-1} c(k) \times \sum_i^n D\xi_i = 0, c(k) > 0, \quad (2.2)$$

where

$$D\xi_i = E(\xi_i - E\xi_i)^2 = E\xi_i^2 - (E\xi_i)^2 \quad (2.2a)$$

determines dispersion of random $\xi_i$.

The correlation starts at the satisfaction condition (2.2a).

The relation of *a priori* and *a posteriori* probabilities ((1.53), Sec.1.5) for the current random events along the observed trajectory evaluates the direct connection with the correlation.

The existence of correlation between random $\xi_m, \xi_n$ establishes coefficient correlation $\mathrm{r}_{mn}$ which defines formulas for the mathematical expectation of random events related to dispersions [1:87].

Then $\mathrm{r}_{mn}$ establishes the ratio of observing time intervals:

$$\mathrm{r}_{mn} = \sqrt{\frac{t_m}{\mathrm{t}_n}}, t_m = (t - t_o)(t_1 - u), \mathrm{t}_n = (u - t_o)(t_1 - t), t_o < t < u < t_1 \quad (2.2b)$$

where $\mathrm{t}_m, \mathrm{t}_n$ are fixed random moments of $\xi_m = \xi_m(t_m), \xi_n = \xi_n(t_n)$ within observing moments $t_o < t < u < t_1$ [18: 32].

From that, it follows

$$c(n) = \sqrt{\frac{t_m}{\mathrm{t}_n}} / (|\xi_m - \xi_n|) > 0 \quad (2.2c)$$

which determines a threshold of starting correlation and observation time $\mathrm{t}_n$.

The starting correlation becomes stable if, at any initial $D\xi_i \ne 0$ (in 2.2a) and restricted (2.2b), it is found such $n$ when condition (2.2) is satisfied.

Stable correlations keep temporal memory.

It is initially assumed that existence of the trajectories of the stochastic process satisfy the limitation [2:44]:

$$\lim_{c \to \infty} \overline{\lim_{t^o \to t}} P\{|\xi(t^o) - \xi(t)|\} > c\sqrt{(t^o - t)} = 0 \quad (2.2d)$$

determines by this probability measure. At conformity of both differences in (2.2d), the first difference is called a Laplace variable [18:22] for which (2.2b) is satisfied. For these variables, the coefficients drift and diffusion in a stochastic process determine the following relations [18:28]:

$$a(t) = \frac{(t_1 - t)\xi_o - (t - t_o)\xi_1}{t_1 - t_o}, \sigma^2(t) = \frac{(t_1 - t) - (t - t_o)}{t_1 - t_o} \quad . \quad (2.2e)$$

*Condition (2.2c) determines starting the virtual observation where interacting impulses begin correlation, which stabilizes condition (2.2) for a stochastic process satisfying* (2.2*d*).

The Markov drift and diffusion connects additive functional (Sec. 1.4), which links to the process correlation matrixes $r_t$:

$$E[a^u(t,\tilde{x}_t)^T (2b(t,\tilde{x}_t))^{-1} a^u(t,\tilde{x}_t)] = 1/2 r_t^{-1} \dot{r}_t \quad . \quad (2.2f)$$

6. Each elementary interaction with opposite actions ↓↑ models Dirac's delta-function, whose impulse's *interactive cut* originates from the step-down and step-up interactive actions within the impulse.
The impulse discrete function, switching the entropy from its minimum to the cutting maximum, and then back from the maximum to the next minimum, provides the maxmin–minimax principle.
The minimax variation principle establishes the *invariance of the impulse entropy measure through the observing process.*
7. Cutting the Markov diffusion process (Sec. 3.4) determines the minimal entropy of step-down interactive action ¼ Nat, the minimal increment between the interactive impulse 1/2 Nats, and the step-up action's entropy ¼ Nat.
An interactive impulse ↓↑ with both step-down and step-up virtual interactive actions carries the entropy 1Nat through the multi-dimensional observing process. For ¼ Nat, as the threshold of minimal entropy increments $S_{ki1} = 1/4$ for a dimension $n = 1$, a minimal increase dimension to $n = 2$ brings minimal increments of the interactive impulse $S_{ki2} = 1/2$ Nat. Correlation within each impulse holds the related time interval $r_{im} = c\sqrt{\tau_{im}}$, which for `each common 1Nat unifies the impulse probability 0 or 1, the time interval, and entropy measures:

$$M_p \to M_{im} = [1]_{\tau_{im}} \to [1]_{Nat}. \quad (2.3)$$

For an impulse with minimal interactive entropy 1/4 Nat, its size square measure time interval $1/2o(\tau_k)$ of that entropy:

$$M_{\tau_k} = [1/2o(\tau_k)]^2 = 1/4o(\tau_k)^2 \ . \quad (2.3a)$$

The impulse, preserving measure (2.3), extends its initial time unit $1/2o(\tau_k)$ to $o(\tau_k) = 2$ for reaching measure

$$M_p = [1/2 \times 2] = [1] \to [1]_{Nat}. \quad (2.3b)$$

The step-down action cuts the correlation which holds the entropy hidden in the cutoff correlation.
If the impulse preserves the invariant maxmin entropy measure, then the impulse's equivalent time and space intervals are connected through imaginary time directly.
That follows from correlation $r_{tj} \to \pm\sqrt{\delta_{ij}}, \delta_{ij} = (t_i - t_j) > 0$ which for an inverse time interval $\delta_{ij} = -\delta_{ji}$ are imaginary. This occurs inside the cutting impulse with the emerging space interval.
8. The opposite Yes-No probability events reveal its hidden correlation, whose posterior correlations automatically increase under Bayesian probabilities.
Assuming each probability 0 or 1 is *a priori* or *a posteriori* accordingly for a virtual impulse, from relation (1.5.13) it follows that each impulse posterior correlation $r_{im}$ increases relatively to the impulse starting auto-correlation $r_{io}$ in ratio

$$r_{im} / r_{io} = 4 \ . \quad (2.4)$$

Such self-growing correlation indicates the *emergence of an elementary virtual Observer, with measure (2.3b) and self-cutting the observing correlations* (originating from the step-down and step-up interactive actions within the impulse).
If an impulse delivers minimal entropy $S_i = 1/2$ to the following impulses, upon reaching this threshold, a self–observing process starts. Its posteriori action virtually coveys the next impulse cutting action, enabling the process to continue through self-support.
This virtual Observer rises as a part of the observing random process with interactive impulses.
9. Growing correlations intensity of entropy per the interval (as entropy density) that increases on each following interval, indicate a shift between the virtual actions, a displacement. The displacement identifies an entropy gap between the invariant impulses. Displacement $a$, starting under physical uncertainty inside of sub-Plank region [91], measures the proportion of the Plank constant to number $N$ : $a = h^o / 2\pi N = \hbar / N$ of the impulse reaching $a$. This ratio evaluates the relative closeness of the displacement to the uncertainty needed to reach the standard Plank edge. The minimal relative displacement evaluates ratio

$a^* / a = 1.000262774 N / N_*$ at $N / N_* = 1$. (2.5)

The relative displacement's distance from its minimal value (2.5) evaluates ratio

$d_a = 1 - 1.000262774 N / N_*$. (2.5a)

Relation (2.5a) measures maximal distance of minimal displacement (2.5) from the Plank edge. The maximal distance estimates the interactive impulse with space measure, which begins forming a minimal volume at that displacement.

The interactive impulse' momentum rotates a shift between the displaced states.

The opposite actions create the shift starting with a finite entropy of the displacement gap.

An extreme entropy for multiple impulses identifies the minimal difference between the opposite actions measured by time shift $\delta_k^{\tau+} / 4$ which evaluates the finite impulse width (before starting the space interval).

The ratio of entropy of the impulse step-down action's width part to entropy 0.25 Nat of that impulse step-down action evaluates the relative width

$\upsilon_o = (0.025 / 0.25) = 0.1$ . (2.5b)

The minimal displacement distance between the invariant impulses, equal to $d_a = 0.1$, can be reached using (2.5b) under a ratio of the numbers of observing impulses:

$N_* / N = 1.111403$.

To reach minimal displacement (2.5b) initial $N = m_o = 8800$, starting the observation, needs to increase up to $N_* \cong 9780$.

The entropy gradient, curving displacement (2.5b), measures the growing entropy force.

Under the growing entropy gradient, the curving displacement estimates its starting radius

$r_{e1} = \sqrt{1 + (0.025 / 0.25)^2} = \mp 1.0049875$ . (2.5c)

That radius defines the verge of the threshold. The curving rotation starts by overcoming it.

Rising virtual Euclid's curvature $K_{e1} = (r_{e1})^{-1}$ estimates this threshold:

$K_{e1} \cong +0.995037$ . (2.5d)

Starting step-down curvature's radius initiates the emerging rotation movement of the impulse, whose trajectory (Fig. 3) follows from the minimax variation principle.

Radius (2.5c) determines the initial angle $\beta$ of the rotation trajectory of cone, Figure 3, from relation

$r_{e1} = \rho = b \sin(\varphi \sin \beta)$ at $\varphi = \pi k / 2, \mathrm{k} = 1, b = 1/4$. (2.5e)

10. The impulse step-up action displaces the time measured virtual impulse's interval through rotation on angle $\varphi = \pi / 2$ . The displacement within the impulse leads to the discrete time-space form of the impulse *preserving* its measure (2.3b) in the emerging time-space coordinate system.

Comments 6.1. Let the rotation start on a spherical surface at conditional probability distribution probabilities for a distance with latitude $\theta : -\pi \le \theta \le \pi$ at given longitude $\psi$ having form [1:75]:

$$P(\theta_1 \le \theta \le \theta_2 \mid \psi) = 1/4 \int_{\theta_1}^{\theta_2} |\cos \theta| \, d\theta \cdot$$

Then this conditional probability distance is irregular.•

That indicates changing an impulse's time interval unit with the appearance of curved impulses, which is extending while curving. With growing probability, the intensity of the entropy force draws together the impulse action and reaction, squeezing the time interval between these actions up to the jump when these actions merge and start the microprocess. The invariant measure is conserved in following time-space movement.

Preserving the impulse $\bar{u}_k$ measure $|M_{io}|=|1|_M\,[\tau]\times[l]$ at $h=2, p=1/2$, $M[\bar{u}_k]=|2\times1/2|\xrightarrow{p[\bar{u}_k]}|1|_M$ leads to space $[l]$ and time $[\tau]$ invariant measures:

$$[l]=\pm[(|M_{io}|/|1|_M)(2/\pi)]^{1/2},[\tau]=\mp[(|M_{io}|/|1|_M)(\pi/2)]^{1/2}\,, \qquad (2.6)$$

and to $|M_{io}|=M[\bar{u}_k]\pi/2\times[l]^2$, which at $p=1/2h, M[\bar{u}_k]=1/2h^2, 1/2h^2[l]^2\pi/2=|M_{io}|$ and $h[l]=2$ holds impulse invariant measure

$$|M_{io}|=\pi. \qquad (2.6a)$$

That impulse's time-space irrational measure preserves the impulse entropy measure, when the virtual Observer is cutting the correlations in the curving rotation.

*Condition (2.6) determines the emerging space-time impulse with measure (2.6a) after overcoming a threshold (2.5c), which defines the starting rotation with* $\varphi=\pi/2$ . The rotating coordinate system of the curved impulse starts angular velocity *c* measured by the rate of changing the angular displacement.

In the rotating space-time an impulse appears, the starting virtual observer's geometrical shape with volume $V_c=2\pi c^3/3(k\pi)^2 tg\psi^o$ [4] determines the initial space angular velocity *c*, the cone geometrical parameter *k*, and the angle at each cone vertex $\psi^o$ (Figure 8).

11. The displacement shift's parameters define the following relations.

The Information analog of Plank constant $\hat{h}$ (at maximal frequency of energy spectrum of Information wave in its absolute temperature) evaluates the maximal Information speed of the observing process:

$$c_{mi}=\hat{h}^{-1}\cong(0.536\times10^{-15})^{-1}Nat/\sec\cong1.86567\times10^{15}Nat/\sec. \qquad (2.7)$$

That value also estimates a minimal time interval corresponding the time shift:

$$\delta t_e\cong1.59459\times10^{-14}\sec\approx1.6\times10^{-14}\sec. \qquad (2.7a)$$

Time shift at maximal light speed $c_o=3\times10^9 m/\sec$ allows the estimate a minimal space shift:

$$\delta_{lo}\approx4.8\times10^{-5}m. \qquad (2.7b)$$

The angular velocity, emerging with maximal linear speed $c_o$, curves length $\delta_{lo}$ to the length

$$\delta_{low}=\pi\delta_{lo}[m],\ \delta_{low}=15\times10^{-5}m\ . \qquad (2.8)$$

Ratio $c_o/\delta_{low}\cong w_o$ approximates a maximal angular velocity for the curved length $\delta_{low}$:

$$w_o\approx0.1989\times10^{14}\sec^{-1}. \qquad (2.8a)$$

Maximal entropy speed can rotate the entropy increments on the starting displacement $\Delta s_{apo}=-\ln(0.8437)\cong0.117Nat$ with maximal entropy angular velocity

$$w_{oe}=0.73\times10^{15}Nat/\sec\ . \qquad (2.8b)$$

12. The microprocess emerges inside a random process, modeling by Markov diffusion process, when the displacement verges at distance (2.5b) reaches the minimal me interval (2.7a) upon merging the nearest impulse's opposite actions.

The optiposite actions $u_-^t$ and $u_+^t$ are fixed variables of the Markov diffusion process, which preserves both their additive and multiplicative functions (Sec. 2.1).

It requires fulfillment

$$u_+^t-u_-^t=u_+^t\times u_-^t \qquad (2.9)$$

which leads to

$u^t_+ / u^t_- = 2$ (2.9a)

if both actions are real. And to functions

$u_{+*} = (j-1), u_{-*} = (j+1)$ (2.9b)

when both actions are complex conjugated. At equal absolute values of actions $|u^t_+| = |u^t_-|$, imaginary functions

$u^t_{+*} = j\sqrt{2}, u^t_{-*} = -j\sqrt{2}$ (2.9c)

satisfy only multiplicative part of (2.9) while the impulse additive measure holds $\mathrm{U}_a = 0$.

Functions (2.9b) fulfill both additive and multiplicative measures equal to $\mathrm{U}_a = U_m = -2$.

When the sub-Markov process gets negative entropy measure $S^*_{\mp a} = -2$ of the impulse actions with relative probability $p_{a\pm} = \exp(-2) = 0.1353$, it starts opposite imaginary actions (2.9b) or (2.9c), initiating the microprocess.

Within the impulse time interval $\tau = 1Nat$, *entanglement starts before its space is formed and ends with the beginning of the space during the reversible relative time interval of* $0.015625\pi$ part of the impulse invariant measure $\pi$.

*Since entanglement has no space measure, the entangled states can be everywhere in a space.*

13. The space interval, beginning the displacement shift, starts within interval of entanglement having the probability $P_{po*} \cong 0.8231$, continues during the shift, and extends to the space part of the impulse multiplicative measure after the displacement ends. That means the displacement widens, extending its ending probability up to the impulse's inner part, where it ends with probability $P^i_n = 0.86$, holding entropy $S_\pm = 0.15$. The end of displacement indicates the formation of a space interval within that impulse. Or *a priori* $i$-probability $P^i_n = 0.86$ is the indicator of the appearance of the first impulse space interval ($n$-from starting observation).

If this impulse's positive curvature interacts with the next impulse's negative curvature, then the interacting part holds the transitional curvature' entropy sum $S_\Delta = 0.5085$ (Sec. 2.6). The difference $S_\pm - S_\Delta \cong 0.01$ estimates the increment of both impulse asymmetries which concurs with estimation (Sec. 2.6). This means the opposite asymmetries of interacting impulses estimates probability $P^i_n = 0.86$. The increment of the probability, starting an external interacting impulse, and the probability of injecting energy evaluates: $\Delta P_{ie} = 0.981699525437 - 0.9855507502 = -0.1118$ holds entropy $\Delta S_{\pm a} = -2.191$. The difference $\delta S_\pm = -0.191$ determines the related increment of entropy within this impulse before the injection of Landauer's minimal energy measure $\ln 2$ within the interval of encoding information $\ln 2$ Nat.

The imaginary microprocess ends with the entangling entropy volume, the Information microprocess emerges with providing energy, killing that entropy and memorizing the classical Bit by the end of external impulse.

Probability $p^*_\pm = \exp(-2h^{o*1}_\alpha) \cong 0.9866617771$ identifies physical structural parameter $h^{o1}_\alpha$ which counts the sub-Plank spot above, resulting from the interactive impulse with this probability during the observation.

On a path from uncertainty to certainty, the increasing number of interacting impulses $N = 8800$ allows the observer closer approach to the gap of reality through decreasing uncertain displacement of the sub-Planck spots.

After entropy volume of the $N+$ impulses increase to overcome uncertain volume (2.5b), the entropy reaches the edge of certainty-reality with increasing probability $p^*_\pm$. Since a Bit is created at the probability approaching 1 with the number of each interaction $N^o_* \cong 8828$, each impulse observation can create the Bit with frequency

$F_{im} = 1/8828 = 10^{-4} \times 1.13276$. (2.10)

Moreover, because each Bit creation needs a final interaction of the impulses with opposite curvatures (Sec. 2.6), such interaction needs $N = 8800$, which evaluates the probability, and the frequency of appearance that impulse

$F_{imo} = 1/8800 = 10^{-4} \times 1.13636$. (2.10a)

Both frequencies evaluate the optimal number of impulses for a single observation.

The Information Bit, as two memorized qubits, can be produced through interaction, which generates the qubits contained by a material or device (a conductor-transmitter) that preserves the curvature of the transitional impulse (Secs.2.5.2,2.5.3) inside a closed device. Memorizing the entangled curvature is the Information "demon cost" for the entangled correlation, which naturally holds its entropy, time, and the curvature of the transitional impulse. •

### 6.3. MATH SUMMARY. INTERACTIVE INTEGRATED INFORMATION DYNAMICS (IIID)

1. Probabilities and conditional entropies of random events.

A *priori* $P^{a}_{s,x}(d\omega)$ and *a posteriori* $P^{p}_{s,x}(d\omega)$ probabilities observe the Markov diffusion process $\tilde{x}_t$ distributions of random variable $\omega$ (events).

For each $i,k$ random event $A_i, B_k$ along the observing process, each conditional *a priori* probability $P(A_i / B_k)$ follows the conditional *a posteriori* probability $P(B_k / A_{i+1})$.

Conditional Kolmogorov probability

$$P(A_i / B_k) = [P(A_i)P(B_k / A_i)] / P(B_k) \quad (3.1)$$

defines the Bayes probability after substituting average probability:

$$P(B_k) = \sum_{i=1}^{n} P(B_k / A_i)P(A_i).$$

Conditional entropy

$$S[A_i / B_k)] = E[-\ln P(A_i / B_k))] = -\ln \sum_{i,k=1}^{n} P(A_i / B_k)]P(B_k) \quad (3.1a)$$

averages the conditional Kolmogorov-Bayes probability for multiple events along the observing process.

Conditional probability satisfies Kolmogorov's 1-0 Law for function $f(x)|\xi$ of an $\xi, x$ infinite sequence of independent random variables:

$$P_\delta(f(x)|\xi) = \begin{cases} 1, & f(x)|\xi) \ge 0 \\ 0, & f(x)|\xi) < 0 \end{cases}. \quad (3.1b)$$

This probability measure has been applied for the impulse probing in observable random process, which holds opposite Yes-No probabilities-as the unit of the probability impulse step-function.

Random current conditional entropy of the finite sequence of the random events is

$$\tilde{\mathrm{S}}_{ik} = -\ln \mathrm{P}(A_i / B_k)\,\mathrm{P}(B_k). \quad (3.1c)$$

Probability density measure on the Markov process trajectories as function of events $\omega$:

$$p(\omega) = \frac{\tilde{P}_{s,x}(d\omega)}{P_{s,x}(d\omega)} = \exp\{-\varphi_s^t(\omega)\}, \quad (3.1d)$$

is connected with this process additive functional

$$\varphi_s^T = 1/2\int_s^T a^u(t,\tilde{x}_t)^T (2b(t,\tilde{x}_t))^{-1} a^u(t,\tilde{x}_t)dt + \int_s^T (\sigma(t,\tilde{x}_t))^{-1} a^u(t,\tilde{x}_t)d\xi(t) \ . \quad (3.1e)$$

This functional is defined through controllable functions drift $a^u(t,\tilde{x}_t)$ and diffusion $b(t,\tilde{x}_t) = 1/2\sigma(t,\tilde{x}_t)\sigma(t,\tilde{x}_t)^T$ of the process, where (3.1e) also describes the transformation of the Markov process's random time traversing the process trajectory.

2. The *integral measure* of the observing *process* trajectories formalizes an *Entropy Functional* (EF), which is expressed through the above functions of Markov diffusion process $\tilde{x}_t$:

$$\Delta S[\tilde{x}_t]|_s^T = 1/2E_{s,x}\{\int_s^T a^u(t,\tilde{x}_t)^T(2b(t,\tilde{x}_t))^{-1}a^u(t,\tilde{x}_t)dt\} = \int_{\tilde{x}(t)\in B} -\ln[p(\omega)]P_{s,x}(d\omega) = -E_{s,x}[\ln p(\omega)], \quad (3.2)$$

and the probability density measure on the process trajectories.

3. Cutting the EF by the impulse delta-function determines the increments of Information for each impulse:

$$\Delta I[\tilde{x}_t]|_{t=\tau_k^{-o}}^{t=\tau_k^{+o}} = \begin{cases} 0, t<\tau_k^{-o} \\ 1/4Nat, t=\tau_k^{-o} \\ 1/4Nat, t=\tau_k^{+o} \\ 1/2Nat, t=\tau_k, \tau_k^{-o}<\tau_k<\tau_k^{+o} \end{cases} \quad (3.3)$$

with total $\sum_{t=\tau_k^{-o}}^{t=\tau_k^{+o}} \Delta I[\tilde{x}_t]_{\delta t} = 1Nat$. (3.3a)

*4. The Information Path Functional* (IPF) unites the Information cutoff contributions $\Delta I[\tilde{x}_t / \varsigma_t]_{\delta_k}$ along $n$-dimensional Markov process impulses during its total time interval $(T-s)$:

$$I[\tilde{x}_t]|_s^{t\to T} = \lim_{k=n\to\infty} \sum_{k=1}^{k=n} \Delta I[\tilde{x}_t / \varsigma_t]_{\delta_k} \to S[\tilde{x}_t], \quad (3.4)$$

which in the limit approaches the EF. The IPF along the cutting time correlations on optimal (extreme) process trajectory $x_t$, in the limit, determines equation

$$I[\tilde{x}_t / \varsigma_t]_{x_t} = -1/8\int_s^T Tr[(r_s \dot{r}_t^{-1}]dt = -1/8Tr[\ln((r(\mathrm{T})/r(s))]. \quad (3.4a)$$

depending on the trace of the relative correraliions.

5. The equation of the EF for a microprocess:

$$\partial S(t^*)/\delta t^* = u_{\pm}^{t1}S(t^*), u_{\pm}^{t1} = [u_+ = \uparrow_{\tau_k^{+o}} (j-1), u_- = \downarrow_{\tau_k^{+o}} (j+1)] \quad (3.5)$$

under inverse actions of function $u_{\pm}^{t1}$, starts the impulse opposite time $t_{\pm}^* = \pm\pi/2t^i$ which measures a space rotating angle relative to the impulse inner time $t^i$.

The equation' solutions for the conjugated entropies $S_+(t_+^*)$, $S_-(t_-^*)$ determine functions

$S_+(t_+^*) = [exp(-t_+^*)(\mathrm{Cos}(t_+^*) - jSin(t_+^*))]|, S_-(t_-^*) = [exp(-t_+^*)(\mathrm{Cos}(-t_+^*) + jSin(-t_+^*))]$ at

$$\begin{aligned} S_{\pm}(t_{\pm}^*) &= 1/2S_+(t_+^*)\times S_-(t_-^*) = 1/2[\exp(-2t_+^*)(\mathrm{Cos}^2(t_+^*) + Sin^2(t_+^*) - 2Sin^2(t_+^*))] = \\ & 1/2[\exp(-2t_+^*)((+1-2(1/2 - Cos(2t_+^*))))] = 1/2\exp(-2t_+^*)Cos(2t_+^*) \end{aligned} \quad (3.5a)$$

Minimal interactive entropy $S_{\pm}(t_{\pm})$ *begins the space measure* during reversible relative time interval $0.015625\pi$ of the impulse invariant measure $\pi$. The running microprocess, overcoming the entropy-Information gap, starts Information Bit and an Observer information macrodynamics.

6. The Information macrodynamic equations:

$$\partial I/\partial x_t = X_t, a_x = \dot{x}_t = I_f, I_f = b_t X_t \quad (3.\ 6)$$

define $X_t$ -a gradient (force) of Information path functional $I$ (3.4) on macroprocess' trajectories $x_t$, $I_f$ - Information flow determined through speed $\dot{x}_t$ of the macroprocess. The flow emerges from drift $a^u(t, \tilde{x}_t)$ being averaged by function $a_x$ along the observing process, and the averaged diffusion $b_t \to b$ for the macroprocess force.

The Information Hamiltonian of the macrodynamics:

$$-\frac{\partial \tilde{S}}{\partial t} = (a^u)^T X + b\frac{\partial X}{\partial x} + 1/2a^u(2b)^{-1}a^u = -\frac{\partial S}{\partial t} = H \, . \qquad (3.7)$$

determines macro equations (3.6) from the minimax variation principle for the EF using Jacobi-Hamiltonian equations.

The impulse interactive observations sequentially join extremals of the multidimensional extreme process discretely changing Hamiltonian which accumulates the growing information.

The discrete Hamiltonian divides irreversible dynamic trajectory on the partial reversible segments, predicting the next emerging Information unit. The joint extremals form borders of the segments.

Equations (3.6) are the Information form of the equations of Irreversible Thermodynamics [60, 26], which the *Information Macrodynamics* generalize.

The flows and forces determine the macroprocess Hamiltonian in the invariant form

$H = X \times I$ .

Information curvature $K_{\alpha}^{m}$, density of Information mass $M_{vm}^{*}$, and effective complexity $MC_m^{\delta e}$ connect equation [ \]:

$$K_m^{\alpha} = M_{vm}^{*} MC_m^{\delta e}, \qquad (3.8)$$

where

$$MC_m^{\delta e} = 3\dot{H}_m^V MC_m \qquad (3.9)$$

includes the differential of Hamiltonian per volume $\dot{H}_m^V$ and the IN cooperative complexity $MC_m$.

*The single Eq. (3.8) at (3.9) encloses all previous Eqs. (3.1-3.7), unifying the formal math description of this approach.*

These new results validate analytical and computer simulations and illustrates the experimental applications [92,93].

The Equations of *Interactive Integrated Information Dynamics* (IIID) (3.2-3.9) formalize the description of information regularities of observing interactive process.

The IIID integrates the Information Path from Randomness and Uncertainty to Information, Thermodynamics, and the emerging Intelligence of Observer.

The interactive action universally arises in all Observers as receivers, transmitters, and the sources of Information which evolve in communications.

Together, such actions are unified in the concept of the Information Observer.

The IIID united information approach [99] shows how:

-information emerges in Observer from observing random process probabilities of interacting impulses, bringing a probabilistic causality of the observations;

-removing the random process' uncertainty converts it to equivalent certainty-information;

-merging action and reaction join probabilistic prior and posterior actions on edge of predictability generating microprocess;

-time of entanglement starts space interval composing two qubits or Bit of reversible logic;

-an entropy-information gap on an edge of reality arises and restriction on overcoming the gap;

-each memorized information binds reversible microprocess with irreversible information macroprocess along multi-dimensional observing process;

-assembling and moving information creates Observer's information process, with information dynamics, information network, its logic and code;

-raising the observer's structures with information regularities, differentiation, evolution, cognition, and intelligence;

-observation process integrates information bits in information process encoding the Information Observer.

Instead of designing the AI manipulators, the IIID focuses on creation an artificial brain in evolving information process of the intelligent Observer.

Since brain neurons operate with information form of different observation, images representations, they recognize the observing information process up to rising the observer cognition and intellect.

At an equal IN's level of AI and human intellect, the difference between AI and a human makes the *personalized history* of memory for each human being.

The *personalized history* structures the IN of personal memory which accumulates the observer's conscience connected with all Observer cognitive observations.

Integrated information theory (IIT) [100, 101] focuses on "system's consciousness determined by causal properties of physical system and each consciousness experience composed of multiple "*phenomenological distinctions*."

The IIID is more general and systemic approach integrating the *observing interactive path* to intelligence including a cognitive conscience.

The IIID integral follows from the observing process properties and is not axiomatic introduced as the IIT. The IIID formalizes Observer's regularities in comprehensive information-physical theory, connecting the quantum world with the physical classic and relativistic world, which the information law can predict.

Moreover, the IIID formalism with DSS reveals how artificially encodes an observer in universal information coding structure like DNA. It involves naturally creating information Bits and natural encoding triplet code enables recognizing other encoded information. These Bits becomes standard units of different information languages in the observer communications.

Natural interactive process, transferring Bits, creates information process self-organizing and encoding the observer' brain logic, cognition, and intelligence with function of consciousness.

Consciousness defines observer ability of self-recognizing where-(in which IN node or domain)-was memorized specific observed information, enables understanding and in creative intercommunication.

Natural origin of information and natural encoding information defines natural observer

That includes

1. Natural emerging subjective bits;
2. Natural creation information logic, cognition, and intelligence;
3. Self-created consciousness.

These are components of the IIID where

1. Natural correlated interactive Yes-No actions provide both Bits and Hidden information between them revealed as Free Information which the IPF integrates. Cooperative attraction, created by Free information, structures (at least 4.48 bits [99]) selects a minimal *subjective* observer-a triplet whose Free information enables adjoins other Bits in a logical structure during the cooperative attraction.

2. Equalizing speeds of the Free Information's attracting movement synchronize the frequencies of the triplet's logic in a resonance which assembles a logical loop. The multiple resonances of the logical loops assemble Information Network of hierarchical nodes of the nested triplet's logic.

Each nested coherent logical loop accepts only such units that each IN node logic recognizes, self-organizing a logical chain of *cognitive functions*, which are distributed along hierarchy of the assembling logical units: triplets, IN nested nodes, and the IN ending nodes.

The Observer cognition emerges as an evolving intentional ability to request, integrate, and predict the needed Observer Information that builds the Observer's growing networks. The hierarchy of loop logic is memorizing and then encoding after coordinated access of external energy to each local logical unit at every hierarchical level.

3.The coordination performs the Observer own time clock with its time course, which commands the hierarchical switching.

4.The Observer hierarchical codes physically organizes the multiple IN, self-encoding their local codes in the Information structure of the *Observer intelligence*.

The self-creating observer's intelligence enables decoding each IN's node memorized information recognizing it in intercommunication associated with self-awareness. That proceeds through the self-creation, which initiates some observed bit and their cooperating triplets (as threshold for the selective observer 4.48 bit.) These bits free information of an attracting logic requests for cooperation some logical information located in the specific IN node.

Thus, the observer intelligence includes function of a self-awareness modeling consciousness.

The last result confirms philosophical [102] conclusion that term "consciousness" had no constant meaning, and should be abandoned, "it does stand for a function". ...

**REFRENCES**